\documentclass[traditabstract]{aa}  
\usepackage[english] {babel}
\usepackage{astron}
\usepackage[dvips]{graphicx}
\usepackage{ae}
\usepackage{color}
\usepackage{amsfonts}
\usepackage{tabularx}
\usepackage{natbib}
\usepackage{amssymb,amsmath}
\usepackage{listliketab}
\usepackage[toc,page]{appendix}
\usepackage{txfonts}
\usepackage{placeins}
\usepackage{float}
\usepackage{morefloats}
\usepackage{afterpage}
\usepackage{url}

\begin{document}
\bibliographystyle{aa} 

   \title{Calibrating the fundamental plane with SDSS DR8 data}


   \author{Christoph Saulder
          \inst{1,2}
          \and
          Steffen Mieske
          \inst{1}
          \and
          Werner W. Zeilinger
          \inst{2}      
          \and
          Igor Chilingarian
          \inst{3,4}              
          }

   \institute{
   European Southern Observatory,
   Alonso de C\'{o}rdova 3107, Vitacura, Casilla 19001, Santiago, Chile\\
    \email{csaulder@eso.org,smieske@eso.org}
    \and
   Department of Astrophysics, University of Vienna,
   T\"urkenschanzstra\ss e 17, 1180 Vienna, Austria\\
   \email{werner.zeilinger@univie.ac.at}
   \and
   Smithsonian Astrophysical Observatory, Harvard-Smithsonian Center for Astrophysics, 60 Garden St. MS09 Cambridge, MA 02138, USA \\
   \email{igor.chilingarian@cfa.harvard.edu}
   \and
   Sternberg Astronomical Institute, Moscow State University, 13 Universitetski prospect, 119992 Moscow, Russia  \\ 
             }

   \date{Received March 13, 2013 ; accepted May 27, 2013}

 
  \abstract
 {We present a calibration of the fundamental plane using SDSS Data Release 8. We analysed about 93000 elliptical galaxies up to $z<0.2$, the largest sample used for the calibration of the fundamental plane so far. We incorporated up-to-date K-corrections and used GalaxyZoo data to classify the galaxies in our sample. We derived independent fundamental plane fits in all five Sloan filters u, g, r, i and z. A direct fit using a volume-weighted least-squares method was applied to obtain the coefficients of the fundamental plane, which implicitly corrects for the Malmquist bias. We achieved an accuracy of 15\% for the fundamental plane as a distance indicator. We provide a detailed discussion on the calibrations and their influence on the resulting fits. These re-calibrated fundamental plane relations form a well-suited anchor for large-scale peculiar-velocity studies in the nearby universe. In addition to the fundamental plane, we discuss the redshift distribution of the elliptical galaxies and their global parameters. }


   \maketitle

\section{Introduction}
The fundamental plane is an empirical relation between three global parameters of elliptical galaxies: the central velocity dispersion $\sigma_{0}$, the physical effective radius $R_{0}$, and the mean surface brightness $\mu_{0}$ within the effective radius. The last parameter is usually expressed as $I_{0}$, which is a renormalised surface brightness (see Equation \ref{logI0}). The functional form of the fundamental plane reads 
\begin{equation}
\textrm{log}_{10}\left(R_{0}\right)=a \cdot \textrm{log}_{10}\left(\sigma_{0}\right) + b \cdot \textrm{log}_{10}\left(I_{0}\right) + c .
\label{fundamentalplane}
\end{equation}
Historically, the fundamental plane of elliptical galaxies was first mentioned in \citet{Terlevich:1981}. It was defined and discussed in more detail in \citet{Dressler:1987} and \citet{Djorgovski:1987}. As part of an extensive study on elliptical galaxies \citep{Bernardi:2003a,Bernardi:2003b,Bernardi:2003c,Bernardi:2003d}, the first work on the fundamental plane using SDSS data was done in Bernardi's paper \citep{Bernardi:2003c}. Afterwards considerable work was done on the fundamental plane by a wide range of scientists e.g.  \citet{DOnofrio:2008}, \citet{LaBarbera:2008}, \citet{Gargiulo:2009}, \citet{Hyde:2009}, \citet{LaBarbera:2010}, \citet{FraixBurnet:2010}, and \citet{Magoulas:2012}.

The central velocity dispersion as well as the mean surface brightness are distance-independent quantities. Consequently, one can use the fundamental plane as a distance indicator by comparing the predicted effective radius with the observed one. We plan to use this standard-candle property of the fundamental plane in future work on the peculiar-velocity field in the nearby universe.

According to \citet{Bernardi:2003c}, a direct fit is the most suitable type of fit to obtain the fundamental plane coefficients if one plans on using them as a distance indicator, because it minimises the scatter in the physical radius $R_{0}$. Other types of fits also have their advantages, when using the fundamental plane for different applications (such as investigating the global properties of elliptical galaxies). In Table \ref{fp_coefficients_by_others}, we collect the results for the fundamental plane coefficient of previous literature work. As \citet{Bernardi:2003c} already pointed out, the coefficients depend on the fitting method. Table \ref{fp_coefficients_by_others} shows that the coefficient $a$ is typically smaller for direct fits than for orthogonal fits.

On theoretical grounds, it is clear that virial equilibrium predicts interrelations between the three parameters $R_{0}$, $\sigma_{0}$, and $I_{0}$. The coefficients of the fundamental plane can be compared with these expectations from virial equilibrium, and a luminosity-independent mass-to-light (M/L) ratio for all elliptical galaxies. Virial equilibrium and constant M/L predicts $a=2$ and $b=-1$. Any deviation (usually lower values for $a$ and higher values for $b$) of these values is referred to as tilt in the literature. From Table~\ref{fp_coefficients_by_others} it is clear that the actual coefficients of the fundamental plane deviate from these simplified assumptions (e.g. a$\sim$1 instead of 2). The physical reasons that give rise to this deviation are obviously a matter of substantial debate in the literature since it provides fundamental information about galaxy evolution \citep{Ciotti:1996,Busarello:1997,Busarello:1998,Graham:1997,Trujillo:2004,DOnofrio:2006,Cappellari:2006} or its environment dependence \citep{Lucey:1991,Jorgensen:1996,Pahre:1998,deCarvalho:1992,LaBarbera:2010b}. The empirical relation as such is very well documented and is often used as a distance indicator.

There are several two-dimensional relations that can be derived from the fundamental plane: the Faber-Jackson relation \citep{FaberJackson} between the luminosity and the velocity dispersion, and the Kormendy relation \citep{Kormendy:1977} between the luminosity and effective radius and the $D-\sigma$-relation \citep{Dressler:1987}, which connects the photometric parameter $D$ with the velocity dispersion $\sigma$.

In this paper, we provide a calibration of the fundamental plane for usage as a distance indicator, using a sample of about 93000 elliptical galaxies from the eighth data release of the Sloan Digital Sky Survey (SDSS DR8) \citep{SDSS}. This doubles the sample of the most extensive FP calibration in the present literature \citep{Hyde:2009}. We assumed a $\Lambda$-CDM cosmology with a relative dark-energy density of $\Omega_{\Lambda}=0.7$ and and a relative matter density of $\Omega_{M}=0.3$ as well as a present-day Hubble parameter of $H_{0}=70\, \textrm{km}\,\, \textrm{s}^{-1}\,\, \textrm{Mpc}^{-1}$. One can use the parameter $h_{70}$ to rescale the results for any other choice of the Hubble parameter.\\
\begin{table*}
\begin{center}
\begin{tabular}{cccccccc}
 band & $a$ & $b$ & $c$ & $\sigma_{\textrm{dist}}$ [\%] & $N_{\textrm{gal}}$ & type of fit & authors \\ \hline 
B & $1.39 \pm 0.14$ & $-0.90 \pm 0.09$ & - & 20 & 106 & 2-step inverse R & \citet{Djorgovski:1987} \\
B & $1.33 \pm 0.05$ & $-0.83 \pm 0.03$ & - & 20 & 97 & inverse R & \citet{Dressler:1987} \\
V+R & $1.43 \pm 0.03$ & $-0.84 \pm 0.02$ & $ -7.995 \pm 0.021$ & 21 & 694 & inverse R & \citet{Smith:2001} \\
R & $1.38 \pm 0.04$ & $-0.82 \pm 0.03$ & - & 21 & 352 & inverse R & \citet{Hudson:1997} \\
R & $1.37 \pm 0.05$ & $-0.84 \pm 0.03$ & - & 21 & 428 & inverse R & \citet{Gibbons:2001} \\
V & $1.26 \pm 0.07$ & $-0.82 \pm 0.09$ & - & 13 & 66 & forward R & \citet{Lucey:1991} \\
V & $1.14$ & $-0.79$ & - & 17 & 37 & forward R & \citet{Guzman:1993} \\
r & $1.24 \pm 0.07$ & $-0.82 \pm 0.02$ & - & 17 & 226 & orthogonal R & \citet{Jorgensen:1996} \\
R & $1.25$ & $-0.87$ & - & 19 & 40 & orthogonal R & \citet{Muller:1998} \\
V & $1.21 \pm 0.05$ & $-0.80 \pm 0.01$ & - & - & - & orthogonal R & \citet{DOnofrio:2008} \\
r & $1.42 \pm 0.05$ & $-0.76 \pm 0.008$ & - & 28 & 1430 & orthogonal R & \citet{LaBarbera:2008} \\
K & $1.53 \pm 0.04$ & $-0.77 \pm 0.01$ & - & 29 & 1430 & orthogonal R & \citet{LaBarbera:2008} \\
R & $1.35 \pm 0.11$ & $-0.81 \pm 0.03$ & - & 21 & 91 & orthogonal R & \citet{Gargiulo:2009} \\
g & $1.40 \pm 0.02$ & $-0.76 \pm 0.02$ & $-8.858 $ & 31 & 46410 & orthogonal R & \citet{Hyde:2009} \\
r & $1.43 \pm 0.02$ & $-0.79 \pm 0.02$ & $-8.898 $ & 30 & 46410 & orthogonal R & \citet{Hyde:2009} \\
i & $1.46 \pm 0.02$ & $-0.80 \pm 0.02$ & $-8.891 $ & 29 & 46410 & orthogonal R & \citet{Hyde:2009} \\
z & $1.47 \pm 0.02$ & $-0.83 \pm 0.02$ & $-9.032 $ & 29 & 46410 & orthogonal R & \citet{Hyde:2009} \\
g & $1.38 \pm 0.02$ & $-0.788 \pm 0.002$ & $-9.13 \pm 0.08 $ & 29 & 4467 & orthogonal R & \citet{LaBarbera:2010} \\
r & $1.39 \pm 0.02$ & $-0.785 \pm 0.002$ & $-8.84 \pm 0.06 $ & 26 & 4478 & orthogonal R & \citet{LaBarbera:2010} \\
i & $1.43 \pm 0.02$ & $-0.780 \pm 0.002$ & $-8.76 \pm 0.05 $ & - & 4455 & orthogonal R & \citet{LaBarbera:2010} \\
z & $1.42 \pm 0.02$ & $-0.793 \pm 0.002$ & $-8.74 \pm 0.07 $ & - & 4319 & orthogonal R & \citet{LaBarbera:2010} \\
Y & $1.47 \pm 0.02$ & $-0.785 \pm 0.002$ & $-8.53 \pm 0.06 $ & - & 4404 & orthogonal R & \citet{LaBarbera:2010} \\
J & $1.53 \pm 0.02$ & $-0.795 \pm 0.002$ & $-8.57 \pm 0.06 $ & 26 & 4317 & orthogonal R & \citet{LaBarbera:2010} \\
H & $1.56 \pm 0.02$ & $-0.795 \pm 0.002$ & $-8.42 \pm 0.08 $ & 27 & 4376 & orthogonal R & \citet{LaBarbera:2010} \\
K & $1.55 \pm 0.02$ & $-0.790 \pm 0.002$ & $-8.24 \pm 0.08 $ & 28 & 4350 & orthogonal R & \citet{LaBarbera:2010} \\
K & $1.53 \pm 0.08$ & $-0.79 \pm 0.03$ & - & 21 & 251 & orthogonal R & \citet{Pahre:1998} \\
V & $1.31 \pm 0.13$ & $-0.86 \pm 0.10$ & - & 14 & 30 & orthogonal R & \citet{Kelson:2000}   \\
R & $1.22 \pm 0.09$ & $-0.84 \pm 0.03$ & - & 20 & 255 & orthogonal ML & \citet{Colless:2001} \\
g & $1.45 \pm 0.06$ & $-0.74 \pm 0.01$ & $ -8.779 \pm 0.029$ & 25 & 5825 & orthogonal ML & \citet{Bernardi:2003c} \\
r & $1.49 \pm 0.05$ & $-0.75 \pm 0.01$ & $ -8.778 \pm 0.020$ & 23 & 8228 & orthogonal ML & \citet{Bernardi:2003c} \\
i & $1.52 \pm 0.04$ & $-0.78 \pm 0.01$ & $ -8.895 \pm 0.021$ & 23 & 8022 & orthogonal ML & \citet{Bernardi:2003c} \\
z & $1.51 \pm 0.04$ & $-0.77 \pm 0.01$ & $ -8.707 \pm 0.023$ & 22 & 7914 & orthogonal ML & \citet{Bernardi:2003c} \\
J & $1.52 \pm 0.03$ & $-0.89 \pm 0.008$ & - & 30 & 8901 & orthogonal ML & \citet{Magoulas:2012} \\
H & $1.47 \pm 0.02$ & $-0.88 \pm 0.008$ & - & 29 & 8568 & orthogonal ML & \citet{Magoulas:2012} \\
K & $1.46 \pm 0.02$ & $-0.86 \pm 0.008$ & - & 29 & 8573 & orthogonal ML & \citet{Magoulas:2012} \\
g & $1.08 \pm 0.05$ & $-0.74 \pm 0.01$ & $ -8.033 \pm 0.024$ & - & 5825 & direct ML & \citet{Bernardi:2003c} \\
r & $1.17 \pm 0.04$ & $-0.75 \pm 0.01$ & $ -8.022 \pm 0.020$ & - & 8228 & direct ML & \citet{Bernardi:2003c} \\
i & $1.21 \pm 0.04$ & $-0.77 \pm 0.01$ & $ -8.164 \pm 0.019$ & - & 8022 & direct ML & \citet{Bernardi:2003c} \\
z & $1.20 \pm 0.04$ & $-0.76 \pm 0.01$ & $ -7.995 \pm 0.021$ & - & 7914 & direct ML & \citet{Bernardi:2003c} \\
g & $1.12 \pm 0.02$ & $-0.74 \pm 0.02$ & $-8.046 $ & - & 46410 & direct R & \citet{Hyde:2009} \\
r & $1.17 \pm 0.02$ & $-0.76 \pm 0.02$ & $-8.086 $ & - & 46410 & direct R & \citet{Hyde:2009} \\
i & $1.20 \pm 0.02$ & $-0.76 \pm 0.02$ & $-8.048 $ & - & 46410 & direct R & \citet{Hyde:2009} \\
z & $1.23 \pm 0.02$ & $-0.78 \pm 0.02$ & $-8.216 $ & - & 46410 & direct R & \citet{Hyde:2009} \\
I & $1.25 \pm 0.02$ & $-0.79 \pm 0.03$ & - & 20 & 109 & direct R & \citet{Scodeggio:1998} \\
R & $1.13 \pm 0.03$ & $-0.84 \pm 0.01$ & $8.53 \pm 0.1$ & - & 699 & direct R & \citet{FraixBurnet:2010} \\
u & $0.798 \pm 0.030$  & $-0.700 \pm 0.008$  & $-7.53 \pm 0.10$ & $16.5$ & 92953 & direct R & this paper\\
g & $0.966\pm \pm 0.030$  & $-0.740 \pm 0.013$  & $-7.75 \pm 0.13$ & $15.6$ & 92953 & direct R & this paper\\
r & $1.034 \pm 0.030$  & $-0.753 \pm 0.013$  & $-7.77 \pm 0.13$ & $15.3$ & 92953 & direct R & this paper\\
i & $1.062 \pm 0.030$  & $-0.757 \pm 0.013$  & $-7.75 \pm 0.13$ & $15.0$ & 92953 & direct R & this paper\\
z & $1.108 \pm 0.030$  & $-0.763 \pm 0.013$  & $-7.81 \pm 0.13$ & $14.8$ & 92953 & direct R & this paper
\end{tabular}
\end{center}
\caption{A list of previous publications (we do not claim completeness) of fundamental plane coefficients, based on the list of \citet{Magoulas:2012}, which is itself based on the lists of \citet{Bernardi:2003c} and \citet{Colless:2001}. It is sorted by method (note: R=regression, ML=maximum likelihood) and date of publication. Some of the values in the list cannot be found in the same form due to slightly different definitions in the referenced papers. We sometimes had to renormalise the coefficient $b$, when the fundamental plane was defined using with the mean surface brightness $\mu_{0}$ instead of the parameter $\textrm{log}_{10}\left(I_{0}\right)$. Furthermore, the coefficient $c$, if available, is always given for a Hubble parameter of $H_{0}=70\, \textrm{km}\,\, \textrm{s}^{-1}\,\, \textrm{Mpc}^{-1}$ here, therefore we had to rescale it if other values of $H_{0}$ were used in the referenced paper. In addition to the fundamental plane coefficients $a$, $b$ and if available $c$, the distance error $\sigma_{\textrm{dist}}$ and the number of galaxies $N_{\textrm{gal}}$ in the sample is given. Furthermore, we also list the type of fit (R = regression, ML = maximum likelihood), which was used to obtain the fundamental plane coefficient, because it is known that the coefficients not only depend on the wavelength, but also on the fitting method.}
\label{fp_coefficients_by_others}
\end{table*}

\section{Sample}
\subsection{Definition}
\label{sample_def}

Our starting sample consisted of 100427 elliptical galaxies from SDSS DR8 \citep{SDSS}. These galaxies were selected by the following criteria, as also summarised in Table \ref{criteria}: 

\begin{table}
\begin{center}
\begin{tabular}{cc}
parameter & condition \\ \hline
 \emph{SpecObj.z} &> 0 \\
 \emph{SpecObj.z} &< 0.5 \\
 \emph{SpecObj.zWarning} &= 0 \\
 \emph{zooVotes.p\_el} &> 0.8 \\
\emph{zooVotes.nvote\_tot} &> 10 \\
 \emph{SpecObj.veldisp} &> 100\\
 \emph{SpecObj.veldisp} &< 420\\
\emph{SpecObj.snMedian} &> 10\\
\emph{SpecObj.class} &='GALAXY'\\
\emph{PhotoObj.deVAB\_u} &> 0.3\\
\emph{PhotoObj.deVAB\_g} &> 0.3\\
\emph{PhotoObj.deVAB\_r} &> 0.3\\
\emph{PhotoObj.deVAB\_i} &> 0.3\\
\emph{PhotoObj.deVAB\_z} &> 0.3\\
\emph{PhotoObj.lnLDeV\_u}&>\emph{PhotoObj.lnLExp\_u}\\
\emph{PhotoObj.lnLDeV\_g}&>\emph{PhotoObj.lnLExp\_g}\\
\emph{PhotoObj.lnLDeV\_r}&>\emph{PhotoObj.lnLExp\_r}\\
\emph{PhotoObj.lnLDeV\_i}&>\emph{PhotoObj.lnLExp\_i}\\
\emph{PhotoObj.lnLDeV\_z}&>\emph{PhotoObj.lnLExp\_z}
\end{tabular}
\end{center}
\caption{Selection criteria given in the language of the SDSS CAS-job queries. As a direct consequence of these requirements, we demand that there must be spectroscopic data for every galaxy in our sample. Hereby, we impose the target limit for galaxy spectroscopy of SDSS on our sample, which is a minimum Petrosian magnitude in the r band of 17.77 mag \citep{SDSS_spectarget}.}
\label{criteria}
\end{table}

\begin{figure}[ht]
\begin{center}
\includegraphics[width=0.16\textwidth]{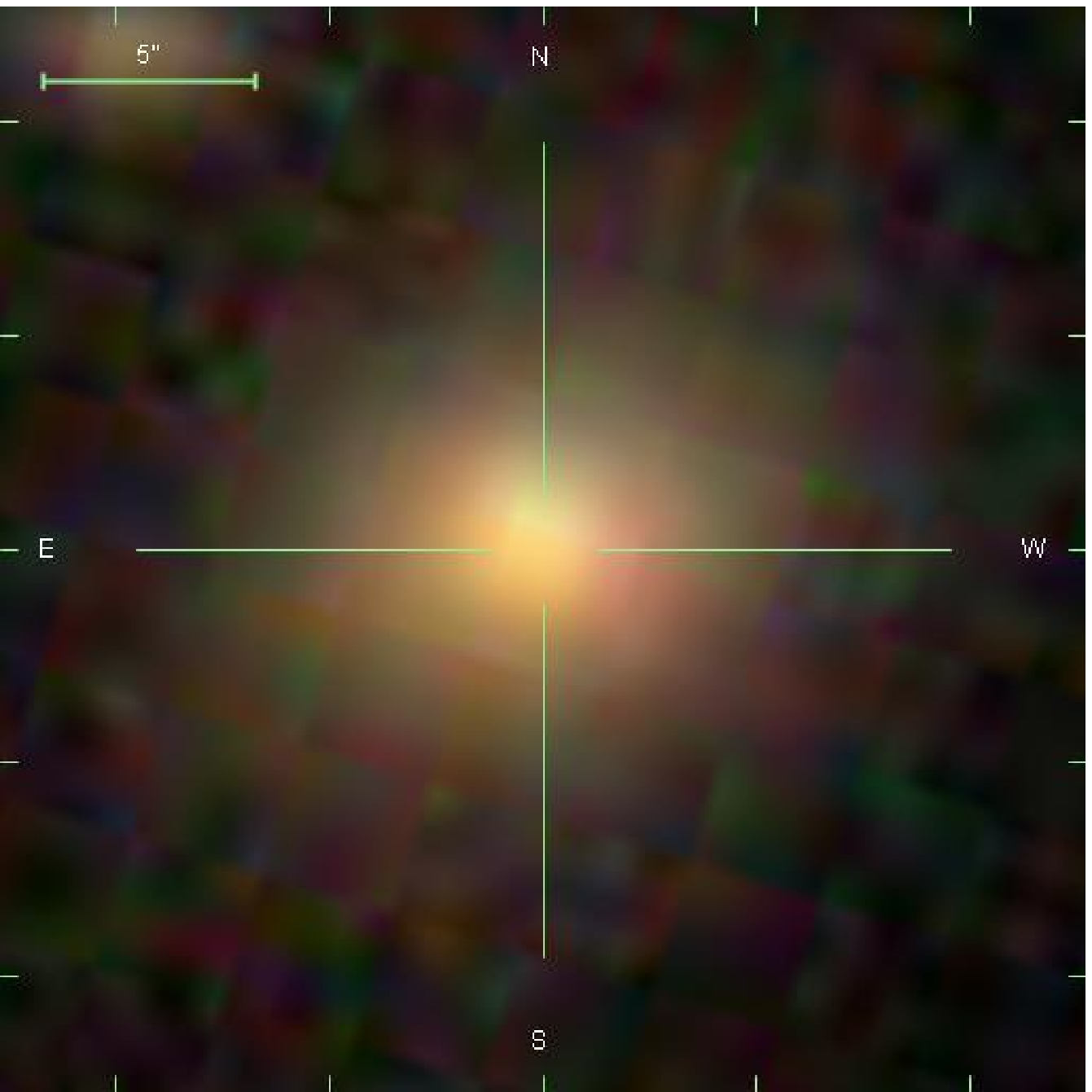}
\includegraphics[width=0.16\textwidth]{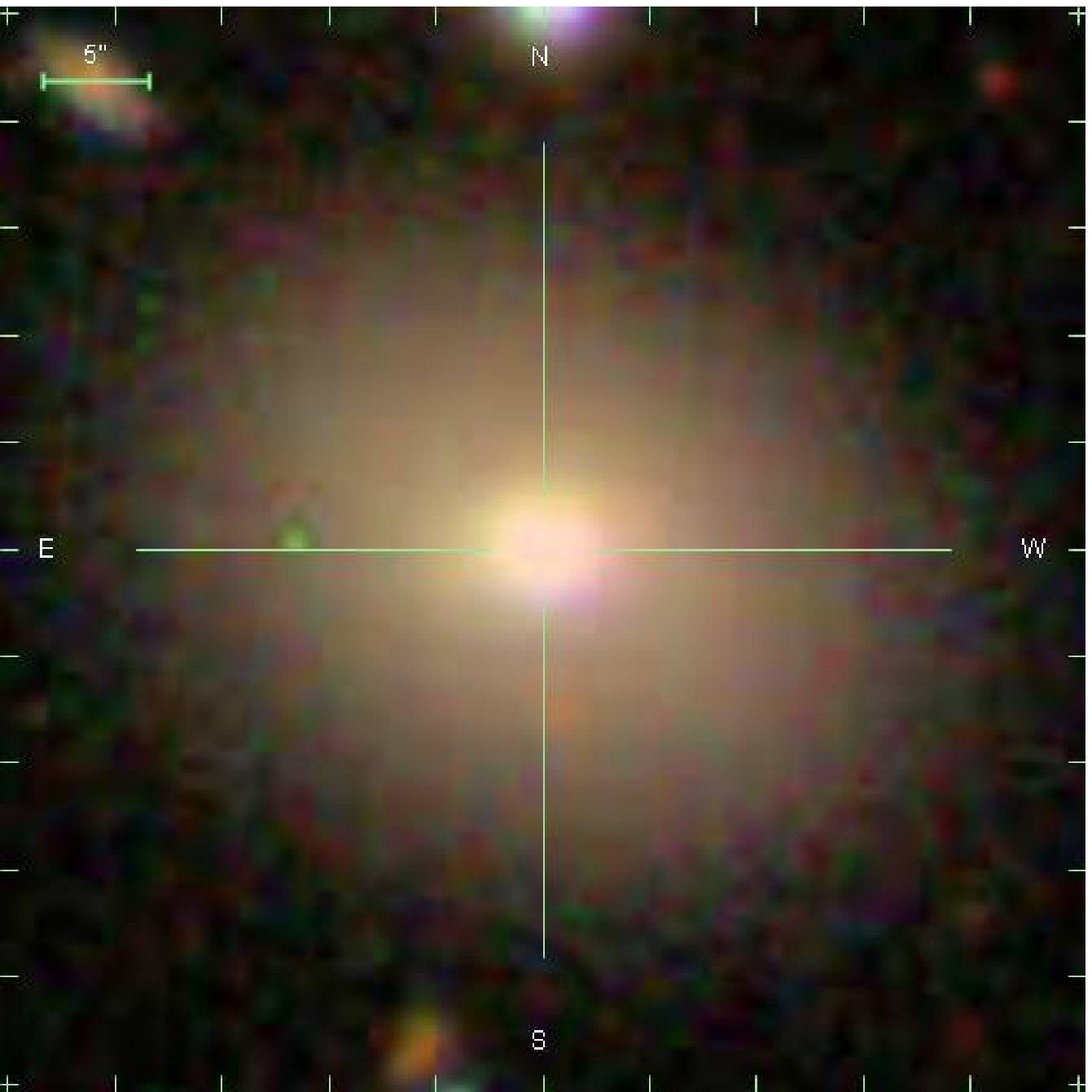}
\includegraphics[width=0.16\textwidth]{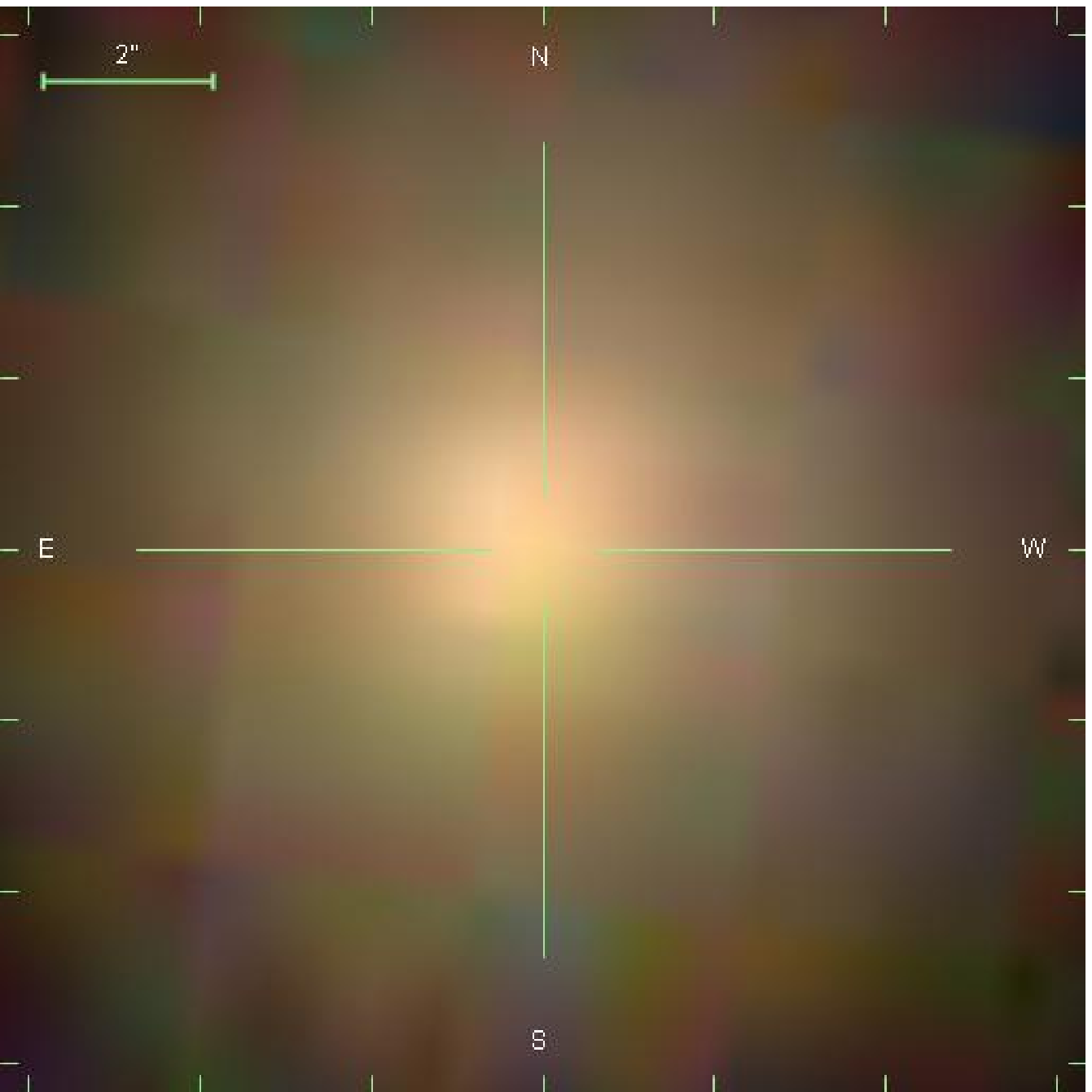}\\
\includegraphics[width=0.16\textwidth]{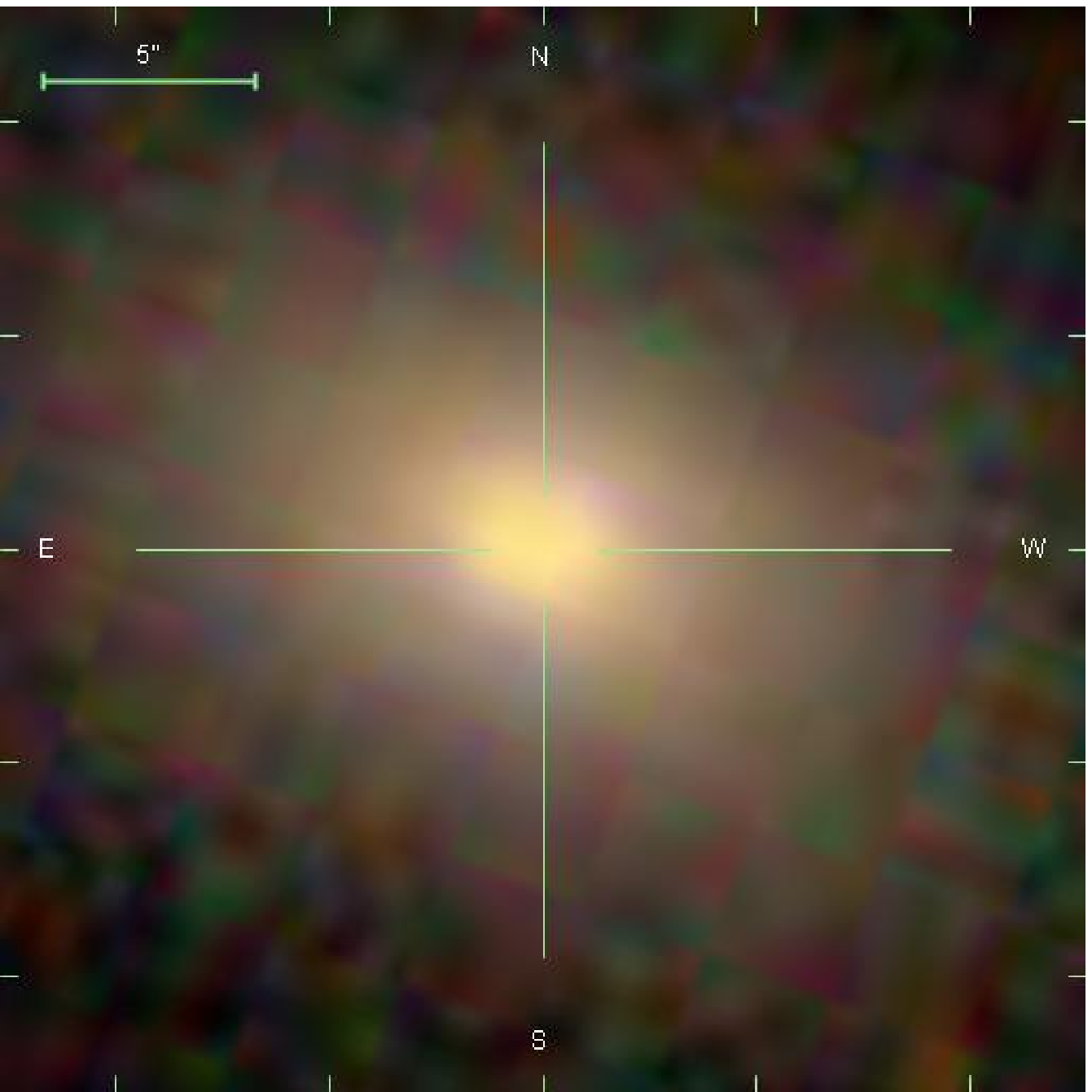} 
\includegraphics[width=0.16\textwidth]{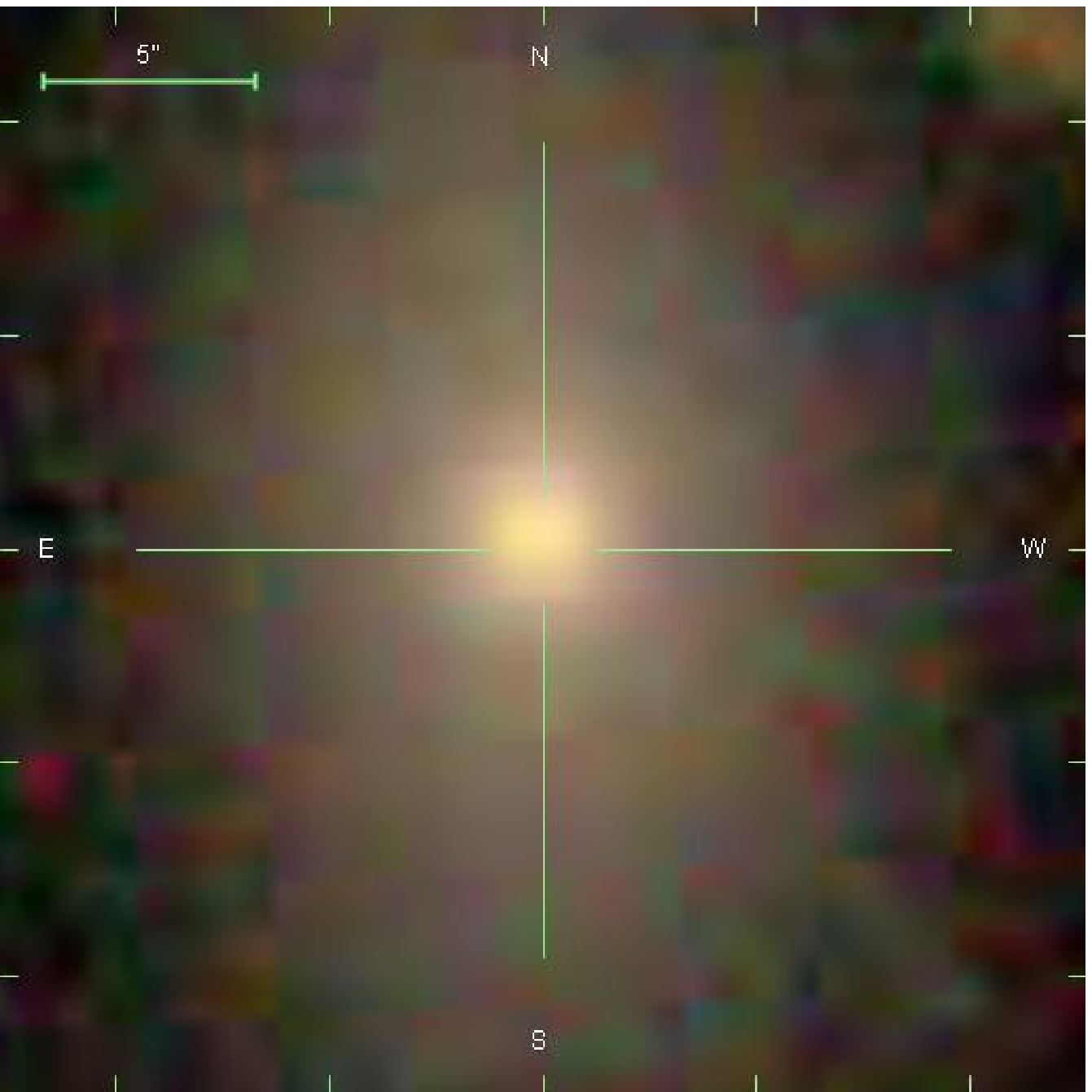}
\includegraphics[width=0.16\textwidth]{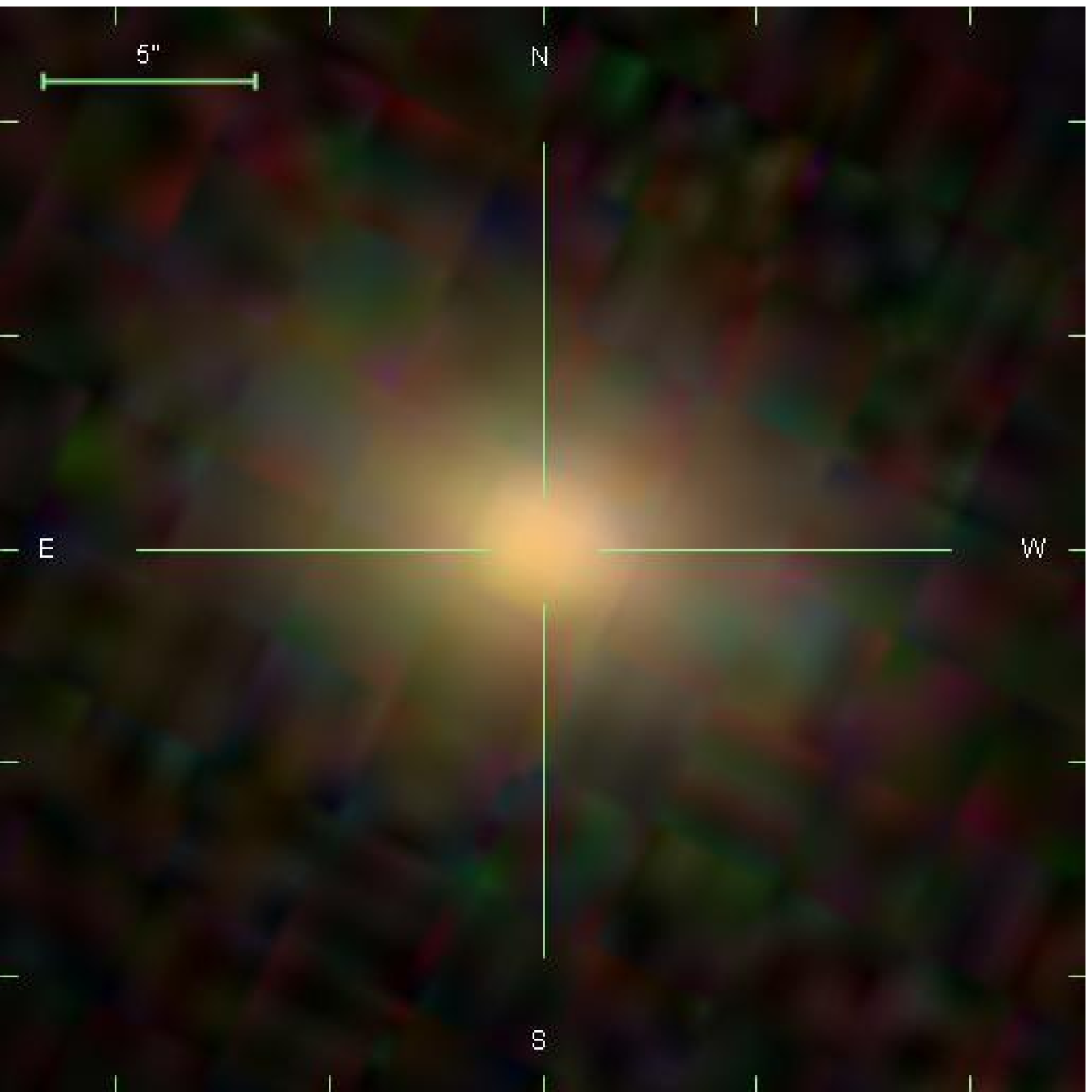}\\
\includegraphics[width=0.16\textwidth]{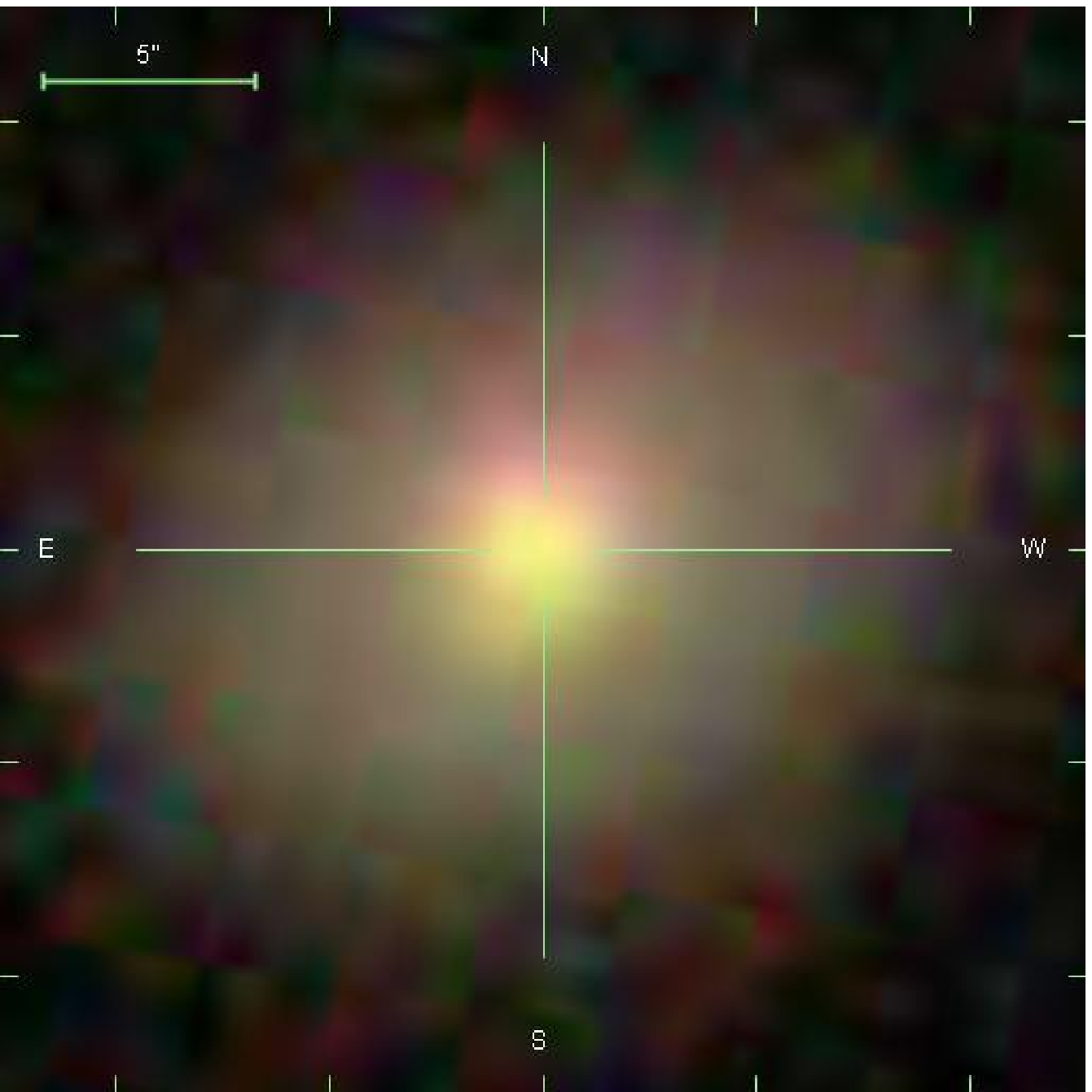}
\includegraphics[width=0.16\textwidth]{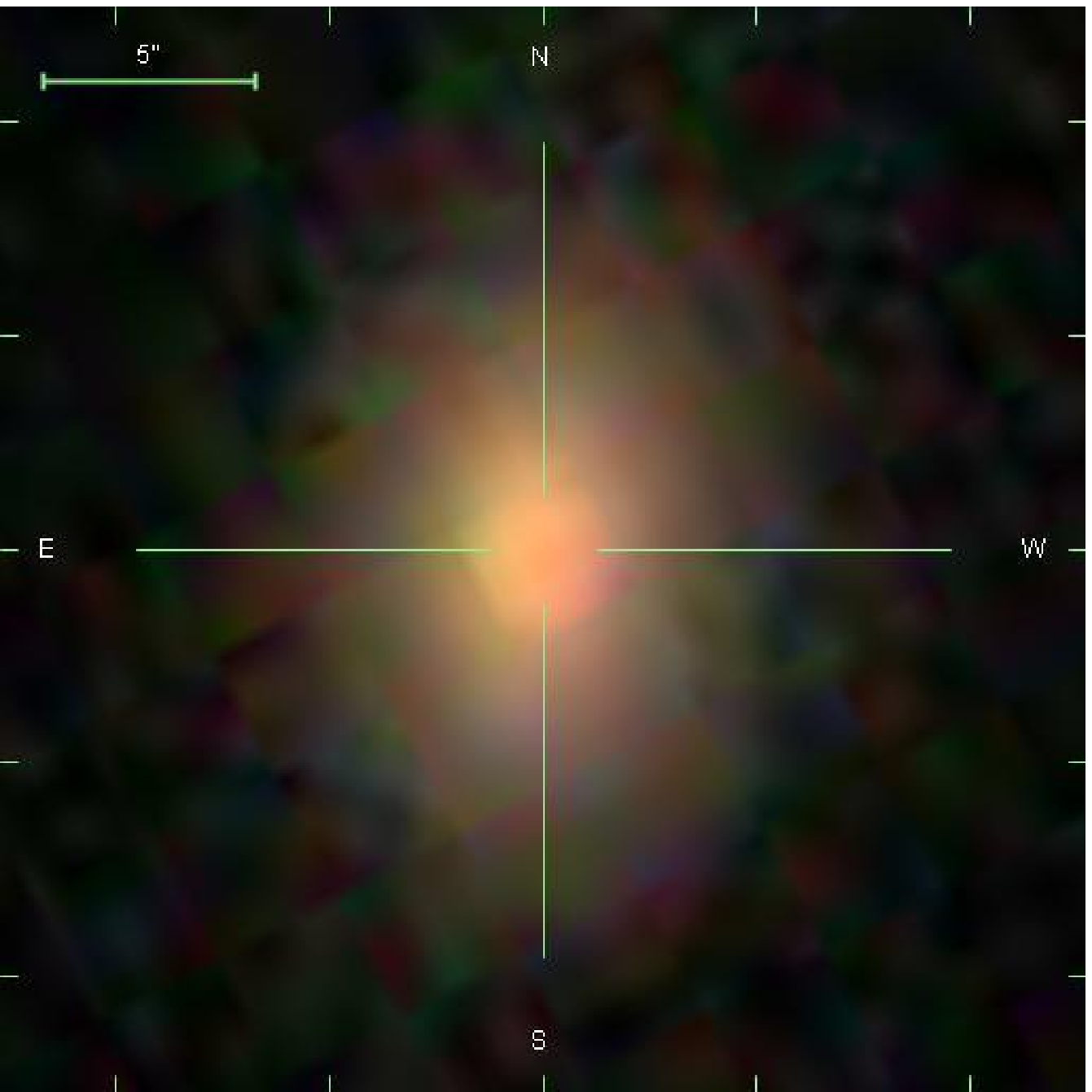}
\includegraphics[width=0.16\textwidth]{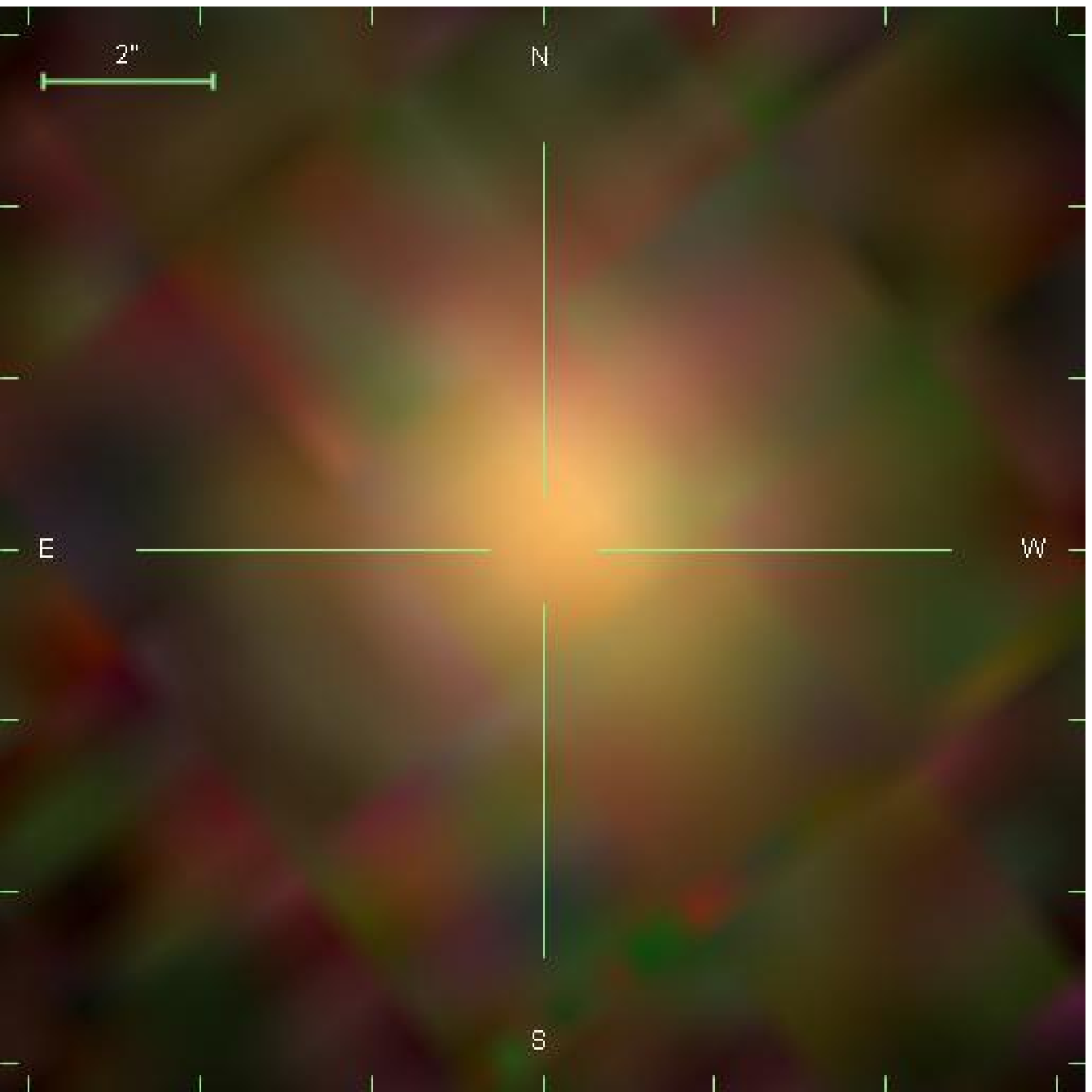}\\
\includegraphics[width=0.16\textwidth]{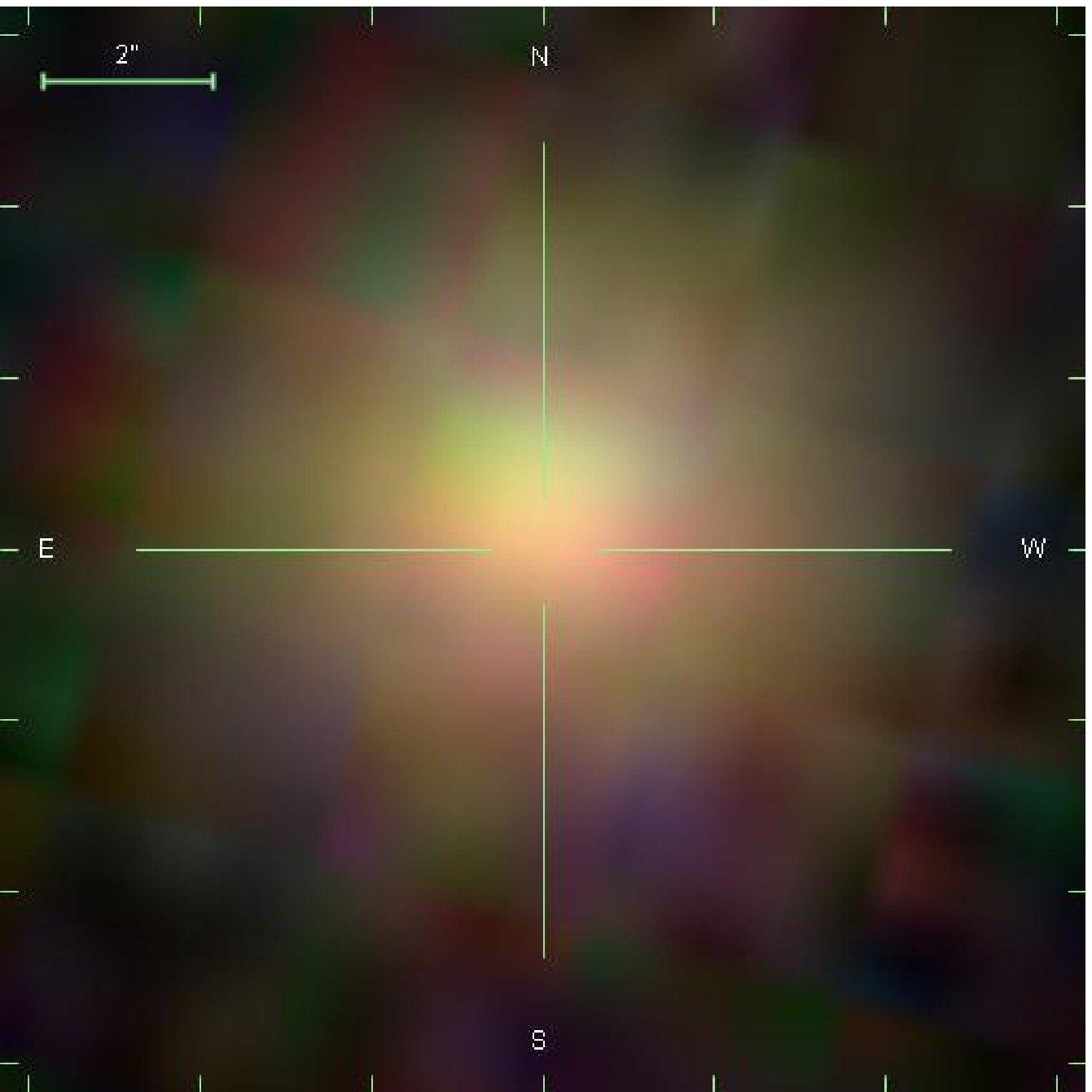}
\includegraphics[width=0.16\textwidth]{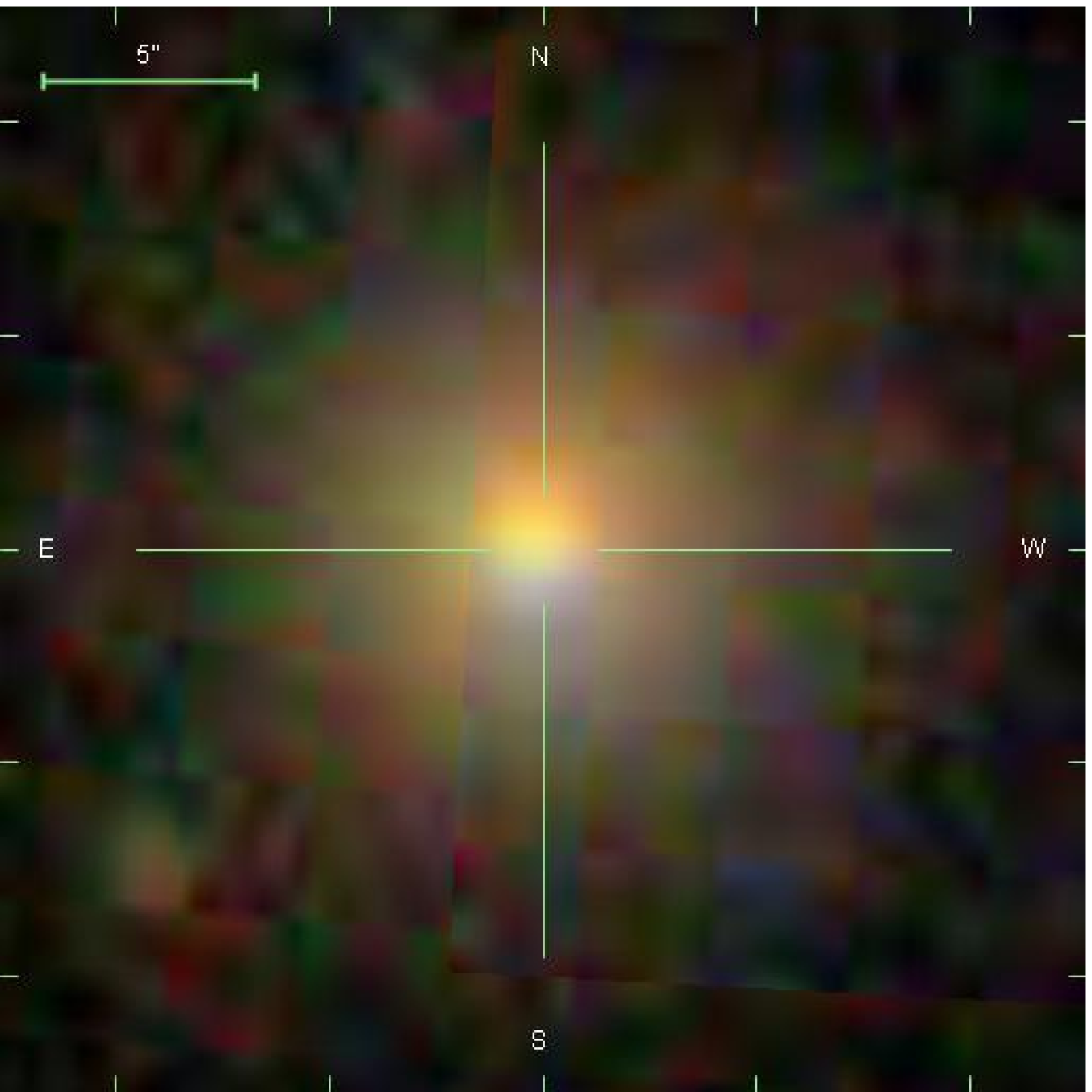}
\includegraphics[width=0.16\textwidth]{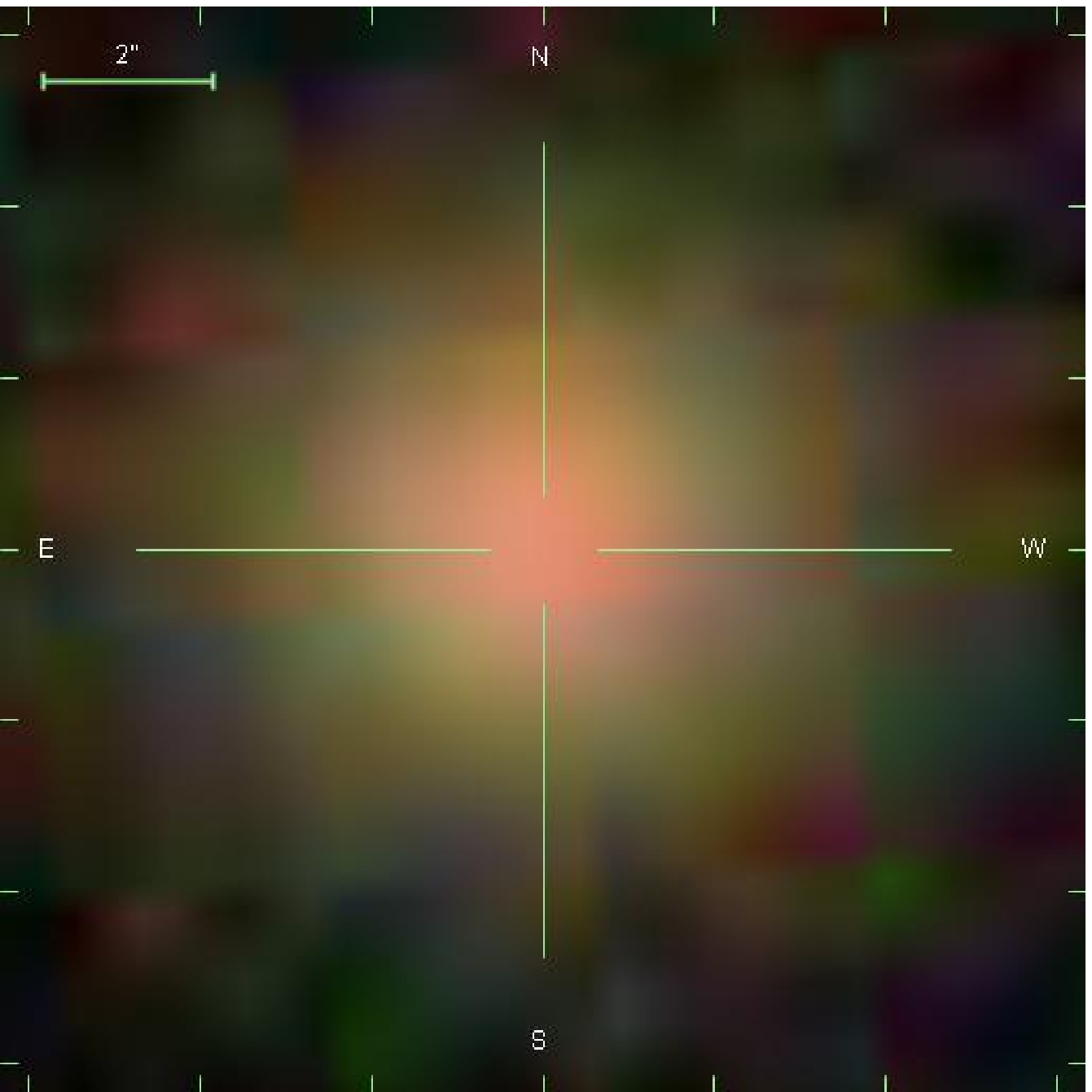}
\caption{Randomly selected subsample of 12 galaxies of our selected sample with a redshift lower than 0.1. All of them were classified to be elliptical galaxies by our selection criteria, which is confirmed by their morphology.}
\label{near_gal}
\end{center}
\end{figure}
\begin{figure}[ht]
\begin{center}
\includegraphics[width=0.16\textwidth]{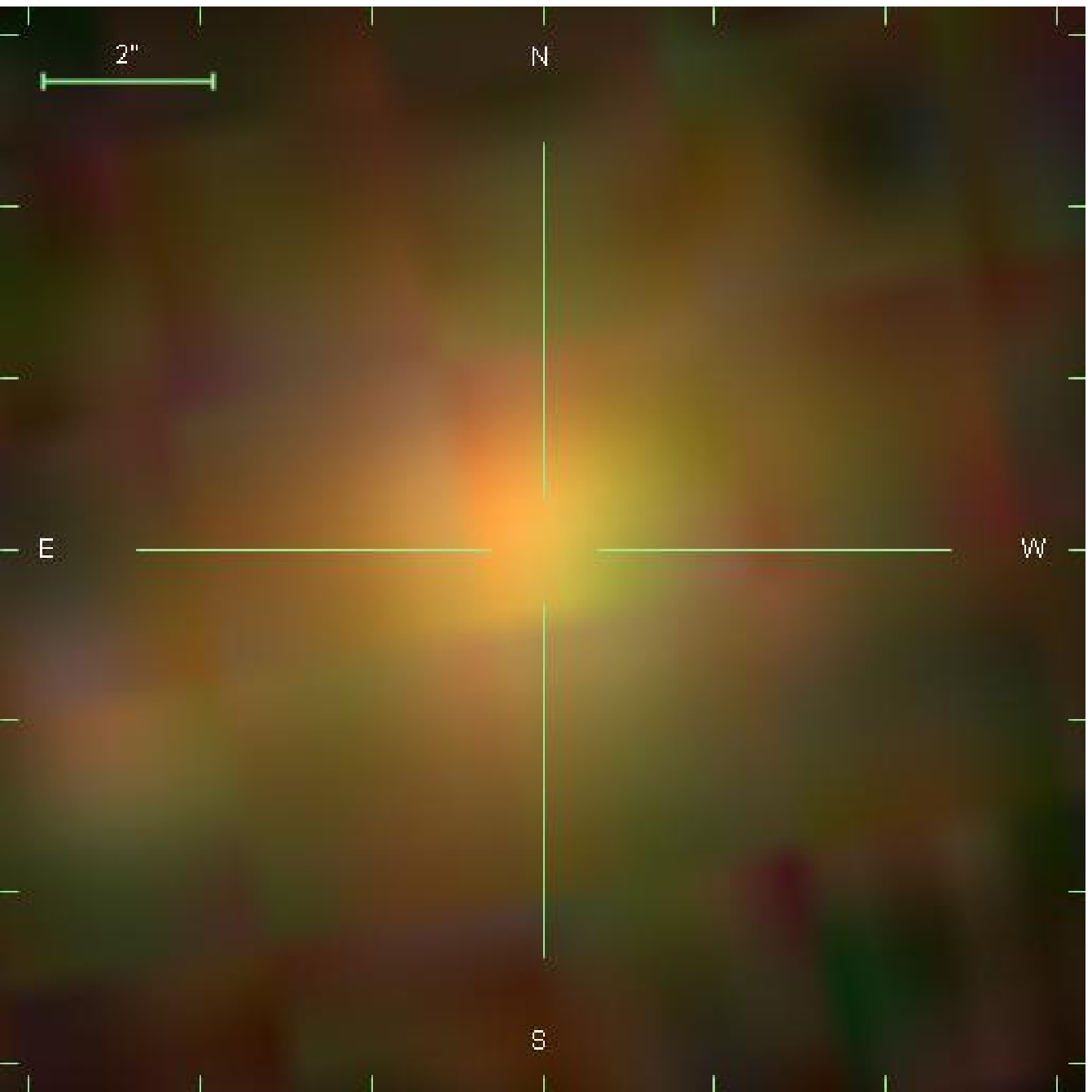}
\includegraphics[width=0.16\textwidth]{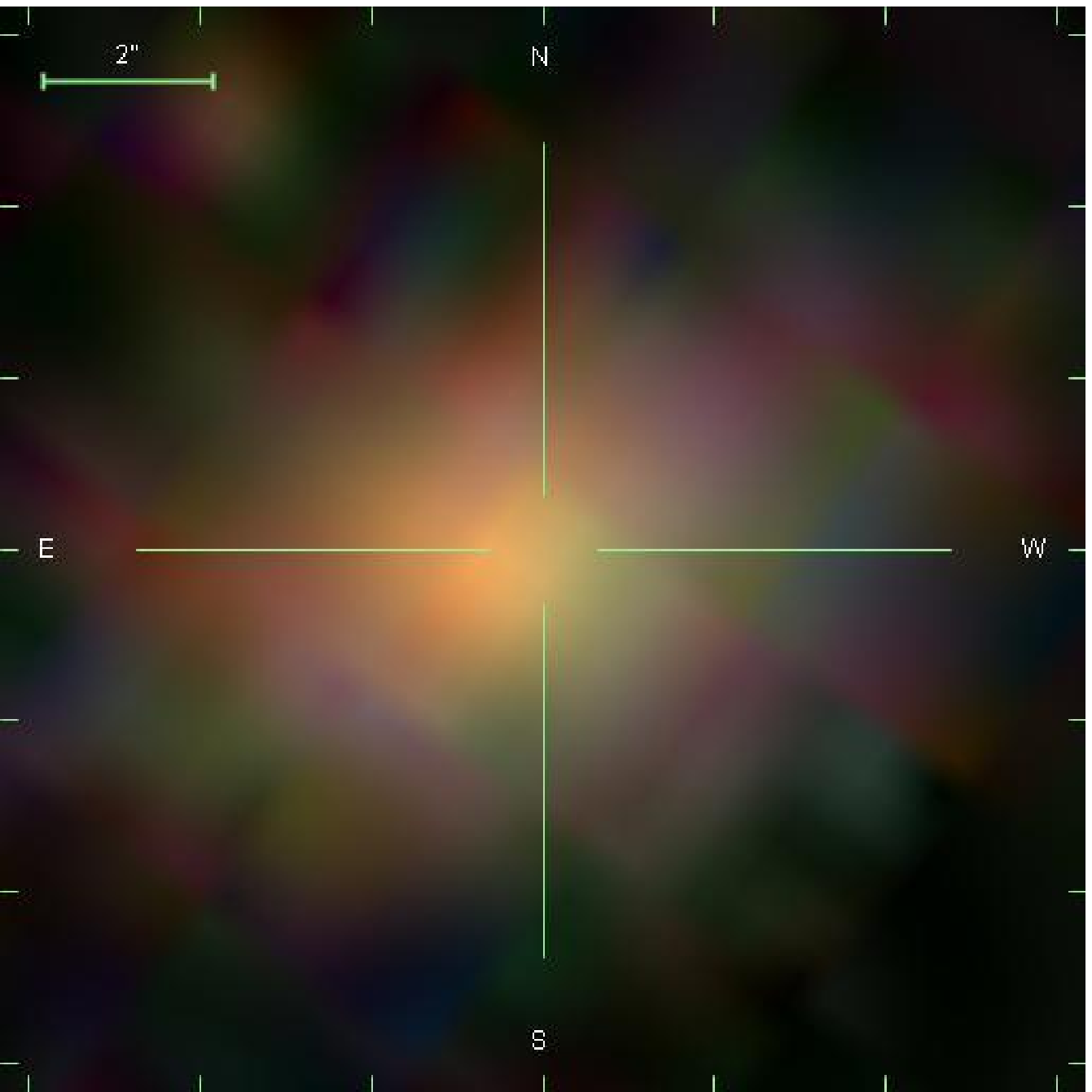}
\includegraphics[width=0.16\textwidth]{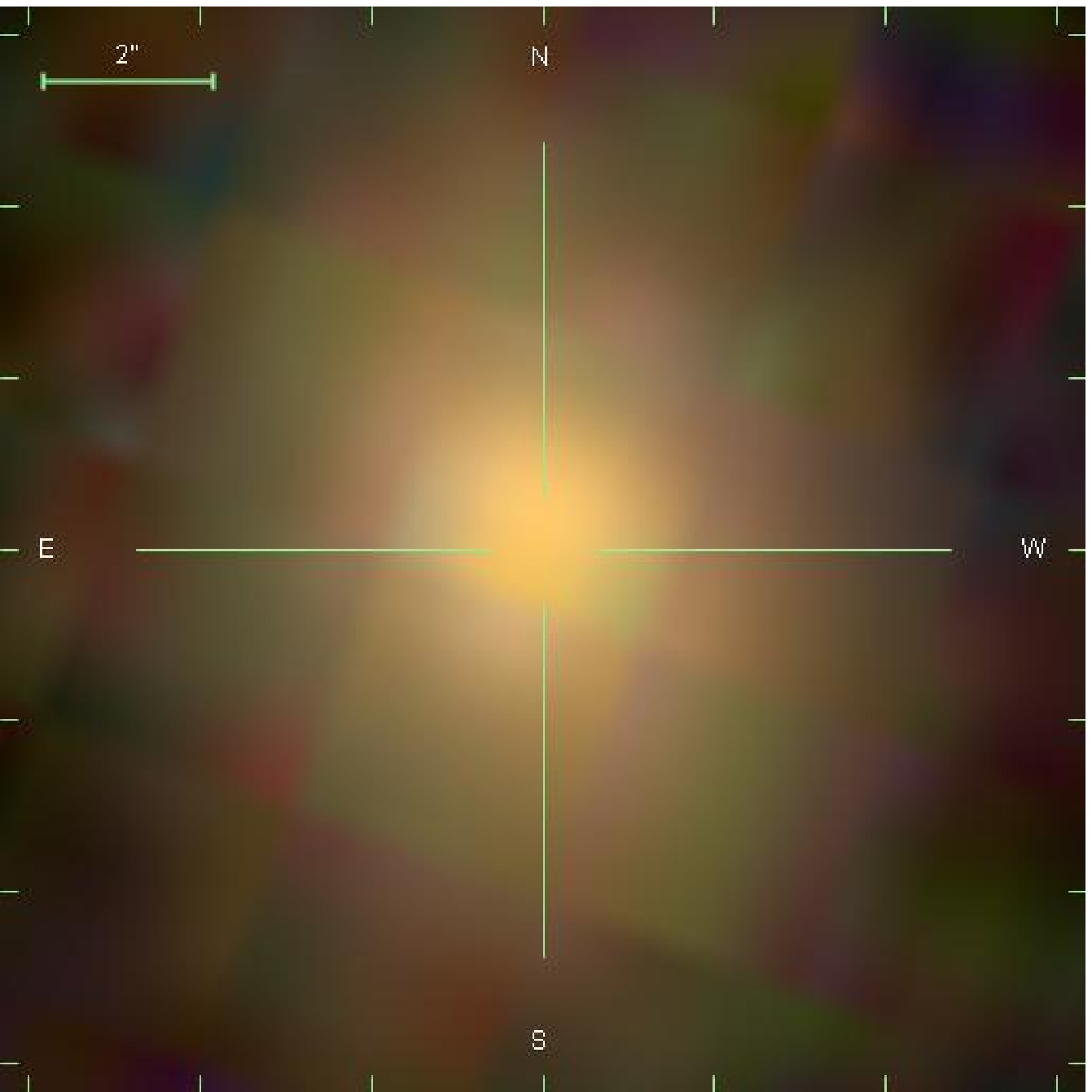}\\
\includegraphics[width=0.16\textwidth]{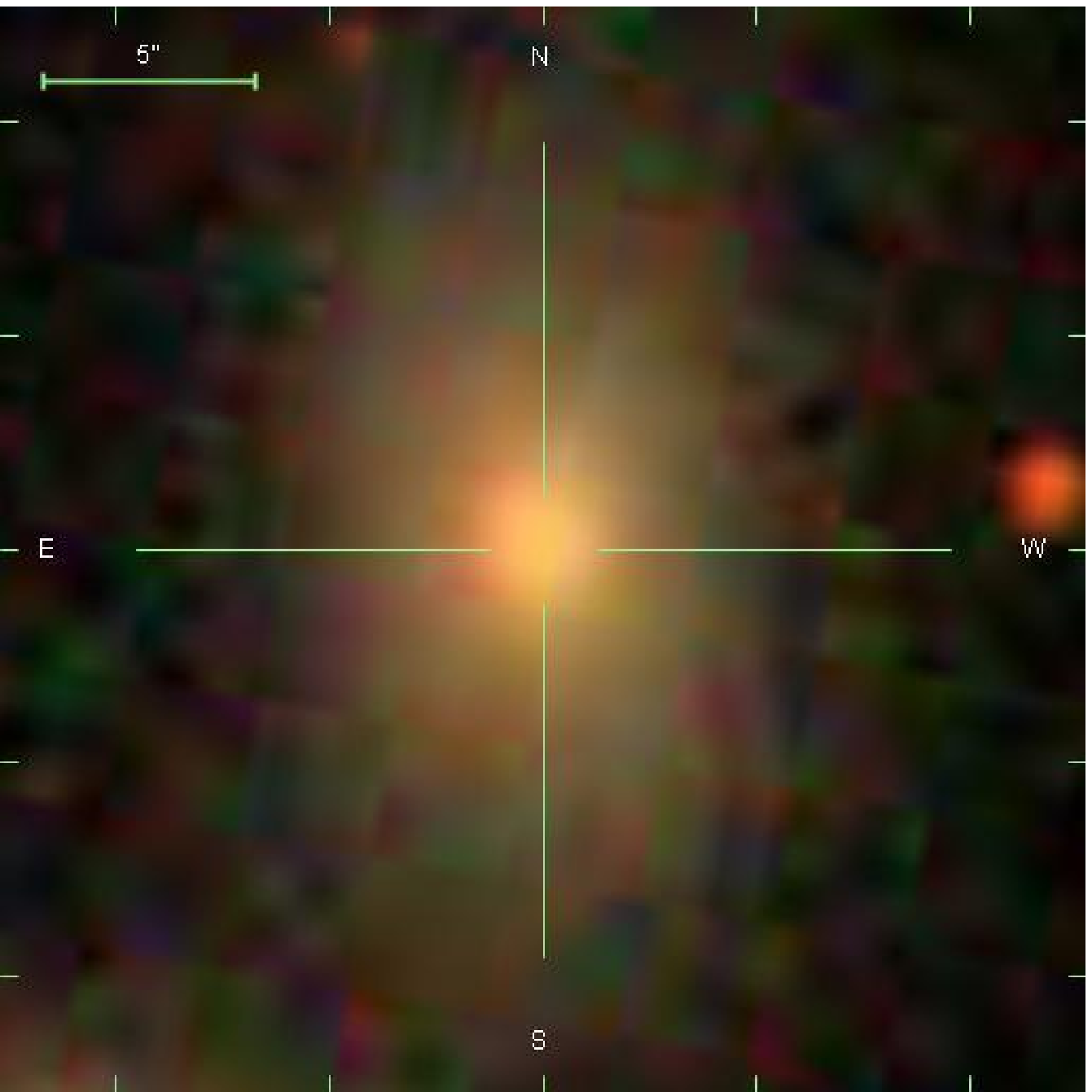} 
\includegraphics[width=0.16\textwidth]{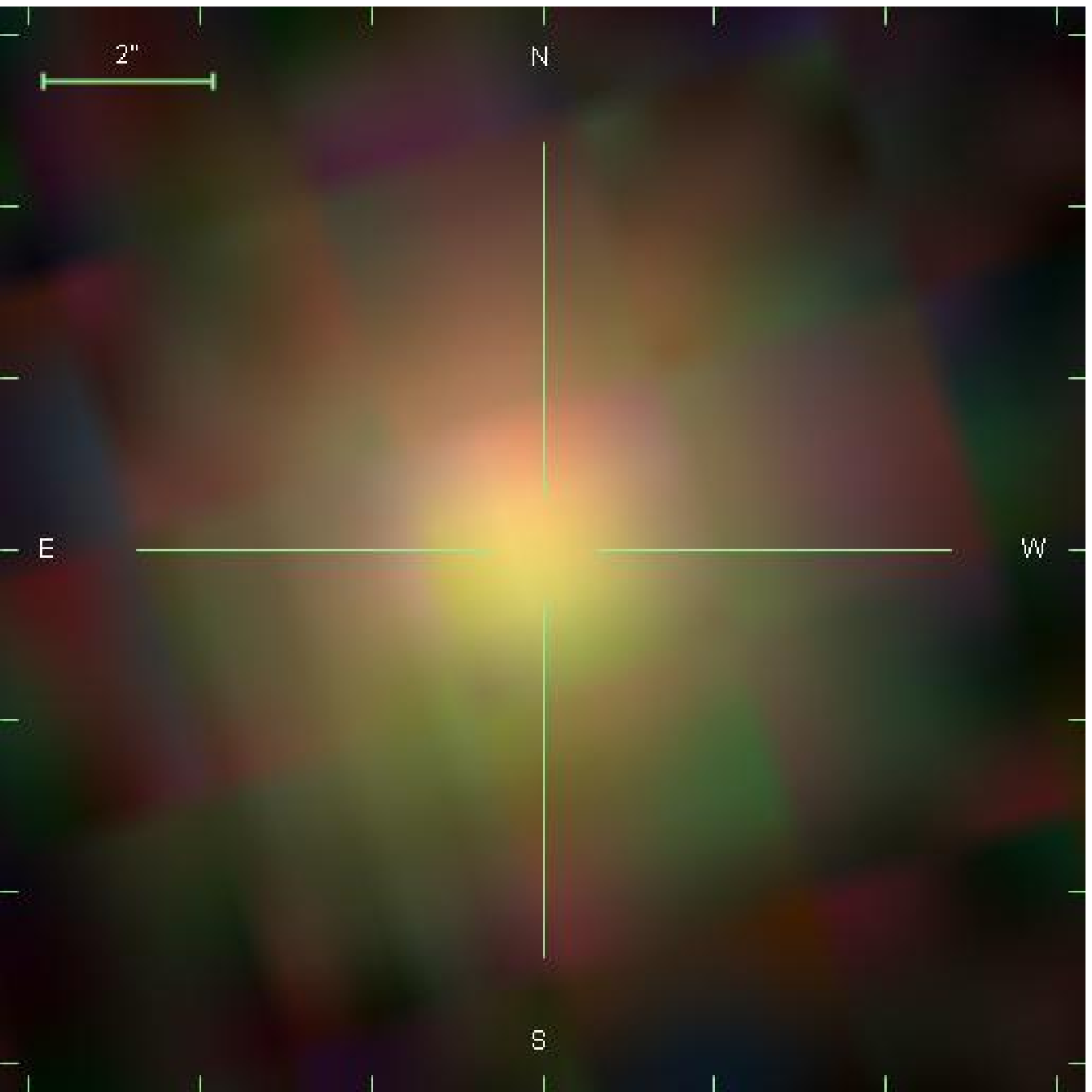}
\includegraphics[width=0.16\textwidth]{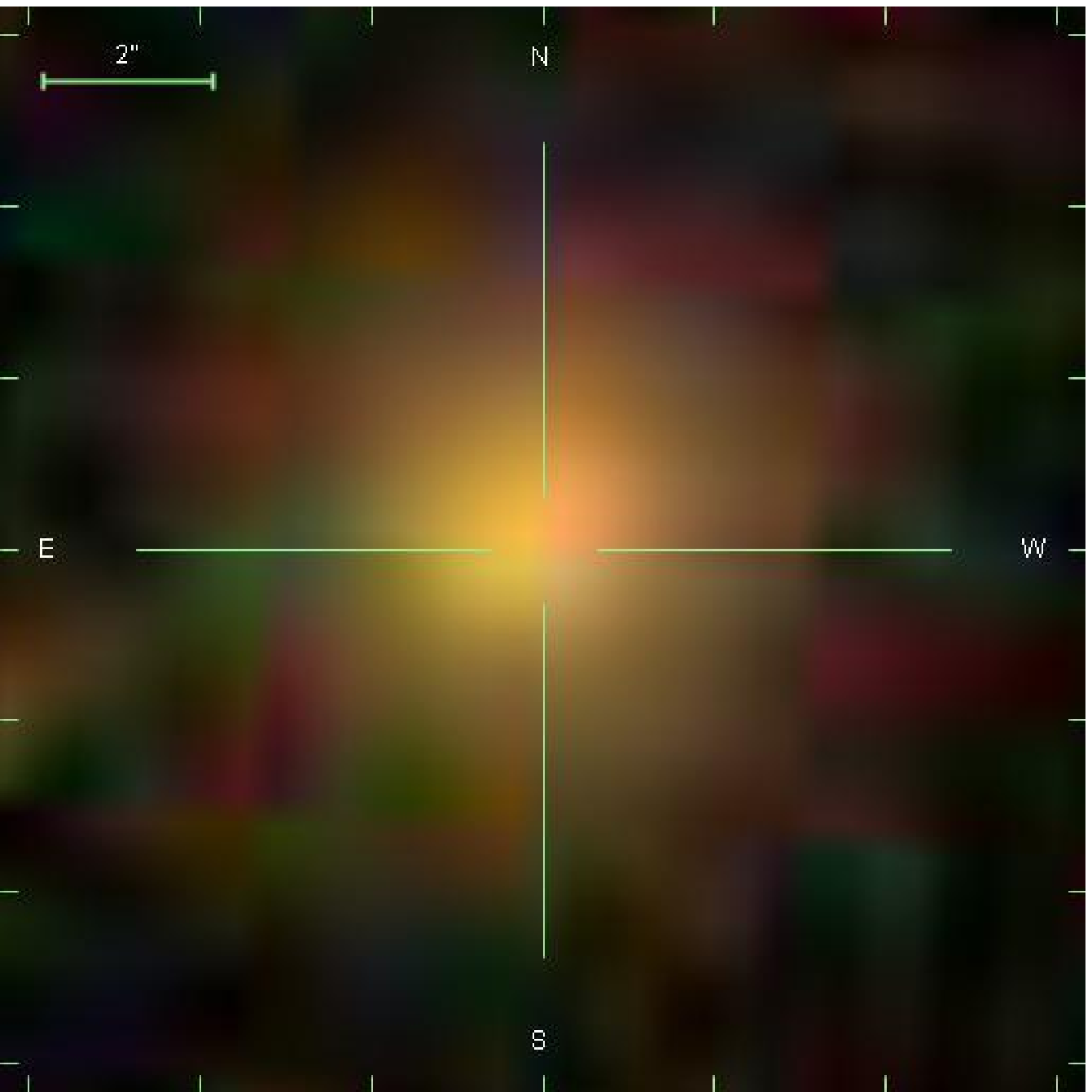}\\
\includegraphics[width=0.16\textwidth]{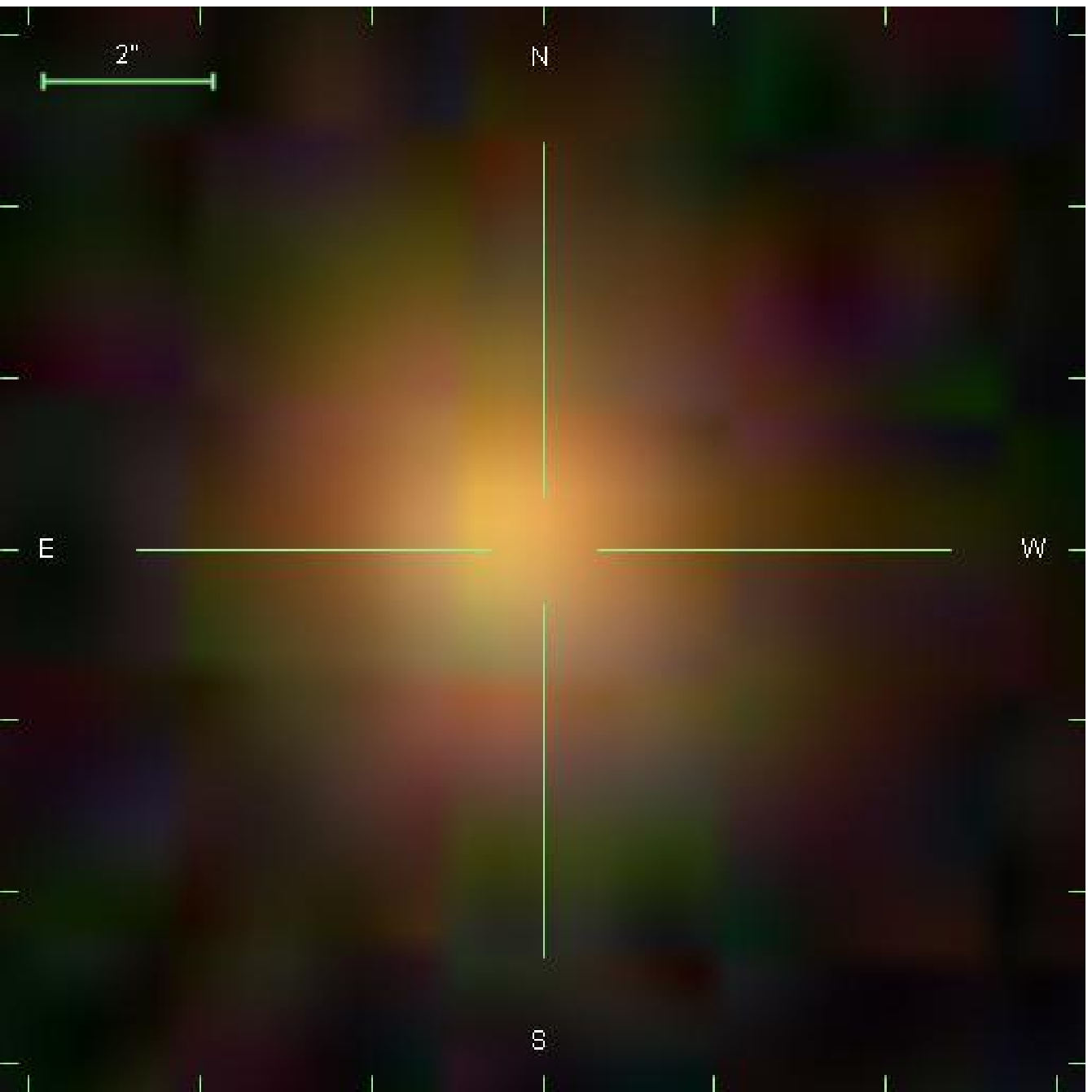}
\includegraphics[width=0.16\textwidth]{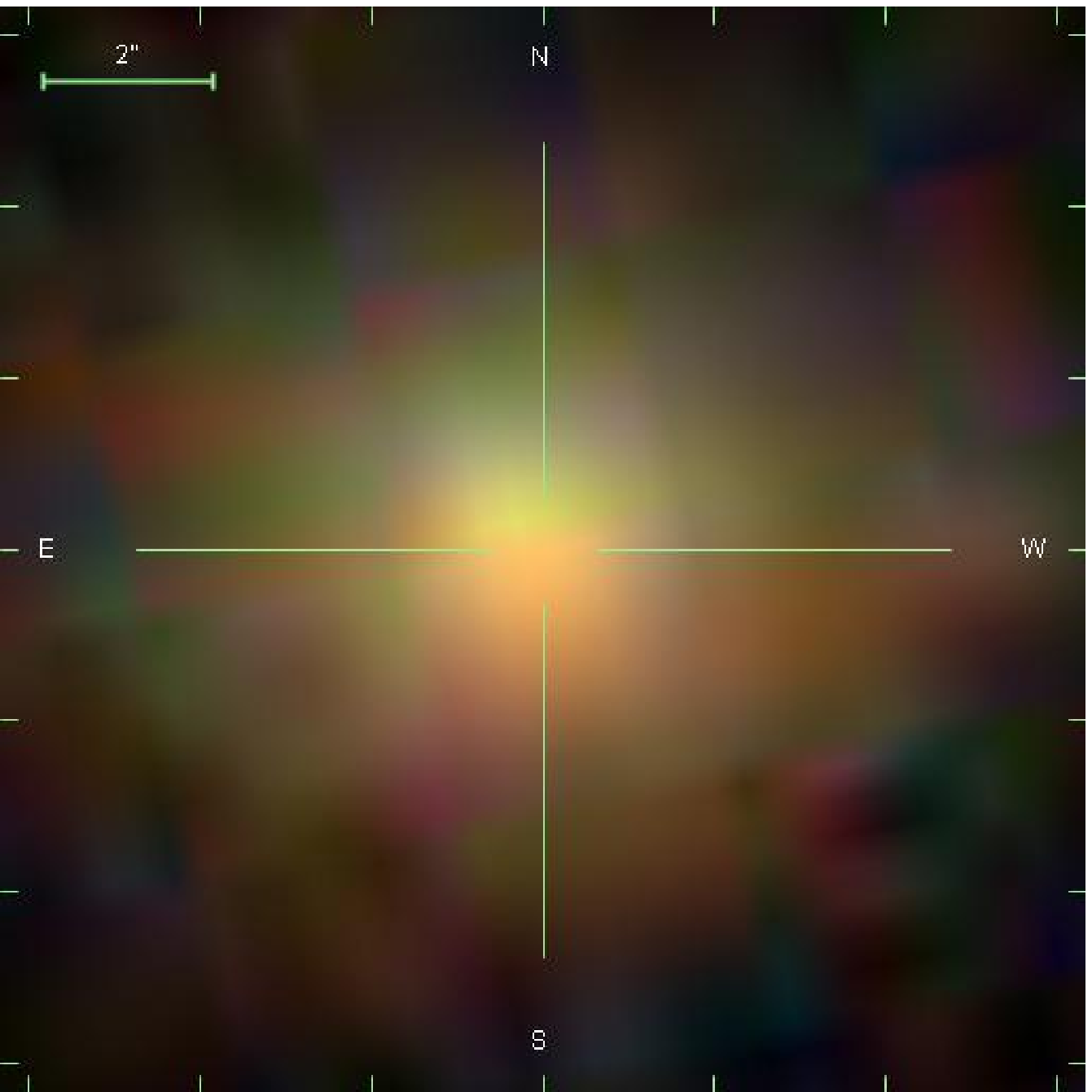}
\includegraphics[width=0.16\textwidth]{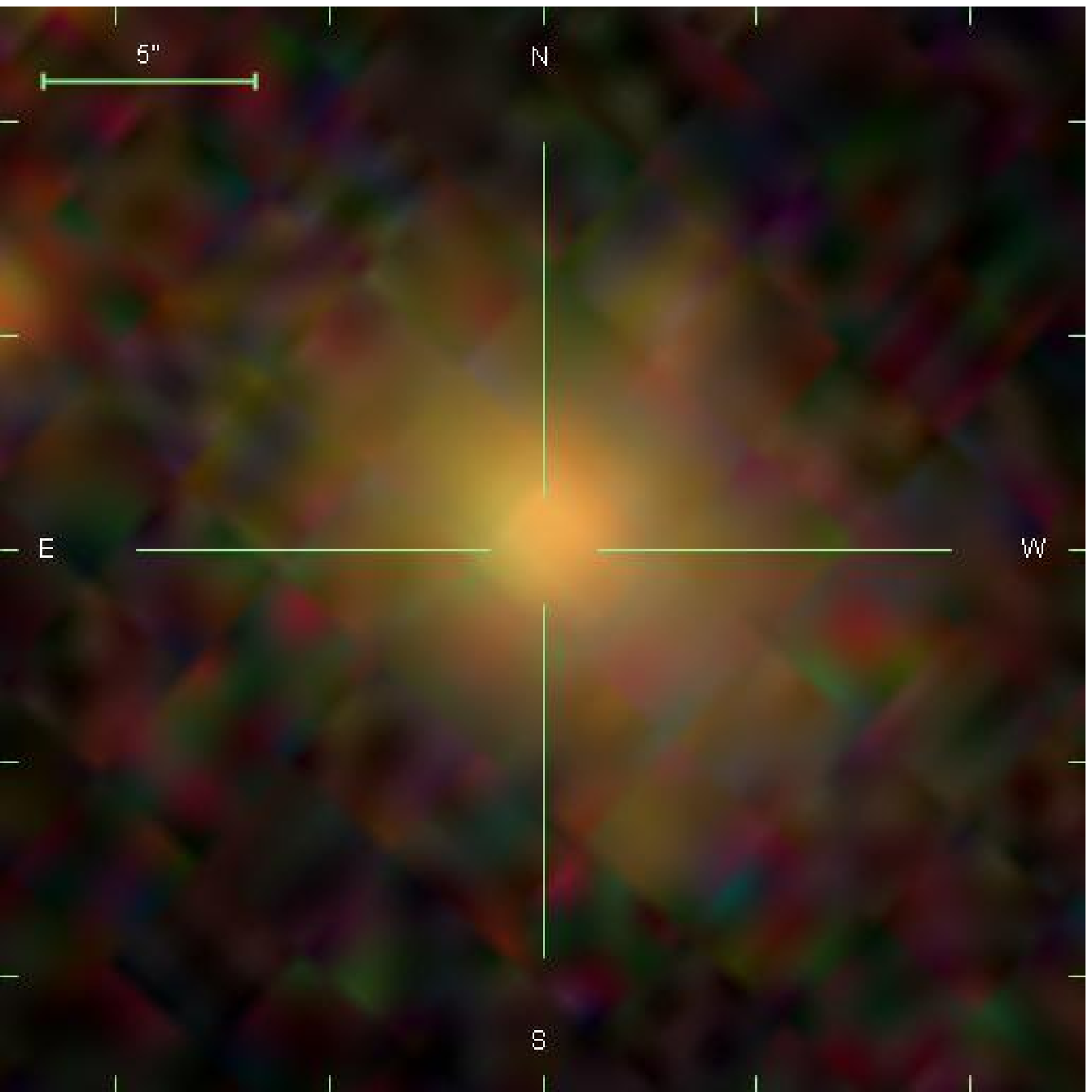}\\
\includegraphics[width=0.16\textwidth]{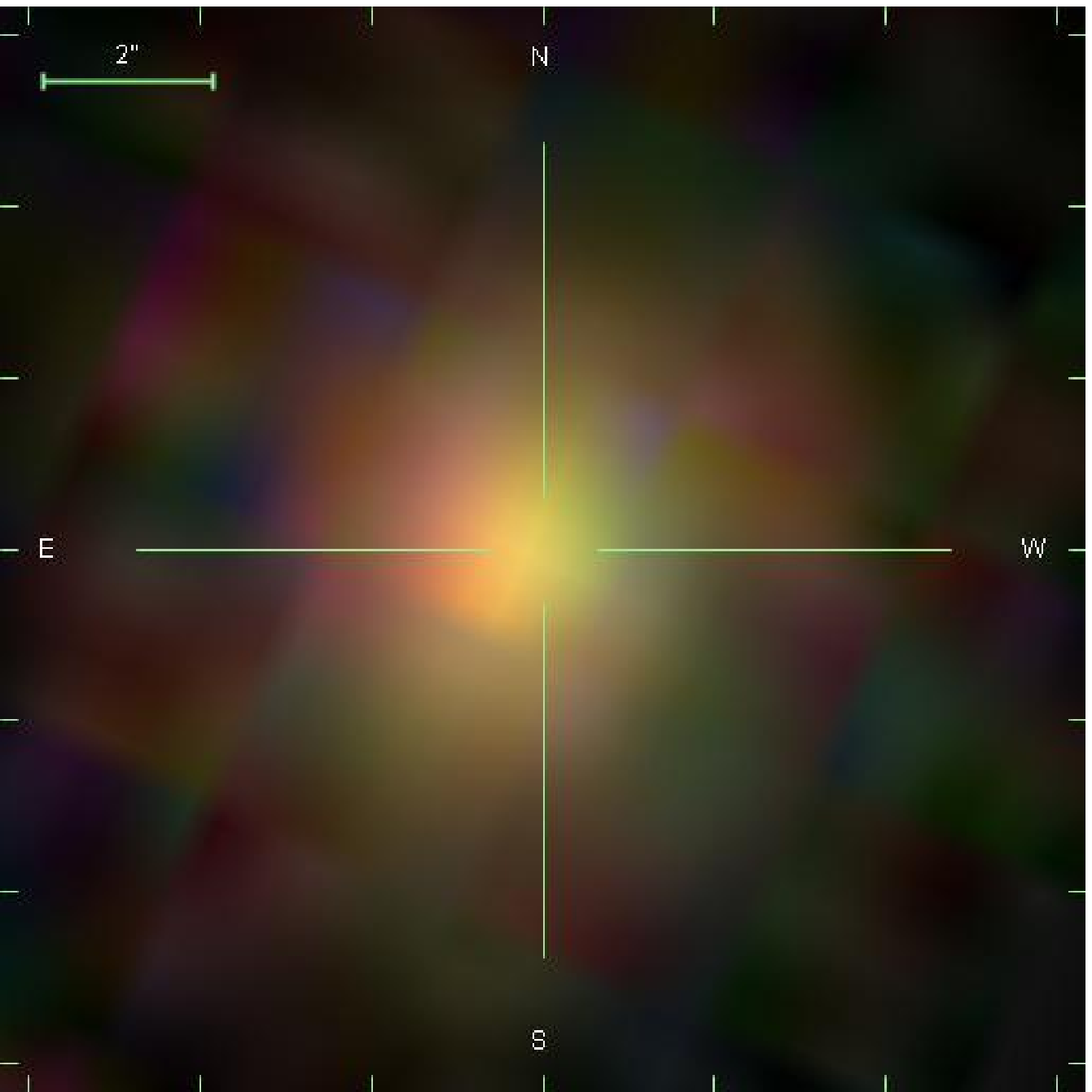}
\includegraphics[width=0.16\textwidth]{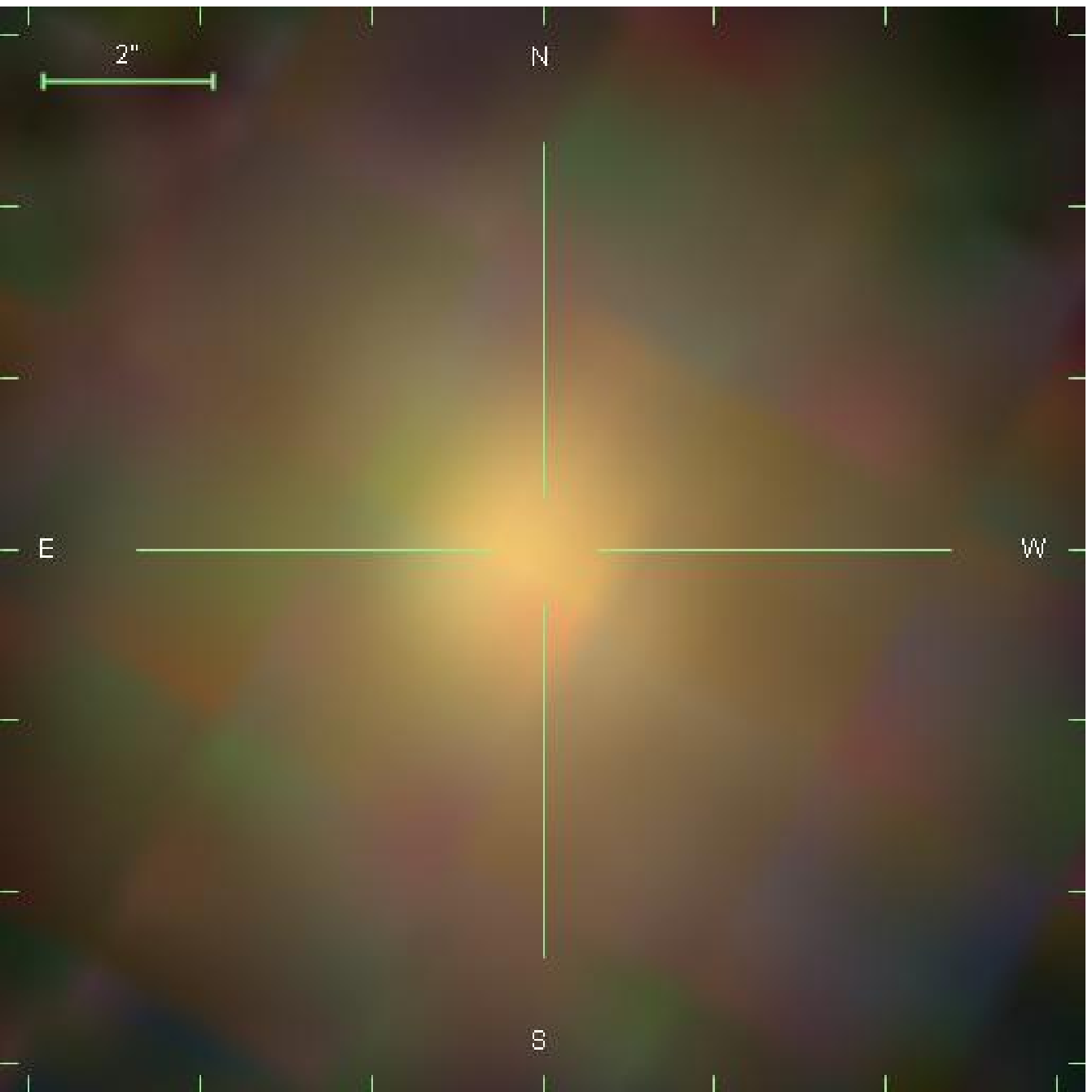}
\includegraphics[width=0.16\textwidth]{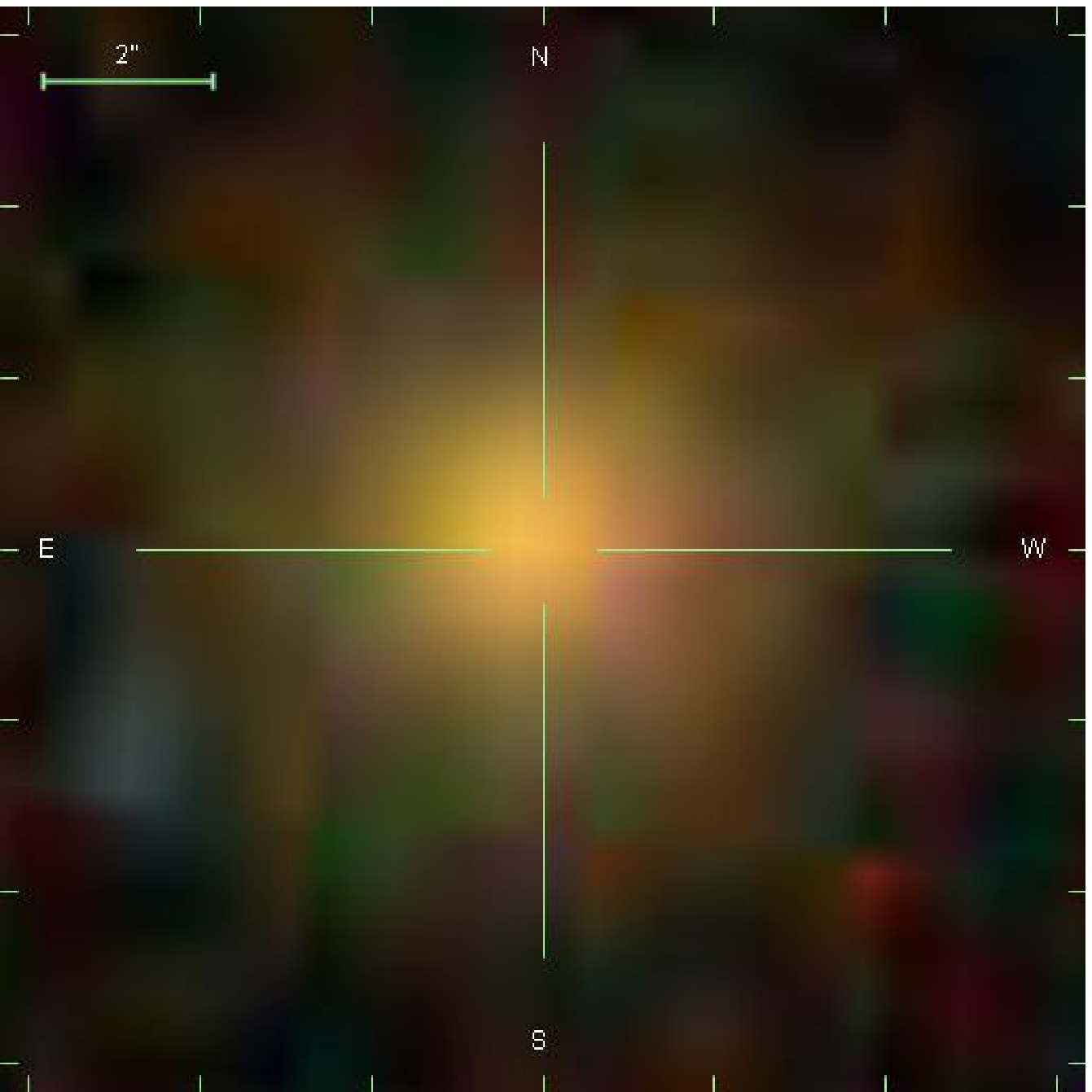}
\caption{Randomly selected subsample of 12 galaxies of our selected sample with a redshift range of [0.1,0.2]. All of them were classified to be elliptical galaxies by our selection criteria, which is confirmed by their morphology.}
\label{middle_gal}
\end{center}
\end{figure}
\begin{figure}[ht]
\begin{center}
\includegraphics[width=0.16\textwidth]{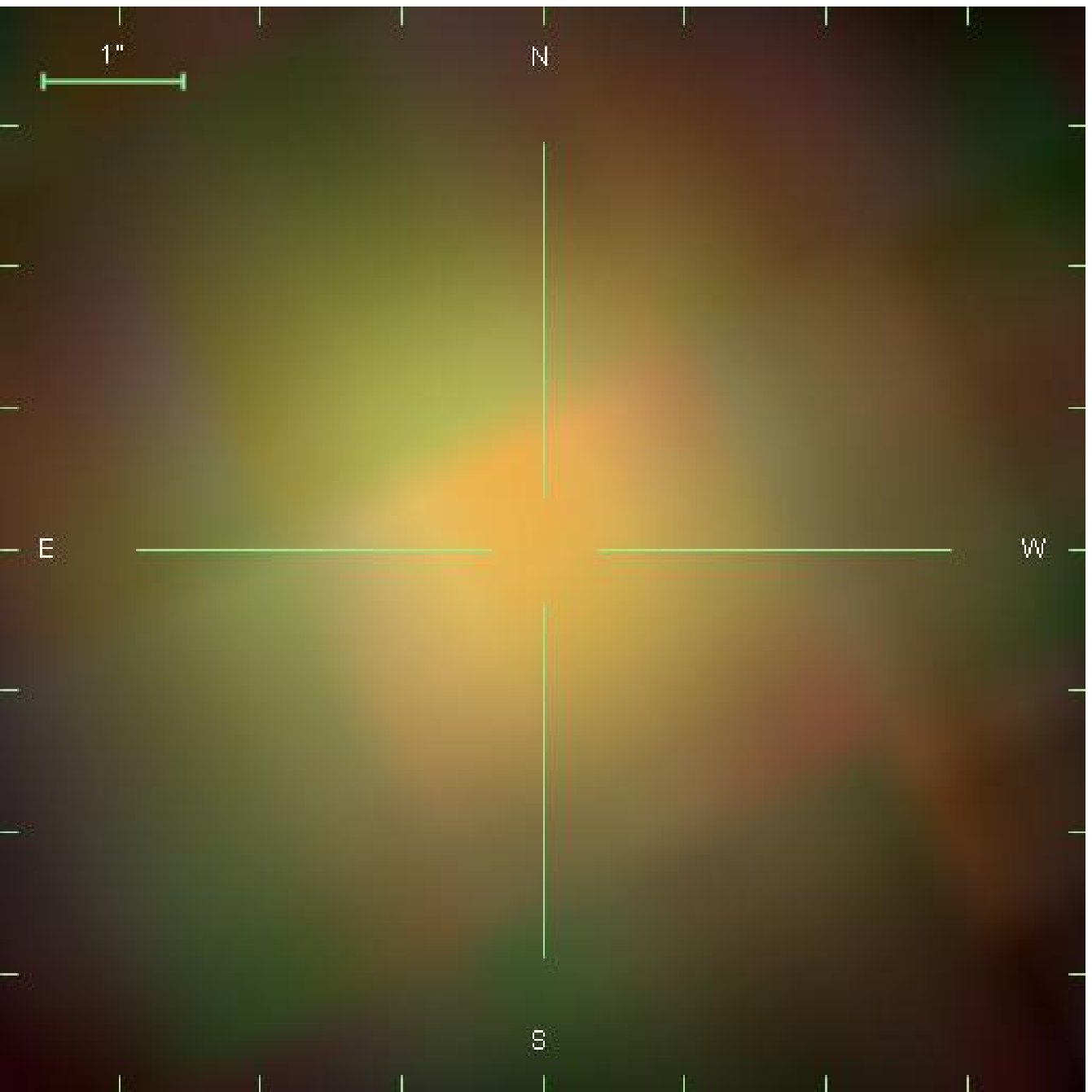}
\includegraphics[width=0.16\textwidth]{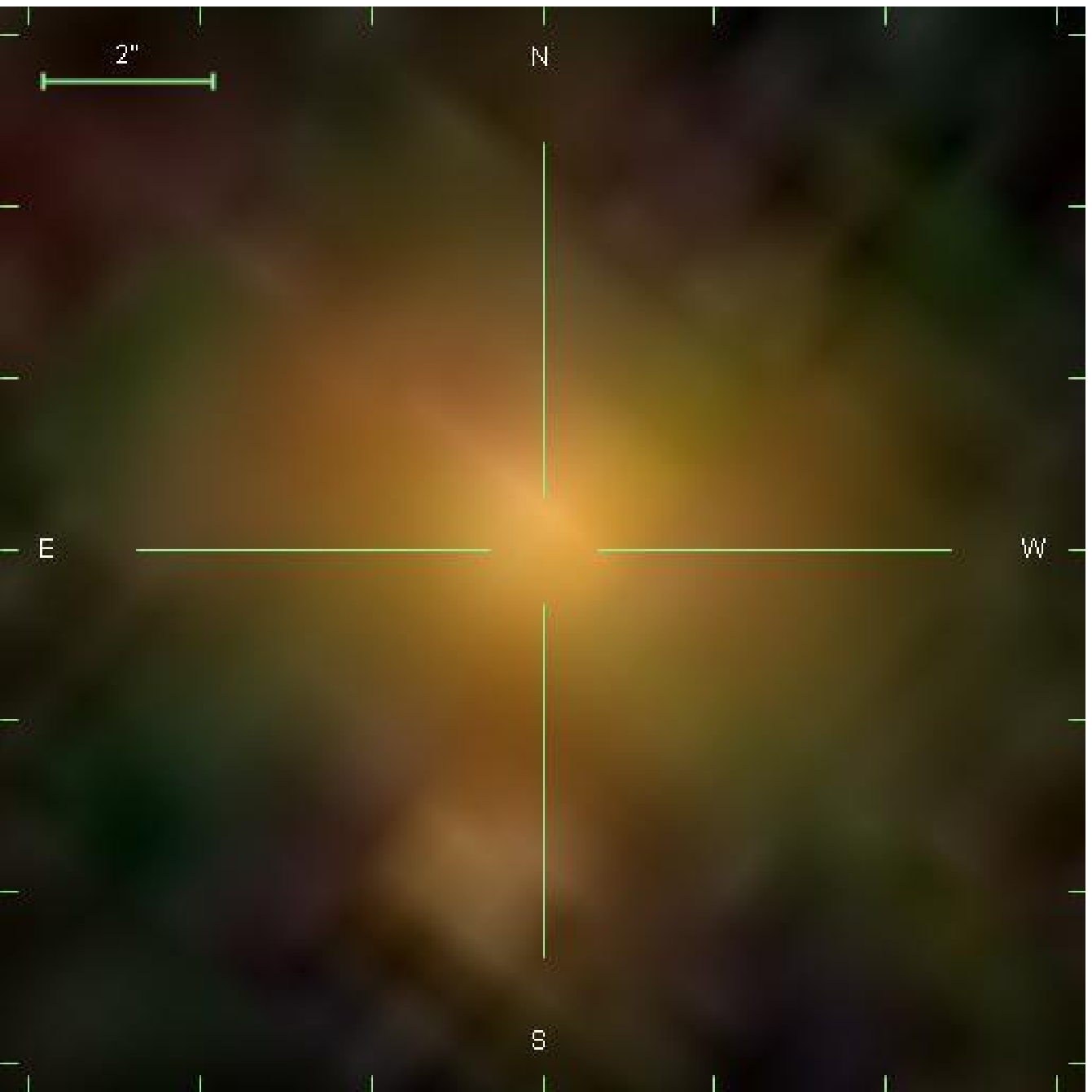}
\includegraphics[width=0.16\textwidth]{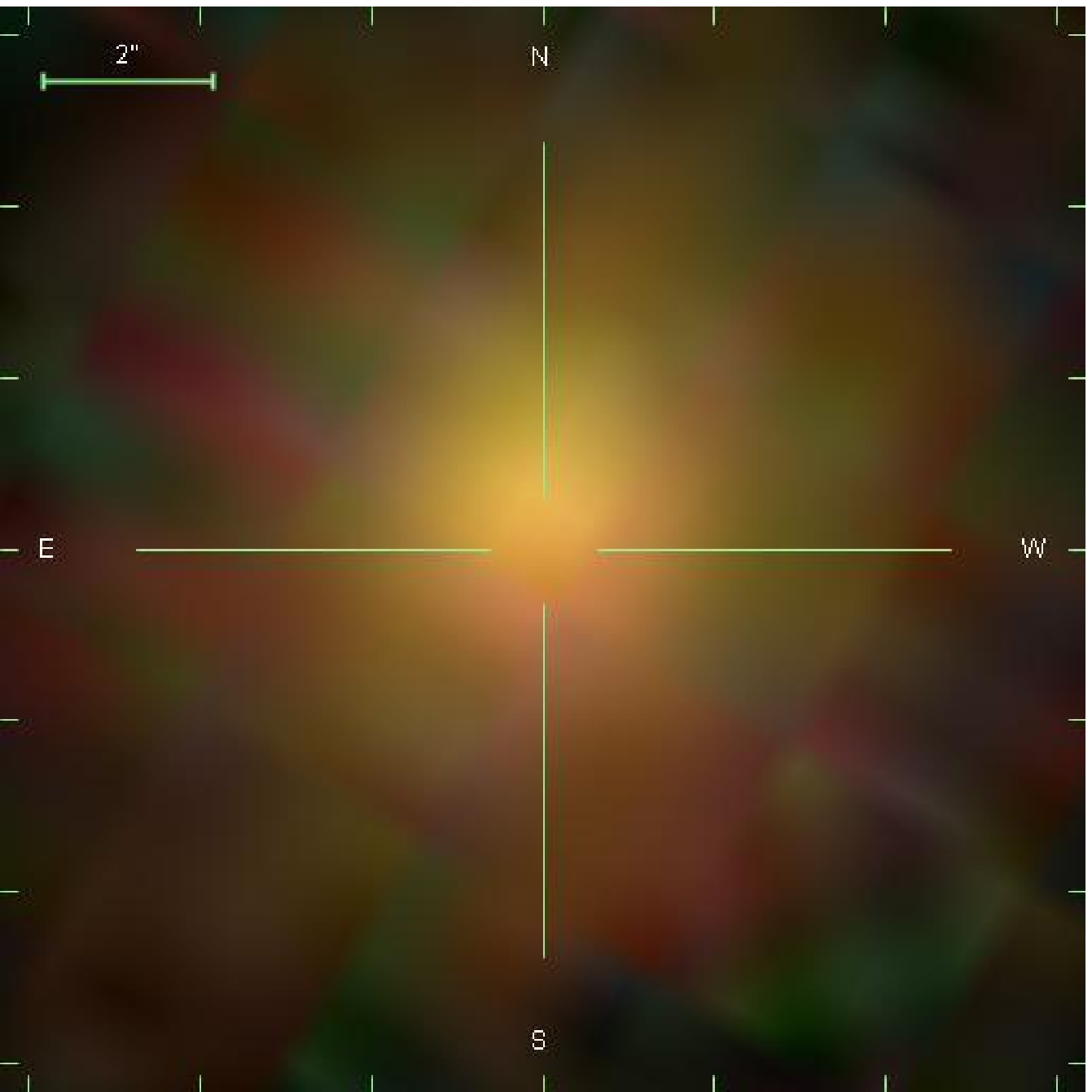}\\
\includegraphics[width=0.16\textwidth]{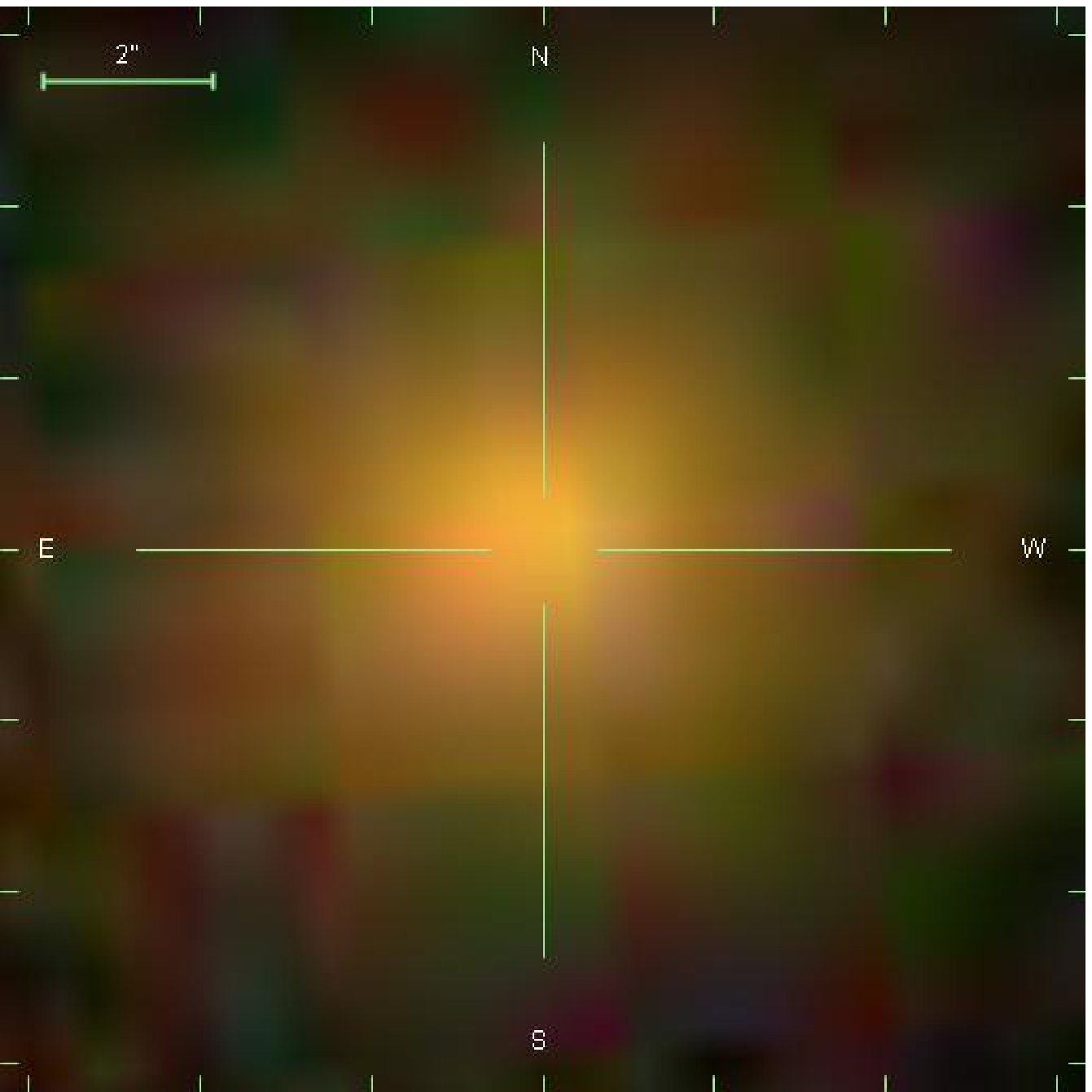} 
\includegraphics[width=0.16\textwidth]{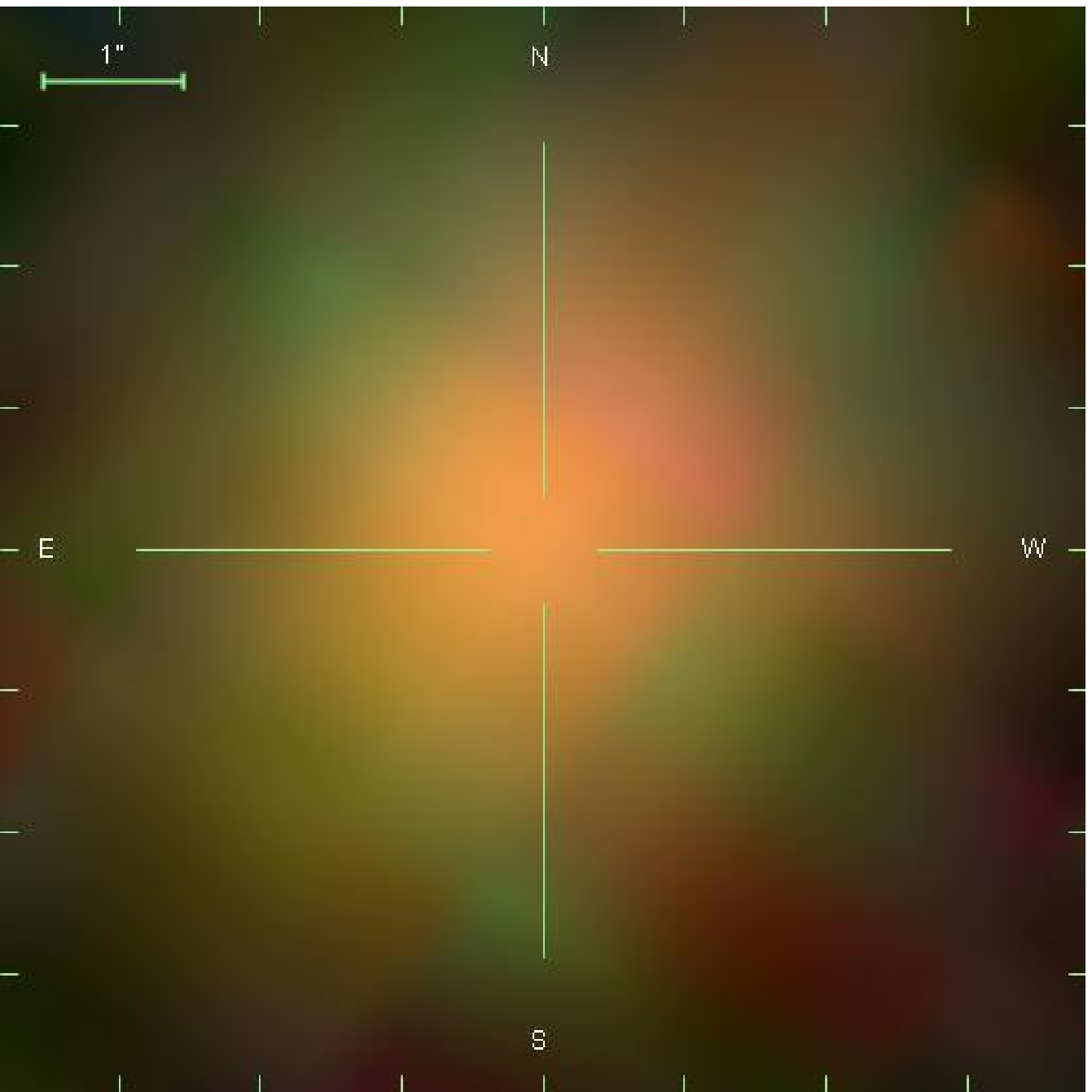}
\includegraphics[width=0.16\textwidth]{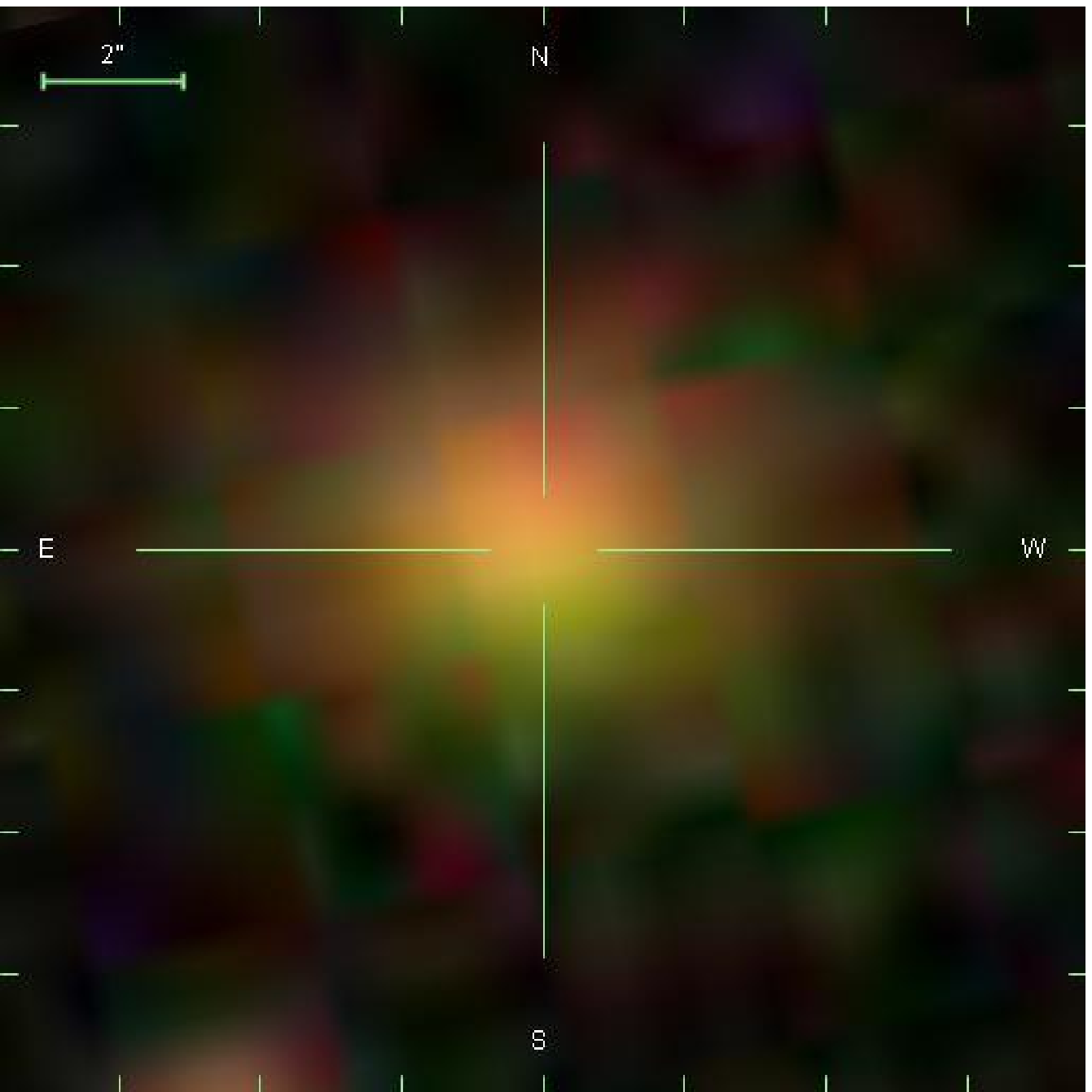}\\
\includegraphics[width=0.16\textwidth]{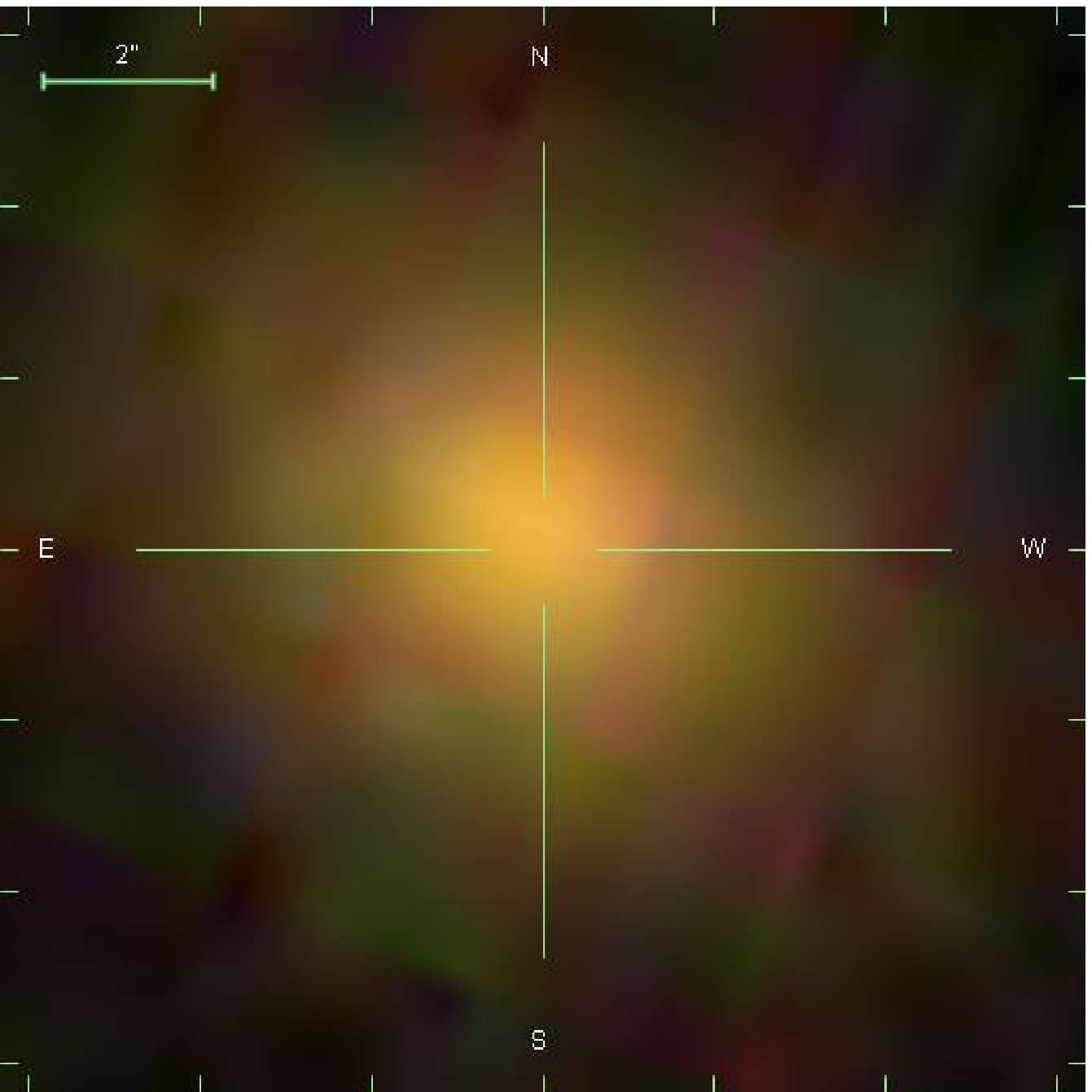}
\includegraphics[width=0.16\textwidth]{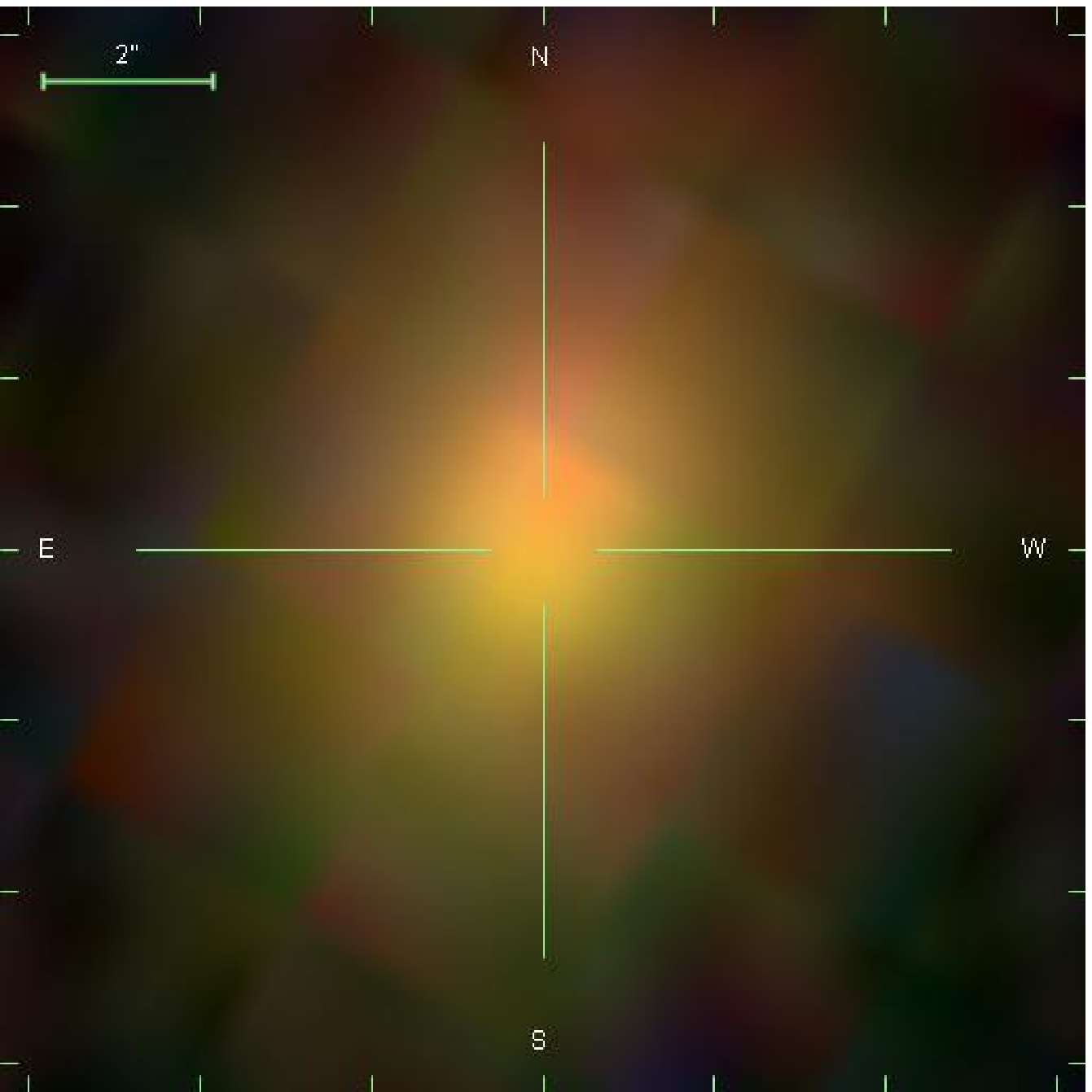}
\includegraphics[width=0.16\textwidth]{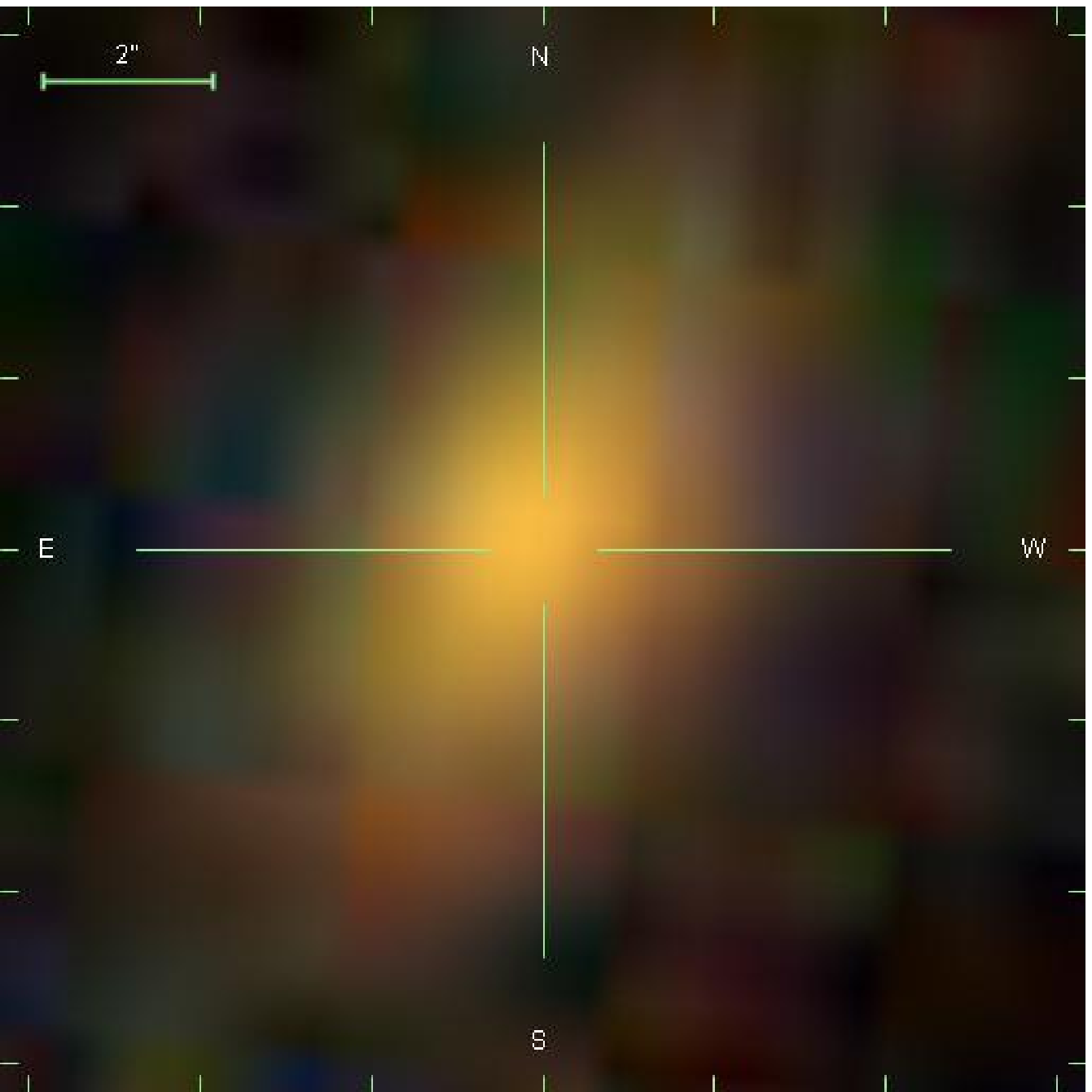}\\
\includegraphics[width=0.16\textwidth]{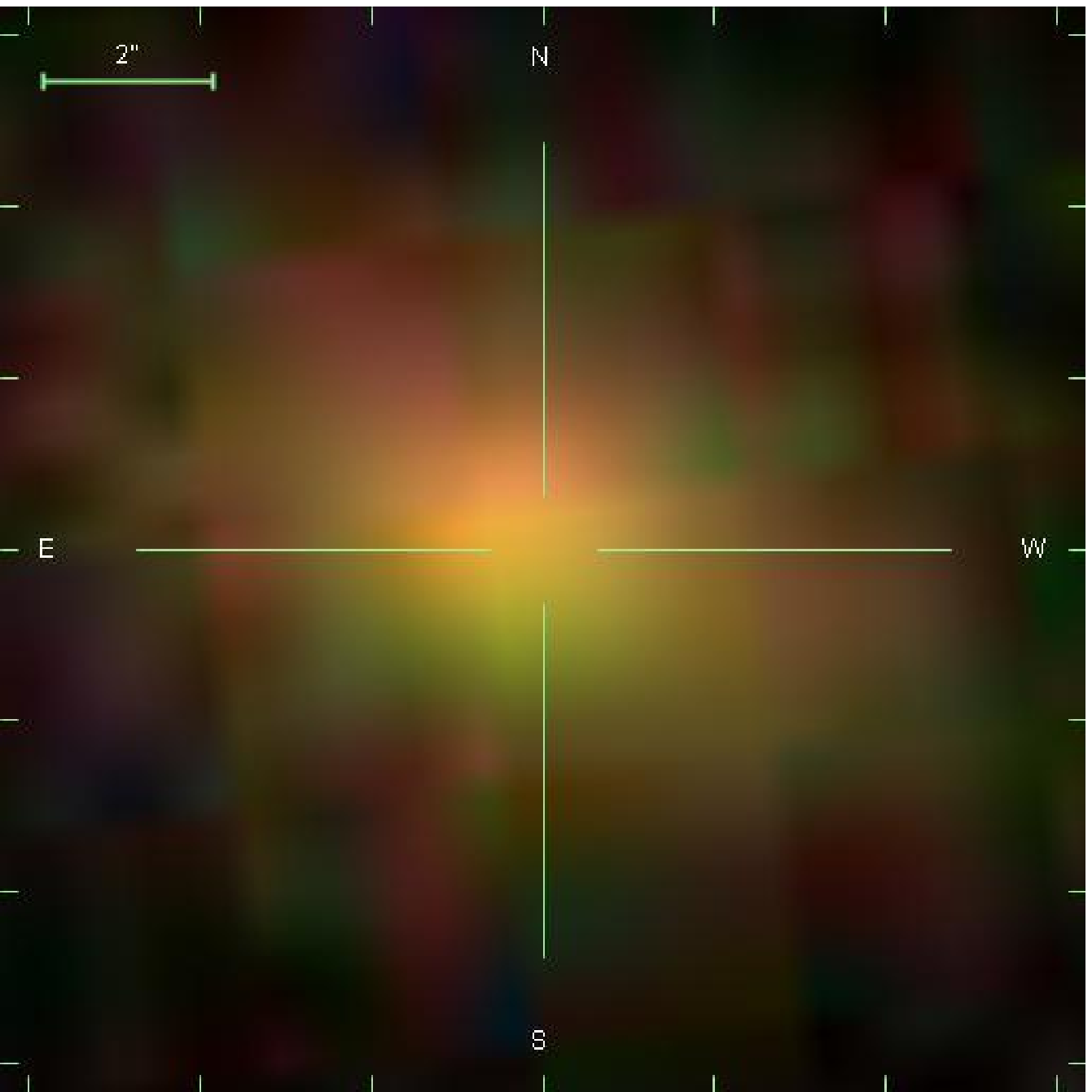}
\includegraphics[width=0.16\textwidth]{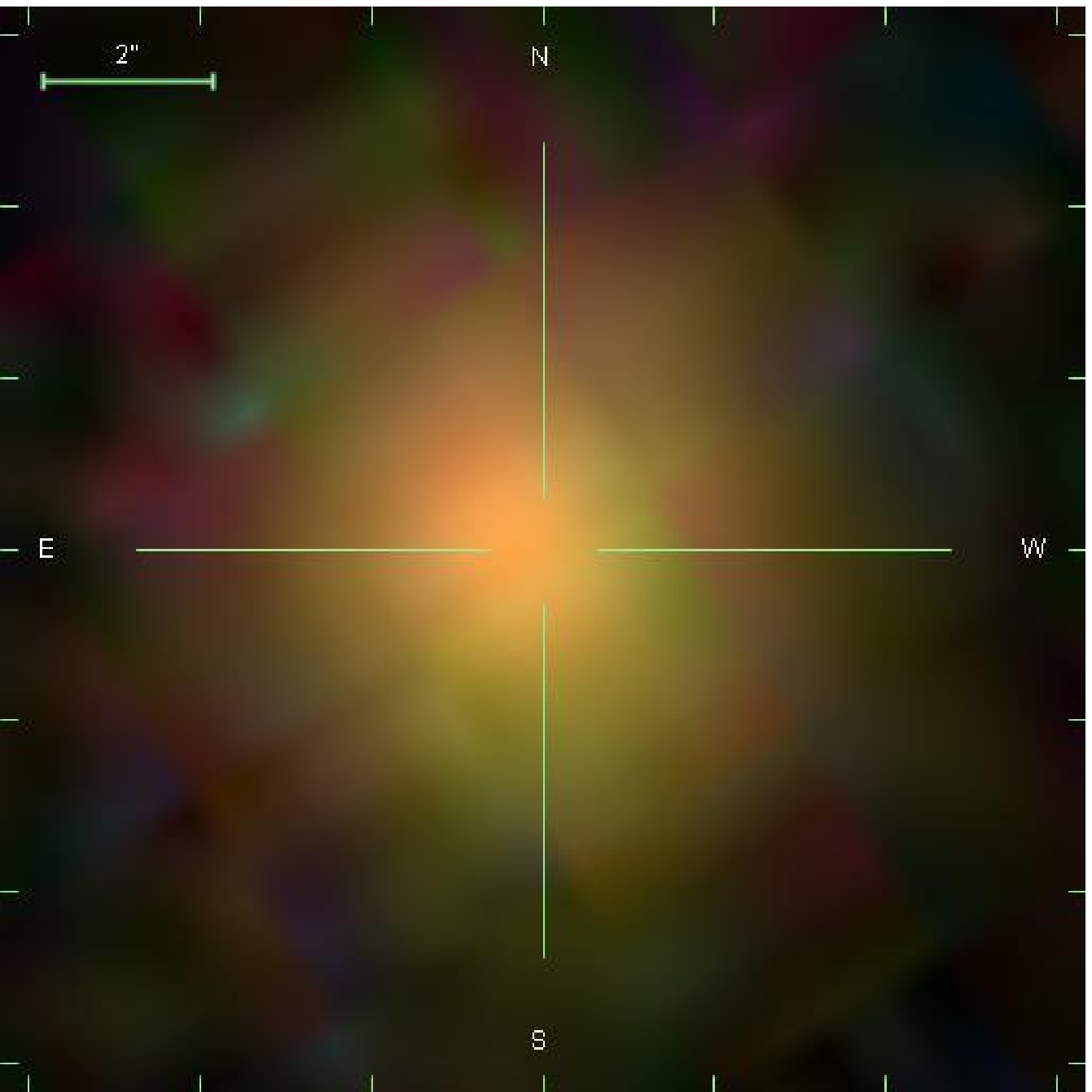}
\includegraphics[width=0.16\textwidth]{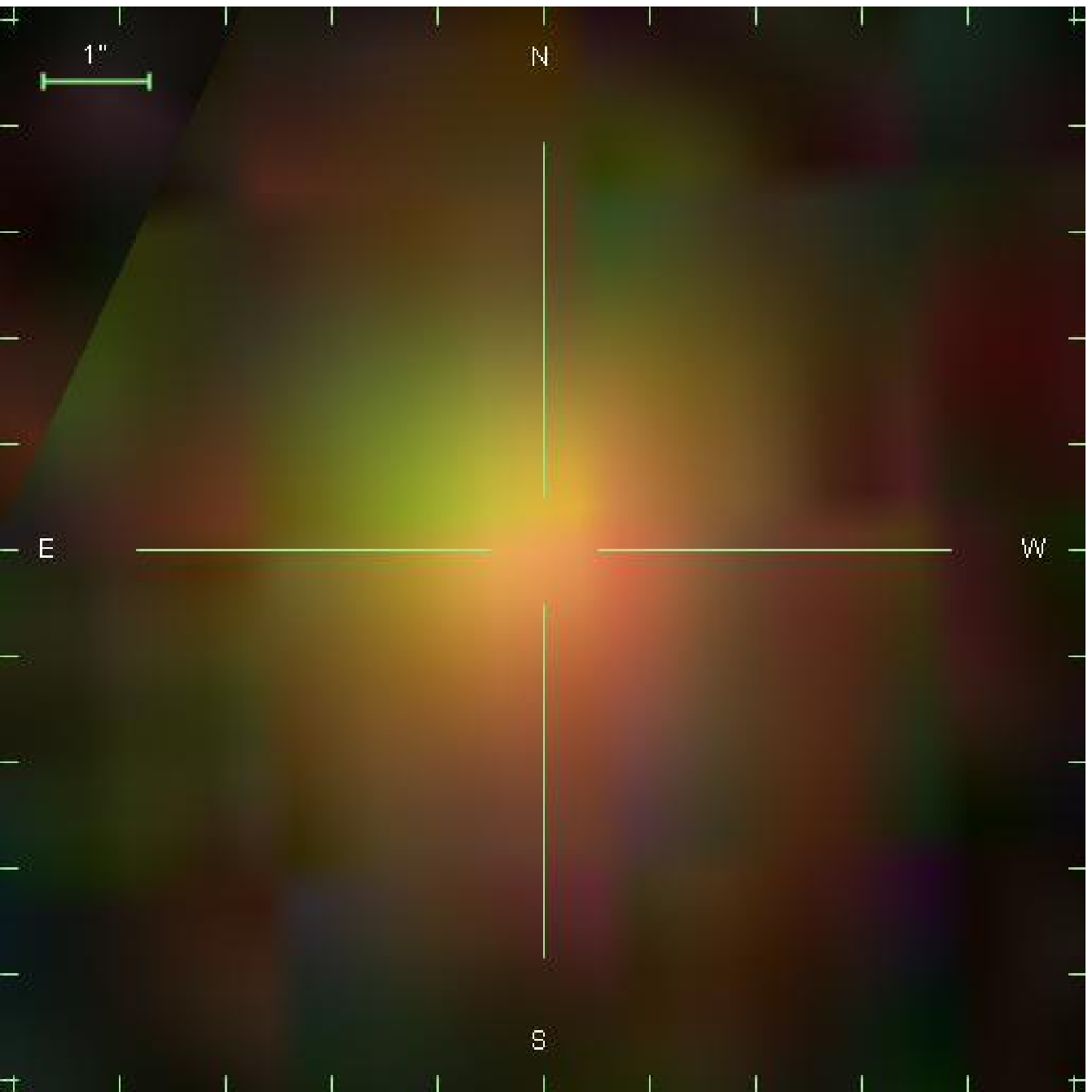}
\caption{Randomly selected subsample of 12 galaxies of our extended sample (limited to a redshift lower than 0.2 for our main analysis) with a redshift higher than 0.2. All of them were classified to be elliptical galaxies by our selection criteria, which is confirmed by their morphology. Note that for some of the sources with the smallest angular extent, the SDSS colour-composite image becomes imprecise, causing the apparent rainbow-like colour structure.}
\label{far_gal}
\end{center}
\end{figure}

The redshift z has to be between 0 and 0.5. Furthermore, to ensure that the redshift measurements were trustworthy, the \emph{SpecObj.zWarning} flag had to be 0. For the morphological selection we made use of the citizen science project GalaxyZoo \citep{GalaxyZoo}, in which volunteers on the internet classify SDSS galaxies in a simplified manner (no scientific background required). The results of these visual classifications \citep{GalaxyZoo_data} were integrated into the SDSS query form. To obtain a reasonable sample of elliptical galaxies with only a small number of misclassification, we demanded that the probability that a galaxy is an elliptical is greater than 0.8 and that at least ten GalaxyZoo users classified it. This probability is the fraction of all users who classified the given galaxy as elliptical. It would make no sense to set the parameter to 1 in the query because this would exclude too many galaxies since many users occasionally misclassify a galaxy by accident or trolling. Based on criterion, GalaxyZoo provided 170234 candidates for elliptical galaxies within the given redshift range of 0 and 0.5.

To reduce the number of misqualifications and to ensure the quality of our data set, we applied the following criteria. The signal-to-noise ratio of the spectra had to be higher than 10 and the central velocity dispersion of every galaxy in our sample had to be higher than 100 km/s and lower than 420 km/s. In general, velocity dispersion measurements of SDSS are only recommended to be used if they are between 70 and 420 km/s. Our choice of 100 km/s as lower limit is an additional precaution to avoid contamination of our sample by misclassification, because we found that a significant number of galaxies with a low central velocity dispersion (<100 km/s) are misclassified as elliptical galaxies by GalaxyZoo, although most of them actually are bulge-dominated spiral galaxies, as we found be taking a small random sample and visually inspecting the imaging and spectroscopic data. Furthermore, the spectrum has to be identified by the SDSS pipeline to be of a galaxy, and the likelihood of a de Vaucouleurs fit on a galaxy has to be higher than the likelihood of an exponential fit, in all five SDSS filters. We also demanded the axis ratio derived from the de Vaucouleurs fit to be higher than 0.3 (which excludes all early-type galaxies later than E7) in all filters, thereby removing very elongated elliptical and lenticular galaxies from our sample. In Figures \ref{near_gal} and \ref{middle_gal}, randomly selected SDSS colour thumbnails of our selected sample are shown. Their morphologies are all consistent with being ellipticals (some artificial apparent green/red granulation can occur in the colour composite), without obvious spiral/disk patterns, and overall smoothness. We thus conclude that our sample is sufficiently clean. For redshifts higher than 0.2, our set of criteria still yields a pretty clean sample as one can see in Figure \ref{far_gal}. However, in Subsection \ref{sample_prop} that our criteria create an additional bias at redshifts higher than 0.2, because an increasing fraction of galaxies is rejected due to uncertain classification.

There are 100427 galaxies in SDSS DR8 that fulfil all these requirements; they form our basic sample \footnote{finer cuts in redshift and colours as well as a rejection of outliers will reduce this to about 93000 galaxies in the end, see next sections}. We downloaded the galactic coordinates \emph{PhotoObj.b} and \emph{PhotoObj.l}, the redshift \emph{SpecObj.z} and its error \emph{SpecObj.zErr}, the central velocity dispersion \emph{SpecObj.veldisp} and its error \emph{SpecObj.veldispErr}, the axis ratio of the de Vaucouleurs fit \emph{PhotoObj.deVAB\_filter}, the scale radius of the de Vaucouleurs fit \emph{PhotoObj.deVRad\_filter} and its error \emph{PhotoObj.deVRadErr\_filter}, the model magnitude of the de Vaucouleurs fit \emph{PhotoObj.deVMag\_filter} and its error \emph{PhotoObj.deVMagErr\_filter}, the magnitude of the composite model fit \emph{PhotoObj.cModelMag\_filter} and its error \emph{PhotoObj.cModelMagErr\_filter}, the scale radius of the Petrosian fit \emph{PhotoObj.petroRad\_filter} and its error \emph{PhotoObj.petroRadErr\_filter}, the model magnitude of the Petrosian fit \emph{PhotoObj.petroMag\_filter} and its error \emph{PhotoObj.petroMagErr\_filter} and the extinction values \emph{PhotoObj.extinction\_filter}, which are based on Schlegel maps \citep{Schlegelmaps}, for all five SDSS filters (if a parameter is available for different filters, the wild card \emph{filter} is placed there, which can stand for either u, g, r, i, or z) and all galaxies in our basic sample. The SDSS filters have a central wavelength of 355.1 nm for u, 468.6 nm for g, 616.6 nm for r, 748.0 nm for i, and 893.2 nm for z \citep{SDSS_early}. 

\begin{figure}[ht]
\begin{center}
\includegraphics[width=0.45\textwidth]{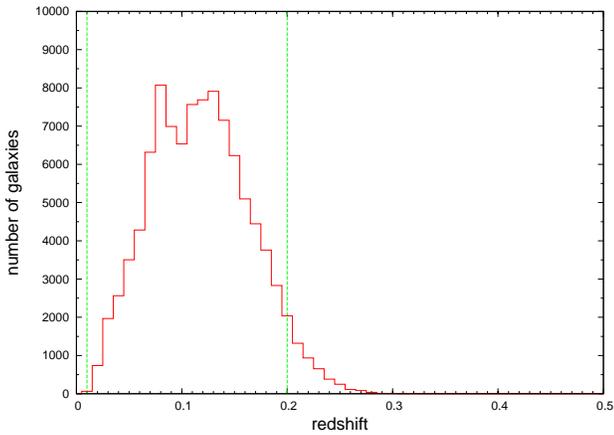}\\ 
\caption{Redshift distribution of the basic sample. The green vertical lines at z=0.01 and z=0.2 indicate the limitation of the selected sample that was used for fitting the fundamental plane.}
\label{zbinned}
\end{center}
\end{figure}

\begin{figure}[ht]
\begin{center}
\includegraphics[width=0.45\textwidth]{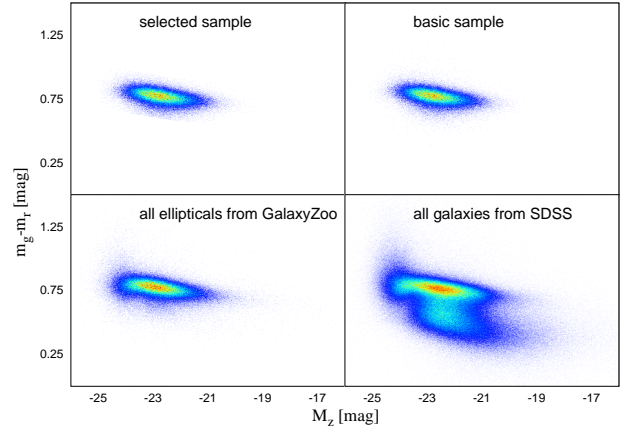}\\ 
\caption{Selection of our sample illustrated using colour-magnitude diagrams. This set of g-r colour versus z band absolute magnitude diagrams shows that with our constraints only the red sequence remains in the selected sample (top left panel) of about 95000 galaxies, which is used to fit the fundamental plane. The basic sample (top right panel), which consists of 100427 galaxies, is already relatively clean by itself. The 170234 candidates of elliptical galaxies from GalaxyZoo can be found in the bottom left panel. The bottom right panel shows 852173 unclassified galaxies with proper spectroscopic data from the SDSS database and in this plot, other structures in the colour magnitude diagram, such as the blue cloud, are clearly visible as well. The colour-coding in the plots is always relative to the maximum number density per pixel. Red indicates the maximum number density (which can vary from figure to figure) per pixel. Colours between red and orange cover number densities between the maximum and 2/3 of the maximum. Shades from yellow to green cover the range between $\sim$2/3 and $\sim$1/3 of the maximum number density. Shades from cyan to blue indicate less than $\sim$1/3 of the maximum number density. White pixels contain no objects. The same colour scheme is used for all other density map plots in this paper as well.}
\label{sample_colours}
\end{center}
\end{figure}

In Figure \ref{zbinned}, we consider the overall redshift distribution of our basic sample. In this Figure \ref{zbinned}, we use redshifts corrected for the Milky Way's motion relative to the CMB, but note that the impact of the correction on the Figure is insignificant. Figure \ref{zbinned} shows that there are no galaxies in our sample with redshifts greater than 0.3. Furthermore, the number density decreases rapidly after a redshift of 0.15. We adopt a final cut at redshift of 0.2, since beyond that the sample is heavily biased towards only the most luminous galaxies. Quantitative motivation for the cut at a redshift of 0.2 comes from considering the Malmquist bias in detail, see Section \ref{sec:Malmquist}. Moreover, we introduced a lower cut at a redshift of 0.01 to remove the galaxies for which peculiar velocities can notably distort the Hubble flow. The limitation of our sample to a redshift interval of [0.01,0.2] reduces the number of galaxies by roughly 5000. Furthermore, excluding some objects, with unreasonably large or small absolute magnitudes or physical radii removes a handful galaxies more. We also introduce a colour cut by demanding that the galaxies in our sample lie on the red sequence \citep{Chilingarian:2012}, which is a narrow region in the colour-magnitude diagram where early-type galaxies are located. Since our sample was already relatively clean at this stage, we fitted a second-order polynomial to it in the colour-magnitude diagram. For this end, we used the g-r colours of the apparent magnitudes and the absolute magnitude in the z band. The results of these fits are shown in Figures \ref{c_mz_vs_g-r_map} to \ref{p_mz_vs_g-r_map} with their fitting parameters in Table \ref{redsequence_fit}. Using these fits, we perform a 3-$\sigma$ clipping to remove outliers. Less than 1 $\%$ (the exact ratio depends on the choice of the photometric fits, i.e., the composite model, de Vaucouleur or Petrosian (see Section \ref{section_results} for details on these fits), of the sample is removed by the colour cut. For comparison, applying the same colour cuts on the sample classified only via GalaxyZoo, about 4.5$\%$ are removed. The remaining $\sim$95000 galaxies were used for the fundamental plane calibrations and form our selected sample. During the fitting process another about 2000 galaxies were excluded as outliers, which leaves a sample of 93000 galaxies for our final analysis. A set of comparative colour-magnitude diagrams (see Figure \ref{sample_colours}) illustrates the cleaning process of our sample all the way from the 852173 SDSS galaxies with proper spectroscopic data to our selected sample of about 95000 galaxies.


\subsection{Substructure in the redshift distribution}
\label{sample_prop}
In this section we discuss the redshift substructure in our sample, identifying three peaks in redshift at which galaxies cluster.

Consider Figure \ref{zbinned} again, in which one may immediately notice two peaks in the galaxy counts. The first one, which appears to be most prominent at z=0.08 in this plot, is associated with the Sloan Great Wall, which is located at a redshift of 0.073 \citep{Gott:2005}. The other peak is around a redshift of 0.13 and has been reported previously in \citet{Bernardi:2003b}, though it was not discussed in detail afterwards.

In the following we describe how we corrected the redshift histogram in Figure \ref{zbinned} for completeness and sampling effects to investigate the redshift substructure of our sample in more detail. First of all, for a volume-limited sample one expects the number of galaxies to increase with the third power of the distance (which is in first-order approximation linearly related to the redshift). Then, due to magnitude limitation, one looses the less luminous part of the sample starting at a certain redshift, and number counts will decrease towards zero at very large distances. The combination of volume sampling and magnitude limitation yield a function that grows as a function of redshift from zero to a peak, and then decreases afterwards to zero again. Our sample in Figure \ref{zbinned} has this overall peak at about z=0.1, close to the two putative sub-peaks.

\begin{figure}[ht]
\begin{center}
\includegraphics[width=0.45\textwidth]{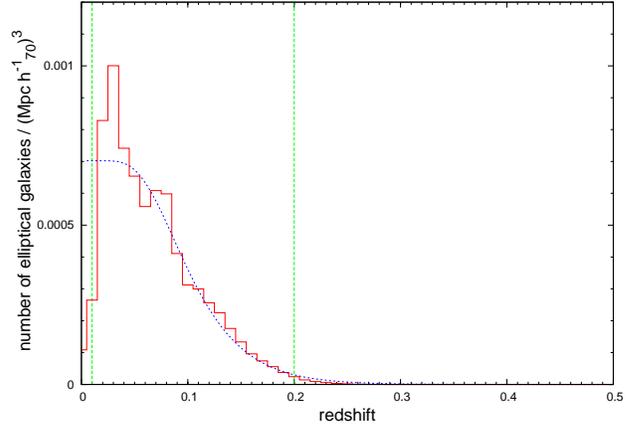}\\ 
\caption{Redshift distribution of the number density of elliptical galaxies in our sample. The measured density of elliptical galaxies is given by the red solid line of the histogram. We divided our sample into redshift bins with a width of 0.01. The blue dashed line represents our best-fit model to the data. The green dashed vertical lines indicate the limits of the sample, which were used for the fundamental plane fitting.}
\label{volumebinned}
\end{center}
\end{figure}

It is therefore important to remove the signature of magnitude limitation from this histogram, which will allow a clean assessment of the existence of those possible sub-peaks. As a first step (Figure \ref{volumebinned}), we thus considered the galaxy count per comoving volume instead of the absolute numbers, dividing the number of galaxies in each redshift bin by the comoving volume of the bin. The following equations define the comoving volume of such a redshift bin $V_{C}(z_{1},z_{2})$:
\begin{equation}
V_{C}(z_{1},z_{2})=\frac{4 \pi}{3} \frac{A_{\textrm{SDSS}}}{A_{\textrm{sky}}} \left(D_{C}^{3}(z_{2})-D_{C}^{3}(z_{1})\right)
\label{comvol}
\end{equation}
\begin{equation}
D_{C}(z)=D_{L}(z) \cdot (1+z)^{-1}
\label{comdist}
\end{equation}
\begin{equation}
D_{L}(z)=\frac{c \cdot z}{H_{0}} \left(1+\left( \frac{z \cdot (1-q_{0})}{\sqrt{1+2 q_{0} \cdot z}+1+q_{0} \cdot z} \right)\right)
\label{lumdist}
\end{equation}
\begin{equation}
q_{0}=\frac{\Omega_{M}}{2}-\Omega_{\Lambda} .
\label{decelpar}
\end{equation}
The comoving volume $V_{C}(z_{1},z_{2})$ is derived from the comoving distance $D_{C}(z)$ and the spectroscopic sky coverage of SDSS DR8 $A_{\textrm{SDSS}}$, which is 9274 $\textrm{deg}^2$, normalised to the total size of the sky $A_{\textrm{sky}}$, which is $\sim 41253$ $\textrm{deg}^2$. The comoving distance $D_{C}$ itself is derived from the luminosity distance $D_{L}$, which is given in Equation \ref{lumdist}. We assumed a Hubble parameter $H_{0}$ of $70\, \textrm{km}\,\, \textrm{s}^{-1}\,\, \textrm{Mpc}^{-1}$ for  our calculations, which can be rescaled to any measured Hubble parameter using $h_{70}$. The subscript $70$ emphasises that this scaling parameter is relative to our chosen value of the Hubble parameter. $c$ denotes the speed of light and $q_{0}$ the deceleration parameter of the universe, which is $-0.55$ for a universe with a relative matter density of $\Omega_{M}=0.3$ and relative dark-energy density of $\Omega_{\Lambda}=0.7$ according to Equation \ref{decelpar}.

In the next step, we corrected for the magnitude limitation. We assumed that we have the same functional shape of the luminosity function of giant elliptical galaxies in every volume element of the universe. For this, we adopted a Gaussian luminosity function with a mean luminosity $\bar{M}$ slowly evolving (evolution parameter $Q$) as linear function of the redshift, and assumed that the standard deviation $\sigma_{M}$ is the same for all redshifts and volume elements. The Gaussian shape of the luminosity function of large elliptical galaxies is discussed in detail in Section \ref{res_lumfct}. Although the bulk of the dwarf elliptical galaxies are already too faint at the minimum redshift of our sample, there are still a few bright dwarf elliptical galaxies that do not fall below the magnitude limit. However, these galaxies are not part of the sample, because their light profiles are more exponential than elliptical and therefore, they are already excluded by the sample selection.

Our sample is limited by a fixed apparent magnitude $m_{\textrm{limit}}$ (basically the spectroscopy limit of SDSS), which will cut into the elliptical galaxy luminosity function more and more as redshift/distance increases, biasing our sample towards intrinsically brighter luminosities at high distances. This effect is also known as Malmquist bias.

We can now express the expected space density of ellipticals in our sample with Equation \ref{surveylimit}:
\begin{equation}
\rho_{\textrm{obs}}(z)=\frac{\rho_{0}}{2}\left(1+\textrm{erf}\left( \frac{ \Delta m + Q \cdot z - 5  \cdot \textrm{log}_{10}(D_{L}/\textrm{pc}) + 5 }{\sqrt{2} \cdot \sigma_{M}} \right)\right) .
\label{surveylimit}
\end{equation}
\begin{equation}
m_{\textrm{app}}-M_{\textrm{abs}}=5 \cdot \textrm{log}_{10}(D_{L}/\textrm{pc})-5
\label{distmod}
\end{equation}
In these equations, we defined a new parameter $\Delta m = m_{\textrm{limit}}-\bar{M}$, which denotes the difference between limit magnitude $m_{\textrm{limit}}$ and the mean luminosity of the luminosity function at redshift zero $\bar{M}$. To represent the Malmquist bias we made use of the distance modulus (see Equation \ref{distmod}), which defines the difference between the apparent magnitude $m_{\textrm{app}}$ and the absolute magnitude $M_{\textrm{abs}}$ in dependence on the luminosity distance $D_{L}$.

We then fitted this 4-parameter function for the observed galaxy density $\rho_{\textrm{obs}}(z)$ to the redshift distribution in Figure \ref{volumebinned}. The four varied parameters are the density of elliptical galaxies $\rho_{0}$, the evolution parameter $Q$, the standard deviation of the luminosity function $\sigma_{M}$, and the parameter $\Delta m$. We used a modified simplex algorithm to perform the first fit to obtain the galaxy densities. After inverting the error function in Equation \ref{surveylimit}, we used a least-squares fit to obtain the other parameters. For mathematical reasons, we had to exclude the first five bins, which form the plateau of the function, for the least-squares fit. However, since the height of the plateau was already fixed by the simplex fit, which uses these bins, no information is lost. The results of this fit are shown as a dashed (green) line in Figure \ref{volumebinned}. Our best fit yields an average density of elliptical galaxies in the universe $\rho_{0}$ (which fulfil the requirements of our sample) of $7 \cdot 10^{-4}$ galaxies per $(\textrm{Mpc} \cdot h_{70}^{-1})^3$. Furthermore, we derived a mild redshift evolution $Q$ of $1.07$ mag (per $z$). Our values for $\Delta m$ and $\sigma_{M}$ are $\left(38.3 - 5\cdot \textrm{log}_{10}(h_{70}) \right)$ mag and $0.89$ mag, respectively.

\begin{figure}[ht]
\begin{center}
\includegraphics[width=0.45\textwidth]{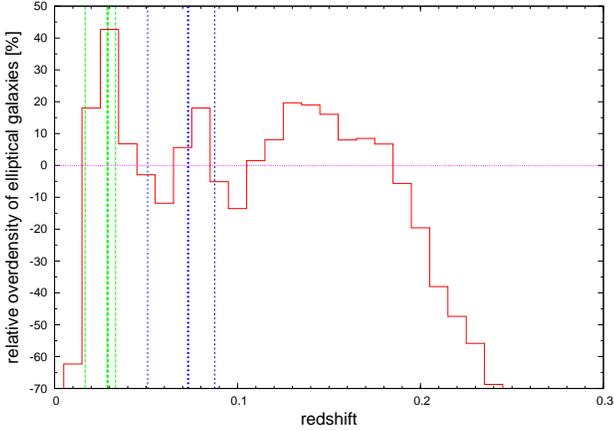}\\ 
\caption{Redshift distribution of the relative number density of elliptical galaxies. The red solid line shows the relative overdensity of the galaxies in our sample compared with the model predictions. Two known large-scale structures are indicated by the green and blue dashed vertical lines. The thin lines are the lower and upper limit of these structures, while the thick line indicates their median redshift. The green lines denote the CfA2 Great Wall and the blue lines denote the Sloan Great Wall.}
\label{overdense}
\end{center}
\end{figure}

To analyse overdensities, we subtracted the fit function from our data in Figure \ref{volumebinned} and normalised it to the fit function's values. The result of this exercise is shown in Figure \ref{overdense}. One notices three overdensities, two of which can be identified with known large-scale structures. There are the CfA2-Great Wall between a redshift of 0.0167 and 0.0333 with a median redshift of 0.029 \citep{Geller:1989} and the Sloan Great Wall between a redshift of 0.0509 and 0.0876 with a median redshift of 0.073 \citep{Gott:2005}. Furthermore, we found another (so far not investigated) overdensity between 0.12 and 0.15 with a peak around 0.13. This overdensity was previously mentioned in \citet{Bernardi:2003b}, who noticed two overdensities in their (SDSS-based) sample of elliptical galaxies: one at a redshift of 0.08 (their paper was published before the Sloan Great Wall was discovered at a similar redshift) and another one at 0.13, which we can now confirm. Identifying any large-scales structures, which might be associated with this over-density, would exceed the scope of this paper and is left open for future investigations.

When looking at Figure \ref{overdense}, one will also notice some significant underdensities at very low and at relatively high redshifts. The underdensity at low redshifts is due to a selection effect in SDSS and GalaxyZoo. Very nearby galaxies are sometimes not included in the SDSS spectroscopic sample because they are too bright. Previously well-classified nearby galaxies were not included in GalaxyZoo as well. These two selection effects cause the apparent deficiency of elliptical galaxies at low redshifts. At high redshifts (z>0.2), we encounter a similar problem. Galaxies at this distance are already rather small and difficult to classify. Therefore, we excepted fewer galaxies that were clearly identified as ellipticals by GalaxyZoo. These two findings strengthen our previous considerations to cut our sample at redshifts lower than 0.01 and higher than 0.2 before using them for the fundamental plane fitting. 

\section{Method}
The fitting procedure of fundamental plane coefficients was performed individually for each SDSS filter to derive independent results. The first matter that needs to be taken into account is galactic extinction:

\begin{equation}
m_{\textrm{extcor}}=m_{\textrm{sdss}} - A_\textrm{Schlegel} .
\label{extinctioncorrection}
\end{equation}

We corrected the SDSS model magnitudes $m_{\textrm{sdss}}$ for extinction $A_\textrm{Schlegel}$ according to the Schlegel maps \citep{Schlegelmaps}. The values for extinction were obtained from the SDSS database, in which they can be found for every photometric object based on its coordinates. 

We also applied a K-correction to $m_{\textrm{extcor}}$, to correct for the effect of the redshift on the spectral profile across the filters, 
\begin{equation}
K(z,m_{f_{1}}-m_{f_{2}})=\sum\limits_{i,j} B_{ij} z^{i} (m_{f_{1}}-m_{f_{2}})^{j}
\label{Kcorrection}
\end{equation}
\begin{equation}
m'_{\textrm{app}} = m_{\textrm{extcor}} - K(z,m_{f_{1}}-m_{f_{2}})  .
\label{apperantmag}
\end{equation}
The K-correction is calculated using the (extinction-corrected) colour $m_{f_{1}}-m_{f_{2}}$ and the redshift $z$ of the galaxy. We used the recent model of \citet{Chilingarian:2010} with updated coefficients as shown in Tables \ref{Kcor_u} to \ref{Kcor_z}. This K-correction model uses a two-dimensional polynomial with coefficients $B_{ij}$ that depend on the filter. One obtains the K-corrected apparent magnitude $m'_{\textrm{app}}$ by a simple subtraction of the K-correction term $K(z,m_{f_{1}}-m_{f_{2}})$. The subscripts $f_{1}$ and $f_{2}$ stand for two different filters, which one can choose for calculating the correction.

The next step was to renormalise the measured model radius from the SDSS data $r_{\textrm{sdss}}$ to take into account the different ellipticities of the elliptical galaxies in our sample,
\begin{equation}
r_{\textrm{cor}} = r_{\textrm{sdss}} \sqrt{q_{b/a}} .
\label{rcor}
\end{equation}
This can be done according to \citet{Bernardi:2003c} by using the minor semi-axis to mayor semi-axis ratio $q_{b/a}$ from SDSS. The corrected radius $r_{\textrm{cor}}$ enables us to directly compare all types of elliptical galaxies. 

The velocity dispersions also need to be corrected for. Because SDSS uses a fixed fibre size, the fibres cover different physical areas of galaxies at different distances. This affects the measured velocity dispersion. Suitable aperture corrections for early-type galaxies were calculated by \citet{Jorgensen:1995} and \citet{Wegner:1999}, see the following equation:
\begin{equation}
\sigma_{0}=\sigma_{\textrm{sdss}} \cdot \left( \frac{r_{\textrm{fiber}}}{r_{\textrm{cor}}/8}  \right)^{0.04} .
\label{sigmacor}
\end{equation}
The radius of the SDSS fibres $r_{\textrm{fiber}}$ is 1.5 arcseconds for all releases up to and including DR8 \footnote{Afterwards the SDSS-telescope was refitted with new smaller (1 arcsecond radius) fibres for BOSS \citep{SDSS_DR9}.}. $\sigma_{0}$ is typically about 10\% higher than the measured value $\sigma_{\textrm{sdss}}$.

We also corrected the measured redshifts $z$ for the motion of our solar system relative to the cosmic microwave background (CMB), because we intend to calculate redshift-based distances afterwards. The corrected redshift $z_{\textrm{cor}}$ for the CMB rest frame can be calculated using some basic mathematics as demonstrated in Appendix \ref{CMB_correction}.

Since we need distances to obtain the physical radii $R_{0}$ of our elliptical galaxies, we calculated angular diameter distances $D_{A}(z_{\textrm{cor}})$: 
\begin{equation}
D_{A}(z_{\textrm{cor}})=D_{L}(z_{\textrm{cor}}) \cdot (1+{z_\textrm{cor}})^{-2} .
\label{angdist}
\end{equation}
They are derived from the luminosity distances $D_{L}$, which have been defined in Equation \ref{lumdist}. The physical radius is given in Equation \ref{realradius}: 
\begin{equation}
R_{0}=D_{A}(z_{\textrm{cor}}) \cdot \textrm{tan}\left(r_{\textrm{cor}}\right) .
\label{realradius}
\end{equation}
Another aspect needs to be considered, the passive evolution of elliptical galaxies
\begin{equation}
m_{\textrm{app}}=m'_{\textrm{app}}+Q \cdot z_{\textrm{cor}} .
\label{evolutioncorrection}
\end{equation}
The K- and extinction-corrected apparent $m'_{\textrm{app}}$ of a sample of elliptical galaxies changes as a function of look-back time due to stellar evolution. We corrected for this effect using an evolutionary parameter $Q$. This parameter was derived by fitting Equation \ref{surveylimit} to the overall redshift distribution of our sample, as done in Subsection \ref{sample_prop}. The fit assumes a passive evolution of the elliptical galaxies that is linear and proportional to the redshift within the sample's redshift range. Using this parameter, one can to calculate the fully (extinction-, K-, and evolution-) corrected apparent magnitude $m_{\textrm{app}}$. 

A final correction was applied to the measured surface brightness:
\begin{equation}
\mu_{0}=m_{\textrm{app}} + 2.5\cdot \textrm{log}_{10}\left( 2\pi \cdot r_{\textrm{cor}}^{2} \right) - 10\cdot \textrm{log}_{10} \left( 1 + z_{\textrm{cor}} \right) .
\label{surfacebrightness}
\end{equation}
The mean surface brightness $\mu_{0}$ within the effective radius $r_{\textrm{cor}}$ is defined by the equation above. The last term corrects for cosmological dimming of surface brightnesses, which is proportional to $(1+z_{\textrm{cor}})^{4}$ in any Friedmann-Lema\^{i}tre-Robertson-Walker metric-based universe \citep{Tolman:1930,Hubble:1935,Sandage:1990a,Sandage:1990b,Sandage:1991,Pahre:1996}.
\begin{equation}
\textrm{log}_{10} \left( I_{0} \right) = -\frac{\mu_{0}}{2.5}
\label{logI0}
\end{equation}
To be consistent with \citet{Bernardi:2003c}, we used $\textrm{log}_{10} \left( I_{0} \right)$ instead of the surface brightness, although they only differ by a factor. 

In the final step, we used the angular diameter distance to determine the physical radius $R_{0}$ of the galaxies in our sample.  With this, we have all parameters at hand that are required for the fundamental plane.

We now briefly discuss our options for fitting it. One first has to take into account is the Malmquist bias. There are several methods to correct for it, and first we tried to use a maximum-likelihood method to fit the fundamental plane, as done in \citet{Bernardi:2003b,Bernardi:2003c}. We found that the extrema landscape of the likelihood function of the multivariate Gaussian (see \citet{Bernardi:2003b} for more details) is unsuitable for the method, since small variations in the initial conditions of our simplex fit led to significantly different results. Other more advanced methods for minimising the likelihood function were considered, but rejected because of their unreasonably high computational costs.

\begin{figure}[ht]
\begin{center}
\includegraphics[width=0.45\textwidth]{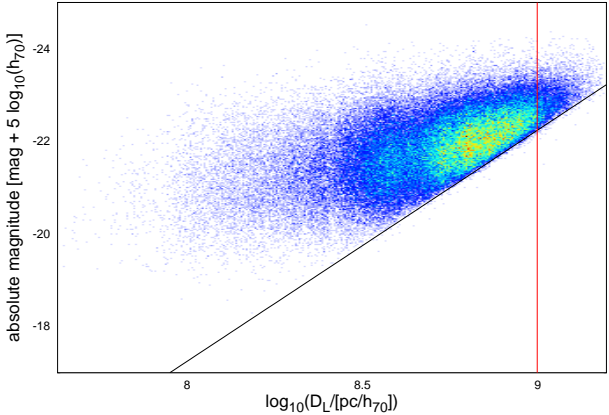}\\ 
\caption{Malmquist bias affecting the magnitude-distance distribution of our sample. Taking the example of the r band for the dV model parameters, one can see a deviation of the pure Malmquist bias (indicated by the black solid line) at large distances. The red solid line indicates a distance corresponding to a redshift of 0.2, which is the limit of our selected sample.}
\label{dV_magabs_vs_dist_in_r_full}
\end{center}
\end{figure}

As a consequence, we decided to use a less complex, yet efficient method to account for the Malmquist bias: volume-weighting \citep{Sheth:2012}. One assigns statistical weights to the galaxies based on the volume in which a galaxy with its luminosity is still visible. To do this, one has to know the exact limits imposed by the bias on one's sample: 
\begin{equation}
\textrm{log}_{10}\left(D_{L,\textrm{limit}}\right)=k_{\textrm{fit}} \cdot M_{\textrm{abs}} + d_{\textrm{fit}}
\label{linfit}
\end{equation}
\begin{equation}
m_{\textrm{limit}}=5 \cdot d_{\textrm{fit}} - 5 .
\label{limit_mag}
\end{equation}
Owing to the nature of the bias, one expects a linear cut in the sample when plotting the logarithm of the luminosity distance $D_{L}$ versus the absolute magnitude $M_{\textrm{abs}}$. The fit parameter $d_{\textrm{fit}}$ is directly connected to the limiting magnitude $m_{\textrm{limit}}$ of the sample. We can perform a simple linear fit (see Equation \ref{linfit}) to this cut, but since we know that it originates in a Malmquist bias, we are able to fix the slope $k_{\textrm{fit}}$ to $-0.2$ and only have to vary the offset $d_{\textrm{fit}}$. This is done in a way that 99.7$\%$ (equivalent to 3$\sigma$) of the data points are located on one side of the fitted line. The fit parameter $d_{\textrm{fit}}$ is directly connected to the limiting magnitude $m_{\textrm{limit}}$ of the sample (equation~\ref{limit_mag}). In the next step, we used this fit to determine the maximum distance $D_{L,\textrm{limit}}$ at which a galaxy with a certain absolute magnitude is still visible. Subsequently, we transformed this luminosity distance into a comoving distance $D_{C}$ (see Equation \ref{comdist}) for which we derived the redshift $z_{\textrm{limit}}$ from the limiting luminosity distance $D_{L,\textrm{limit}}$. An inversion of Equation \ref{lumdist} yields,
\begin{align}
&z=\frac{1}{c^{2}} \Bigg(c^{2} q_{0} - c^{2} + c\, D_{L} H_{0} q_{0} + \\
&\sqrt{c^{2} q_{0}^2 - 2 c^{4} q_{0} + c^{4} + 2 c^{3} D_{L} H_{0} q_{0}^{2} - 4 c^{3} D_{L} H_{0} q_{0} + 2 c^{3} D_{L} H_{0}}\Bigg)\nonumber ,
\label{invlumdist}
\end{align}
which can be used to derive the limiting redshift to determine the limiting comoving distance and consequently the corresponding comoving volume $V_{C}$ (see Equation \ref{comvol}), 
\begin{equation}
w_{\textrm{vol},i}=\frac{\left(V_{C,i}\right)^{-1}}{\sum\limits_{j} \left(V_{C,j}\right)^{-1}} .
\label{volweight}
\end{equation}
The normalised volume weights $w_{\textrm{vol},i}$ for every galaxy $i$ are defined by Equation \ref{volweight} according to \citet{Sheth:2012}. With the volume weights at hand, one can compute the coefficients $a$, $b$, and $c$ of the fundamental plane (see Equation \ref{fundamentalplane}) using a multiple regression based on weighted least-squares. These volume weights correct for the Malmquist bias (see Figure \ref{dV_magabs_vs_dist_in_r_full}), which affects our magnitude-limited sample. One has to solve the following set of equations:
\begin{equation}
\begin{pmatrix} A_{11}
&  A_{12}
& A_{13}\\ 
A_{12}  
&  A_{22}
&  A_{23}  \\ 
A_{13}
& A_{23}
& A_{33} 
\end{pmatrix} \cdot \begin{pmatrix} a  \\ b  \\ c   \end{pmatrix} = \begin{pmatrix} 
 V_{1}\\ 
 V_{2}\\ 
 V_{3}
 \end{pmatrix} ,
\label{leastsquare}
\end{equation}
with
\begin{align}
A_{11} =& \sum\limits_{i=1}^{n}  \left( \left( \textrm{log}_{10}(\sigma_{0,i}) \right)^{2} \cdot \bar{w}_{\textrm{vol},i} \right)\\
A_{12} =& \sum\limits_{i=1}^{n}  \left( \textrm{log}_{10}(\sigma_{0,i} \cdot \textrm{log}_{10}(I_{0,i}) \cdot \bar{w}_{\textrm{vol},i} \right)\\
A_{13} =& \sum\limits_{i=1}^{n}  \left(  \textrm{log}_{10}(\sigma_{0,i})  \cdot \bar{w}_{\textrm{vol},i} \right) \\
A_{22} =& \sum\limits_{i=1}^{n}  \left( \left( \textrm{log}_{10}(I_{0,i}) \right)^{2} \cdot \bar{w}_{\textrm{vol},i} \right) \\
A_{23} =& \sum\limits_{i=1}^{n}  \left(  \textrm{log}_{10}(I_{0,i})  \cdot \bar{w}_{\textrm{vol},i} \right) \\
A_{33} =& n\\
V_{1} =& \sum\limits_{i=1}^{n} \left(\textrm{log}_{10}(R_{0,i})\cdot \textrm{log}_{10}(\sigma_{0,i}) \cdot \bar{w}_{\textrm{vol},i} \right)\\
V_{2} =& \sum\limits_{i=1}^{n} \left(\textrm{log}_{10}(R_{0,i})\cdot \textrm{log}_{10}(I_{0,i}) \cdot \bar{w}_{\textrm{vol},i} \right)\\
V_{3} =& \sum\limits_{i=1}^{n} \left(\textrm{log}_{10}(R_{0,i})\cdot \bar{w}_{\textrm{vol},i} \right) ,
\end{align}
which is done using Cramer's rule. It should be noted that $\bar{w}_{\textrm{vol},i} = w_{\textrm{vol},i} \cdot n $ are renormalized volume weights that were only multiplied by the number of galaxies $n$ used for the fit.
\begin{equation}
s_{\varepsilon} = \sqrt{\frac{ \sum\limits_{i=1}^{n} \bar{w}_{\textrm{vol},i} \left(a \,\, \textrm{log}_{10}\left(\sigma_{0,i}\right) + b \,\, \textrm{log}_{10}\left(I_{0,i}\right) + c - \textrm{log}_{10}\left(R_{0,i}\right) \right)^{2} }{n}}
\label{rms_fp}
\end{equation}
\begin{align}
\sigma_{a}=&\sqrt{\frac{A_{22} A_{33}-(A_{23})^{2}}{\textrm{det}(A)}}\\
\sigma_{b}=&\sqrt{\frac{A_{11} A_{33}-(A_{13})^{2}}{\textrm{det}(A)}}\\
\sigma_{c}=&\sqrt{\frac{A_{11} A_{22}-(A_{12})^{2}}{\textrm{det}(A)}}
\label{err_coeff}
\end{align}
We also computed the root mean square $s_{\varepsilon}$ and the standard errors $\sigma_{a}$, $\sigma_{b}$, and $\sigma_{c}$ of the coefficients $a$, $b$, and $c$, where $A$ denotes the matrix from Equation \ref{leastsquare}. 

We performed an iterative $3\sigma$-clipping after the fitting process, which was repeated until all outliers were eliminated. With the entire set of calibration and tools at hand, we then determined the coefficients of the fundamental plane. 
\section{Results}
\label{section_results}
For the photometric parameters of our model galaxies we used the three available sets of models in SDSS: the c model, the dV model, and the p model. The c model uses \emph{cModelMag}\footnote{Actually \emph{PhotoObj.cModelMag\_filter}, to be consistent with Subsection \ref{sample_def}, but we use this short notation and similar abbreviations here.} and \emph{deVRad} (since SDSS does not provide a composite scale radius fit). The dV model uses \emph{deVMag} and \emph{deVRad}. The p model uses \emph{petroMag} and \emph{petroRad}, and all filters independently.

The following equations define the various parameters:
\begin{equation}
I(r)= I_{0} \cdot \textrm{exp}\left(-7.67 \left(\frac{r}{r_{e}}\right)^\frac{1}{4} \right)
\label{deVaucouleur}
\end{equation}
\begin{equation}
I(r)= I_{0} \cdot \textrm{exp}\left(-1.68 \frac{r}{r_{e}} \right)
\label{exponential}
\end{equation}
\begin{equation}
\mathcal{R}_{P}(r)=\frac{\int_{0.8 r}^{1.25 r} d\bar{r} \,\, \bar{r}\cdot I(\bar{r}) }{(1.25^{2}-0.8^{2})\cdot \int_{0}^{r} d\bar{r} \,\, \bar{r}\cdot I(\bar{r})}
\label{PetrosianRatio}
\end{equation}
\begin{equation}
F_{P}=\int_{0}^{N_{P} \cdot r_{P}} d\bar{r}\,\, 2\pi \cdot \bar{r}\cdot I(\bar{r}) .
\label{PetrosianFlux}
\end{equation}
The \emph{cModelMag} are based on a simple weighted adding (depending on the likelihood of the two fits) of the fluxes from de Vaucouleurs fits (see Equation \ref{deVaucouleur}) and the pure exponential fit (see Equation \ref{exponential}). The \emph{deVMag} are the magnitudes derived directly from the de Vaucouleurs fit given in Equation \ref{deVaucouleur}. The Petrosian magnitudes \emph{petroMag} are slightly more complicated. Firstly, one has to calculate the Petrosian ratio $\mathcal{R}_{P}(r)$ according to Equation \ref{PetrosianRatio}, where $I(r)$ stands for the azimuthally averaged surface brightness profile. The Petrosian radius $r_{P}$, which is denoted with \emph{petroRad}, is the radius for which the Petrosian ratio $\mathcal{R}_{P}(r_{p})$ is equal to a defined value (0.2 for the SDSS). The Petrosian flux $F_{P}$ is given by Equation \ref{PetrosianFlux}, where the parameter $N_{P}$ is defined to be 2 (for the SDSS). SDSS also provides \emph{fibreMag}, but since they are by definition calculated for fixed apertures (diameter of the SDSS-fibre), we found them not to be useful for studying a sample of galaxies at different distances. Therefore, we did not construct a model based on them. 

\subsection{Malmquist bias}
\label{sec:Malmquist}

The first stop on our way to proper results is quantifying the Malmquist bias in our sample (see Figure \ref{dV_magabs_vs_dist_in_r_full}). Owing to the physical nature of this bias, it is best to plot the logarithm of the luminosity distance $D_{L}$ versus the absolute magnitude for the sample. Then one fits a straight line to the cut, which is introduced by the Malmquist bias, in the distribution (see Figures \ref{c_magabs_vs_dist_in_u} to \ref{p_magabs_vs_dist_in_z}). This fit is given by Equation \ref{linfit}, and its result is listed in Table \ref{malmquist_fit_maglimit}. Since the parameter $d_{\textrm{fit}}$ is directly connected to the limiting magnitude $m_{\textrm{limit}}$, which has a higher physical significance than the fit parameter itself, by Equation \ref{limit_mag}, we display this limiting magnitude in Table \ref{malmquist_fit_maglimit}. It is assuring to see that the limiting magnitude is almost independent of the model and only depends on the filter. Because spectroscopic data are required for every galaxy in our sample, the limiting magnitudes from Table \ref{limit_mag} are driven by the spectroscopic limit of SDSS, which is the (uncorrected) Petrosian magnitude in the r band of 17.77 \citep{SDSS_spectarget}. This value is almost the same limiting magnitude for the same model and filter, as we found. It is not surprising that the limiting magnitude is fainter in the bluer filter than in the redder ones, since elliptical galaxies are more luminous in the red, as one can see in Subsection \ref{res_lumfct}.

The fit results shown in Table \ref{malmquist_fit_maglimit} were used to calculate the volume weights (see Equation \ref{volweight}) to correct for the Malmquist bias in our analysis. We found that our sample is affected by an additional bias for redshifts z$\gtrsim 0.2$, which is consistent with our previous findings in Subsection \ref{sample_prop} (especially Figures \ref{volumebinned} and \ref{overdense}): when we extend our plots beyond the luminosity distance corresponding to a redshift of 0.2, there is slight shortage of galaxies just above the fitted line (see Figure \ref{dV_magabs_vs_dist_in_r_full}). This happens for all filters and all models, which is another motivation for removing galaxies above redshift of 0.2 from our sample. A useful review on the Malmquist bias in general can be found in \citet{Butkevich:2005}.

\begin{table}
\begin{center}
\begin{tabular}{c|ccccc}
 models & $m_{\textrm{limit,u}}$ & $m_{\textrm{limit,g}}$ & $m_{\textrm{limit,r}}$ & $m_{\textrm{limit,i}}$ & $m_{\textrm{limit,z}}$ \\ \hline
c model & 20.61 & 18.54 & 17.78 & 17.46 & 17.22\\
dV model & 20.59 & 18.54 & 17.78 & 17.46 & 17.22 \\
p model & 20.77 & 18.59 & 17.80 & 17.47 & 17.23
\end{tabular}
\end{center}
\caption{Limiting magnitude $m_{\textrm{limit,filter}}$, which was derived from the coefficients $d_{\textrm{fit,filter}}$ of the fit on the Malmquist bias according to Equation \ref{limit_mag}, for every filter and for every model.}
\label{malmquist_fit_maglimit}
\end{table}

\subsection{Luminosity function}

\label{res_lumfct}
\begin{figure}[ht]
\begin{center}
\includegraphics[width=0.45\textwidth]{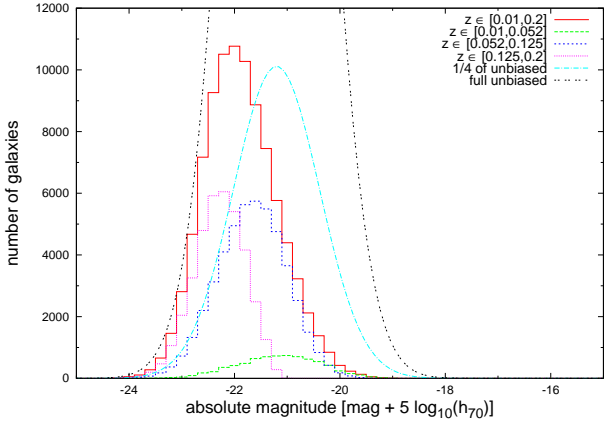}\\ 
\caption{Luminosity function for our sample in the r band for the dV model. We split it into different subsamples (within different redshift bins) and compared the expected unbiased luminosity function and the total observed luminosity function. Our sample is complete at the luminous end, but we miss many of the fainter galaxies due to the Malmquist bias.}
\label{dV_magabs_binned_r}
\end{center}
\end{figure}

Additional information obtained from the preparations of the calibration of the fundamental plane are the luminosity functions of the galaxies in our sample. We note that the faint-luminosity limit of our sample of $M_r = (-18.66$ + $5\,\, \textrm{log}_{10}(h_{70}))$ mag (corresponding to a 3$\sigma$ of the mean luminosity of the sample's galaxies in this filter, see Table \ref{mean_lumiosity}) is brighter than the apparent spectroscopic limit of the SDSS, which is 17.77 Petrosian magnitudes in the r band \citep{SDSS_spectarget}, at the lower redshift limit z=0.01 (distance modulus $m_{\textrm{app}}-M_{\textrm{abs}}$ = (33.16 - $5\,\, \textrm{log}_{10}(h_{70})$) mag). The reason for this is the overall surface brightness limit of SDSS (omitting dwarf galaxies), and the restriction of our sample to galaxies that are better fit by a de Vaucouleurs profile than an exponential. The former profile is characteristic of giant ellipticals, whereas an exponential profile corresponds to fainter, mostly dwarf, galaxies. We hereby conclude that our sample is not contaminated by dwarf ellipticals. 

We calculated the absolute magnitudes of the galaxies using the distance modulus (see Equation \ref{distmod}) and redshift-based distances (see Equation \ref{lumdist}). Since the sample is affected by a Malmquist bias, the luminosity function in different redshift bins is not the same, but shifted to higher luminosity with higher redshift. To analyse the luminosity function, we split the data into 0.2 mag bins. We show the resulting luminosity distributions in Figure \ref{dV_magabs_binned_r} here as well as in Figures \ref{c_magabs_binned_u} to \ref{p_magabs_binned_z}. 

We also calculated the mean luminosity and its standard deviation for every model and filter. We used these values and the mean galaxy density $\rho_{0}$ obtained in Subsection \ref{sample_prop} to calculate the expected luminosity function assuming that our sample would not suffer from a Malmquist bias (see Figures \ref{dV_magabs_binned_r} and \ref{c_magabs_binned_u} to \ref{p_magabs_binned_z}). In this case, we would have almost 416000 galaxies in our selected sample (between a redshift of 0.01 and 0.2), compared to the about 95 000 that we actually found. The magnitude limitation of the SDSS data set thus reduced the number of galaxies by about 80\% for the full redshift range $0.01<z<0.2$, compared to an extrapolated volume limited sample. With the volume weights, we then calculated a bias-corrected mean absolute magnitude $\bar{M}_{\textrm{filter}}$ and a standard deviation $\sigma_{\textrm{filter}}$ for every filter of every model. The results are listed in Table \ref{mean_lumiosity}. The standard deviations are very similar between all filters and models. The mean absolute magnitude also depends almost entirely on the filter. This again shows that the completeness is entirely constrained by the spectroscopic survey limit, not by photometric limits. 

\begin{table*}
\begin{center}
\begin{tabular}{c|cc|cc|cc|cc|cc}
models & $\bar{M}_{u} $ & $\sigma_{u}$ & $\bar{M}_{g}$  & $\sigma_{g}$  & $\bar{M}_{r}$  & $\sigma_{r}$ & $\bar{M}_{i}$  & $\sigma_{i}$  & $\bar{M}_{z}$  & $\sigma_{z}$ \\ 
& [mag + & [mag] & [mag +   & [mag]  & [mag +   & [mag] & [mag +   & [mag]  & [mag +   & [mag] \\ 
&  $5\,\, \textrm{log}_{10}(h_{70})$] &  &  $5\,\, \textrm{log}_{10}(h_{70})$]  &   &  $5\,\, \textrm{log}_{10}(h_{70})$]  &  & $5\,\, \textrm{log}_{10}(h_{70})$]  &   &  $5\,\, \textrm{log}_{10}(h_{70})$]  &  \\ 
 \hline
 c model & -18.82 & 0.79 & -20.47 & 0.80 & -21.20 & 0.82 & -21.55 & 0.82 & -21.78 & 0.83 \\
dV model & -18.84 & 0.80 & -20.48 & 0.80 & -21.20 & 0.82 & -21.55 & 0.82 & -21.79 & 0.83 \\
 p model & -18.55 & 1.04 & -20.38 & 0.80 & -21.12 & 0.82 & -21.48 & 0.82 & -21.74 & 0.83
\end{tabular}
\end{center}
\caption{Bias- and evolution-corrected absolute magnitudes $\bar{M}_{\textrm{filter}}$ and the corresponding standard deviation $\sigma_{\textrm{filter}}$ can be found in this table for all models and all filters. }
\label{mean_lumiosity}
\end{table*}

\subsection{Parameter distribution}
In this subsection, we discuss the properties of the different parameters that define the fundamental plane and their observables. The parameters of the fundamental plane are derived from three observables: the apparent corrected radius $r_{\textrm{cor}}$, the extinction- and K-corrected apparent magnitude $m_{\textrm{app}}$, and the central velocity dispersion $\sigma_{0}$, which is already corrected for the fixed aperture size of the SDSS fibres. The three parameters of the fundamental plane are the logarithm of the physical radius $\textrm{log}_{10}(R_{0})$, the logarithm of the central velocity dispersion $\textrm{log}_{10}(\sigma_{0})$, and the mean surface brightness $\mu_{0}$ in lieu of which as a convention, the parameter $\textrm{log}_{10}(I_{0})$ is used in the fitting process, which only differs by a factor of $-2.5$.

\begin{figure}[ht]
\begin{center}
\includegraphics[width=0.45\textwidth]{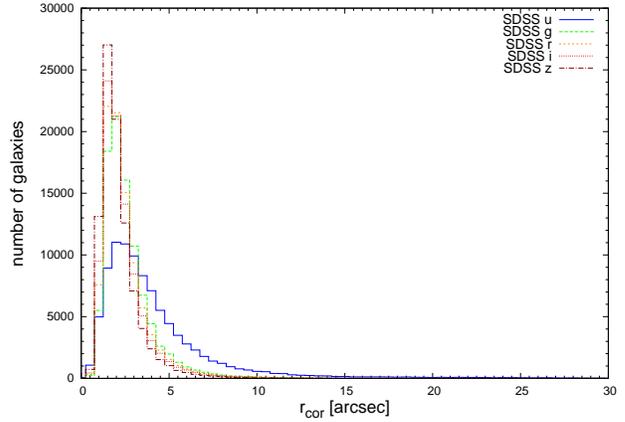}\\ 
\caption{Distribution of the apparent corrected radius $r_{\textrm{cor}}$ is displayed in different filters for the dV model and the c model (the radii are the same is both models). The vast majority of the measured radii are smaller than 5 arcseconds.}
\label{c_rcor_binned}
\end{center}
\end{figure}

The distribution of the radii $r_{\textrm{cor}}$ is shown in Figures \ref{c_rcor_binned} and \ref{p_rcor_binned}. The apparent radius is typically in the order of couple of arcseconds (with its peak about 1.5-2 arcseconds for the c and dV models and aroung 3-4 arcseconds for the p model), but in the case of the SDSS u band, they can be moved to much larger radii (especially for the Petrosian model) because of the known problems with this filter (see the SDSS website\footnote{\url{http://www.sdss.org/dr7/start/aboutdr7.html#imcaveat}}). It has been reported in \citet{Fathi:2010} that the SDSS fitting algorithm tends to prefer certain sets of values for de Vaucouleurs and exponential fits. We found that this happens for the Petrosian fits as well and that it even is much more prominent there, especially in the u and z band. One can already see some grouping in the plots of the corrected apparent radius $r_{\textrm{cor}}$ against the apparent magnitudes $m_{\textrm{app}}$ (see Figures \ref{p_r_cor_vs_m_app_u} and \ref{p_r_cor_vs_m_app_z} and compare them with Figure \ref{p_r_cor_vs_m_app_r}, which is for the r band and does not show any peculiarities). It becomes more prominent in plots of the $\textrm{log}_{10}(R_{0})$ versus redshift, in which one can clearly see band-like structures (see Figures \ref{p_z_dependence_logR0_u} and \ref{p_z_dependence_logR0_z}). Furthermore, the average apparent radii in all filters are larger for the Petrosian model than for the de Vaucouleurs model (the c model also uses de Vaucouleurs radii). There are some tiny differences (too small to be seen in a plot) between the de Vaucouleurs model and the composite model, which are created by the selection of galaxies (because of limits in the magnitudes and the 3-$\sigma$ clipping).

In Table \ref{additional_parameters}, the averages and standard deviations of all previously mentioned parameters are displayed for all filters and all models. For the p model, there is a suspiciously high standard deviation for the u band and for the z band though to a smaller extent there. There is obviously a problem with the measured radii in the u band, which is most likely due to the known problem of scattered light in this filter, and it is much worse for Petrosian fits, for which the z band is also affected. The distribution of the apparent magnitude $m_{\textrm{app}}$ for different models is displayed in Figures \ref{c_mag_app_binned} to \ref{p_mag_app_binned}. The distributions show very steep cut-offs around the limiting magnitudes, which is exactly the expected behaviour for a magnitude-limited sample. 

In Figures \ref{c_sigma_cor_binned} and \ref{p_sigma_cor_binned}, one can see that the corrected central velocity dispersion barely depends on the filter. This is not surprising since the correction (see Equation \ref{sigmacor}) only mildly depends on the apparent corrected radius $r_{\textrm{cor}}$, which is different for different filters. Therefore, as a spectroscopic observable, one will not notice any significant differences depending on the model either.

\begin{figure}[ht]
\begin{center}
\includegraphics[width=0.45\textwidth]{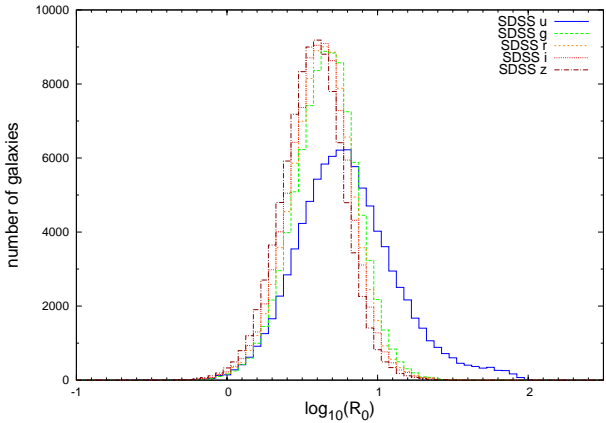}\\ 
\caption{Distributions of the logarithm of the physical radius $\textrm{log}_{10}(R_{0})$ in different filters for the dV model can be well described by sharp Gaussian with their peaks almost exactly at the same value. Only the u band shows some digressive behaviour. In this case the peak is smaller and set apart from the other. Furthermore, the distribution is wider and it shows a small bump at the larger end.}
\label{dV_logR0_binned}
\end{center}
\end{figure}

The three parameters $\textrm{log}_{10}(R_{0})$, $\textrm{log}_{10}(\sigma_{0})$, and $\textrm{log}_{10}(I_{0})$ enter directly into the fit of the fundamental plane. Therefore, the distribution of these parameter is especially important. The distribution of the logarithm of the physical radius $R_{0}$ is shown in Figures \ref{dV_logR0_binned}, \ref{c_logR0_binned}, and \ref{p_logR0_binned}. They show for almost all filters in all models sharp Gaussians with their peaks very close together for almost all filters and with comparable standard deviations (see Table \ref{additional_parameters} for details). Nevertheless, the problems in u band, which we already encountered for the apparent corrected radius $r_{\textrm{cor}}$, are propagated and are even more striking here. In the case of the c model (Figure \ref{c_logR0_binned}) and the dV model (Figure \ref{dV_logR0_binned}), the distribution of $\textrm{log}_{10}(R_{0})$ is widened in the u band compared with the other filters. Moreover, its peak is shifted and the distribution shows a clear deviation from a Gaussian shape at its high end. For the p model (Figure \ref{p_logR0_binned}), one can clearly see a two peaked distribution for the u band and also for the z band to some smaller extent. For these particular models and filters, one has to introduce a cut (or another method) to handle the second peak during the least-squares fitting to avoid unwanted offsets. 

\begin{figure}[ht]
\begin{center}
\includegraphics[width=0.45\textwidth]{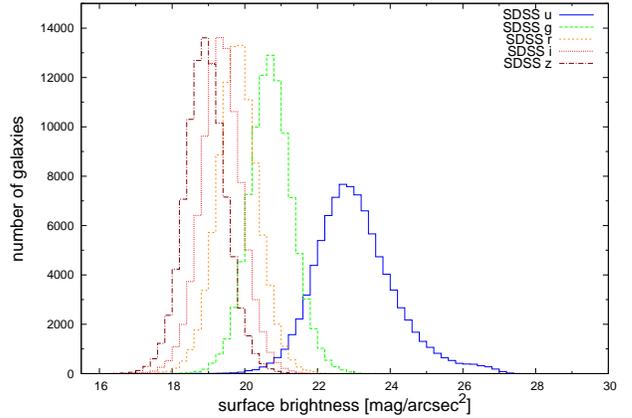}\\ 
\caption{Distribution of the surface brightness $\mu_{0}$ in different filters for the dV model is close to a Gaussian shape. For the u band, the distribution is wider and shows a small bump at the faint end.}
\label{dV_sb_binned}
\end{center}
\end{figure}

Since the mean surface brightnesses $\mu_{0}$ are derived from $r_{\textrm{cor}}$ (see Equation \ref{surfacebrightness}), one expects similar problems from them. Indeed, there are similar sharp Gaussians for all filters, but for the u for all models and for the p model, one may notice a double-peaked distribution for the z band as well (see Figures \ref{dV_sb_binned}, \ref{c_sb_binned} and \ref{p_sb_binned}). 

\begin{figure}[ht]
\begin{center}
\includegraphics[width=0.45\textwidth]{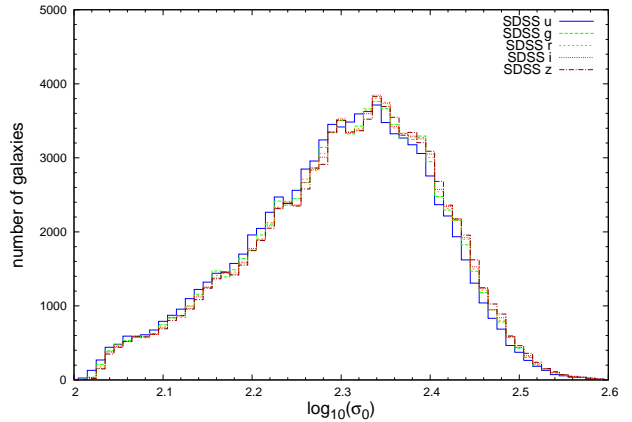}\\ 
\caption{Distributions of the logarithm of the central velocity dispersion $\textrm{log}_{10}(\sigma_{0})$ in different filters for the c model and the dV model are almost exactly the same for all filters.}
\label{c_logsigma_binned}
\end{center}
\end{figure}

The logarithm of the central velocity dispersion $\sigma_{0}$ shows the same behaviour as the central velocity dispersion itself, as one can see in Figures \ref{c_logsigma_binned} and \ref{p_logsigma_binned}. The shape of the distribution of $\textrm{log}_{10}(\sigma_{0})$ is not a perfect Gaussian, in contrast to the distributions of the previous parameters, but this does not have to be the case for the velocity dispersion distribution. Yet, the distributional shape is sufficiently regular (no double peaks or other strange features) to be used in a least-squares fit without concerns.

\subsection{Coefficients of the fundamental plane}
\begin{table*}
\begin{center}
\begin{tabular}{c|ccccc}
 models and filters & $a$ & $b$ & $c$ & $s_{\varepsilon}$ & $\bar{\sigma}_{\textrm{dist}}$ [\%] \\ \hline \hline
  c model &  &  &  & \\ \hline
 u & $  0.809 \pm  0.030 $  & $ -0.696 \pm   0.008 $  & $  -7.53 \pm   0.10 $ & $  0.0947 $ & $   16.5  $\\
 g & $  0.975  \pm   0.030 $  & $ -0.736 \pm  0.013 $  & $  -7.73\pm   0.13  $ & $  0.0933 $ & $   15.6 $\\
 r & $   1.041  \pm   0.030 $  & $ -0.750 \pm  0.013 $  & $  -7.76 \pm   0.13  $ & $  0.0933 $ & $   15.3 $\\
 i & $   1.068 \pm   0.030 $  & $ -0.755  \pm   0.013 $  & $  -7.75 \pm   0.13 $ & $  0.0918 $ & $   15.0 $\\
 z & $   1.113 \pm   0.030 $  & $ -0.760 \pm   0.013 $  & $  -7.80 \pm   0.13 $ & $  0.0947 $ & $   14.8 $\\
 \hline \hline
  dV model &  &  &  & \\ \hline
 u & $  0.798 \pm   0.030 $  & $ -0.700  \pm   0.008 $  & $  -7.53 \pm   0.10 $ & $  0.0941 $ & $   16.5   $\\
 g & $  0.966\pm   0.030$  & $ -0.740       \pm   0.013 $  & $  -7.75 \pm   0.13 $ & $  0.0934 $ & $   15.6  $\\
 r & $   1.034  \pm   0.030 $  & $ -0.753 \pm   0.013 $  & $  -7.77 \pm   0.13   $ & $  0.0933 $ & $   15.3 $\\
 i & $   1.062   \pm   0.030 $  & $ -0.757   \pm   0.013 $  & $  -7.75 \pm   0.13  $ & $  0.0918 $ & $   15.0     $\\
 z & $   1.108   \pm   0.030 $  & $ -0.763 \pm   0.013 $  & $  -7.81 \pm   0.13 $ & $  0.0941 $ & $   14.8$\\
 \hline \hline
  p model &  &  &  & \\ \hline
 u & $  0.852    \pm   0.030 $  & $ -0.550  \pm 0.005 $  & $  -6.36\pm   0.08 $ & $  0.1098 $ & $   19.4      $\\
 g & $  0.987 \pm   0.030 $  & $ -0.697  \pm   0.013 $  & $  -7.58  \pm   0.13 $ & $  0.0970 $ & $   16.5 $\\
 r & $   1.055  \pm  0.030 $  & $ -0.718  \pm   0.013 $  & $  -7.69  \pm   0.14 $ & $  0.0956 $ & $   16.1  $\\
 i & $   1.080    \pm   0.030 $  & $ -0.711  \pm  0.013 $  & $  -7.58  \pm   0.13$ & $  0.0946 $ & $   15.9    $\\
 z & $   1.106 \pm  0.030 $  & $ -0.638  \pm   0.010 $  & $  -6.96 \pm   0.11 $ & $  0.1098 $ & $   16.7 $\\
 \hline
 u (cut) & $  0.849 \pm   0.034 $  & $ -0.539 \pm   0.009 $  & $  -6.25 \pm   0.12 $ & $  0.1110$ & $   19.5  $\\
 z (cut) & $   1.126 \pm  0.031 $  & $ -0.688  \pm   0.012 $  & $  -7.41  \pm   0.13 $ & $  0.1102 $ & $   16.0 $
\end{tabular}
\end{center}
\caption{The results of the best fits for the fundamental-plane coefficients in all filters and for all models using redshift evolution, volume weights, and 3-$\sigma$ clipping.}
\label{fitparameters}
\end{table*}

We performed a volume-weighted least-squares fit in three dimensions (see Equation \ref{leastsquare}) to obtain the coefficients $a$, $b$, and $c$ of the fundamental plane. The results for all parameters, filters, and models, are shown in Table~\ref{fitparameters}. By definition (see Equation \ref{fundamentalplane}), the fundamental plane relates the logarithm of the physical radius $\textrm{log}_{10}(R_{0})$, the logarithm of the central velocity dispersion $\textrm{log}_{10}(\sigma_{0})$, and the renormalized surface brightness $\textrm{log}_{10}(I_{0})$ (see Equation \ref{logI0} for the relation between $I_{0}$ and the surface brightness $\mu_{0}$). We determined its coefficients and their standard errors as well as the root mean square $s_{\varepsilon}$. Furthermore, we determined an upper limit $\bar{\sigma}_{\textrm{dist}}$ for the average distance error $\sigma_{\textrm{dist}}$ by comparing the distances obtained with the fundamental plane to the redshift-based calibration distances. 

The error obtained by this comparison is a combination of the true average distance error $\sigma_{\textrm{dist}}$, a scatter afflicted by peculiar motions, and the finite measurement precision. Consequently, $\bar{\sigma}_{\textrm{dist}}$ is an upper limit to the average distance error, with the true distance error expected to be up to a few percent lower. To estimate the contribution of peculiar motions to the average distance error, we made use of additional data. The catalogue of \citet{Tempel:2012} provides redshifts to galaxies in groups and clusters, which are corrected for the Finger-of-God effect. We picked a subsample that overlaps with our sample and in which every (elliptical) galaxy that we used, is in a group of at least 20 members to have a solid corrected redshift. By comparing the average distance errors of this subsample of 5013 galaxies, once using the redshifts of \citet{Tempel:2012} and once the ones from SDSS, we noticed that there is no difference in the relevant digits between the fits using the Finger-of-God corrected redshifts and those from SDSS. This agrees with a simple estimate one can make using the mean redshift of our entire sample, which corresponds to a velocity of about 34000 km/s. The typical peculiar velocities of galaxies are on the order of 400 km/s \citep{Masters:2006}. The average scatter on the sample inflicted due to peculiar motions is on the order of 1\%. Using the propagation of uncertainty, this 1\% does not significantly contribute to the overall $\sim 15$\% error in the distance measurement. 

For the fitting we used a recursive 3-$\sigma$ clipping to optimise the results. This reduced the number of galaxies in our selected sample by about 2000 galaxies to 92994 for the c Model, to 92953 for the dV Model and to 92801 for the p model. Because of the problem with double peak in the distribution of the physical radii in the u and z band for the p model, we refitted the coefficients using a cut at $\textrm{log}_{10}(R_{0})$ of $1.5$. This value corresponds to a physical radius $R_{0}$ of slightly more than $30$ kpc $h_{70}^{-1}$ and is motivated by the bimodal distribution in Figure \ref{p_logR0_binned}. The results of these refits are given in Table \ref{fitparameters}. As a consequence of this cut, we only used 73914 galaxies for the u band and 91187 galaxies for the z band.

\begin{figure}[ht]
\begin{center} 
\includegraphics[width=0.45\textwidth]{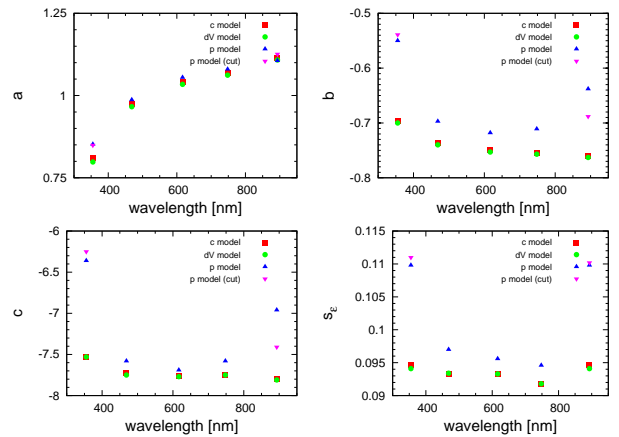}\\ 
\caption{Comparison of the coefficients $a$ (upper left panel), $b$ (upper right panel), and $c$ (lower left panel) and the root mean square $s_{\varepsilon}$ (lower right panel) of the fundamental plane for all models and wavelengths. There are two additional data points for the p model, because we included the refitted coefficients after a cut in $\textrm{log}_{10}(R_{0})$ to remove wrong data points caused by problems discussed earlier in this paper. In general, the behaviour of the coefficients is similar for all models and only depends on the wavelength (notable exception the z band of the p model).}
\label{compare_parameters}
\end{center}
\end{figure}

\begin{figure}[ht]
\begin{center}
\includegraphics[width=0.45\textwidth]{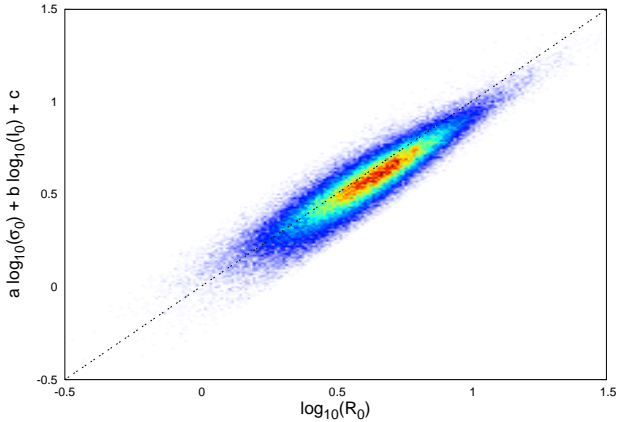}\\ 
\caption{Projection of the fundamental plane for the i band of the dV model. The root mean square is smallest for this particular filter and model.}
\label{dV_fp_i_poster}
\end{center}
\end{figure}

\begin{figure}[ht]
\begin{center}
\includegraphics[width=0.45\textwidth]{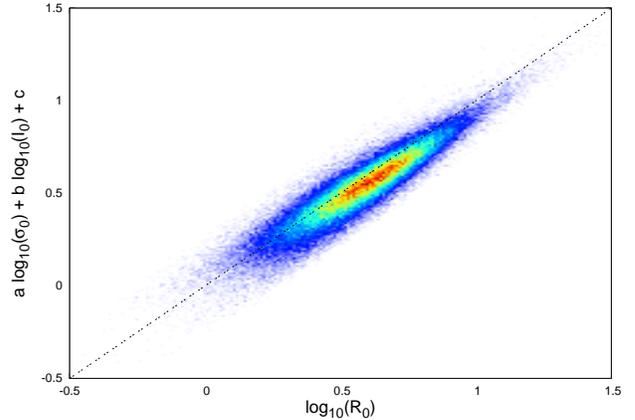}\\ 
\caption{Projection of the fundamental plane for the z band of the dV model. The averaged distance error is smallest for this particular filter and model.}
\label{dV_fp_z_poster}
\end{center}
\end{figure}

A general comparison of the fit parameters and the distance error in Table \ref{fitparameters} shows that the c model and the dV model are clearly better than the p model. Moreover, the upper limit of the average distance error $\bar{\sigma}_{\textrm{dist}}$ is almost exactly the same for the c model and the dV model, and so are the fitting parameters $a$, $b$, and $c$. We found only small differences beyond the relevant digits. Furthermore, we found that though the root mean square $s_{\varepsilon}$ is smallest in the i band, the average distance error decreases with longer wavelengths and is smallest in the z band. In addition to that, the coefficients of the fundamental plane show clear tendencies correlated with the wavelength. The coefficient $a$ increases with long wavelengths, while the other coefficients $b$ and $c$ in general decrease with longer wavelengths. These dependencies are illustrated for all models in Figure \ref{compare_parameters} and hold quite well, save for the two problematic filters u and z in the p model. Figures \ref{dV_fp_i_poster}, \ref{dV_fp_z_poster}, and \ref{c_fp_u_poster} to \ref{p_fp_z_poster} show projections of the fundamental plane for all filters and all models. In Figure \ref{p_fp_u_poster}, one can see a split distribution of two clouds, which is also due to the problems with the p model and the u filter as well. A similar, but smaller, problem occurs for the z filter of the p model (see Figure \ref{p_fp_z_poster}), too. 


\section{Discussion}

\subsection{Comparison with alternative fits}

\begin{figure*}[ht]
\begin{center}
\includegraphics[width=0.90\textwidth]{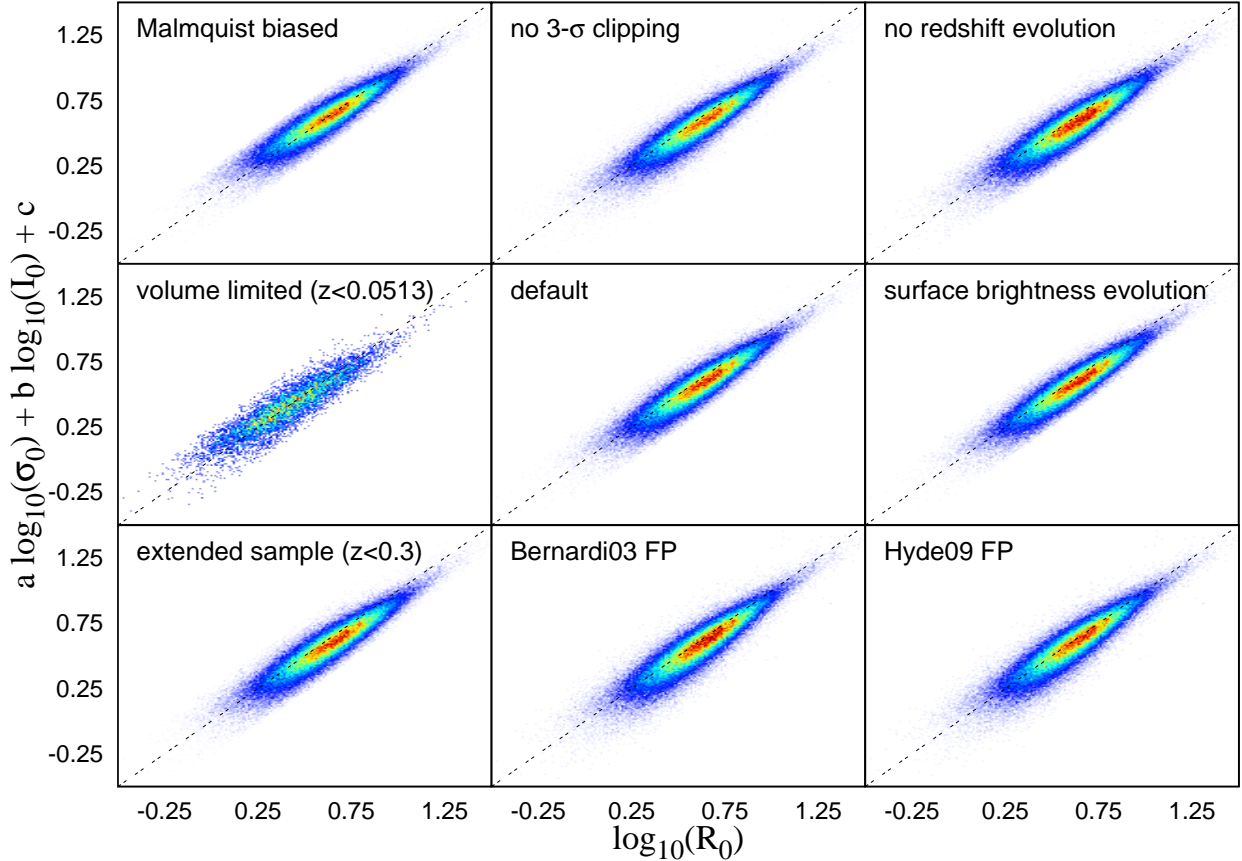}\\ 
\caption{Results for the fundamental plane in the i band for the dV model using our alternatives fits. The plot in the top-left panel does not include the Malmquist bias. We did not perform a 3-$\sigma$ clipping for the plot in the top-middle panel. The plot in the top-right panel excludes the redshift evolution. The results of the volume-limited sample ($z<0.0513$) can be found in the central-left panel. The central-middle panel contains a plot of the default i band fit for the dV model for comparison. We are considering the surface brightness evolution instead of the redshift evolution derived form galaxy number densities in the central-right panel. In the bottom-left panel, the results are shown for an extended sample up to $z=0.3$. The fundamental plane plotted using the coefficients of \citet{Bernardi:2003c}, but with our sample data is displayed in the bottom-middle panel. A similar plot using the coefficients of \citet{Hyde:2009} can be found in the bottom-right panel.}
\label{compare_fp_i}
\end{center}
\end{figure*}

In addition to the main fit (for the resulting coefficients see Table \ref{fitparameters}), which considered redshift evolution and made use of volume weights to correct for the Malmquist bias and a recursive 3-$\sigma$ clipping, we performed additional fits to test features of the code and assumptions we made. A visual comparison of the different fundamental plane fits of the i band for the dV model is shown in Figure \ref{compare_fp_i}.

We provide in Table \ref{fitparameters_noclipping} the fitting results obtained without a 3-$\sigma$ clipping. The change in values of the coefficient is marginal, and the quality of the fit (not surprisingly) is a little poorer than with clipping. Removing outliers is important for the calibrations, because we do not want our coefficients to be influenced by them. 

We also considered the case without corrections for the Malmquist bias: Table \ref{fitparameters_novolumeweight} shows the results of the fitting process with the volume weights turned off. Although the root mean square decreases and the fundamental plane appears to be narrower, its quality as a distance indicator clearly decreases when not correcting for the Malmquist bias. It can be seen in Figure \ref{compare_fp_i} that the fit of the fundamental plane lies directly in the centre of the cloud of data points, if the Malmquist bias is not corrected for by volume weights, while in all other cases (except in the volume-limited sample, which is not affected by a Malmquist bias by definition) the best fit is always slightly above the centre of the cloud. 

To test the Malmquist-bias correction further, we calculated the fundamental-plane coefficients for a volume-limited subsample that does not require any corrections. Making use of Equation \ref{surveylimit} and the results of the corresponding fit, we calculated the redshift distance for which our sample is still to $\sim$95.45\% complete (corresponding to 2-$\sigma$). This is the case up to a redshift of $z=0.0513$, which significantly reduces the number of galaxies in the sample. The c model contains only 7259 galaxies after fitting the fundamental-plane coefficients, the dV model only 7257, and the p model only 7267. We note that the average distance error is by about 2.5 percentage point larger than for the main fit. Although the best fit is going through the centre of the cloud of data points in the same way as for the Malmquist-biased fit (see Figure \ref{compare_fp_i}), the coefficients (see Table \ref{fitparameters_volumelimited}) are clearly less tilted than those of the Malmquist biased fit (see Table \ref{fitparameters_novolumeweight}).  In fact, the coefficients of the volume-limited sample are relatively close (though in general slightly less tilted) than those of the main fit using the Malmquist bias correction (see Table \ref{fitparameters}). This correspondence suggests that the Malmquist bias is handled well by the correction, and therefore our magnitude-limited sample can be used with the correction like a volume-limited sample. 

In contrast to the smaller volume-limited sample, we extended our sample to a redshift of $0.3$, despite the additional bias beyond $z=0.2$, which we found in Subsection \ref{sec:Malmquist}. We used 97341 galaxies for the c model, 97309 for the dV model, and 97050 for the p model. The results of the fit are listed in Table \ref{fitparameters_z03}. We found that the quality of this fit is only marginally poorer than of the main fit (sometimes beyond the relevant digits). However, since the number of additional galaxy in the redshift range between $0.2$ and $0.3$ is rather small (slightly more than 4000) compared with the sample size, and because these galaxies are the most luminous part of the sample and therefore have relatively small statistic weights, it is not surprising that the differences between the main fit and this are so small, in spite of the additional bias. 

The correction for the redshift evolution that we used for the main fit simply takes the evolutionary parameter $Q$, whose value was derived in Subsection \ref{sample_prop}. This is $1.07$ mag per $z$, and this value is only based on the redshift distribution of the observed galaxy number density. Therefore, it is independent of any filter. However, we investigated how the coefficients of the fundamental plane and the quality of the fit changed without considering redshift evolution. The results are listed in Table \ref{fitparameters_noevolution}. We found that the upper limit of the average distance error $\bar{\sigma}_{\textrm{dist}}$ is about one percentage point higher for the non-evolution-corrected fit than for the main fit. However, the coefficients of the fundamental plane are less tilted for the non-evolution-corrected fit, therefore it is possible that the details of handling the passive evolution of elliptical galaxies might contribute to the slightly less tilted coefficients (compared with our main fit) in the literature (see Table \ref{fp_coefficients_by_others}).

We also considered an alternative method of deriving the redshift evolution. For this, we analysed the redshift distribution of the surface brightness in our sample. The surface brightness (if properly corrected for the cosmological dimming) should be a distance, and consequently redshift-independent quantity. However, if the galaxies evolve, one expects different mean surface brightnesses in different redshift bins. We performed a Malmquist-bias-corrected fit to the redshift distribution of the (non-evolution-corrected) surfaces brightness in all filters and for all models. The results are listed in Table \ref{z_evolution}, and a set of graphic examples of the redshift evolution (for the dV model) is shown in Figures \ref{dV_z_dependence_sb_u} to \ref{dV_z_dependence_sb_z}. These evolutionary parameters are surface brightnesses per redshift and not magnitudes per redshift, but they enter the calibration at a point at which these two are mathematically equivalent. The numeric value of these new $Q$ parameters is at least twice as high as of the one derived by galaxy number densities, since this evolution does not only take into account changes of the absolute magnitude of the galaxies over time, but also possible changes in the radial extension of the galaxies. Furthermore, these values are different for every filter, which is more realistic because one may expect some changes not only in the luminosities, but in the colours of the galaxies. The results of the fundamental plane fit using these $Q$ parameters from Table \ref{z_evolution} can be found in Table \ref{fitparameters_filterevolution}. We found that the coefficient of this fit indicates a slightly more tilted fundamental plane than for the main fit. However, the average distance error is smaller by about half a percentage point for all filters and models for the surface brightness evolution fit.

\begin{figure}[ht]
\begin{center} 
\includegraphics[width=0.45\textwidth]{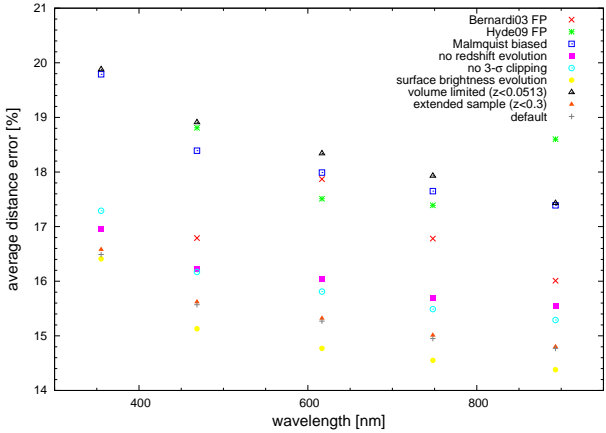}\\ 
\caption{Comparison of the upper limits of the average distance error for all performed fits for all filters of the dV model.}
\label{compare_disterr}
\end{center}
\end{figure} 
 
\subsection{Comparison with the literature}

We compared our results with those from the literature. To this end, we used our selected sample of about 95000 galaxies and the derived fundamental-plane parameters $\textrm{log}_{10}(R_{0})$, $\textrm{log}_{10}(\sigma_{0})$, and $\textrm{log}_{10}(I_{0})$ and see how well the fundamental plane with coefficients from the literature fits them. We used the direct-fit coefficients from the two works that best match our own, which are \citet{Bernardi:2003c} and \citet{Hyde:2009}, and derived the root mean square $s_{\varepsilon}$ and the distance error $\bar{\sigma}_{\textrm{dist}}$ from our sample and their coefficients. The results of this analysis can be found in Table \ref{accuracy_Bernardi}. Our newly derived coefficients are shown to provide a by a distance estimate couple of percentage points better than the previous ones. We point out that we used the same redshift evolutions for their samples as were given in the references. Furthermore, one can see in Figure \ref{compare_fp_i} that the location of the fundamental plane in \citet{Bernardi:2003c} and \citet{Hyde:2009} is similarly slightly above the centre of the cloud of data points, as in our fits. As already mentioned before, this is due to the Malmquist-bias correction. 

An overall visual comparison between the different fits of the fundamental plane performed by us can be found in Figure \ref{compare_fp_i}. Furthermore, we compare the upper limit of the average distance error of each fit or recalculation (\citet{Bernardi:2003c} and \citet{Hyde:2009} coefficients) in Figure \ref{compare_disterr}.

In addition to the fits of the fundamental plane, we studied the redshift distribution of the elliptical galaxies. We found a comoving number density of elliptical galaxies in the universe of $7 \cdot 10^{-4}$ per $(\textrm{Mpc} \cdot h_{70}^{-1})^3$, which is about 35\% of the value derived in \citet{Bernardi:2003b}, who reported $(2.0 \pm 0.1) \cdot 10^{-3}$ per $(\textrm{Mpc} \cdot h_{70}^{-1})^3$. We attribute this difference to the different underlying selection functions of the two samples. If we were to accept all 170962 candidates for elliptical galaxies obtained from GalaxyZoo as true ellipticals, our average comoving number density would increase by about 85\% to $1.3 \cdot 10^{-3}$ per $(\textrm{Mpc} \cdot h^{-1})^3$ in Figure \ref{volumebinned}, which is closer to, albeit still clearly below, the value of \citet{Bernardi:2003b}. When comparing the fractions of SDSS galaxies with spectroscopic data that are classified as ellipticals, our overall selection criteria including GalaxyZoo are slightly stricter than those of \citet{Bernardi:2003a}. Our selected sample consists of $\sim$95000 galaxies taken from a total of $\sim$852000 SDSS DR8 galaxies with proper spectroscopic data. This yields a fraction of about 11\%, somewhat lower than the about 14\% obtained by \citet{Bernardi:2003a}, who classified $\sim$9000 galaxies as ellipticals of a total of $\sim$65000 galaxies, for which spectroscopic and photometric data was provided by SDSS at that time. We thus conclude that differences in selection strictness work towards closing the gap between the higher density estimate by \citet{Bernardi:2003b} and our data, even though they cannot directly explain the full difference. We note that our value is consistent with the luminosity function analysis in Subsection \ref{res_lumfct}.

We found the same overdensities in the number counts by redshift as \citet{Bernardi:2003b}. We identify one as being associated with the CfA2-Great Wall around a redshift of 0.029 \citep{Geller:1989} and another related to the Sloan Great Wall around a redshift of 0.073 \citep{Gott:2005}. In addition to that, we confirm a peak in the number count of elliptical galaxies around 0.13, which has previously been reported in \citet{Bernardi:2003b}, but was not investigated in detail. Another result we obtained by analysing the redshift distribution of sample is the evolution parameter $Q$. The parameter derived just from the galaxy number densities within the sample should be an averaged estimate for all bands. Our fit yields a $Q$ of $1.07$ mag (per $z$), which is similar to the values of \citet{Bernardi:2003b}. In an alternative approach on deriving the redshift evolution, we used the redshift distribution of the surface brightnesses of the galaxies in our sample, which yielded significantly higher values (see Table \ref{z_evolution}) than our first approach and the values of \citet{Bernardi:2003b} and \citet{Hyde:2009}.

\begin{figure}[ht]
\begin{center}
\includegraphics[width=0.45\textwidth]{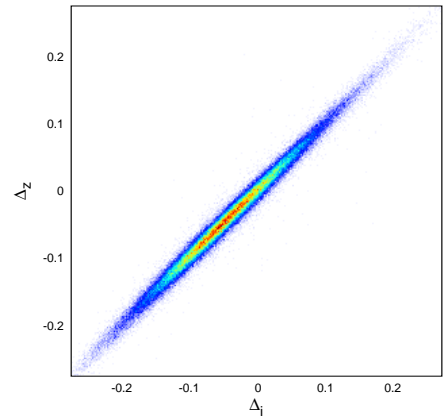}\\ 
\caption{Tight correlation between the residuals of the fundamental plane in the i band $\Delta_{i}$ and of those in the z band $\Delta_{z}$ can be easily seen in this plot. This plot uses the fundamental-plane fit for the dV model.}
\label{dV_residuals_map_iz}
\end{center}
\end{figure}

\subsection{Correlations of the residuals}

The fundamental plane was fitted independently in every filter. One does not necessarily expect any correlation between the residuals $\Delta_{f}$ (see Equation \ref{residuals} for the definition) of the fundamental plane for different filters $f$ due to our methodology.  However, a significant correlation between the residuals of different filters may still exist if the thickness of the fundamental plane is mainly caused by real deviations of the parameters and not by errors in the measurement of these parameters,
\begin{equation}
\Delta = a \cdot \textrm{log}_{10}\left(\sigma_{0}\right) + b \cdot \textrm{log}_{10}\left(I_{0}\right) + c - \textrm{log}_{10}\left(R_{0}\right) .
\label{residuals}
\end{equation}
The strength of the correlation between the residuals of the fundamental plane in different filters is strikingly important, especially when one plans on using the fundamental plane as distance indicator. If the correlation between the residuals in different filters is found to be low, one will be able to treat the distances obtained by the fundamental plane in different filters as independent measurements and will be able to achieve a better distance estimate by combining them than by using just data from one filter. We can quantify the correlation strength by calculating the linear correlation coefficients $r_{f_{1},f_{2}}$ for all possible combinations of filters.
\begin{equation}
r_{f_{1},f_{2}} = \sqrt{\frac{\left(\left(\sum\limits_{j}\left(\Delta_{f_{1}}-\bar{\Delta}_{f_{1}} \right)\right)\cdot\left(\sum\limits_{j} \left(\Delta_{f_{2}}-\bar{\Delta}_{f_{2}} \right)\right)\right)^{2}}{\left(\sum\limits_{j} \left(\Delta_{f_{1}}-\bar{\Delta}_{f_{1}}\right)^{2}\right)\cdot\left(\sum\limits_{j} \left(\Delta_{f_{2}}-\bar{\Delta}_{f_{2}} \right)^{2}\right)}} .
\label{correlation_coeff}
\end{equation}
The linear correlation coefficient $r_{f_{1},f_{2}}$ for the two filters $f_{1}$ and $f_{2}$ can be calculated using the residuals of the fundamental plane $\Delta_{f}$ and their averages $\bar{\Delta}_{f}$ of the corresponding filters $f$. For perfectly linearly correlated data points, the coefficient $r_{f_{1},f_{2}}$ is equal to one and for totally uncorrelated data points, it is zero. We found that the residuals of the fundamental plane of different filters are strongly correlated, as illustrated in Figure \ref{dV_residuals_map_iz} using the example of the i and z band residuals of the dV model. The corresponding plots for all other combinations of filters for the same model are displayed in Figures \ref{dV_residuals_map_rz} to \ref{dV_residuals_map_ug}. The linear correlation coefficients for all filters and models are listed in Table \ref{tab_residuals}. The values are very close to one, which indicates tight correlations. The correlations are slightly weaker, yet very strong for the p model, because the parameters of the fundamental plane have a larger scatter in this model and consequently the larger random errors dampen the correlation somewhat. The same is true for the correlation between the u band and any other filter. Therefore, relative values of the linear correlation coefficients agree with our previous findings about the u band and the p model parameters. 

Furthermore, the correlation coefficients decrease with greater differences in the wavelength of two filters, which is not surprising, because one may expect lower correlation the more different the bands are. We found that the correlation between the filters is too strong to enhance the quality of the distance measurement by combining different filters. In fact, every possible combination of the fundamental plane distances obtained by two and more filters yields a greater average distance error than using the best filter (z band) alone. This means that on average, if the parameters of a galaxy are located away from the fundamental plane in one filter, they are most likely similarly displaced in all other filters as well. This shows that the width of the fundamental plane does not primarily originate in measurement uncertainties of the required parameter, but in the intrinsic properties of the elliptical galaxies. The origin of this scatter is widely discussed in the literature. The theoretical derivation of the fundamental plane assumes a primary pressure-supported system. Nevertheless, it is known that elliptical galaxies are partially rotation-supported \citep{Burkert:2008}, too, and a variation in the fraction of pressure and rotational support can lead to inaccuracies in the mass estimates, which finally manifest themselves in the scatter of the fundamental plane. However, this effect alone does not seem to be sufficient to explain the entire scatter \citep{Prugniel:1996,Onorbe:2005}. Other explanations or contributing factors are the age \citep{Forbes:1998} and variations in the stellar population parameters \citep{Gargiulo:2009}, which are also considered as an explanation for the tilt of the fundamental plane by some authors \citep{LaBarbera:2008,Trujillo:2004}, and the merger history \citep{Hopkins:2008}. An extensive study on the details of the origin of the scatter and the tilt of the fundamental plane is beyond the scope of this paper.

\begin{table}
\begin{center}
\begin{tabular}{c|ccc}
 filters & c model & dV model & p model \\ \hline 
 $r_{i,z}$& 0.9909 & 0.9918 & 0.9259 \\
 $r_{r,z}$& 0.9836 & 0.9852 & 0.9268 \\
 $r_{g,z}$& 0.9635 & 0.9653 & 0.9140 \\
 $r_{u,z}$& 0.8584 & 0.8637 & 0.8190 \\
 $r_{r,i}$& 0.9933 & 0.9939 & 0.9812 \\
 $r_{g,i}$& 0.9788 & 0.9798 & 0.9643 \\
 $r_{u,i}$& 0.8768 & 0.8823 & 0.7979 \\
 $r_{g,r}$& 0.9904 & 0.9905 & 0.9808 \\
 $r_{u,r}$& 0.8865 & 0.8913 & 0.8029 \\
 $r_{u,g}$& 0.9099 & 0.9155 & 0.8191 \\
\end{tabular}
\end{center}
\caption{Linear correlation coefficients of the fundamental-plane residuals for all possible combinations of the five SDSS filters.}
\label{tab_residuals}
\end{table}

\section{Summary and Conclusions}
We analysed a sample of about 93000 elliptical galaxies taken from SDSS DR8. It forms the largest sample used for the calibration of the fundamental plane so far (roughly twice as large as the previous largest sample of \citet{Hyde:2009}). Furthermore, we used the high-quality K-corrections by \citet{Chilingarian:2010}. We also used GalaxyZoo data \citep{GalaxyZoo_data} to classify SDSS galaxies. A direct fit using a volume-weighted least-squares method was applied to obtain the coefficients of the fundamental plane because we plan on using the fundamental plane as a distance indicator in the subsequent work. We achieved an accuracy in the distance measurement of about 15\%. In addition to the fundamental plane, we studied the redshift distribution of the elliptical galaxies and the distribution of their global parameters such as the luminosity function.

We found a comoving number density of $7 \cdot 10^{-4}$ per $(\textrm{Mpc} \cdot h_{70}^{-1})^3$ for elliptical galaxies that qualify for our sample. Furthermore, in the analysis of the redshift distribution of the galaxies in our sample, we detected the same overdensities in the number counts by redshift as \citet{Bernardi:2003b}. One was identified as being associated with the CfA2-Great Wall \citep{Geller:1989} and another is related to the Sloan Great Wall \citep{Gott:2005}. In addition to these two well-known overdensities, we confirm a peak in the number count of elliptical galaxies around 0.13, which has previously been reported in \citet{Bernardi:2003b}, but was not investigated in detail. Moreover, we derived an evolution parameter $Q$ for elliptical galaxies of $1.07$ mag (per $z$), which is similar to the values of \citet{Bernardi:2003b}.

In addition to the results of our main fit, which are listed in Table \ref{fitparameters}, we provided a detailed analysis of the calibrations we made and their influence on the quality of the fitting process. We studied the effects of neglecting the Malmquist-bias correction, the 3-$\sigma$ clipping, or the redshift evolution correction. We also investigated changes in the parameters after using an alternative redshift evolution, a volume-limited sample, or an extended sample.  

To compare our calibrations with the literature, we calculated the root mean square and the upper limit of the average distance error using the coefficients and evolution parameters of \citet{Bernardi:2003c} and \citet{Hyde:2009}, but with the galaxies and parameters of our sample. We picked these two papers, because their work is the most similar to our own. The results can be found in Table \ref{accuracy_Bernardi}, and one can easily see that our main fit (see Table \ref{fitparameters}) provides a better distance indicator by a couple of percentage points. 

We investigated the correlation between the fundamental-plane residuals and found that they correlated too strongly to use a combination of the five independent fits (one for every filter) to reduce the overall scatter by combining two or more of them.

We found that in general the quality of the fundamental plane as a distance indicator increases with the wavelength, although the root mean square has its minimum in the SDSS i band. The upper limit of the average distance error is in general lowest in the z band, as one can see in Figure \ref{compare_disterr}. Furthermore, we found that the tilt of the fundamental plane (for the c and dV model) becomes smaller in the redder filters, as illustrated in Figure \ref{compare_parameters}. In our analysis, we learned that the dV model did best when considering the root mean square and the average distance error. It uses the pure de Vaucouleurs-magnitudes and radii. The c model (using composite magnitudes of a de Vaucouleurs and an exponential fit) only performed insignificantly worse, which indicates that the galaxies in our sample are very well described by de Vaucouleurs profiles. This finding is an expected feature of elliptical galaxies and tells us that our sample is very clean (the contamination by non-elliptical galaxies is insignificantly low). By comparing them to the results of the p model, we can instantly see that the Petrosian magnitudes and radii in SDSS provide poorer fits and cause a larger scatter. Therefore, we recommend only using the pure de Vaucouleurs magnitudes and radii together with our coefficients for them (see Table \ref{fitparameters}) and, if possible, the z or the i band for applications of the fundamental plane. Moreover, we strongly discourage the use of the u band due to known problems and the resulting lower quality of the results for this filter. 

We also found that our coefficients are similar to other direct fits of the fundamental plane of previous authors (see Table \ref{fp_coefficients_by_others}) (though the $a$ coefficient is slightly lower in our case, therefore the fundamental plane is more tilted), but due to our larger sample, we managed to achieve a yet unmatched accuracy.

In future work, we plan on using the fundamental plane to obtain redshift-independent distances for a large sample of elliptical galaxies from the SDSS. We will use those distances in combination with redshift data to derive peculiar velocities, which will form the basis of a cosmological test outlined in \citet{Saulder:2012}. We will investigate the dependence of the Hubble parameter of individual galaxies or clusters on the line of sight mass density towards these objects, and compare this with predictions of cosmological models. 
\section*{Acknowledgments}
Funding for SDSS-III has been provided by the Alfred P. Sloan Foundation, the Participating Institutions, the National Science Foundation, and the U.S. Department of Energy Office of Science. The SDSS-III web site is \url{http://www.sdss3.org/}.

SDSS-III is managed by the Astrophysical Research Consortium for the Participating Institutions of the SDSS-III Collaboration including the University of Arizona, the Brazilian Participation Group, Brookhaven National Laboratory, University of Cambridge, Carnegie Mellon University, University of Florida, the French Participation Group, the German Participation Group, Harvard University, the Instituto de Astrofisica de Canarias, the Michigan State/Notre Dame/JINA Participation Group, Johns Hopkins University, Lawrence Berkeley National Laboratory, Max Planck Institute for Astrophysics, Max Planck Institute for Extraterrestrial Physics, New Mexico State University, New York University, Ohio State University, Pennsylvania State University, University of Portsmouth, Princeton University, the Spanish Participation Group, University of Tokyo, University of Utah, Vanderbilt University, University of Virginia, University of Washington, and Yale University. 

CS acknowledges the support from an ESO studentship. 

IC acknowledges the support from the Russian Federation President's grant MD-3288.2012.2.
 
This research made use of the ``K-corrections calculator'' service available at \url{http://kcor.sai.msu.ru/}. 

IC acknowledges kind support from the ESO Visitor Program.

The publication is supported by the Austrian Science Fund (FWF).
\appendix
\FloatBarrier
\section{Redshift correction for the motion relative to the CMB}
\label{CMB_correction}
The observed redshift $z$ is in the rest frame of our solar system, but for cosmological and extragalactic application, one requires a corrected redshift $z_{\textrm{cor}}$, which is in the same rest frame as the CMB.
\begin{align}
z_{\textrm{cmb},x} &= \frac{v_{\textrm{cmb}}}{c} \textrm{cos}\left( b_{\textrm{cmb}} \right)  \textrm{cos}\left( l_{\textrm{cmb}} \right)\nonumber\\
z_{\textrm{cmb},y} &= \frac{v_{\textrm{cmb}}}{c} \textrm{cos}\left( b_{\textrm{cmb}} \right)  \textrm{sin}\left( l_{\textrm{cmb}} \right)\\
z_{\textrm{cmb},z} &= \frac{v_{\textrm{cmb}}}{c} \textrm{sin}\left( b_{\textrm{cmb}} \right) \nonumber
\label{CMB1}
\end{align}
The solar system moves into the direction of $l_{\textrm{cmb}}= 263.99^{\circ} \pm 0.14^{\circ} $ $b_{\textrm{cmb}}= 48.26^{\circ} \pm 0.03^{\circ}$ (galactic coordinates) with a velocity of $v_{\textrm{cmb}}= (369.0 \pm 0.9) \, \textrm{km} \, \textrm{s}^{-1}$ \citep{WMAP_5}. The first step required for this correction is to calculate the redshift space vector of our motion relative to the CMB $\vec{z_{\textrm{cmb}}} = \left( z_{\textrm{cmb},x}, z_{\textrm{cmb},y}, z_{\textrm{cmb},z} \right)$.
\begin{align}
z_{x} &= z \textrm{cos}\left( b \right)  \textrm{cos}\left( l \right)\nonumber\\
z_{y} &= z \textrm{cos}\left( b \right)  \textrm{sin}\left( l \right)\\
z_{z} &= z \textrm{sin}\left( b \right) \nonumber
\label{CMB2}
\end{align}
Then we translate the coordinates ($l$,$b$,$z$) of the observed galaxies into Cartesian coordinates into redshift space $\vec{z} = \left( z_{x}, z_{y}, z_{z} \right)$. In the next step, we perform a vector addition,
\begin{equation}
\vec{z_{\Sigma}}=\vec{z}+\vec{z_{\textrm{cmb}}} .
\label{CMB3}
\end{equation}
\begin{equation}
z_{\textrm{cor}}=z_{\Sigma,x}\,\textrm{cos}\left( b \right)  \textrm{cos}\left( l \right) + z_{\Sigma,y}\, \textrm{cos}\left( b \right)  \textrm{sin}\left( l \right) + z_{\Sigma,z}\, \textrm{sin}\left( b \right)
\label{CMB4}
\end{equation}
Now we project the vector $\vec{z_{\Sigma}} = \left( z_{\Sigma,x}, z_{\Sigma,y}, z_{\Sigma,z} \right)$ onto the line of sight and obtain the corrected (for our motion relative to the CMB) redshift $z_{\textrm{cor}}$. The corrected redshifts $z_{\textrm{cor}}$ are in the same rest frame as the CMB and can be used to calculate distances using the Hubble relation. 
\section{Additional figures}

\begin{figure}[H]
\begin{center}
\includegraphics[width=0.45\textwidth]{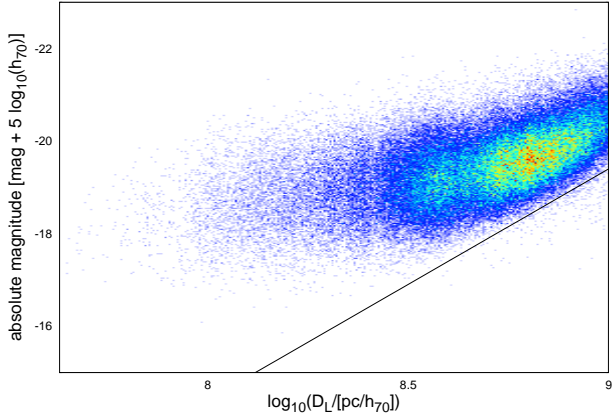}\\ 
\caption{Malmquist bias in the u band for the c model parameters is indicated by the black solid line of our fit. Due to the larger scatter in the u band, the fit is not as tight as for the other filters.}
\label{c_magabs_vs_dist_in_u}
\end{center}
\end{figure}
\begin{figure}[H]
\begin{center}
\includegraphics[width=0.45\textwidth]{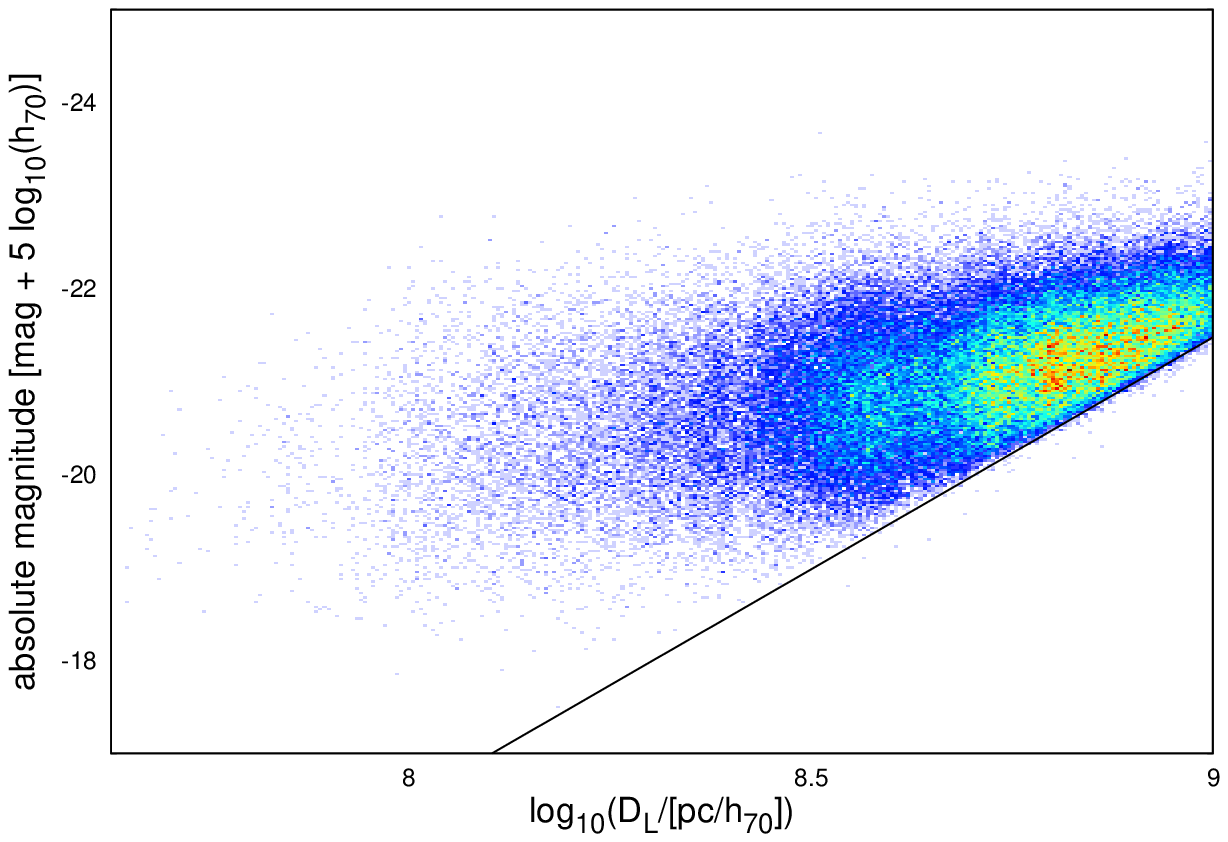}\\ 
\caption{Malmquist bias in the g band for the c model parameters is indicated by the black solid line of our fit.}
\label{c_magabs_vs_dist_in_g}
\end{center}
\end{figure}
\begin{figure}[H]
\begin{center}
\includegraphics[width=0.45\textwidth]{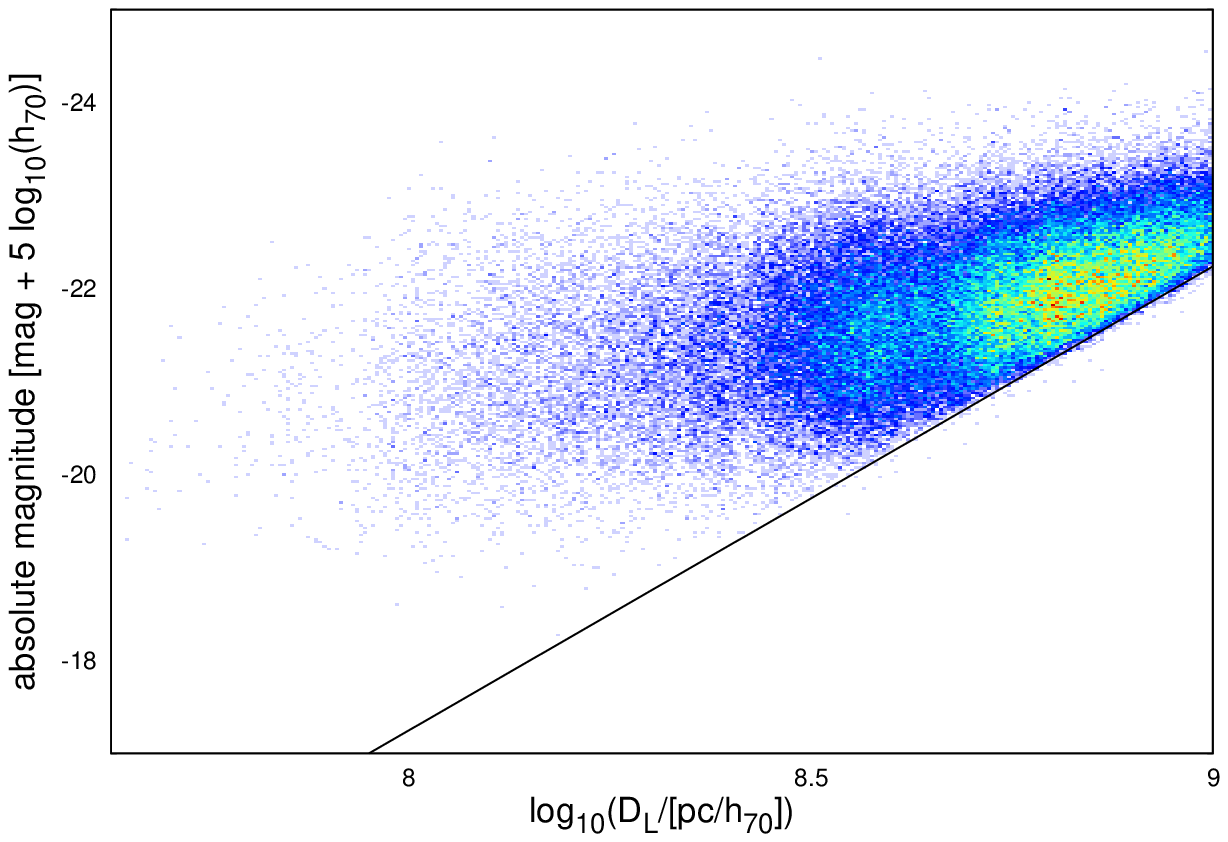}\\ 
\caption{Malmquist bias in the r band for the c model parameters is indicated by the black solid line of our fit.}
\label{c_magabs_vs_dist_in_r}
\end{center}
\end{figure}
\begin{figure}[H]
\begin{center}
\includegraphics[width=0.45\textwidth]{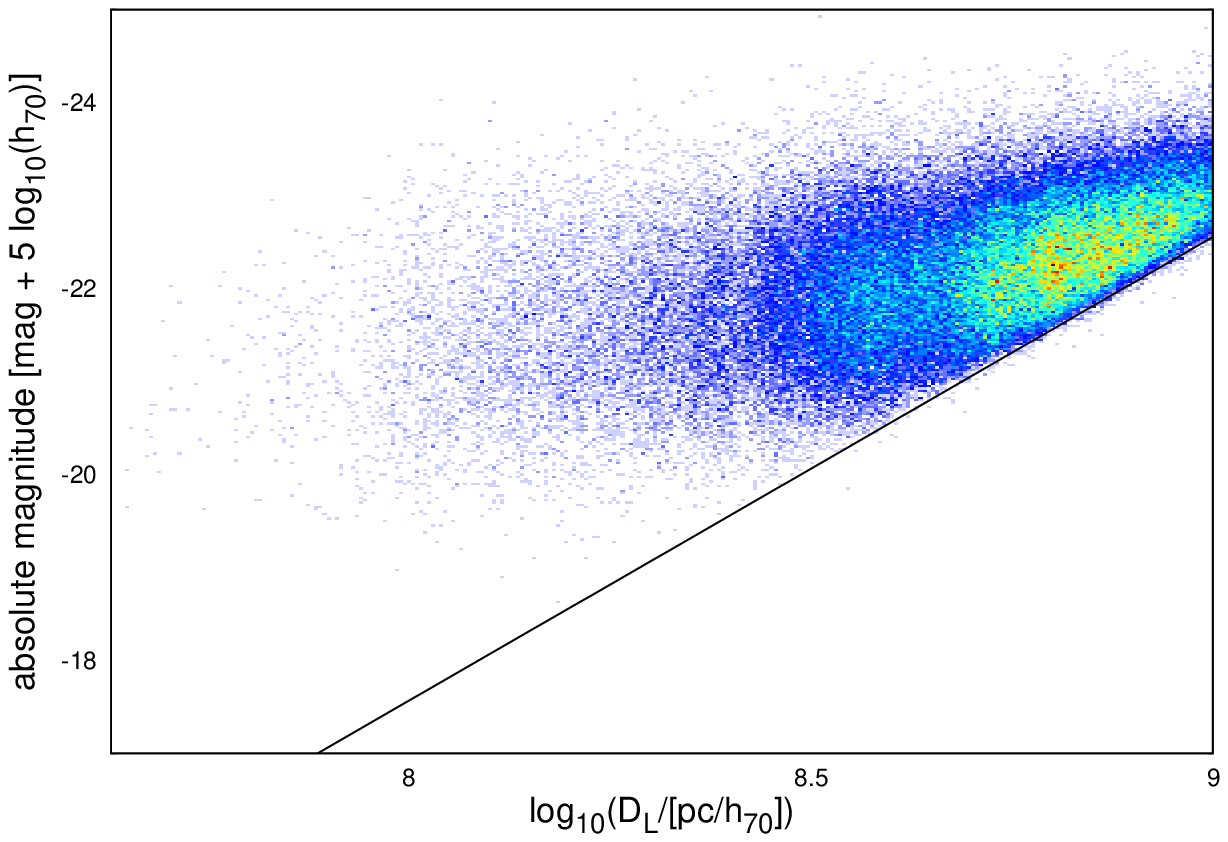}\\ 
\caption{Malmquist bias in the i band for the c model parameters is indicated by the black solid line of our fit.}
\label{c_magabs_vs_dist_in_i}
\end{center}
\end{figure}
\begin{figure}[H]
\begin{center}
\includegraphics[width=0.45\textwidth]{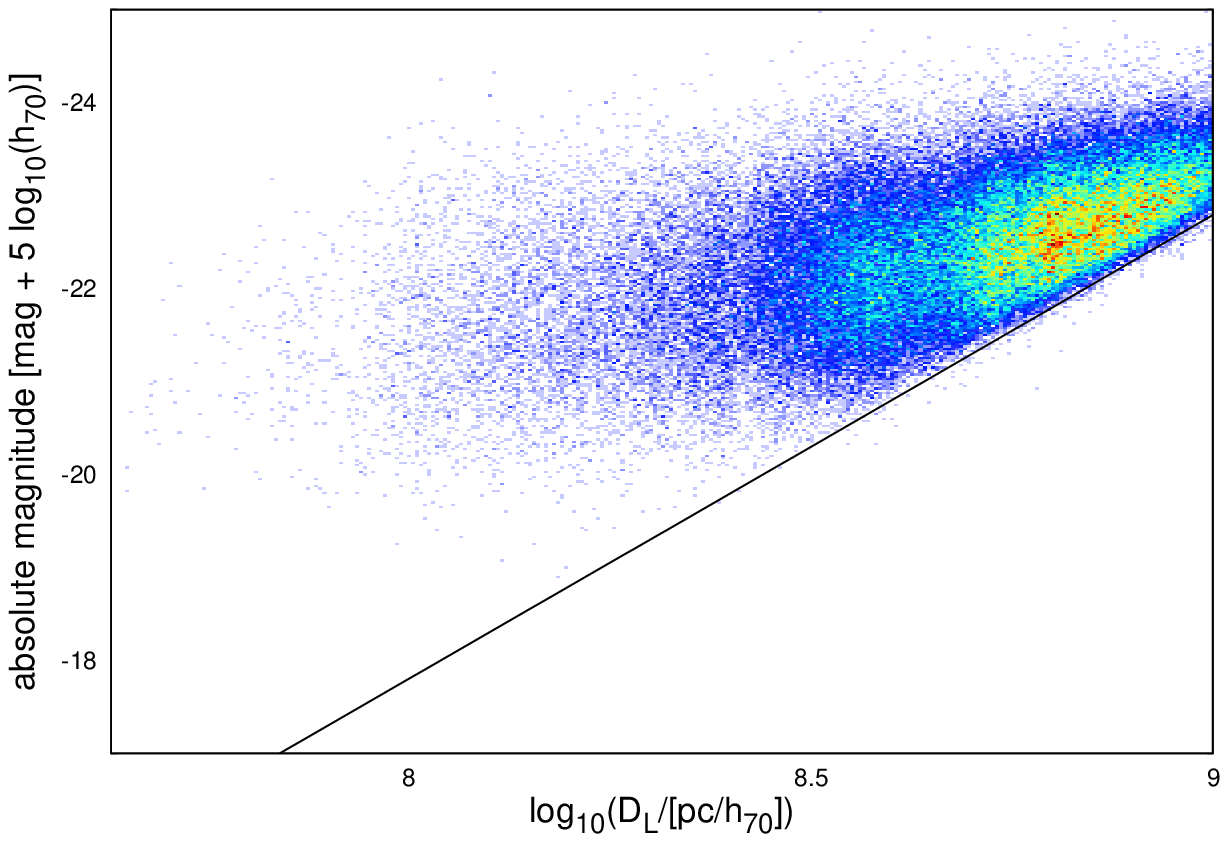}\\ 
\caption{Malmquist bias in the z band for the c model parameters is indicated by the black solid line of our fit.}
\label{c_magabs_vs_dist_in_z}
\end{center}
\end{figure}

\begin{figure}[H]
\begin{center}
\includegraphics[width=0.45\textwidth]{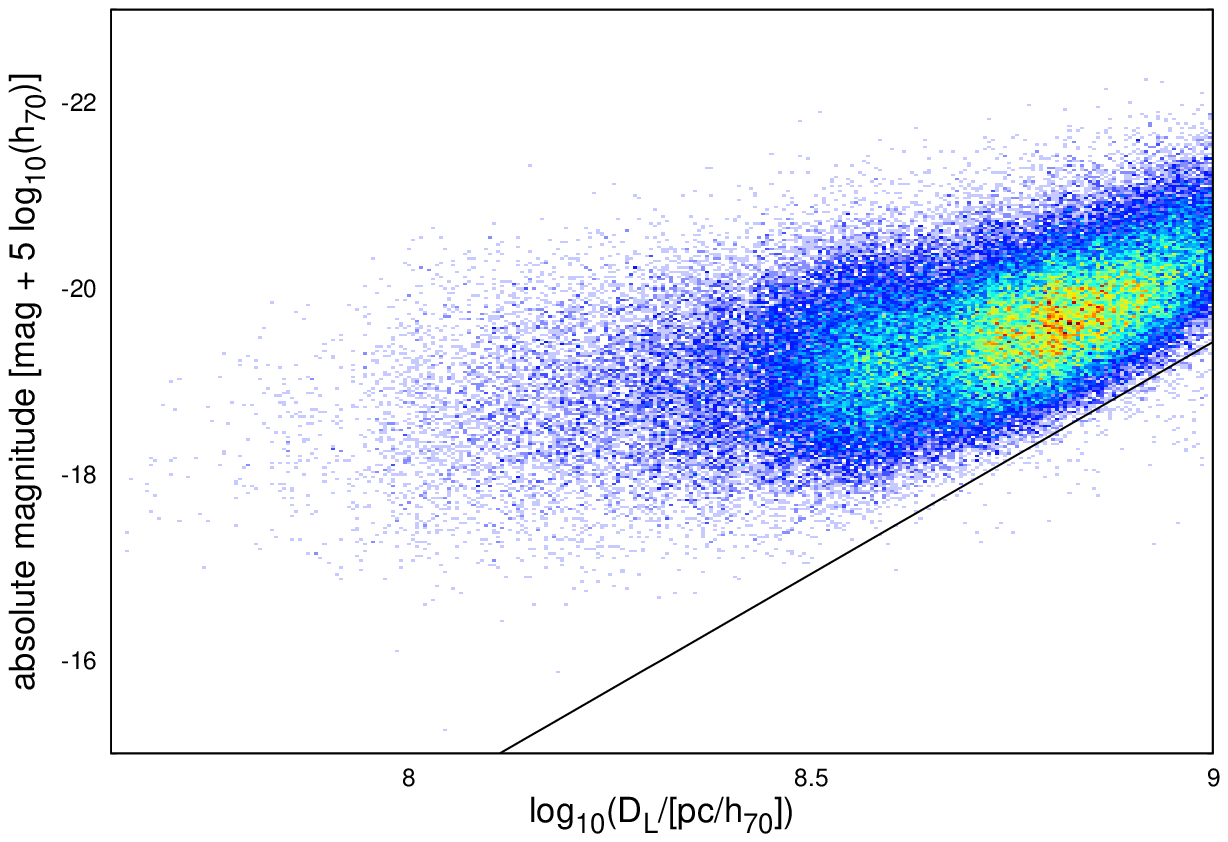}\\ 
\caption{Malmquist bias in the u band for the dV model parameters is indicated by the black solid line of our fit. Due to the larger scatter in the u band, the fit is not as tight as for the other filters.}
\label{dV_magabs_vs_dist_in_u}
\end{center}
\end{figure}
\begin{figure}[H]
\begin{center}
\includegraphics[width=0.45\textwidth]{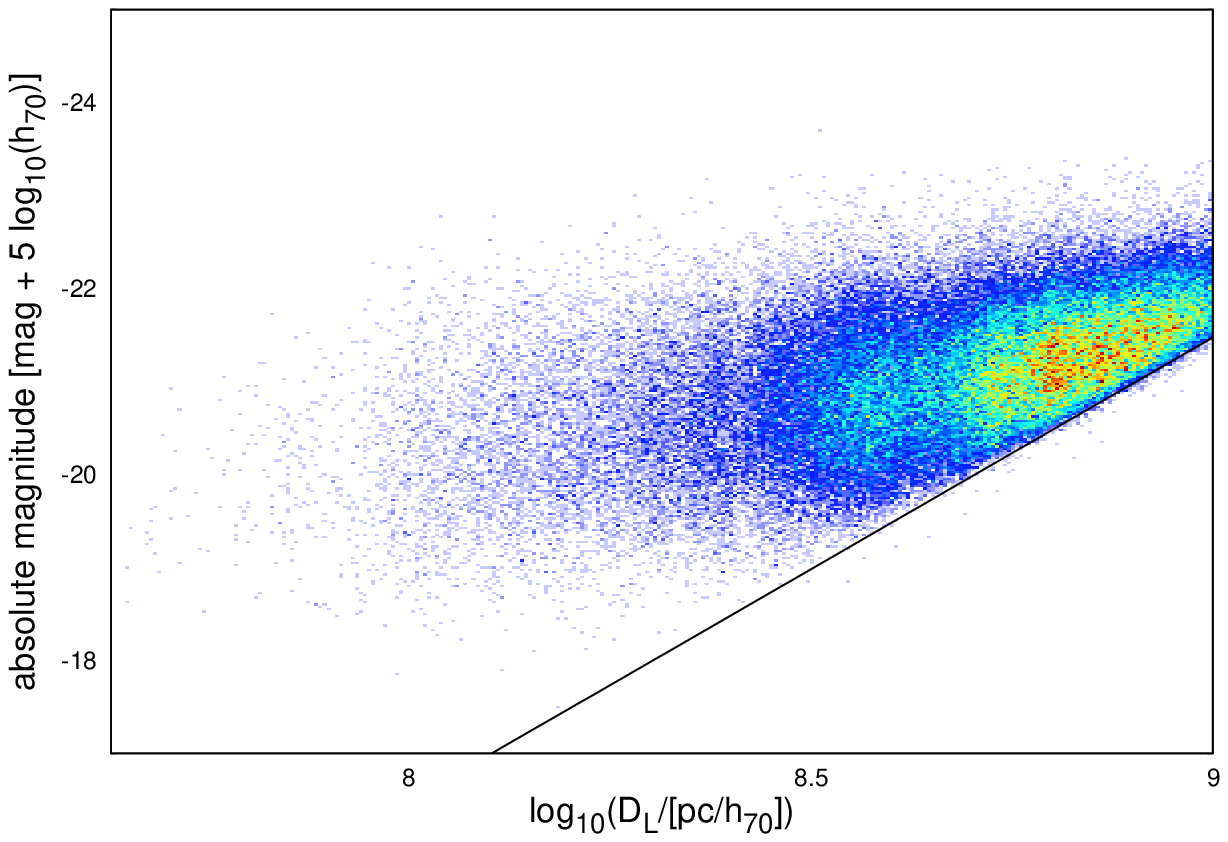}\\ 
\caption{Malmquist bias in the g band for the dV model parameters is indicated by the black solid line of our fit.}
\label{dV_magabs_vs_dist_in_g}
\end{center}
\end{figure}
\begin{figure}[H]
\begin{center}
\includegraphics[width=0.45\textwidth]{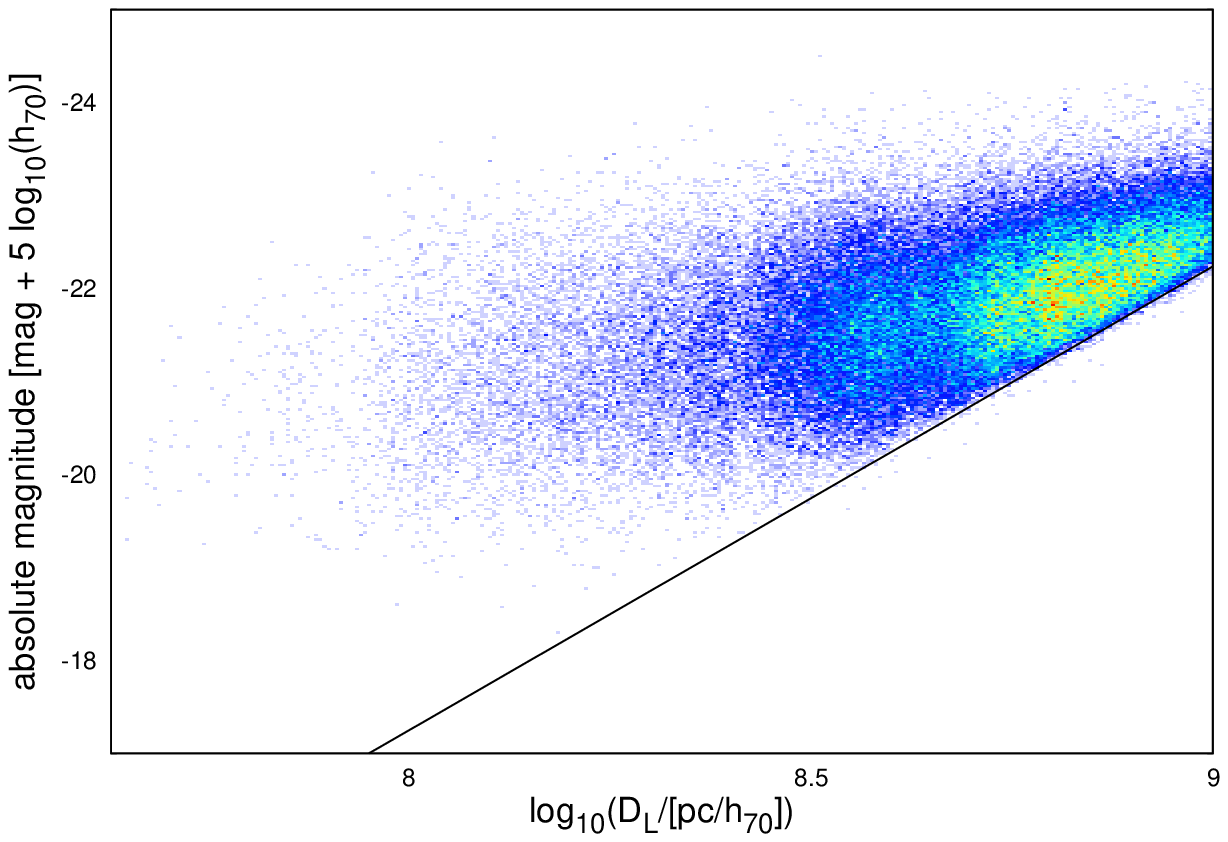}\\ 
\caption{Malmquist bias in the r band for the dV model parameters is indicated by the black solid line of our fit.}
\label{dV_magabs_vs_dist_in_r}
\end{center}
\end{figure}
\begin{figure}[H]
\begin{center}
\includegraphics[width=0.45\textwidth]{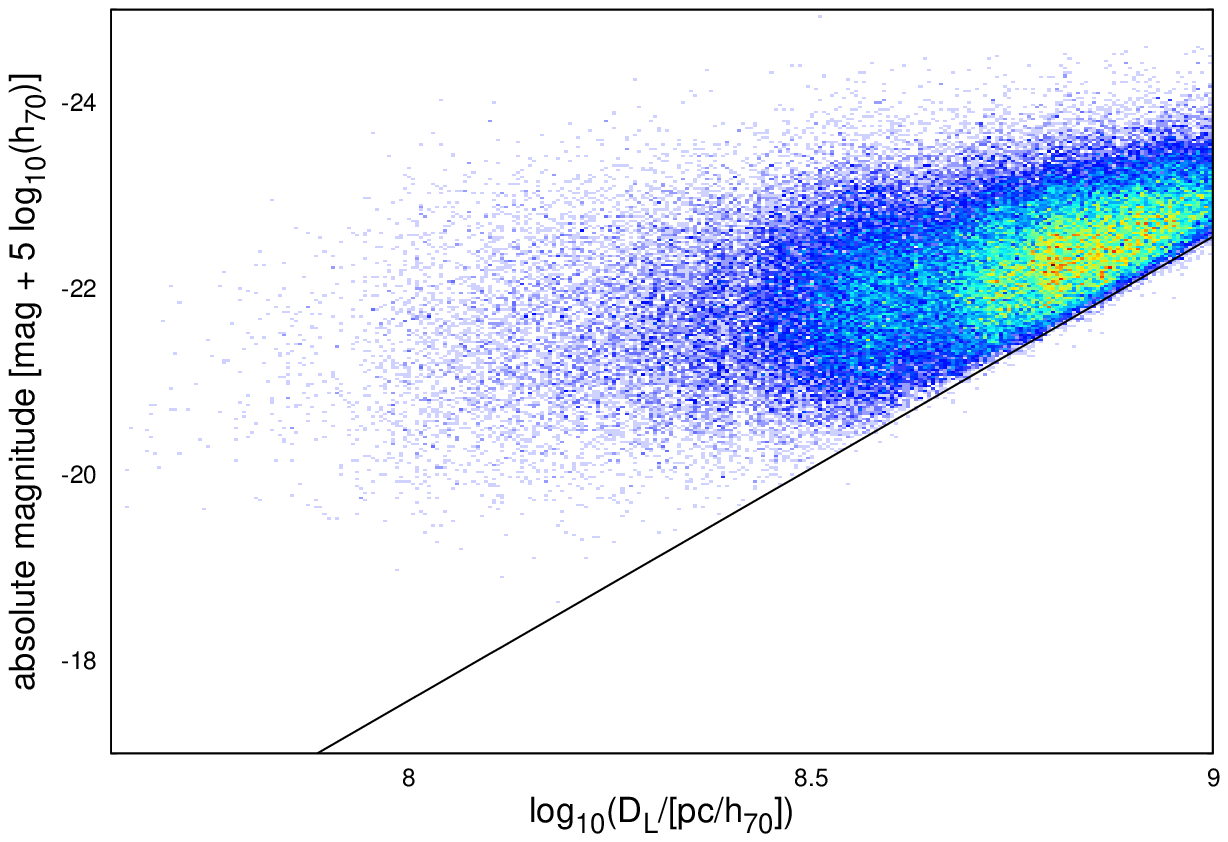}\\ 
\caption{Malmquist bias in the i band for the dV model parameters is indicated by the black solid line of our fit.}
\label{dV_magabs_vs_dist_in_i}
\end{center}
\end{figure}
\begin{figure}[H]
\begin{center}
\includegraphics[width=0.45\textwidth]{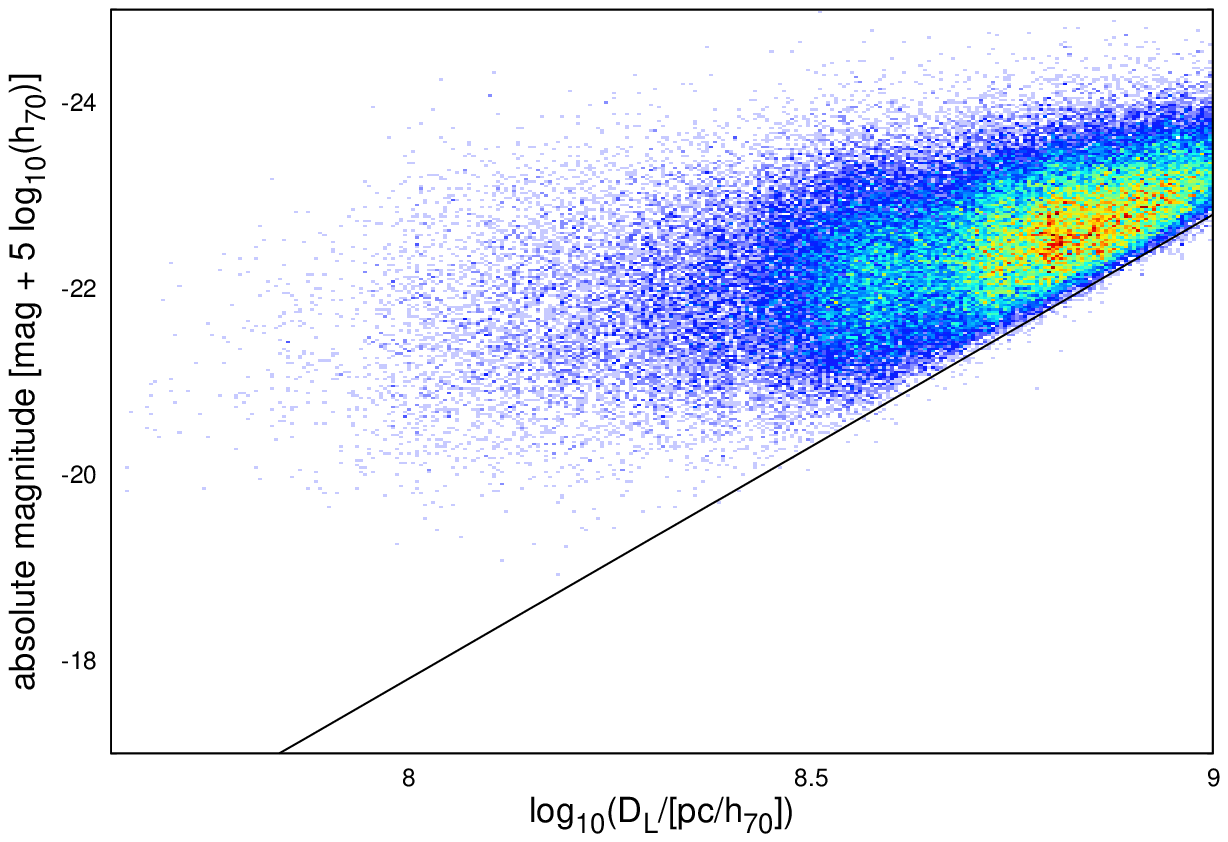}\\ 
\caption{Malmquist bias in the z band for the dV model parameters is indicated by the black solid line of our fit.}
\label{dV_magabs_vs_dist_in_z}
\end{center}
\end{figure}

\begin{figure}[H]
\begin{center}
\includegraphics[width=0.45\textwidth]{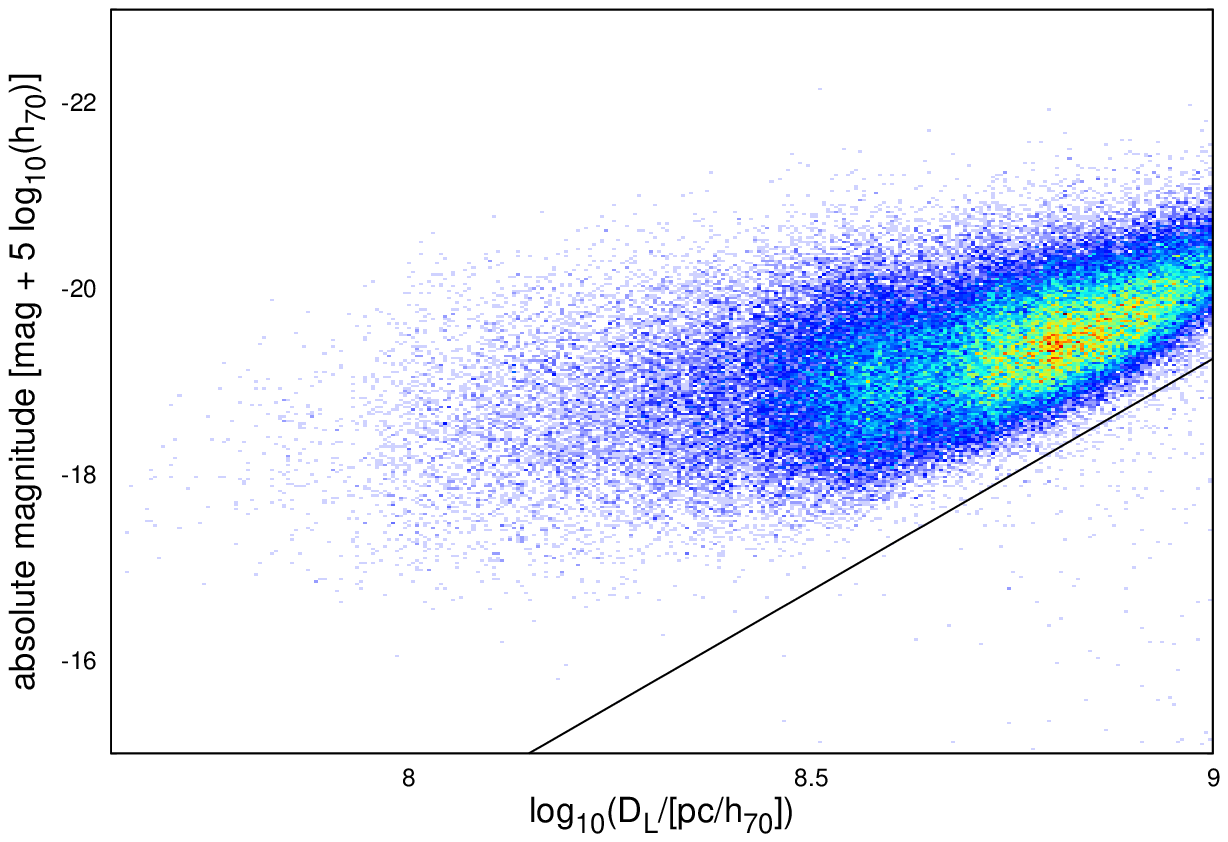}\\ 
\caption{Malmquist bias in the u band for the p model parameters is indicated by the black solid line of our fit. Due to the larger scatter in the u band, the fit is not as tight as for the other filters.}
\label{p_magabs_vs_dist_in_u}
\end{center}
\end{figure}
\begin{figure}[H]
\begin{center}
\includegraphics[width=0.45\textwidth]{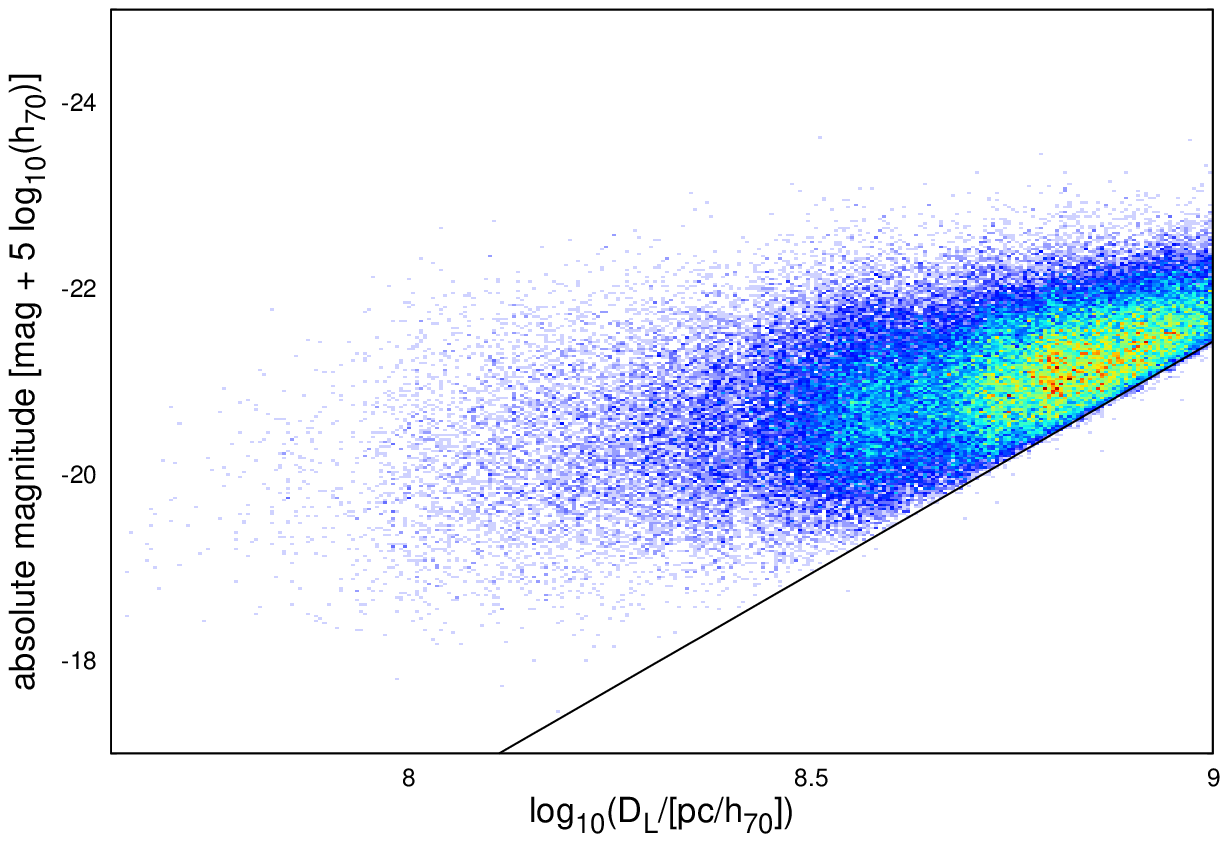}\\ 
\caption{Malmquist bias in the g band for the p model parameters is indicated by the black solid line of our fit.}
\label{p_magabs_vs_dist_in_g}
\end{center}
\end{figure}
\begin{figure}[H]
\begin{center}
\includegraphics[width=0.45\textwidth]{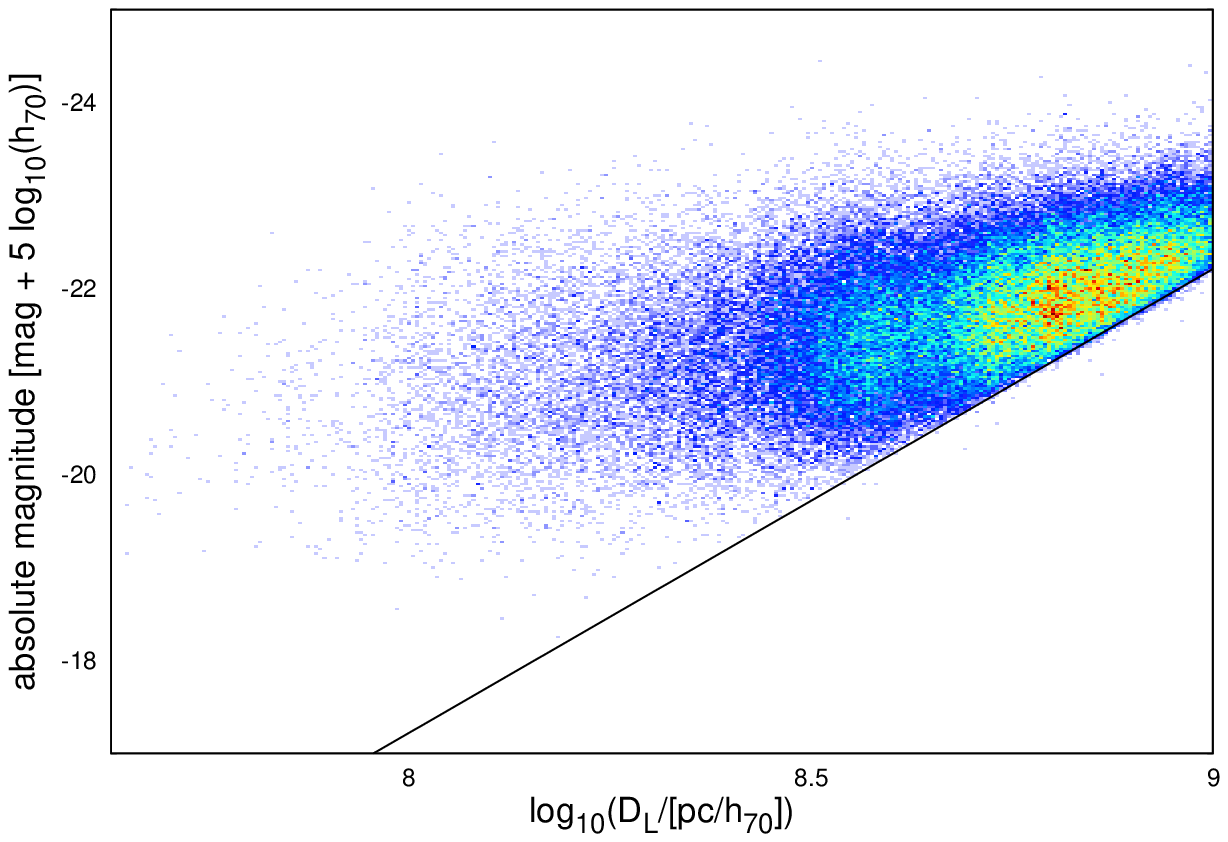}\\ 
\caption{Malmquist bias in the r band for the p model parameters is indicated by the black solid line of our fit.}
\label{p_magabs_vs_dist_in_r}
\end{center}
\end{figure}
\begin{figure}[H]
\begin{center}
\includegraphics[width=0.45\textwidth]{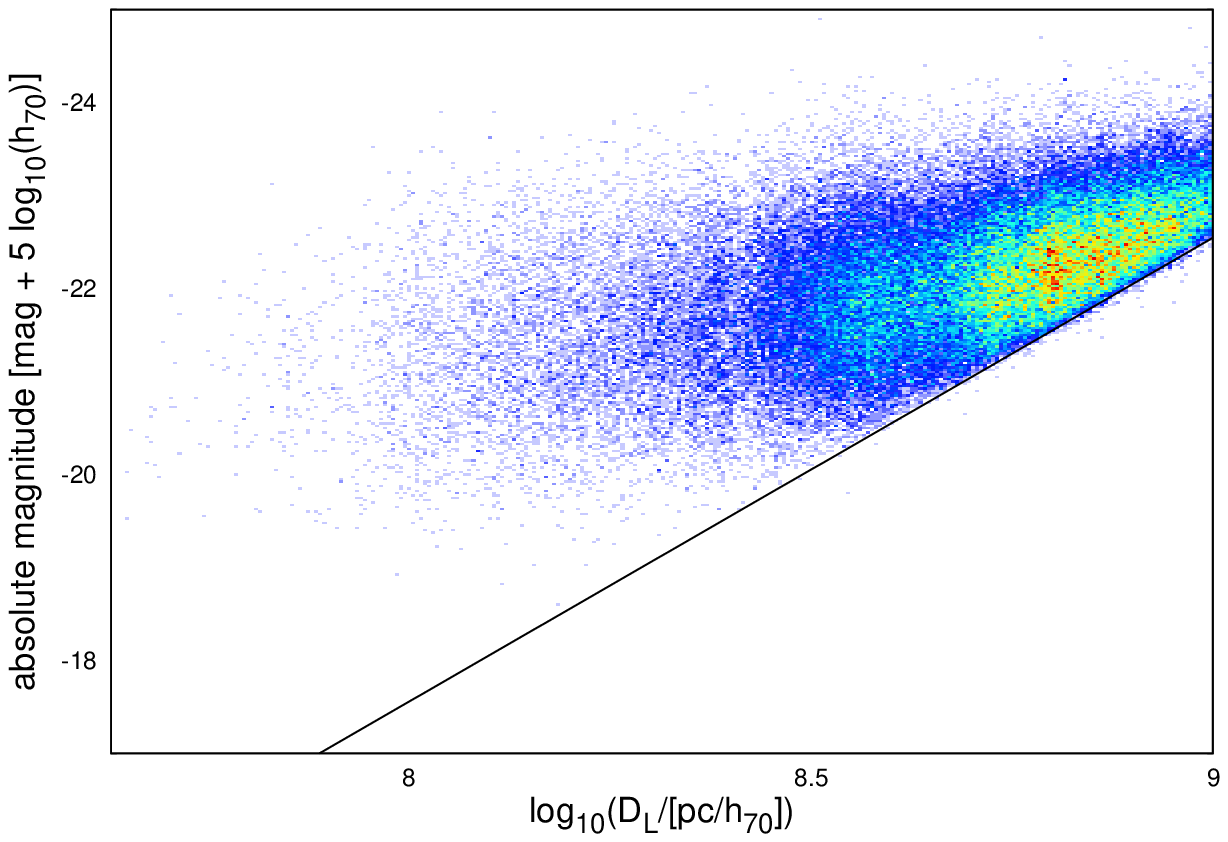}\\ 
\caption{Malmquist bias in the i band for the p model parameters is indicated by the black solid line of our fit.}
\label{p_magabs_vs_dist_in_i}
\end{center}
\end{figure}
\begin{figure}[H]
\begin{center}
\includegraphics[width=0.45\textwidth]{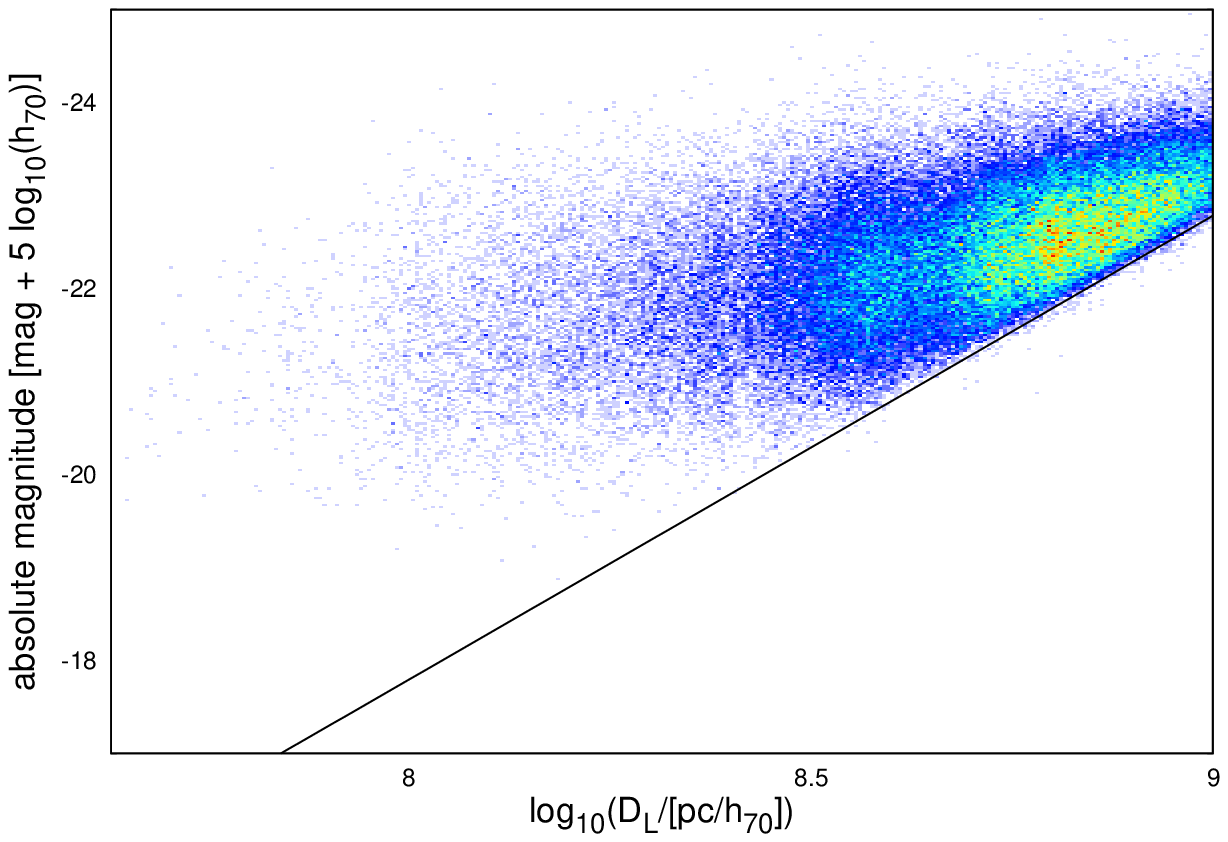}\\ 
\caption{Malmquist bias in the z band for the p model parameters is indicated by the black solid line of our fit.}
\label{p_magabs_vs_dist_in_z}
\end{center}
\end{figure}

\begin{figure}[H]
\begin{center}
\includegraphics[width=0.45\textwidth]{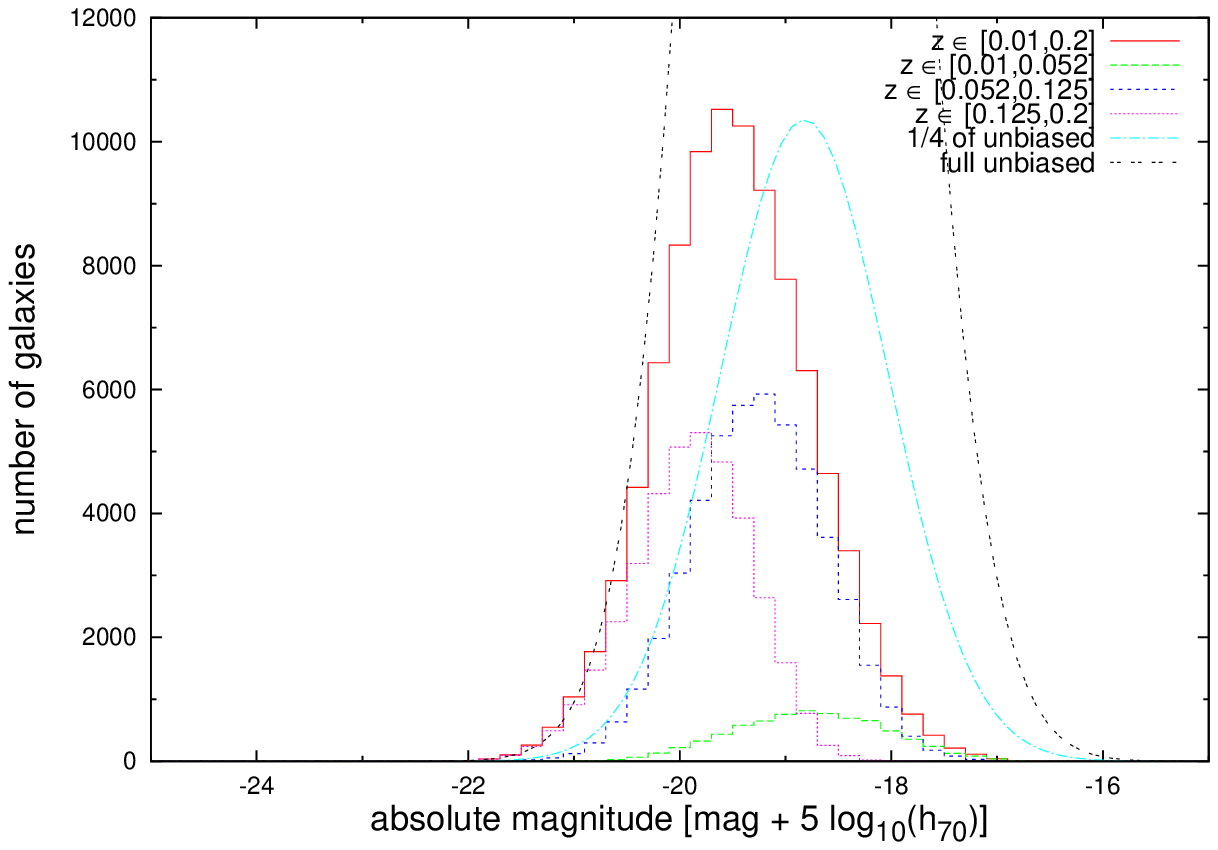}\\ 
\caption{Luminosity function for our sample in the u band for the c model. We split it into different subsamples (within different redshift bins) and compared the expected unbiased luminosity function and the total observed luminosity function. Our sample is almost complete at the luminous end, but we are missing many of the fainter galaxies due to the Malmquist bias.}
\label{c_magabs_binned_u}
\end{center}
\end{figure}
\begin{figure}[H]
\begin{center}
\includegraphics[width=0.45\textwidth]{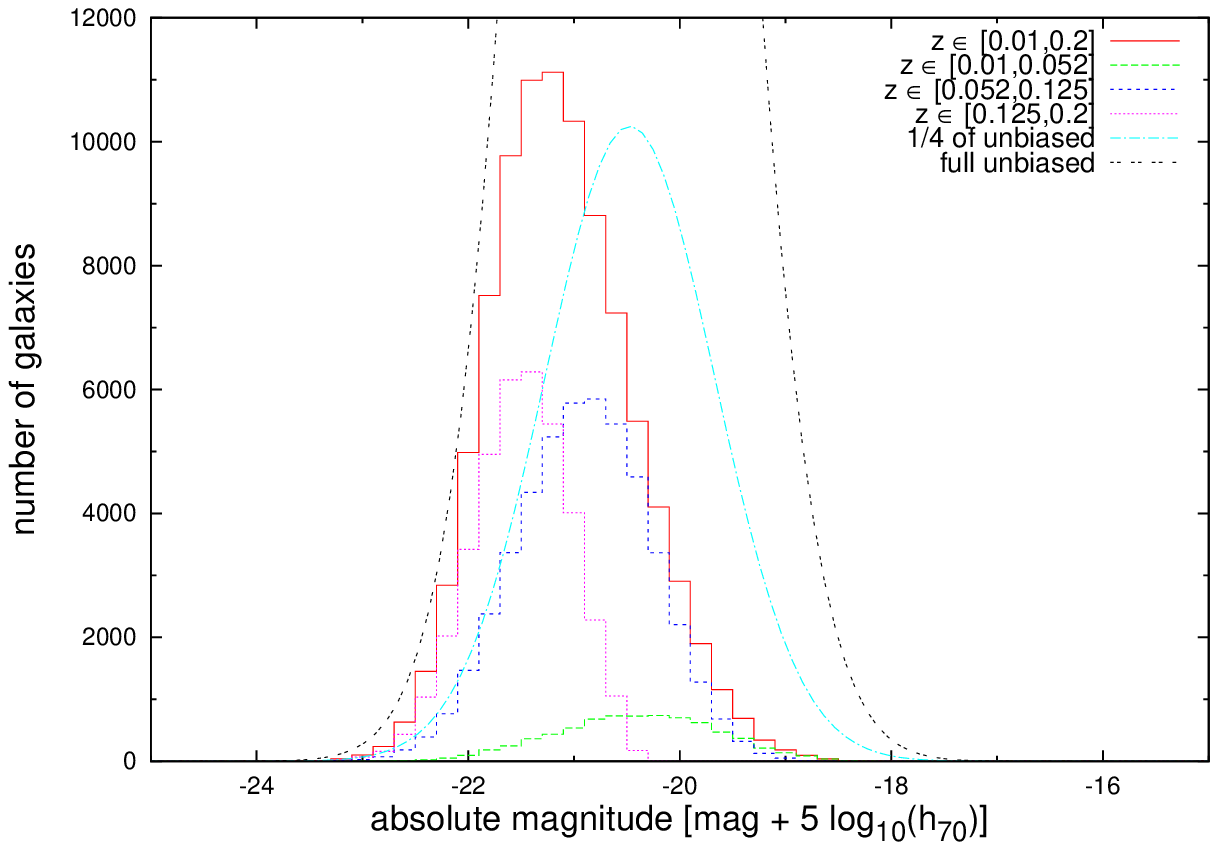}\\ 
\caption{Luminosity function for our sample in the g band for the c model. We split it into different subsamples (within different redshift bins) and compared the expected unbiased luminosity function and the total observed luminosity function. Our sample is almost complete at the luminous end, but we are missing many of the fainter galaxies due to the Malmquist bias.}
\label{c_magabs_binned_g}
\end{center}
\end{figure}
\begin{figure}[H]
\begin{center}
\includegraphics[width=0.45\textwidth]{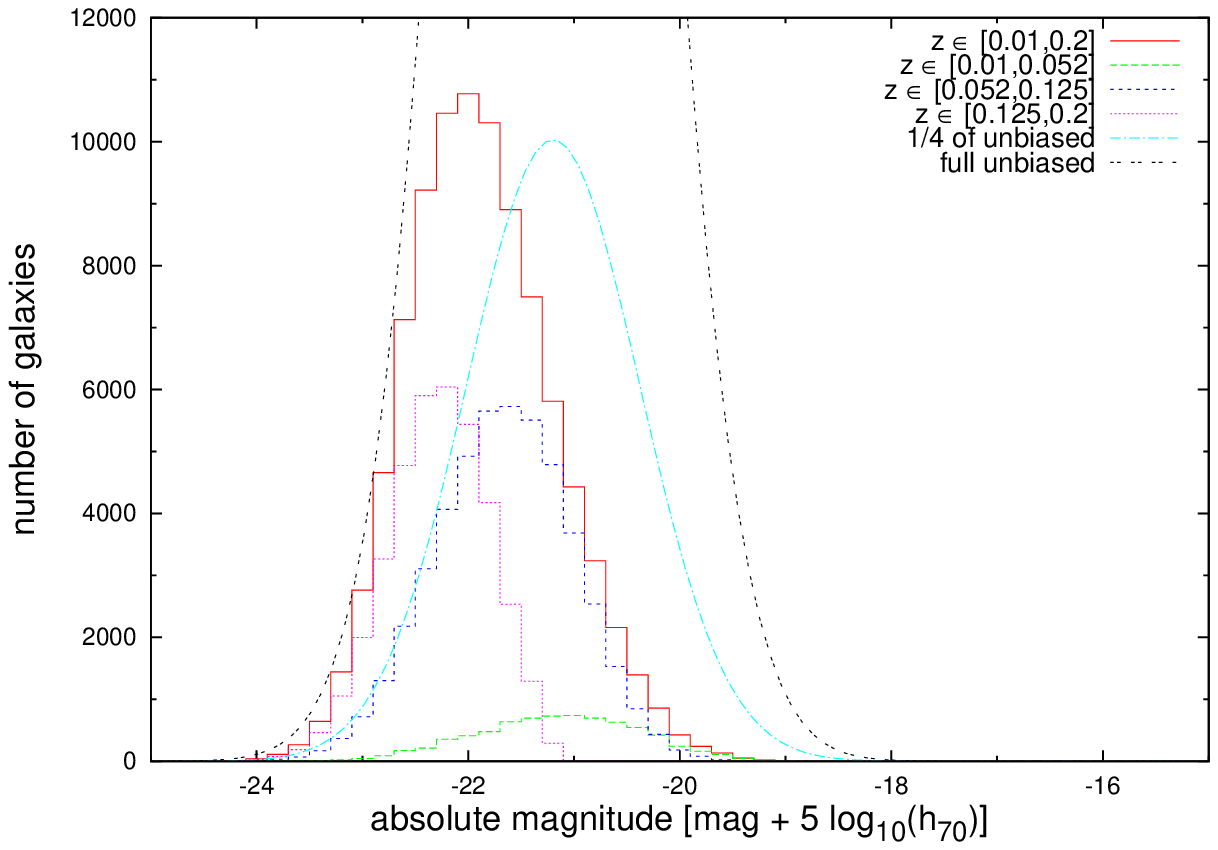}\\ 
\caption{Luminosity function for our sample in the r band for the c model. We split it into different subsamples (within different redshift bins) and compared the expected unbiased luminosity function and the total observed luminosity function. Our sample is almost complete at the luminous end, but we are missing many of the fainter galaxies due to the Malmquist bias.}
\label{c_magabs_binned_r}
\end{center}
\end{figure}
\begin{figure}[H]
\begin{center}
\includegraphics[width=0.45\textwidth]{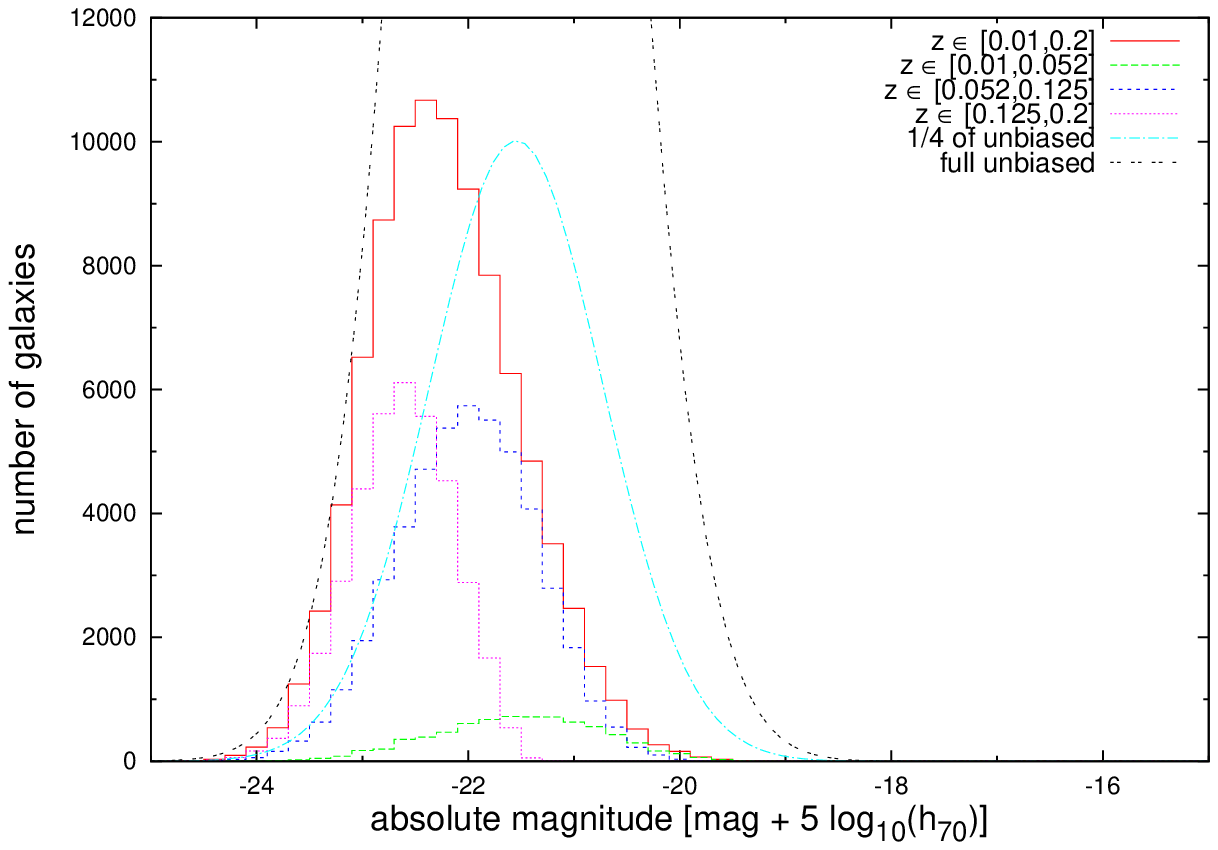}\\ 
\caption{Luminosity function for our sample in the i band for the c model. We split it into different subsamples (within different redshift bins) and compared the expected unbiased luminosity function and the total observed luminosity function. Our sample is almost complete at the luminous end, but we are missing many of the fainter galaxies due to the Malmquist bias.}
\label{c_magabs_binned_i}
\end{center}
\end{figure}
\begin{figure}[H]
\begin{center}
\includegraphics[width=0.45\textwidth]{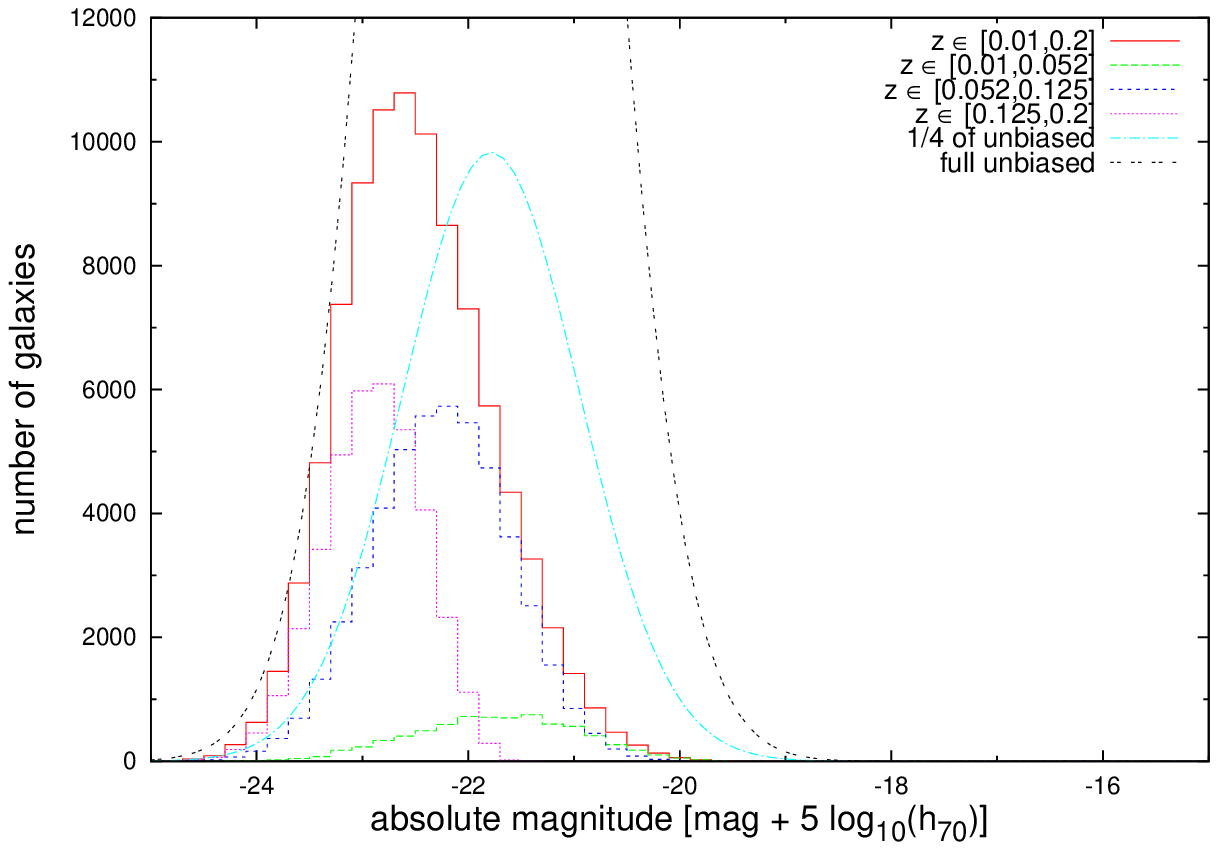}\\ 
\caption{Luminosity function for our sample in the z band for the c model. We split it into different subsamples (within different redshift bins) and compared the expected unbiased luminosity function and the total observed luminosity function. Our sample is almost complete at the luminous end, but we are missing many of the fainter galaxies due to the Malmquist bias.}
\label{c_magabs_binned_z}
\end{center}
\end{figure}

\begin{figure}[H]
\begin{center}
\includegraphics[width=0.45\textwidth]{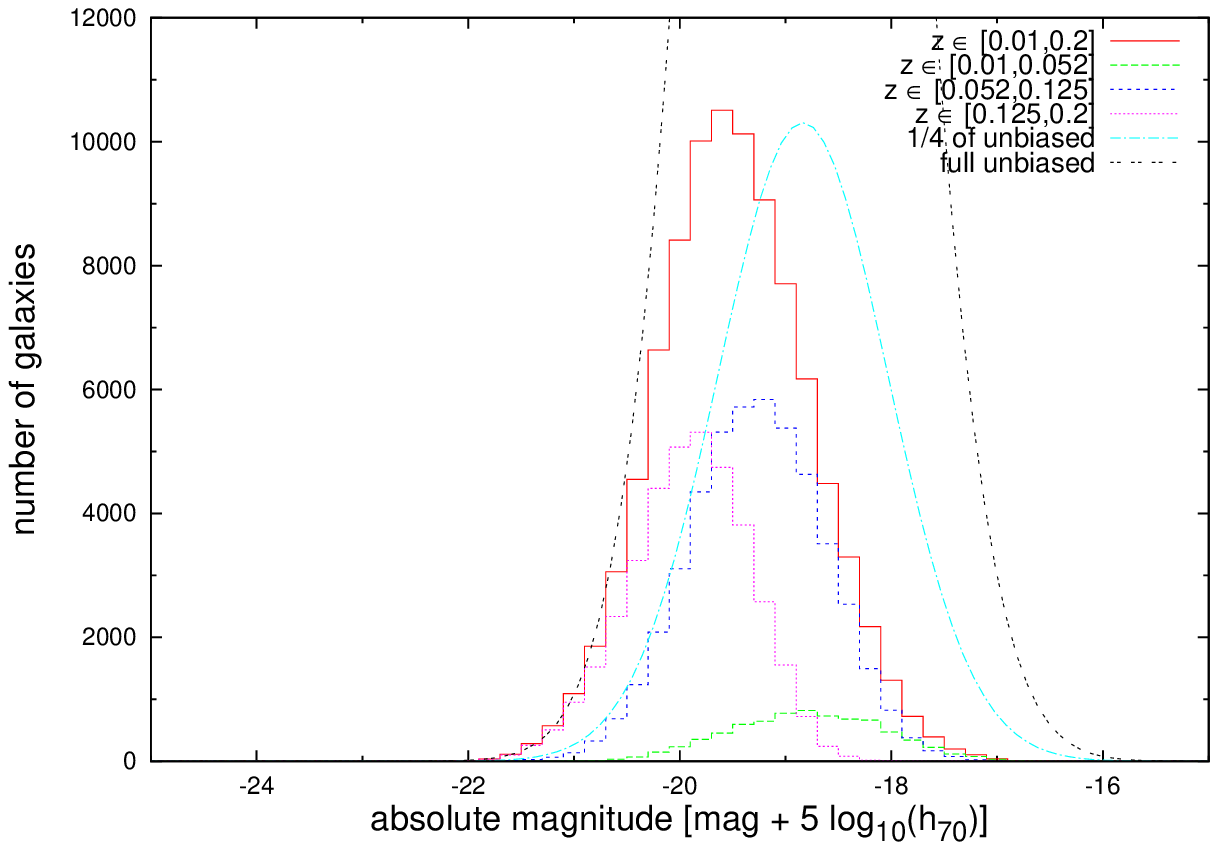}\\ 
\caption{Luminosity function for our sample in the u band for the dV model. We split it into different subsamples (within different redshift bins) and compared the expected unbiased luminosity function and the total observed luminosity function. Our sample is almost complete at the luminous end, but we are missing many of the fainter galaxies due to the Malmquist bias.}
\label{dV_magabs_binned_u}
\end{center}
\end{figure}
\begin{figure}[H]
\begin{center}
\includegraphics[width=0.45\textwidth]{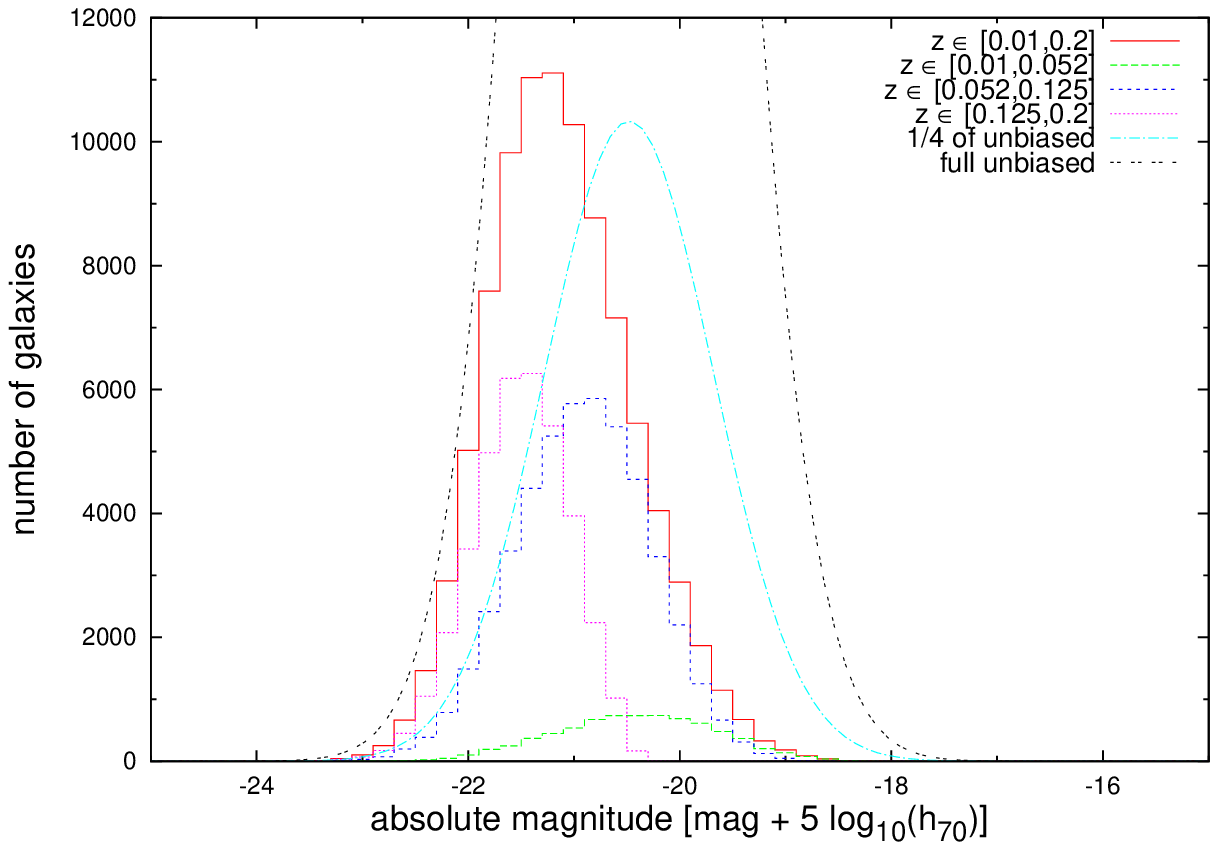}\\ 
\caption{Luminosity function for our sample in the g band for the dV model. We split it into different subsamples (within different redshift bins) and compared the expected unbiased luminosity function and the total observed luminosity function. Our sample is almost complete at the luminous end, but we are missing many of the fainter galaxies due to the Malmquist bias.}
\label{dV_magabs_binned_g}
\end{center}
\end{figure}
\begin{figure}[H]
\begin{center}
\includegraphics[width=0.45\textwidth]{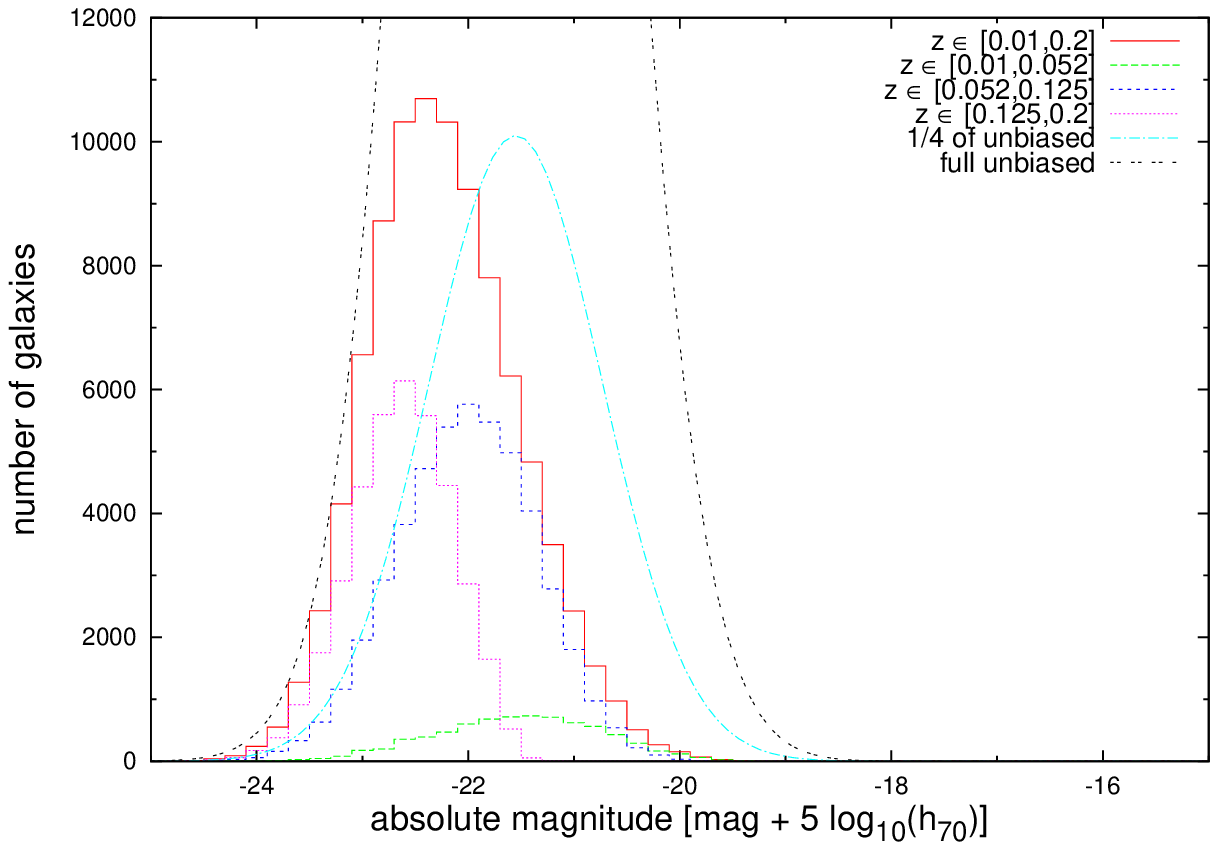}\\ 
\caption{Luminosity function for our sample in the i band for the dV model. We split it into different subsamples (within different redshift bins) and compared the expected unbiased luminosity function and the total observed luminosity function. Our sample is almost complete at the luminous end, but we are missing many of the fainter galaxies due to the Malmquist bias.}
\label{dV_magabs_binned_i}
\end{center}
\end{figure}
\begin{figure}[H]
\begin{center}
\includegraphics[width=0.45\textwidth]{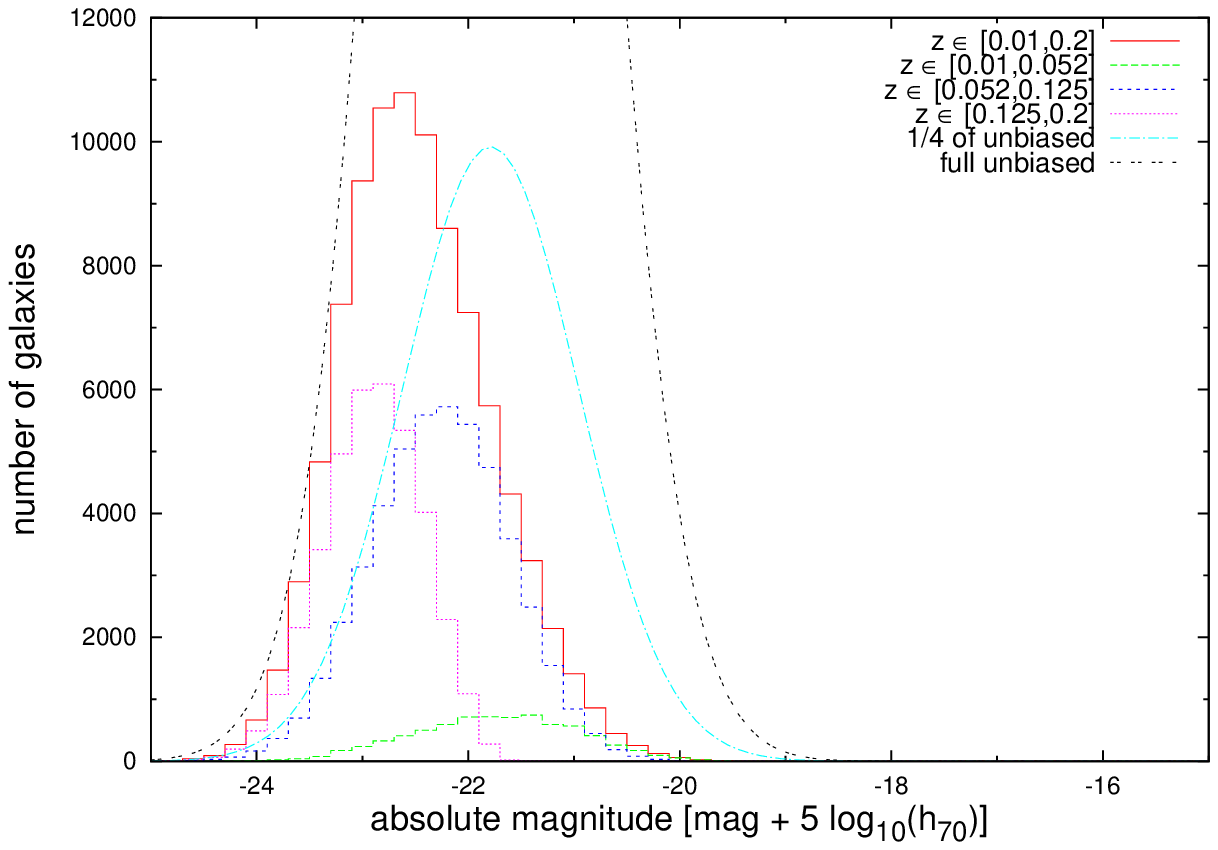}\\ 
\caption{Luminosity function for our sample in the z band for the dV model. We split it into different subsamples (within different redshift bins) and compared the expected unbiased luminosity function and the total observed luminosity function. Our sample is almost complete at the luminous end, but we are missing many of the fainter galaxies due to the Malmquist bias.}
\label{dV_magabs_binned_z}
\end{center}
\end{figure}

\begin{figure}[H]
\begin{center}
\includegraphics[width=0.45\textwidth]{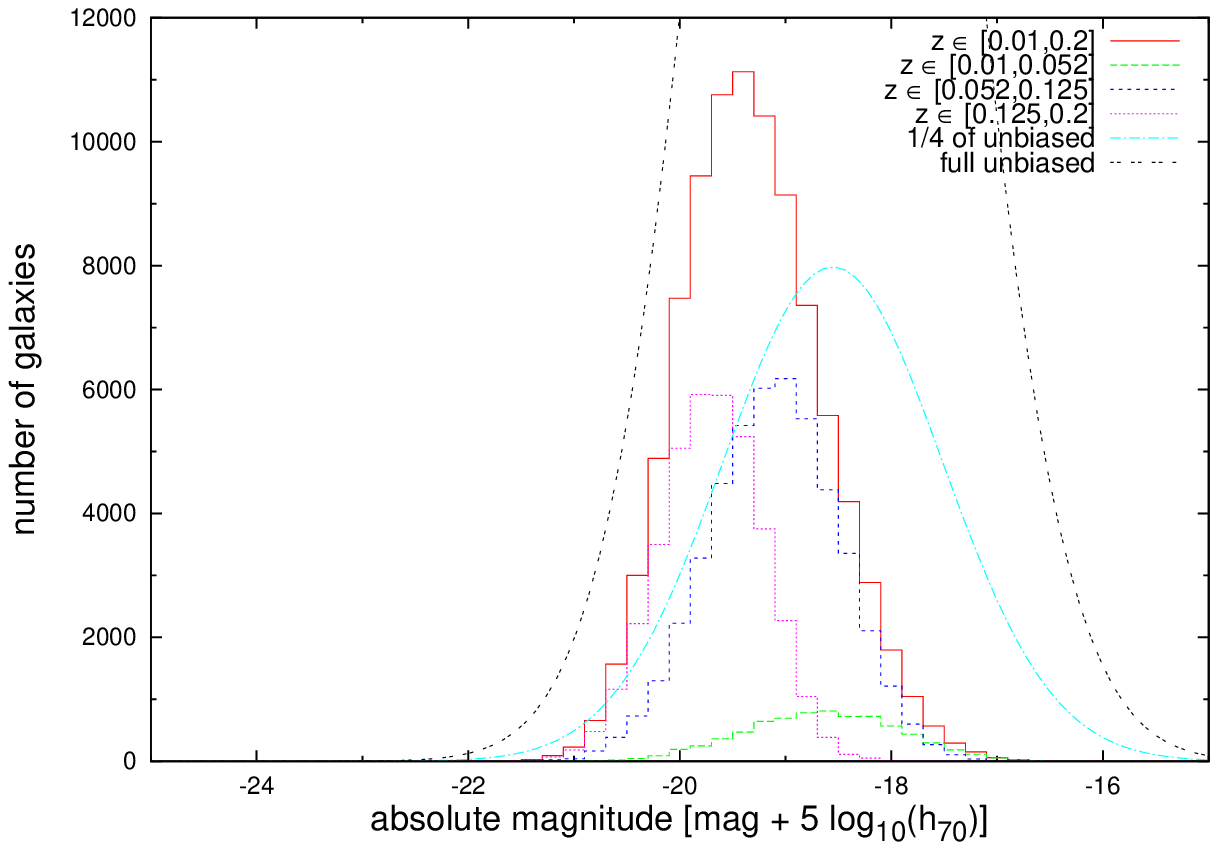}\\ 
\caption{Luminosity function for our sample in the u band for the p model. We split it into different subsamples (within different redshift bins) and compared the expected unbiased luminosity function and the total observed luminosity function. Our sample is almost complete at the luminous end, but we are missing many of the fainter galaxies due to the Malmquist bias.}
\label{p_magabs_binned_u}
\end{center}
\end{figure}
\begin{figure}[H]
\begin{center}
\includegraphics[width=0.45\textwidth]{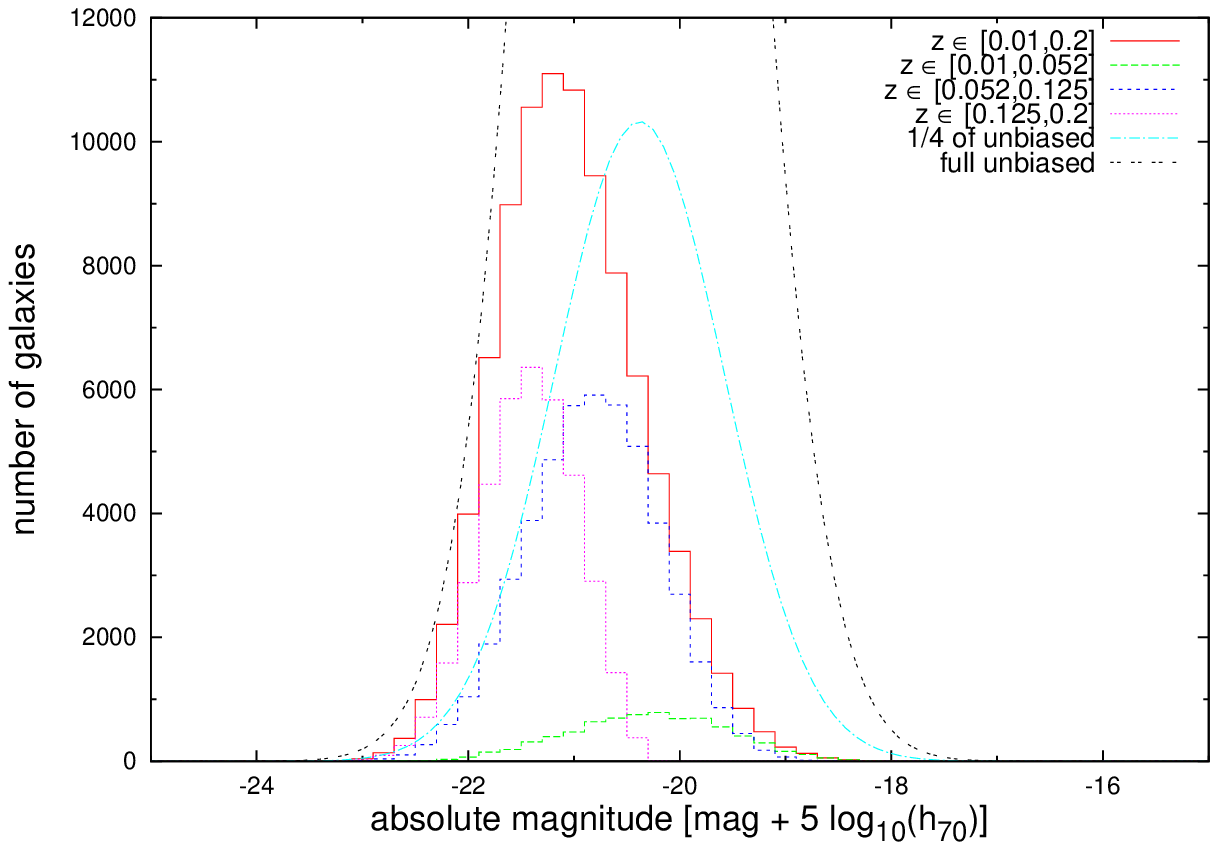}\\ 
\caption{Luminosity function for our sample in the g band for the p model. We split it into different subsamples (within different redshift bins) and compared the expected unbiased luminosity function and the total observed luminosity function. Our sample is almost complete at the luminous end, but we are missing many of the fainter galaxies due to the Malmquist bias.}
\label{p_magabs_binned_g}
\end{center}
\end{figure}
\begin{figure}[H]
\begin{center}
\includegraphics[width=0.45\textwidth]{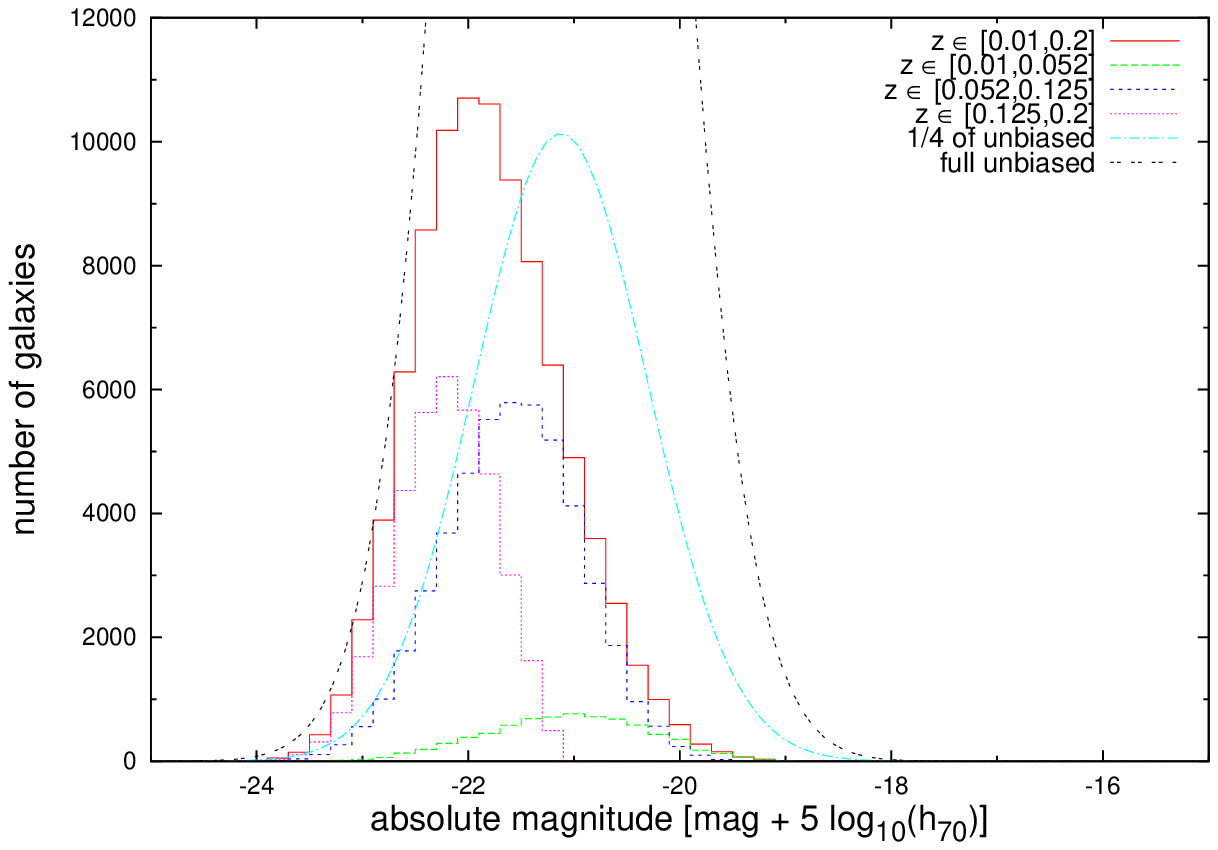}\\ 
\caption{Luminosity function for our sample in the r band for the p model. We split it into different subsamples (within different redshift bins) and compared the expected unbiased luminosity function and the total observed luminosity function. Our sample is almost complete at the luminous end, but we are missing many of the fainter galaxies due to the Malmquist bias.}
\label{p_magabs_binned_r}
\end{center}
\end{figure}
\begin{figure}[H]
\begin{center}
\includegraphics[width=0.45\textwidth]{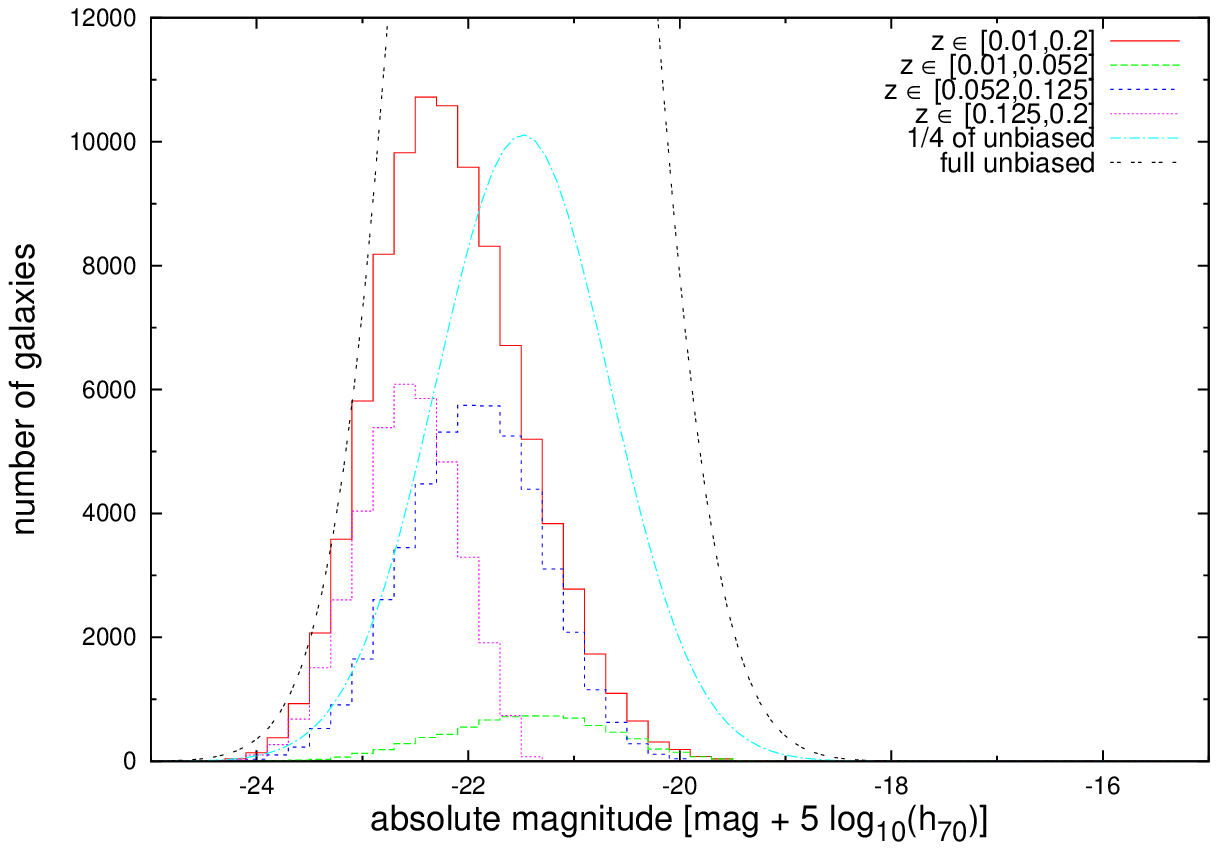}\\ 
\caption{Luminosity function for our sample in the i band for the p model. We split it into different subsamples (within different redshift bins) and compared the expected unbiased luminosity function and the total observed luminosity function. Our sample is almost complete at the luminous end, but we are missing many of the fainter galaxies due to the Malmquist bias.}
\label{p_magabs_binned_i}
\end{center}
\end{figure}
\begin{figure}[H]
\begin{center}
\includegraphics[width=0.45\textwidth]{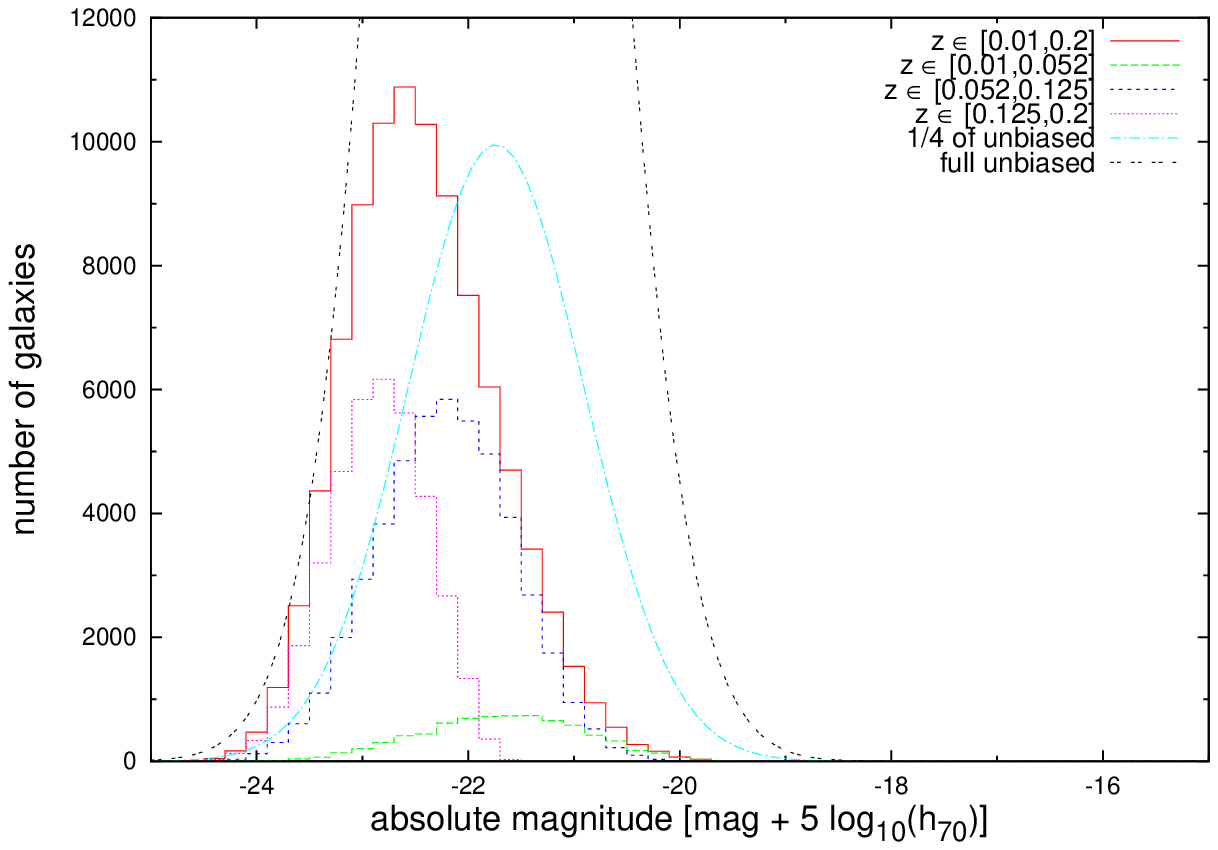}\\ 
\caption{Luminosity function for our sample in the z band for the p model. We split it into different subsamples (within different redshift bins) and compared the expected unbiased luminosity function and the total observed luminosity function. Our sample is almost complete at the luminous end, but we are missing many of the fainter galaxies due to the Malmquist bias.}
\label{p_magabs_binned_z}
\end{center}
\end{figure}

\begin{figure}[H]
\begin{center}
\includegraphics[width=0.45\textwidth]{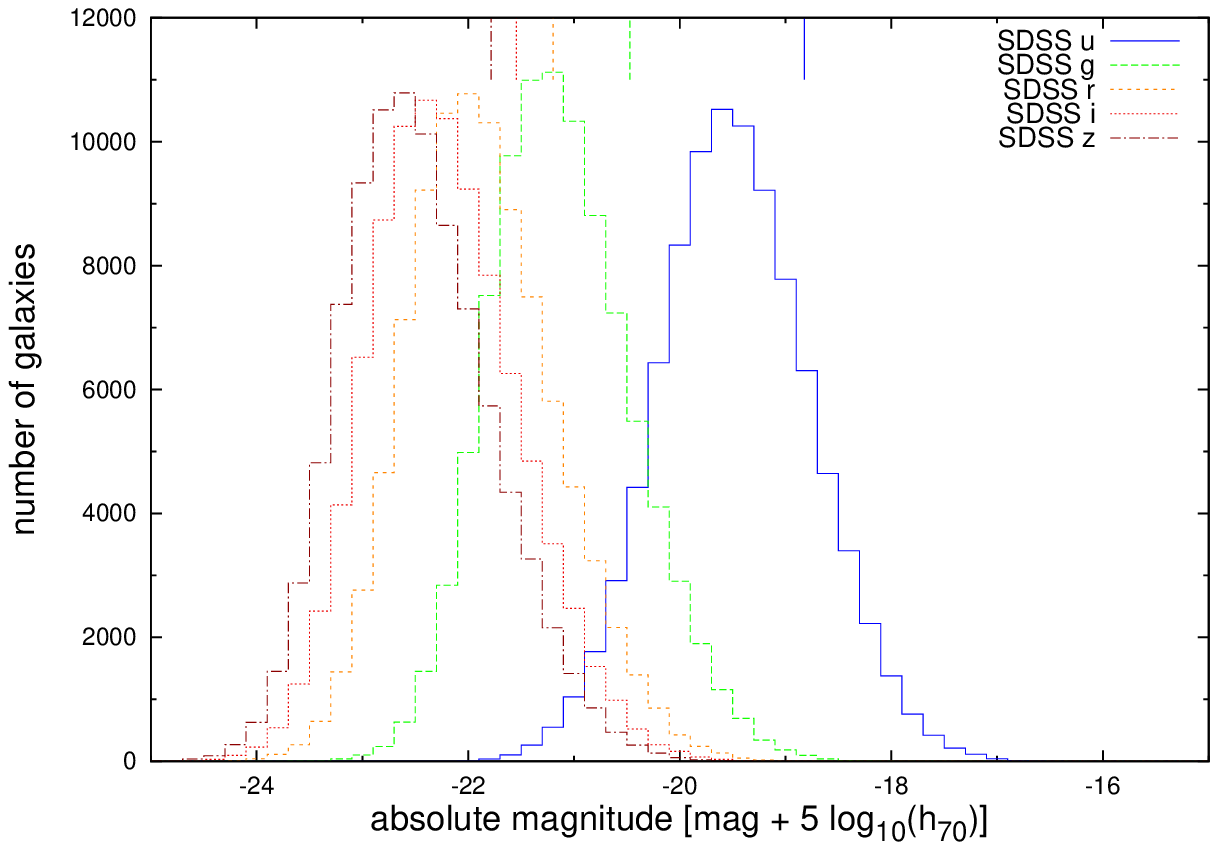}\\ 
\caption{Comparison of the luminosity functions in different filters for the c model. Although their shapes stay approximately the same, the peaks move to higher luminosities with redder filters. The short lines in the upper part of the plot indicate the Malmquist-bias-corrected mean magnitudes of our sample in the corresponding filters.}
\label{c_magabs_binned}
\end{center}
\end{figure}
\begin{figure}[H]
\begin{center}
\includegraphics[width=0.45\textwidth]{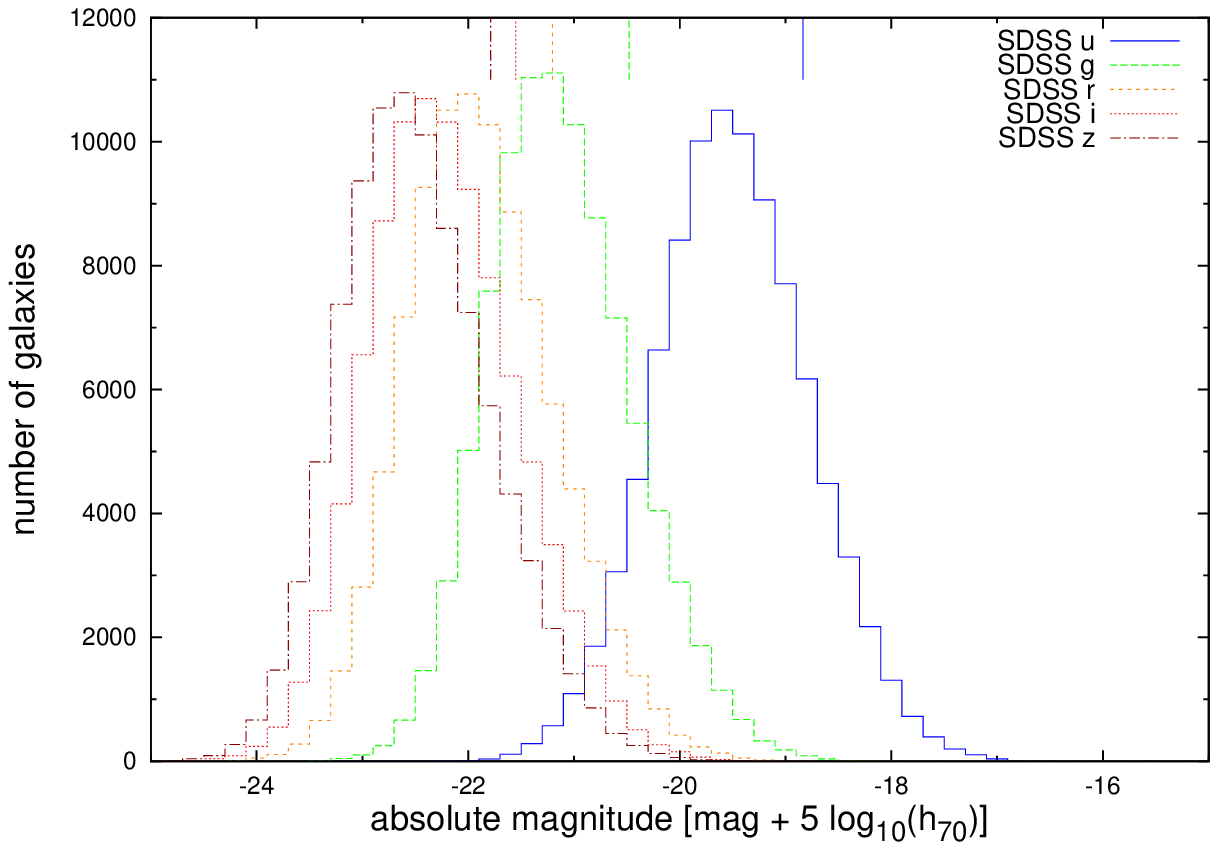}\\ 
\caption{Comparison of the luminosity functions in different filters for the dV model. Although their shapes stay approximately the same, the peaks move to higher luminosities with redder filters. The short lines in the upper part of the plot indicate the Malmquist-bias-corrected mean magnitudes of our sample in the corresponding filters.}
\label{dV_magabs_binned}
\end{center}
\end{figure}
\begin{figure}[H]
\begin{center}
\includegraphics[width=0.45\textwidth]{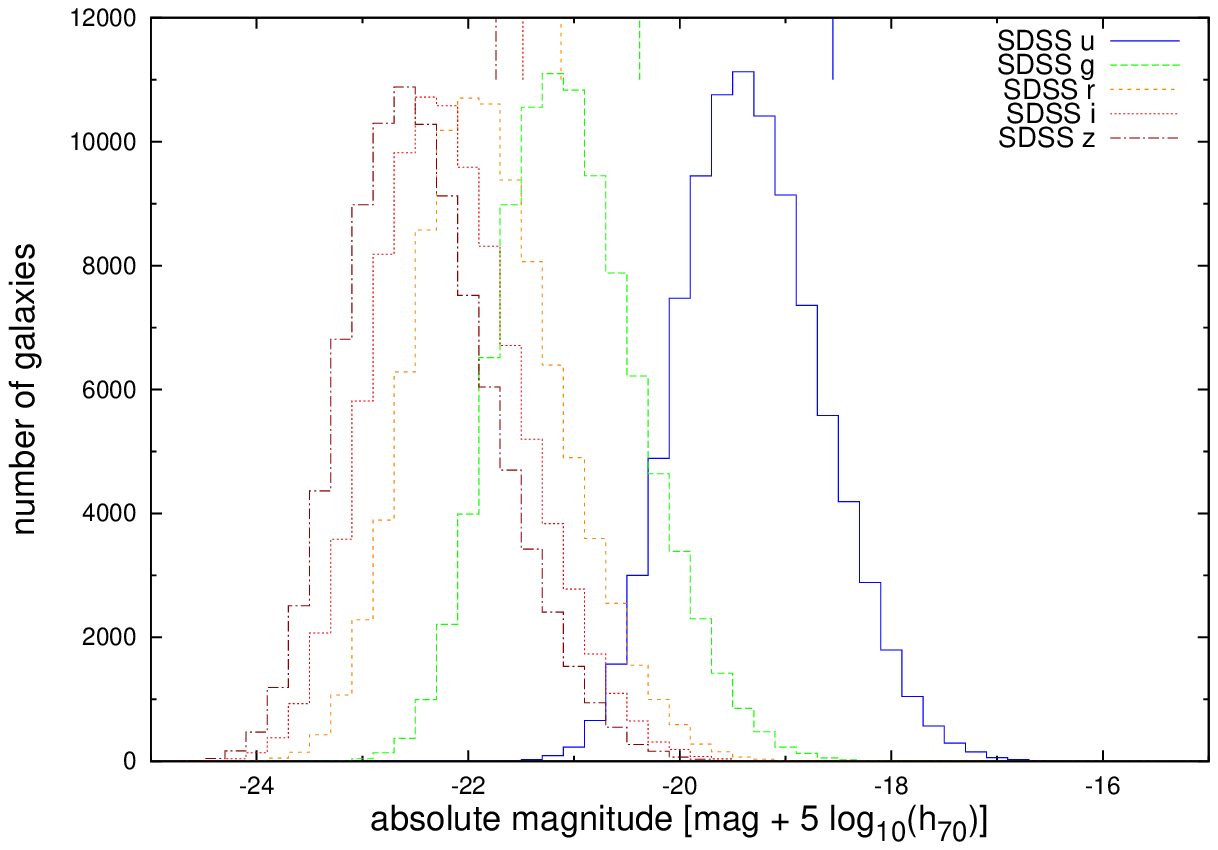}\\ 
\caption{Comparison of the luminosity functions in different filters for the p model. Although their shapes stay approximately the same, the peaks move to higher luminosities with redder filters. The short lines in the upper part of the plot indicate the Malmquist-bias-corrected mean magnitudes of our sample in the corresponding filters.}
\label{p_magabs_binned}
\end{center}
\end{figure}

\begin{figure}[H]
\begin{center}
\includegraphics[width=0.45\textwidth]{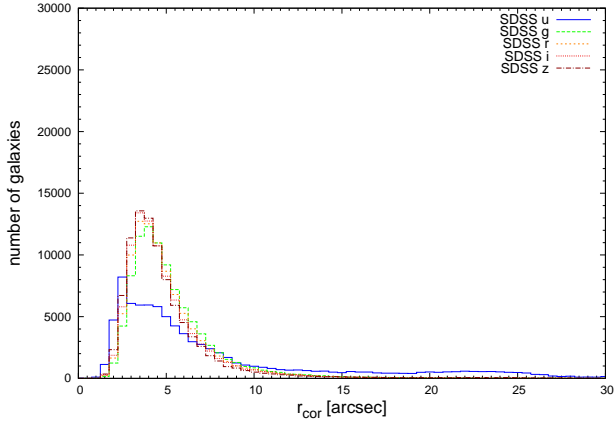}\\ 
\caption{Distribution of the apparent corrected radius $r_{\textrm{cor}}$ is displayed in different filters for the p model. The measured radii of this model are clearly larger than those of the dV model. Furthermore, the distribution is extremely spread out in the u band due to known problems in this filter.}
\label{p_rcor_binned}
\end{center}
\end{figure}

\begin{figure}[H]
\begin{center}
\includegraphics[width=0.45\textwidth]{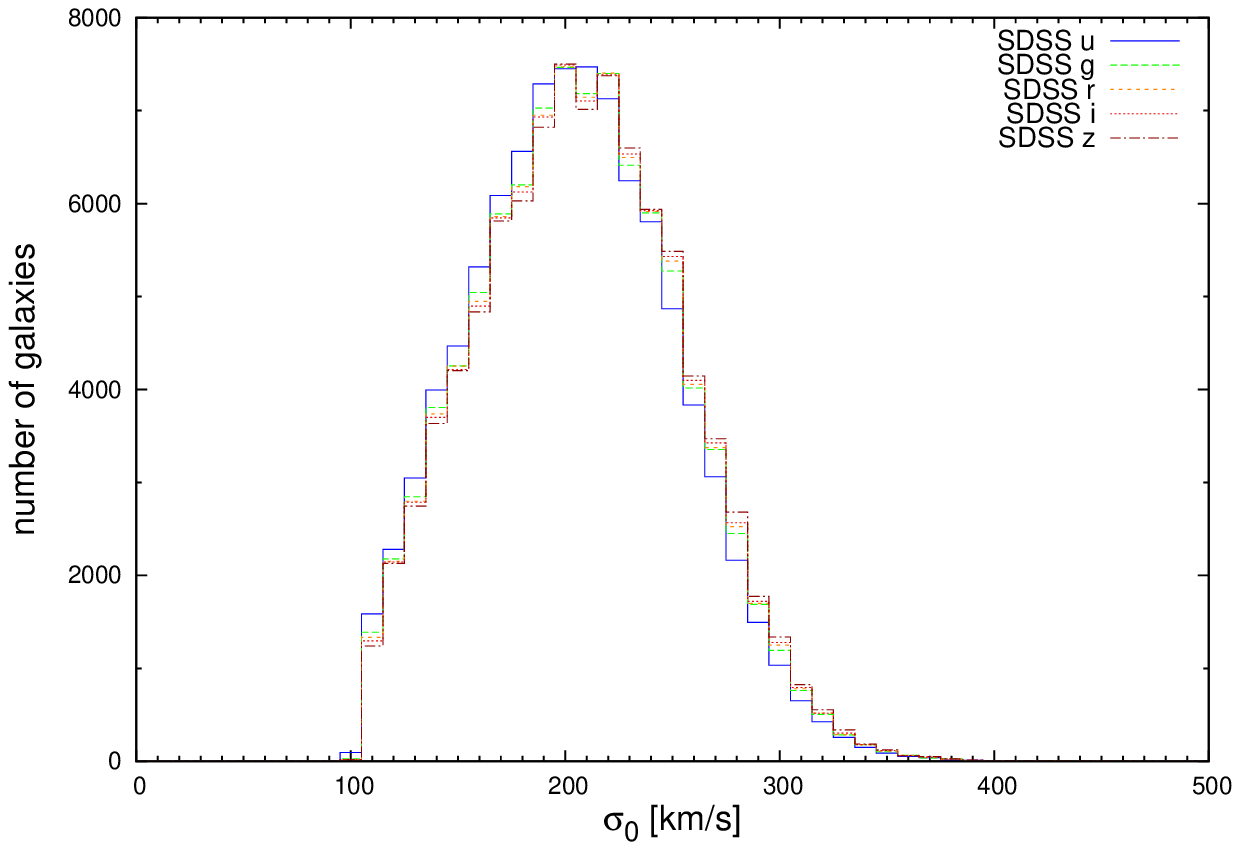}\\ 
\caption{Central velocity dispersion $\sigma_{0}$ for different filters (only slightly different in all of them due to the small correction for the fixed fibre diameters) for the c and the dV model. One can clearly see the cut-off of at 100 km/s, which has been introduced to avoid the contamination of our sample.}
\label{c_sigma_cor_binned}
\end{center}
\end{figure}
\begin{figure}[H]
\begin{center}
\includegraphics[width=0.45\textwidth]{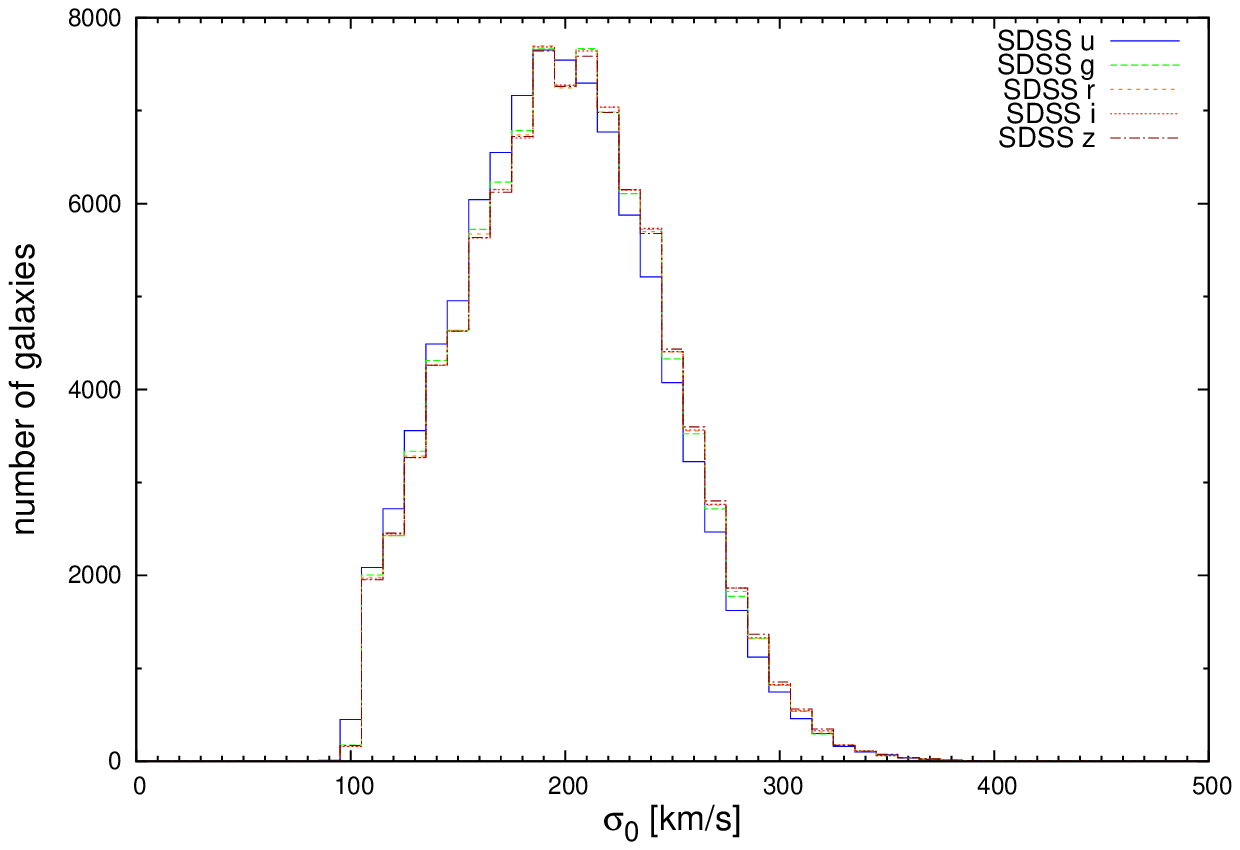}\\ 
\caption{Central velocity dispersion $\sigma_{0}$ for different filters (only slightly different in all of them due to the small correction for the fixed fibre diameters) for the p model. One can clearly see the cut-off of at 100 km/s, which has been introduced to avoid the contamination of our sample.}
\label{p_sigma_cor_binned}
\end{center}
\end{figure}

\begin{figure}[H]
\begin{center}
\includegraphics[width=0.45\textwidth]{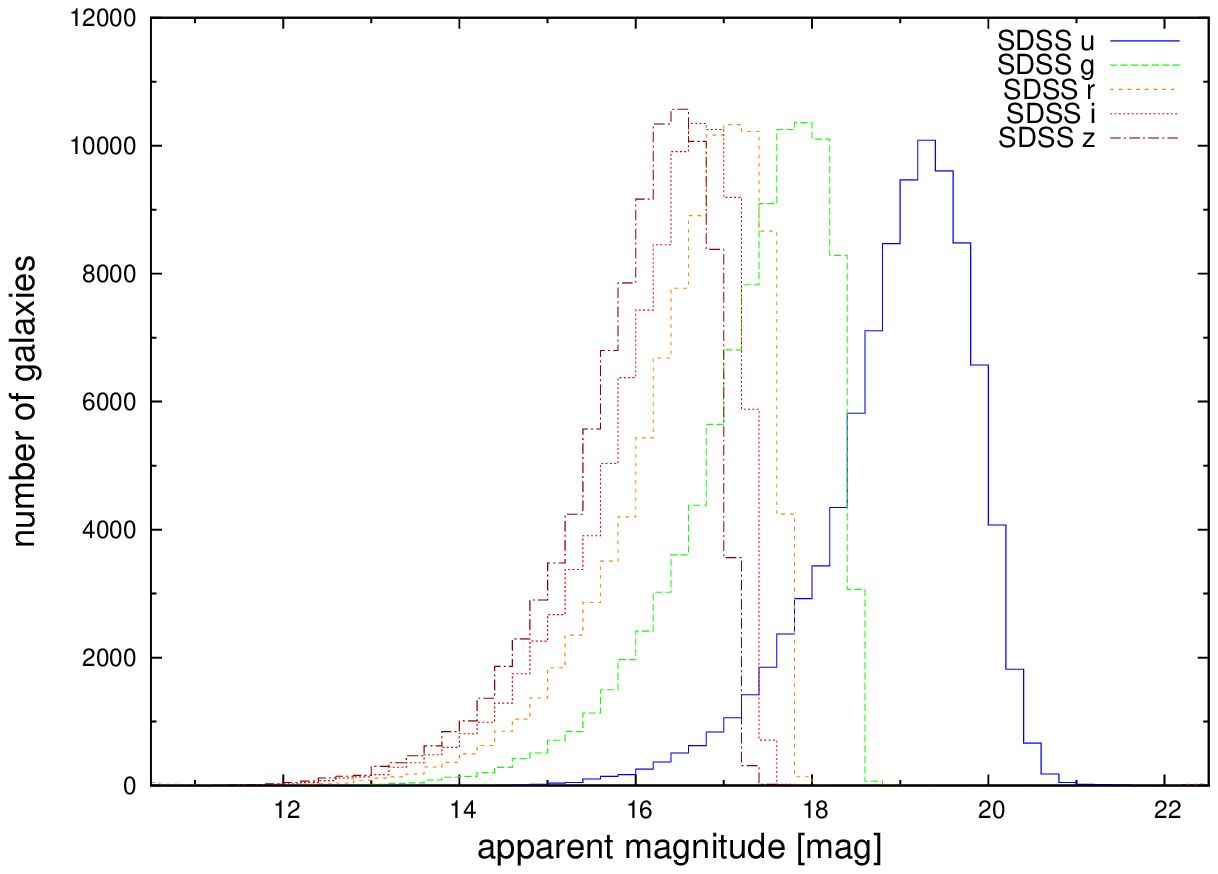}\\ 
\caption{Distribution of extinction- and K-corrected apparent magnitudes $m_{\textrm{app}}$ in different filters for the c model, showing a steady increase in numbers until the steep cut-off at the sample's limiting magnitudes, which are listed in Table \ref{malmquist_fit_maglimit}.}
\label{c_mag_app_binned}
\end{center}
\end{figure}
\begin{figure}[H]
\begin{center}
\includegraphics[width=0.45\textwidth]{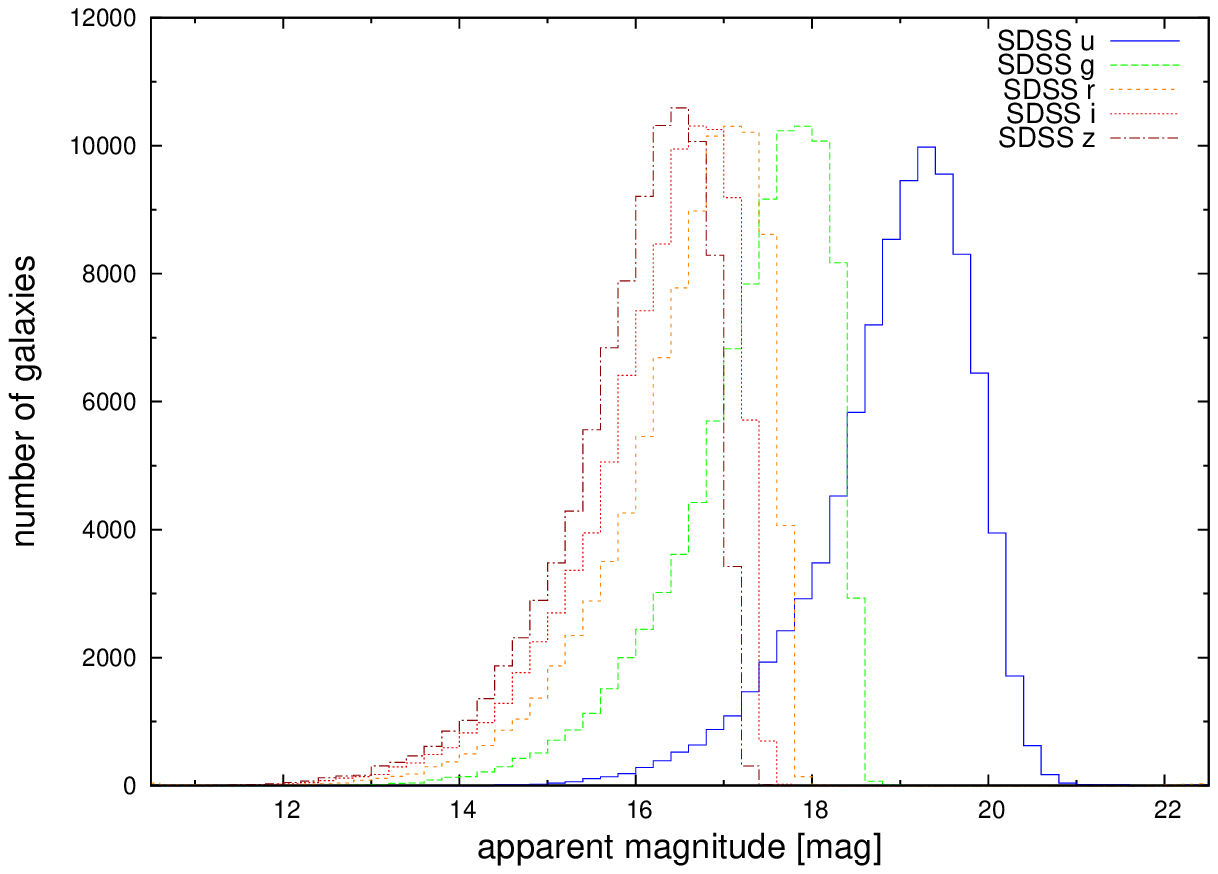}\\ 
\caption{Distribution of extinction- and K-corrected apparent magnitudes $m_{\textrm{app}}$ in different filters for the dV model, showing a steady increase in numbers until the steep cut-off at the sample's limiting magnitudes, which are listed in Table \ref{malmquist_fit_maglimit}.}
\label{dV_mag_app_binned}
\end{center}
\end{figure}
\begin{figure}[H]
\begin{center}
\includegraphics[width=0.45\textwidth]{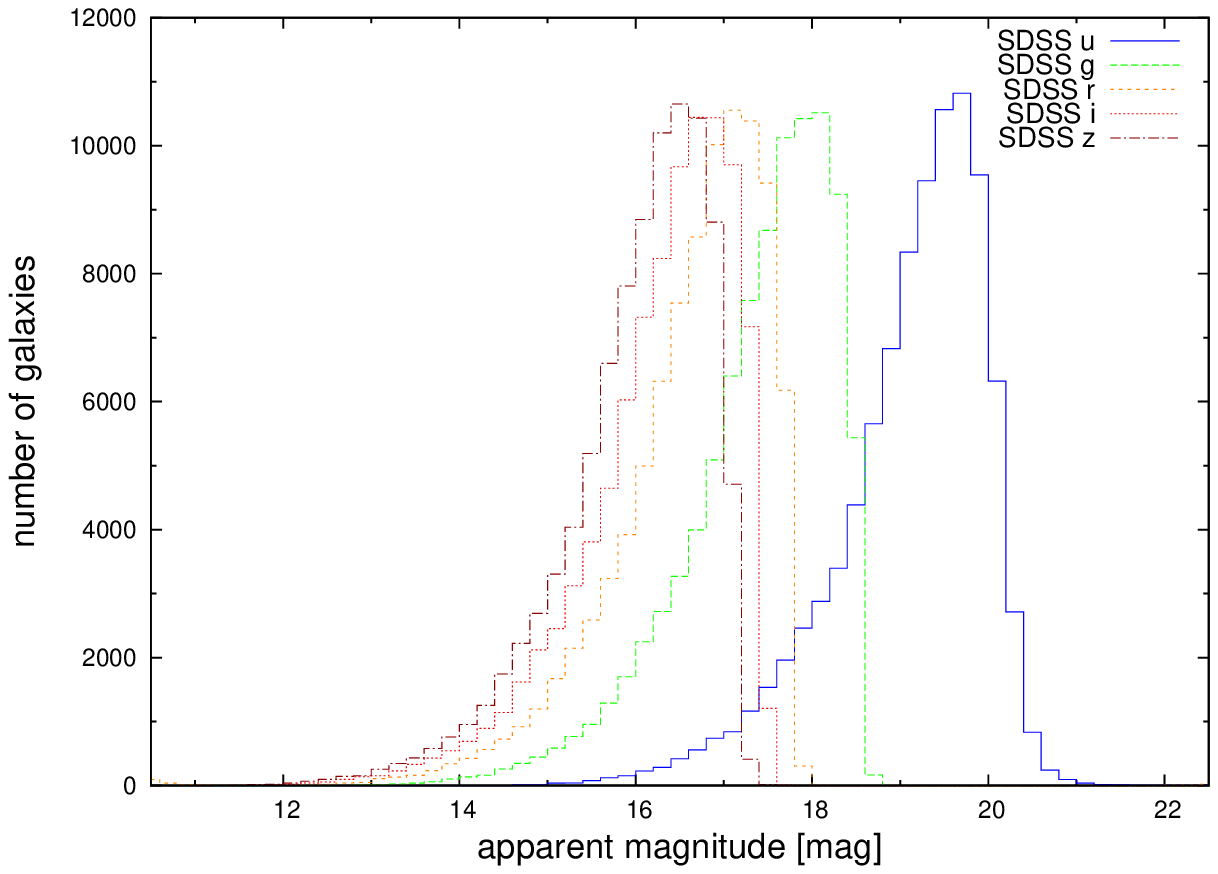}\\ 
\caption{Distribution of extinction- and K-corrected apparent magnitudes $m_{\textrm{app}}$ in different filters for the p model, showing a steady increase in numbers until the steep cut-off at the sample's limiting magnitudes, which are listed in Table \ref{malmquist_fit_maglimit}.}
\label{p_mag_app_binned}
\end{center}
\end{figure}

\begin{figure}[H]
\begin{center}
\includegraphics[width=0.45\textwidth]{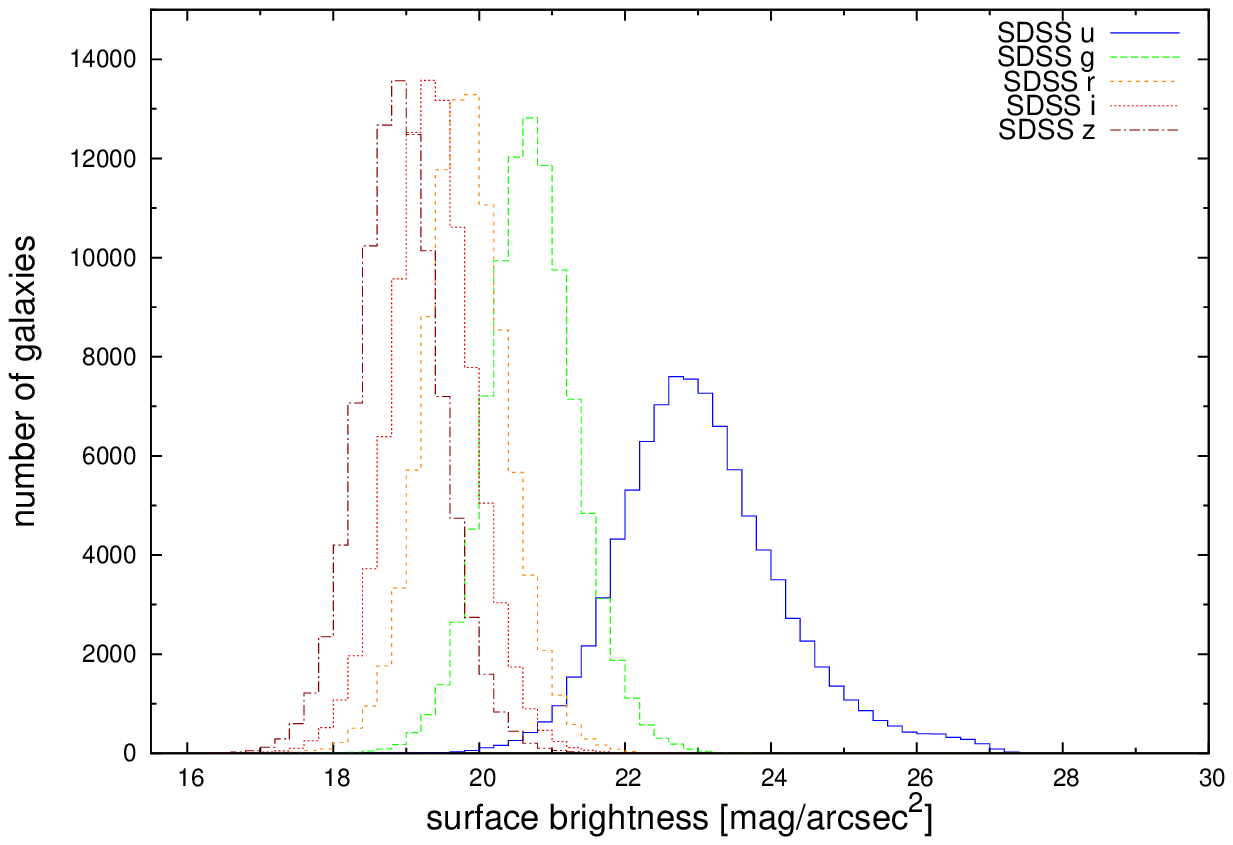}\\ 
\caption{Distribution of the surface brightness $\mu_{0}$ in different filters for the c model showing an almost Gaussian shape. For the u band, the distribution is wider and shows a small bump at the faint end.}
\label{c_sb_binned}
\end{center}
\end{figure}

\begin{figure}[H]
\begin{center}
\includegraphics[width=0.45\textwidth]{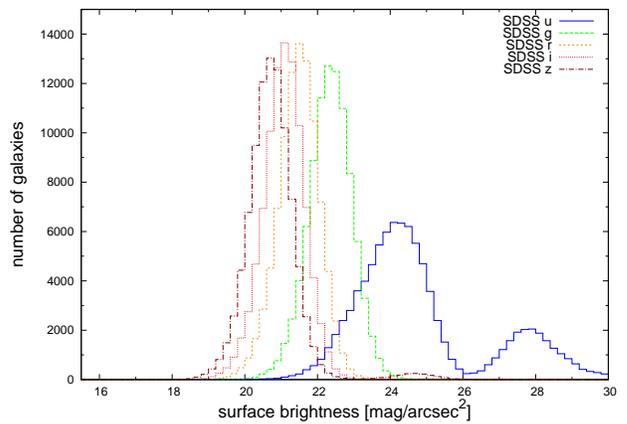}\\ 
\caption{Distribution of the surface brightness $\mu_{0}$ in different filters for the p model shows some peculiar features in the u band and to some smaller extent in z band as well. In these two filters, one can see a clear second peak on the faint side of the main Gaussian.}
\label{p_sb_binned}
\end{center}
\end{figure}

\begin{figure}[H]
\begin{center}
\includegraphics[width=0.45\textwidth]{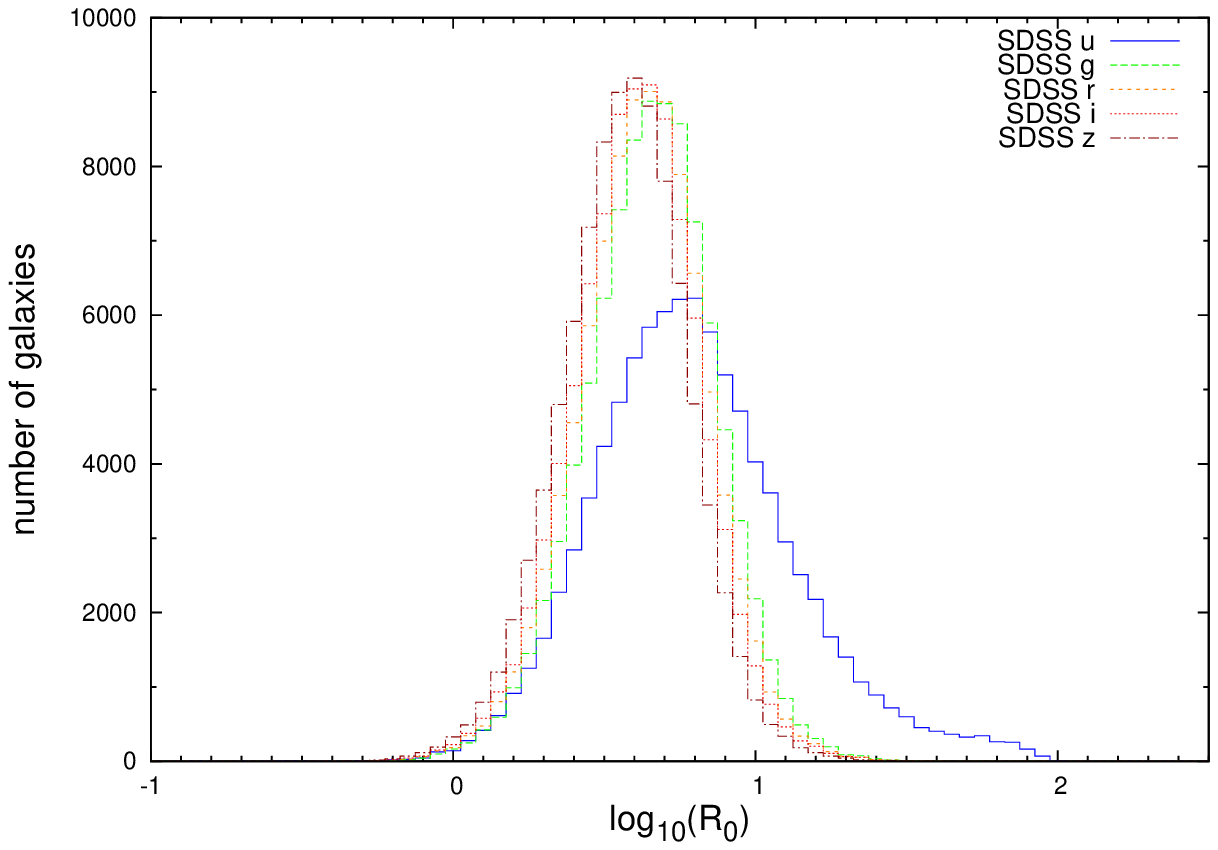}\\ 
\caption{Distributions of the logarithm of the physical radius $\textrm{log}_{10}(R_{0})$ in different filters for the c model are well described by sharp Gaussian with their peaks almost exactly at the same value. Only the u band shows some digressive behaviour. In this case the peak is smaller and set apart from the other. Furthermore, the distribution is wider and shows a small bump at the larger end.}
\label{c_logR0_binned}
\end{center}
\end{figure}

\begin{figure}[H]
\begin{center}
\includegraphics[width=0.45\textwidth]{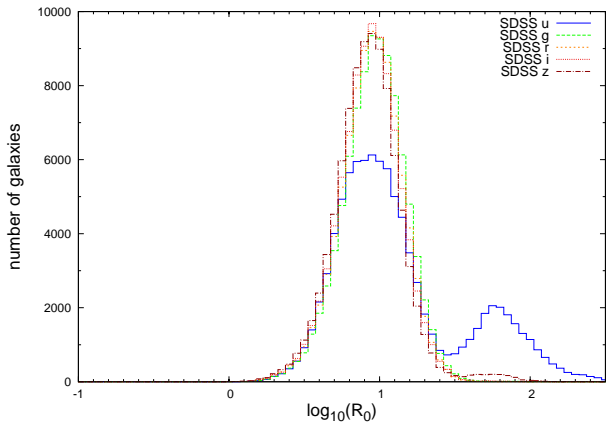}\\ 
\caption{Distributions of the logarithm of the physical radius $\textrm{log}_{10}(R_{0})$ in different filters for the p model are well described by sharp Gaussian with their peaks almost exactly at the same value. However, the u band shows a peculiar second peak aside the consequently smaller (in comparison to the other filters) main one. In addition to this deviation, the z band distribution has a small bump at its larger end.}
\label{p_logR0_binned}
\end{center}
\end{figure}

\begin{figure}[H]
\begin{center}
\includegraphics[width=0.45\textwidth]{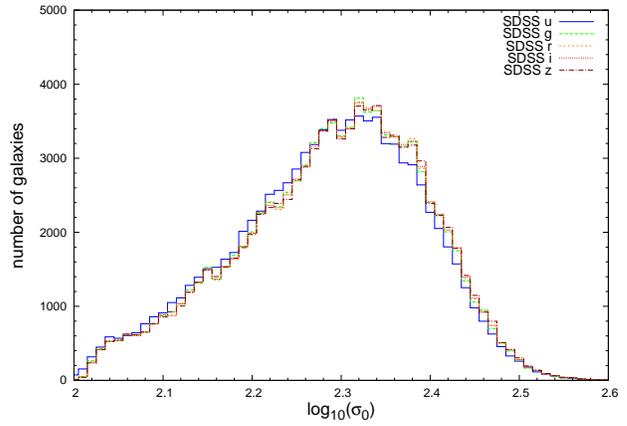}\\ 
\caption{Distributions of the logarithm of the central velocity dispersion $\textrm{log}_{10}(R_{0})$ in different filters for the p model are almost exactly the same for all filters. They show an general abundance (compared with a perfect Gaussian) of galaxies at the lower end, which might indicate some residual contamination of the sample.}
\label{p_logsigma_binned}
\end{center}
\end{figure}

\begin{figure}[H]
\begin{center}
\includegraphics[width=0.45\textwidth]{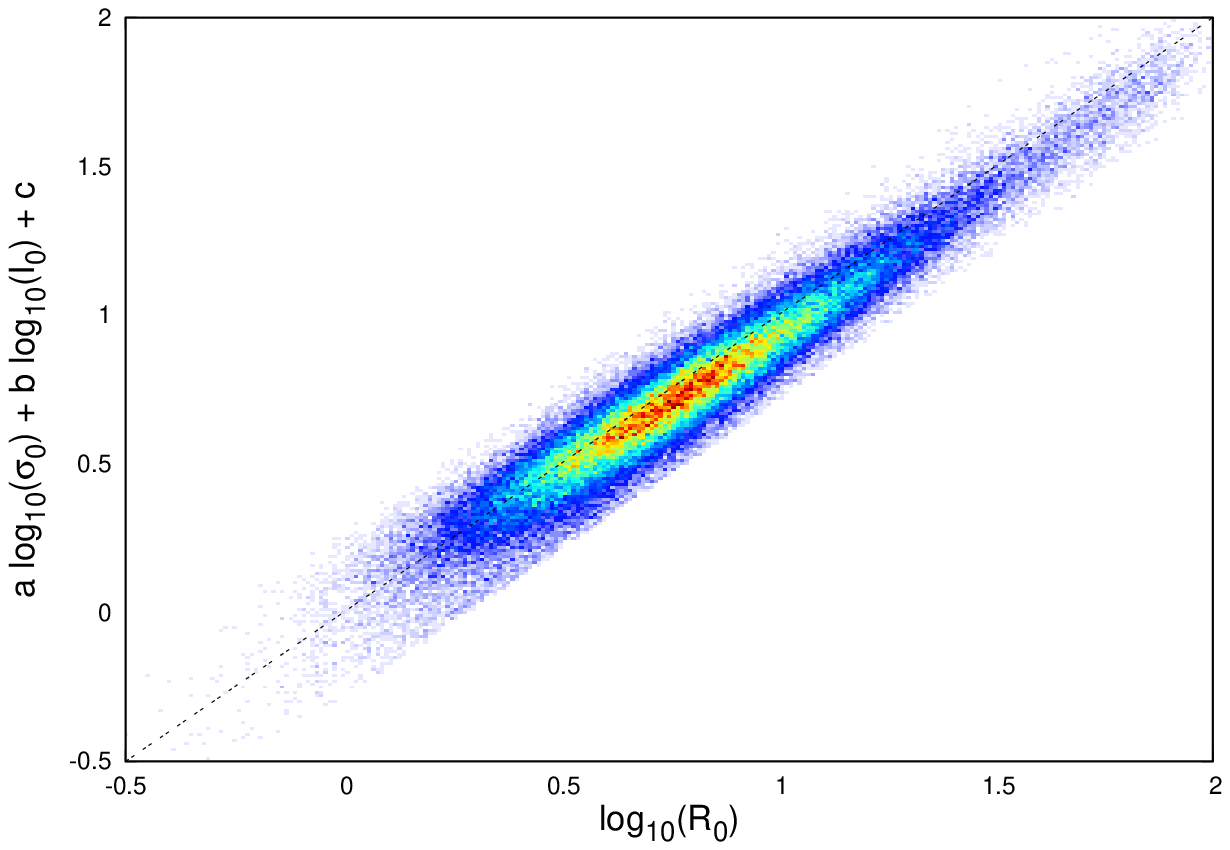}\\ 
\caption{Projection of the fundamental plane for the u band of the c model.}
\label{c_fp_u_poster}
\end{center}
\end{figure}
\begin{figure}[H]
\begin{center}
\includegraphics[width=0.45\textwidth]{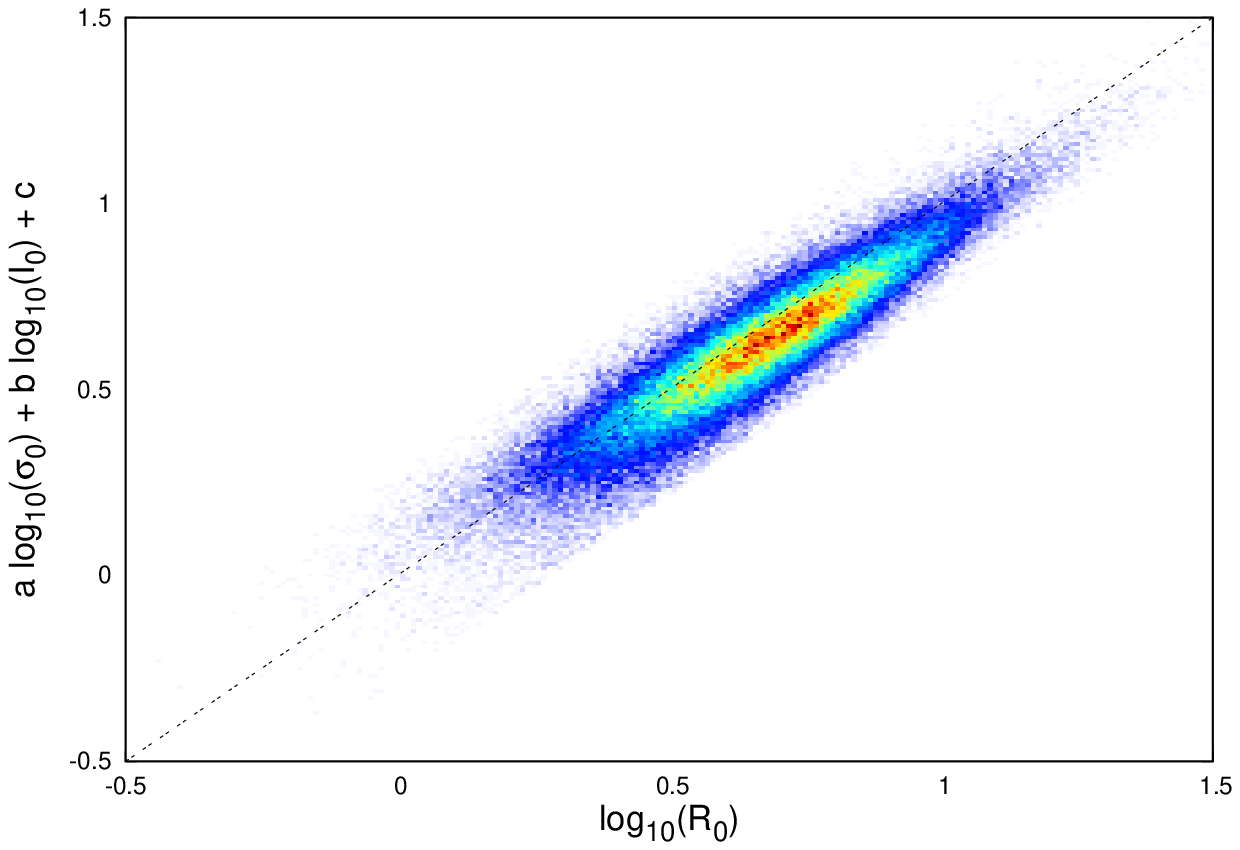}\\ 
\caption{Projection of the fundamental plane for the g band of the c model.}
\label{c_fp_g_poster}
\end{center}
\end{figure}
\begin{figure}[H]
\begin{center}
\includegraphics[width=0.45\textwidth]{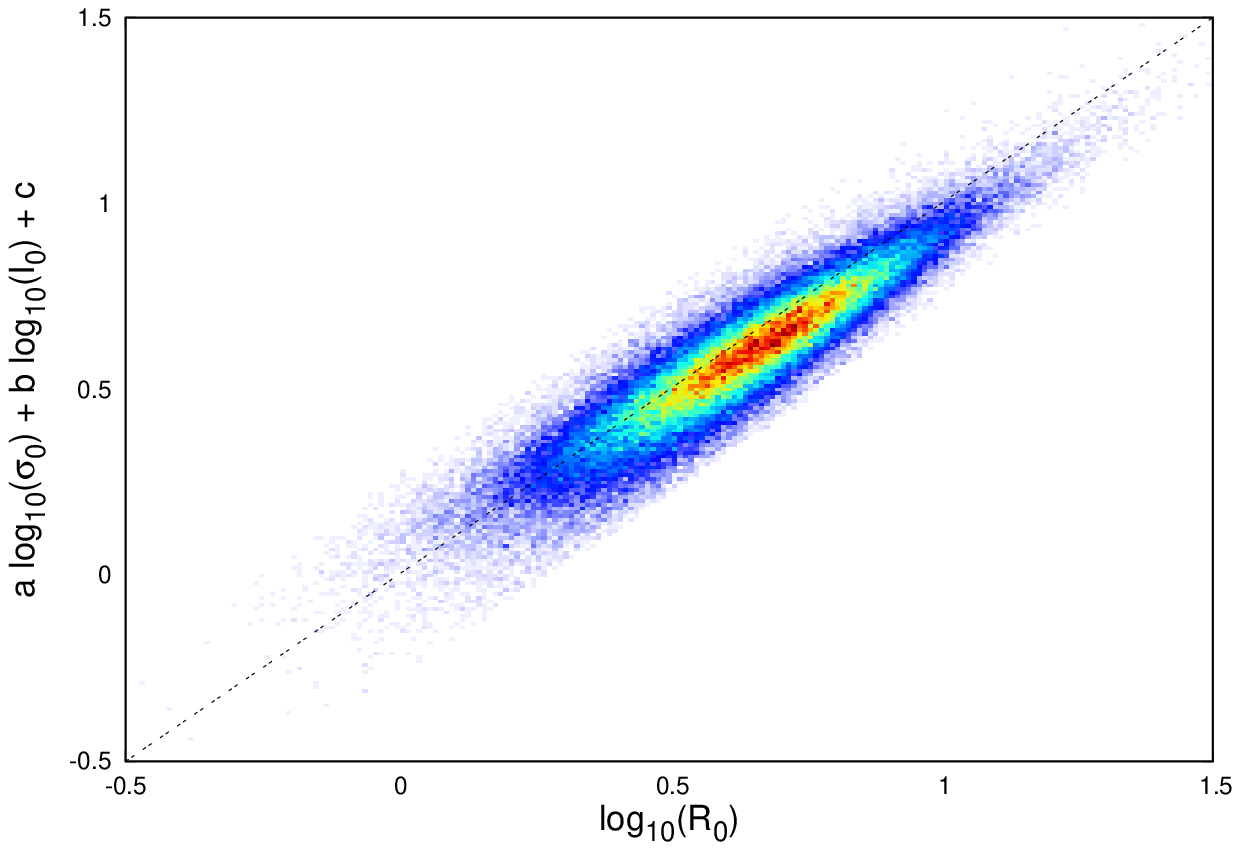}\\ 
\caption{Projection of the fundamental plane for the r band of the c model.}
\label{c_fp_r_poster}
\end{center}
\end{figure}
\begin{figure}[H]
\begin{center}
\includegraphics[width=0.45\textwidth]{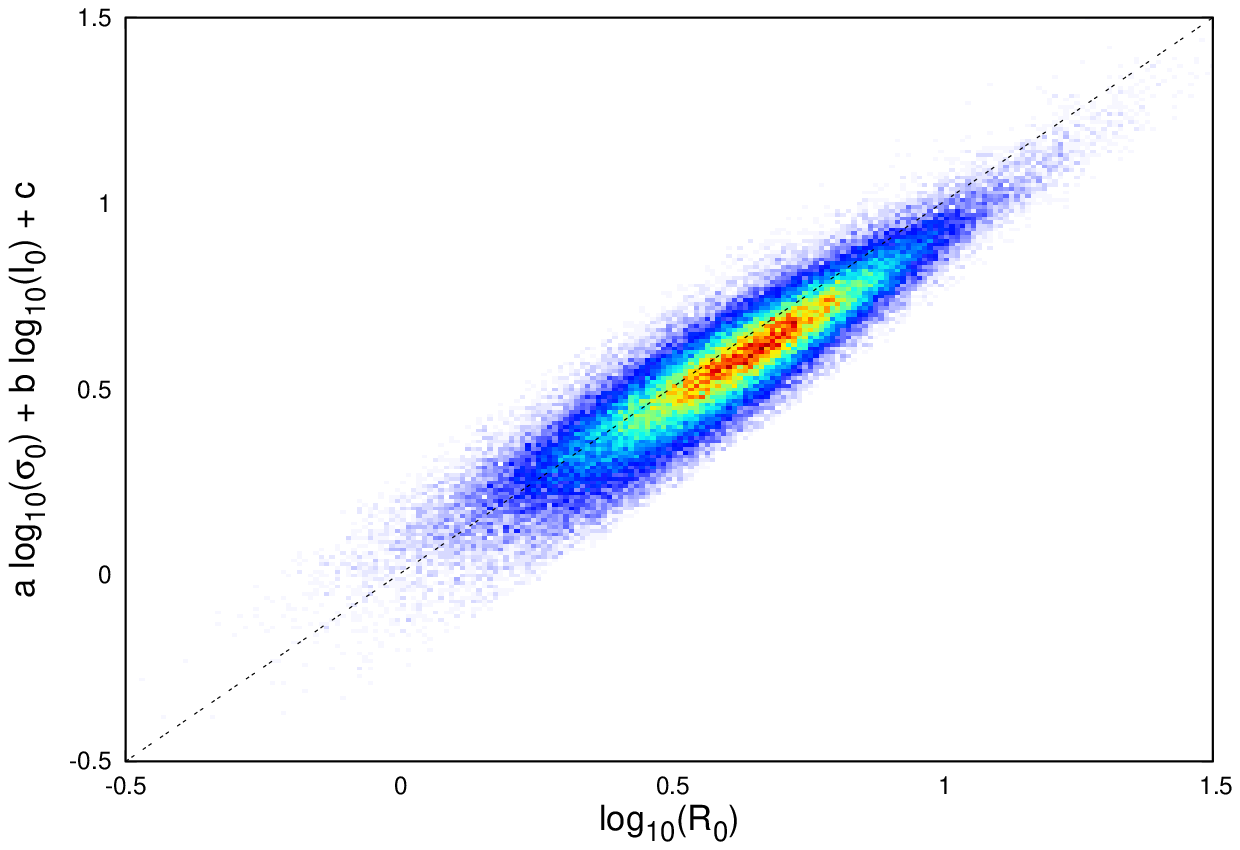}\\ 
\caption{Projection of the fundamental plane for the i band of the c model.}
\label{c_fp_i_poster}
\end{center}
\end{figure}
\begin{figure}[H]
\begin{center}
\includegraphics[width=0.45\textwidth]{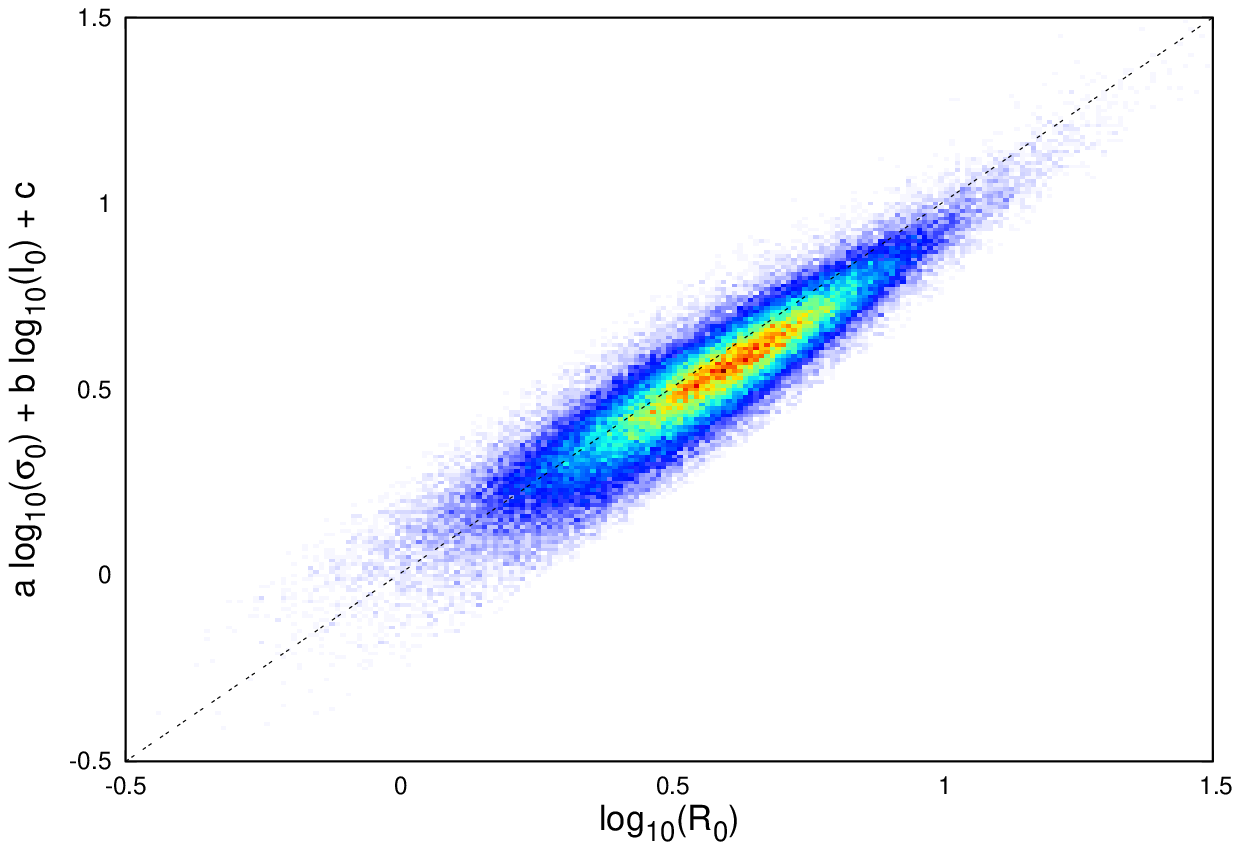}\\ 
\caption{Projection of the fundamental plane for the z band of the c model.}
\label{c_fp_z_poster}
\end{center}
\end{figure}

\begin{figure}[H]
\begin{center}
\includegraphics[width=0.45\textwidth]{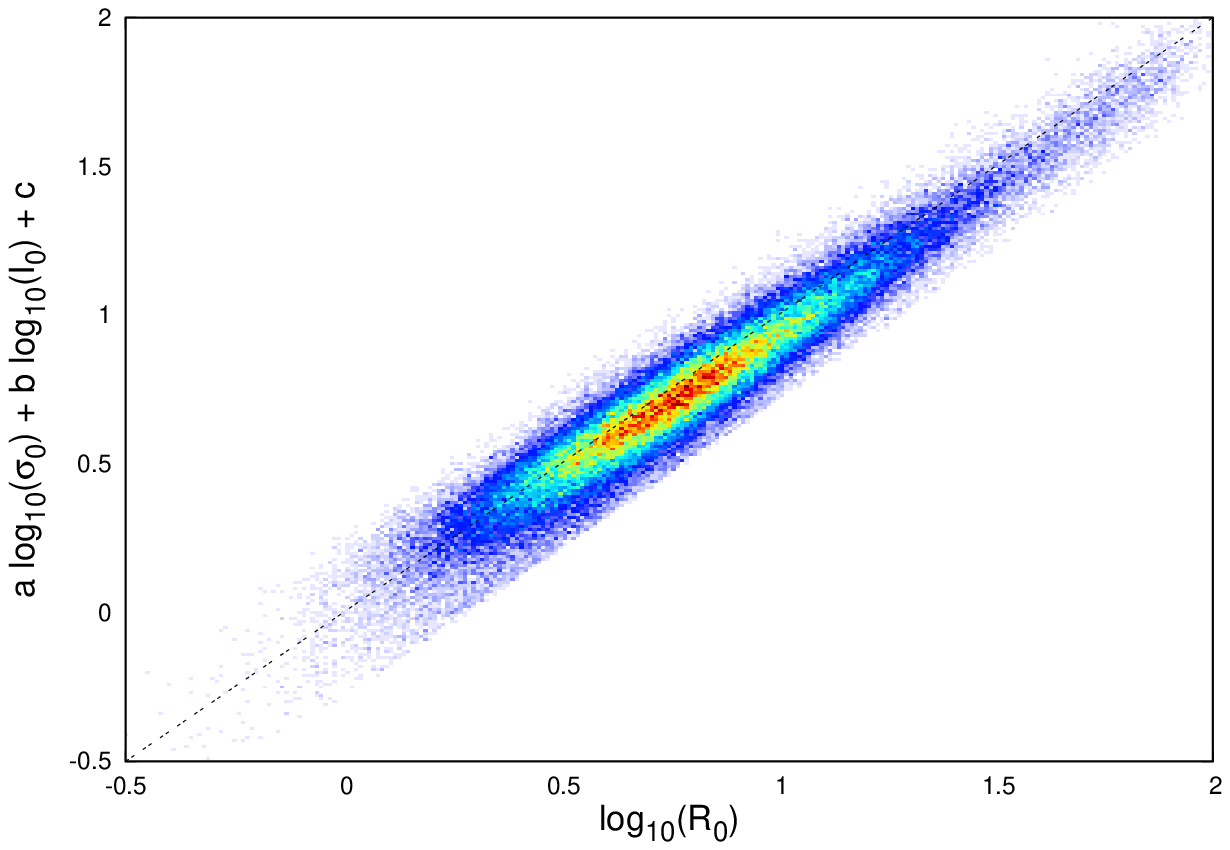}\\ 
\caption{Projection of the fundamental plane for the u band of the dV model.}
\label{dV_fp_u_poster}
\end{center}
\end{figure}
\begin{figure}[H]
\begin{center}
\includegraphics[width=0.45\textwidth]{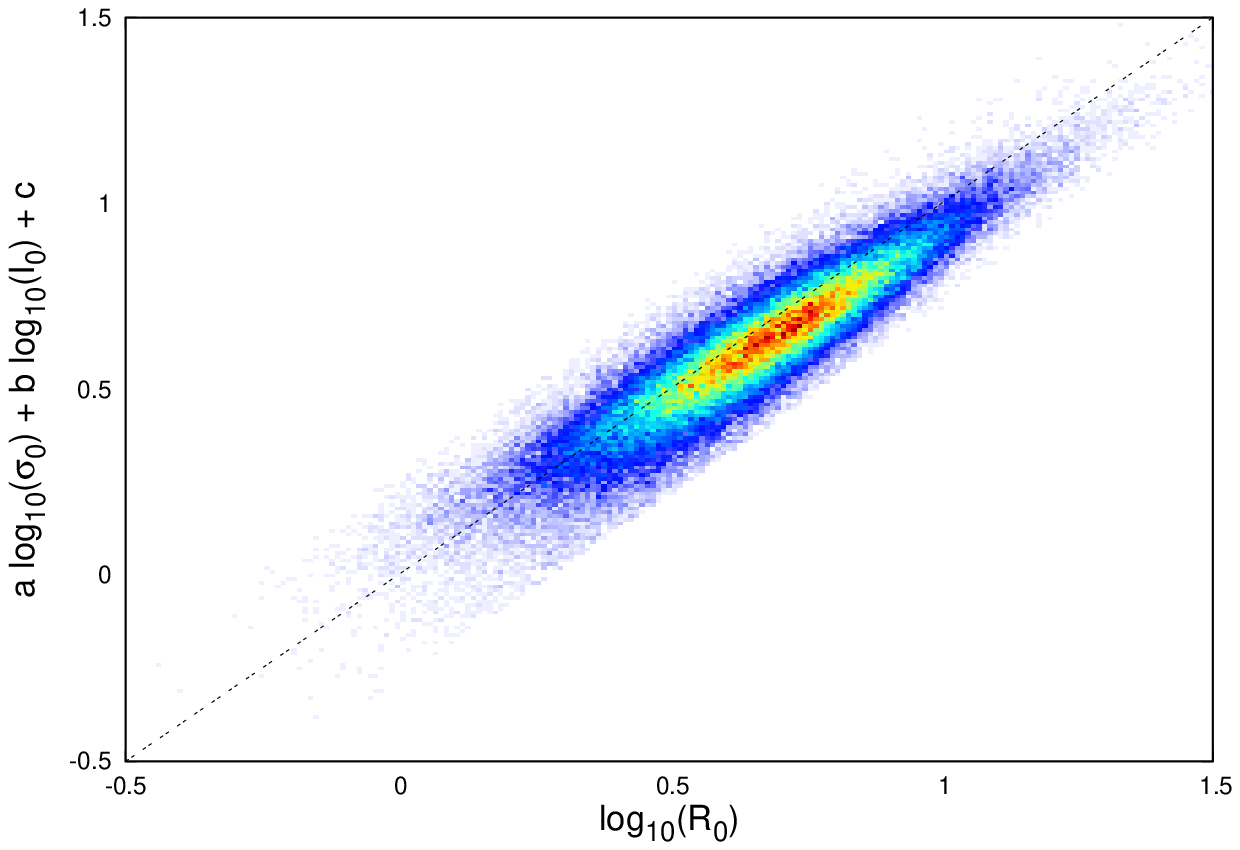}\\ 
\caption{Projection of the fundamental plane for the g band of the dV model.}
\label{dV_fp_g_poster}
\end{center}
\end{figure}
\begin{figure}[H]
\begin{center}
\includegraphics[width=0.45\textwidth]{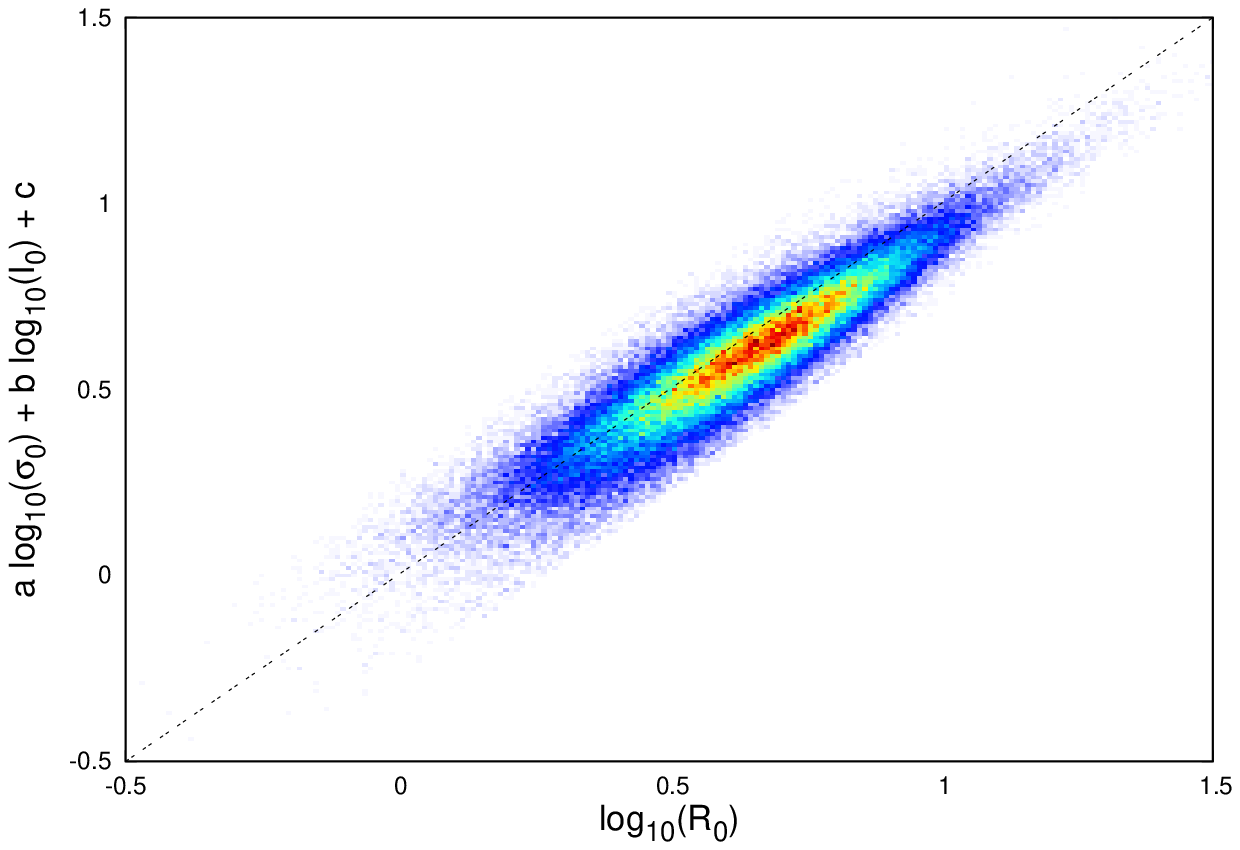}\\ 
\caption{Projection of the fundamental plane for the r band of the dV model.}
\label{dV_fp_r_poster}
\end{center}
\end{figure}

\begin{figure}[H]
\begin{center}
\includegraphics[width=0.45\textwidth]{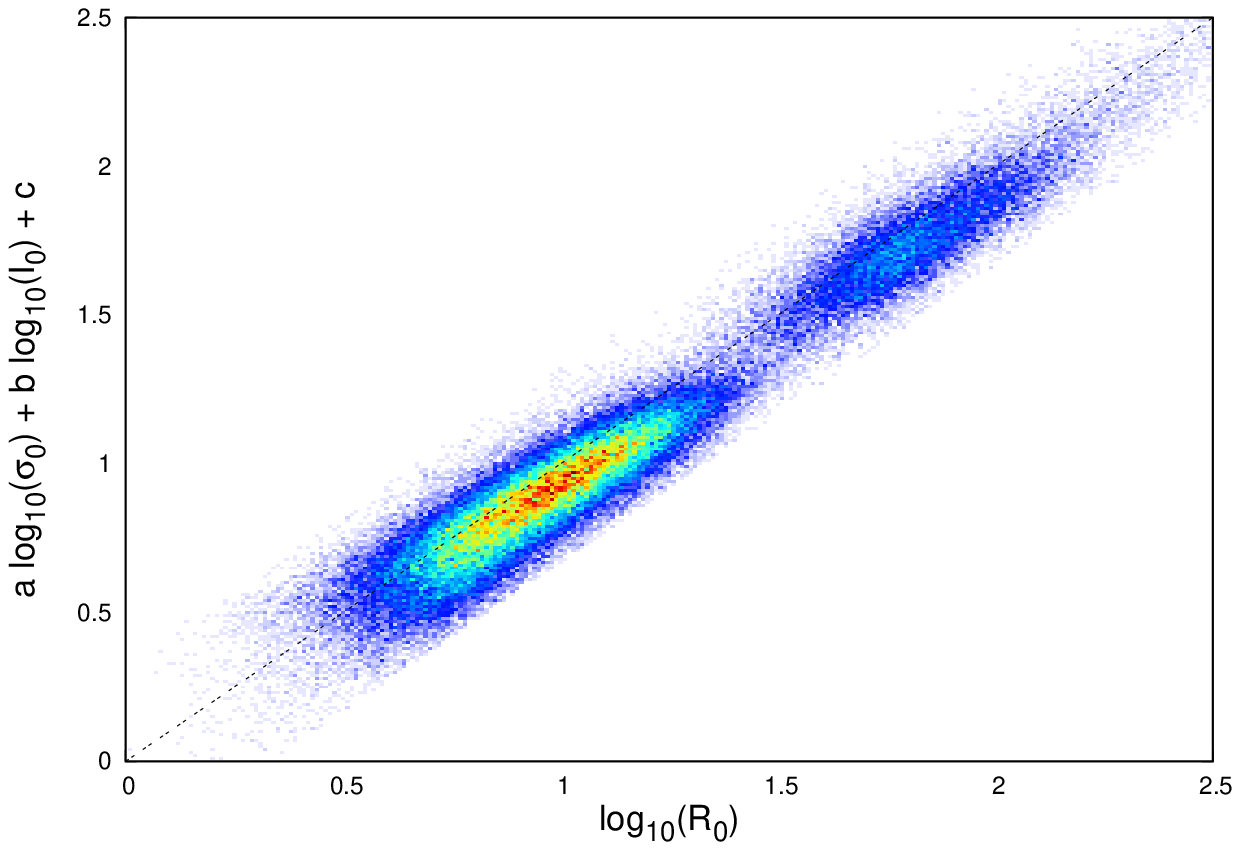}\\ 
\caption{Projection of the fundamental plane for the u band of the p model.}
\label{p_fp_u_poster}
\end{center}
\end{figure}
\begin{figure}[H]
\begin{center}
\includegraphics[width=0.45\textwidth]{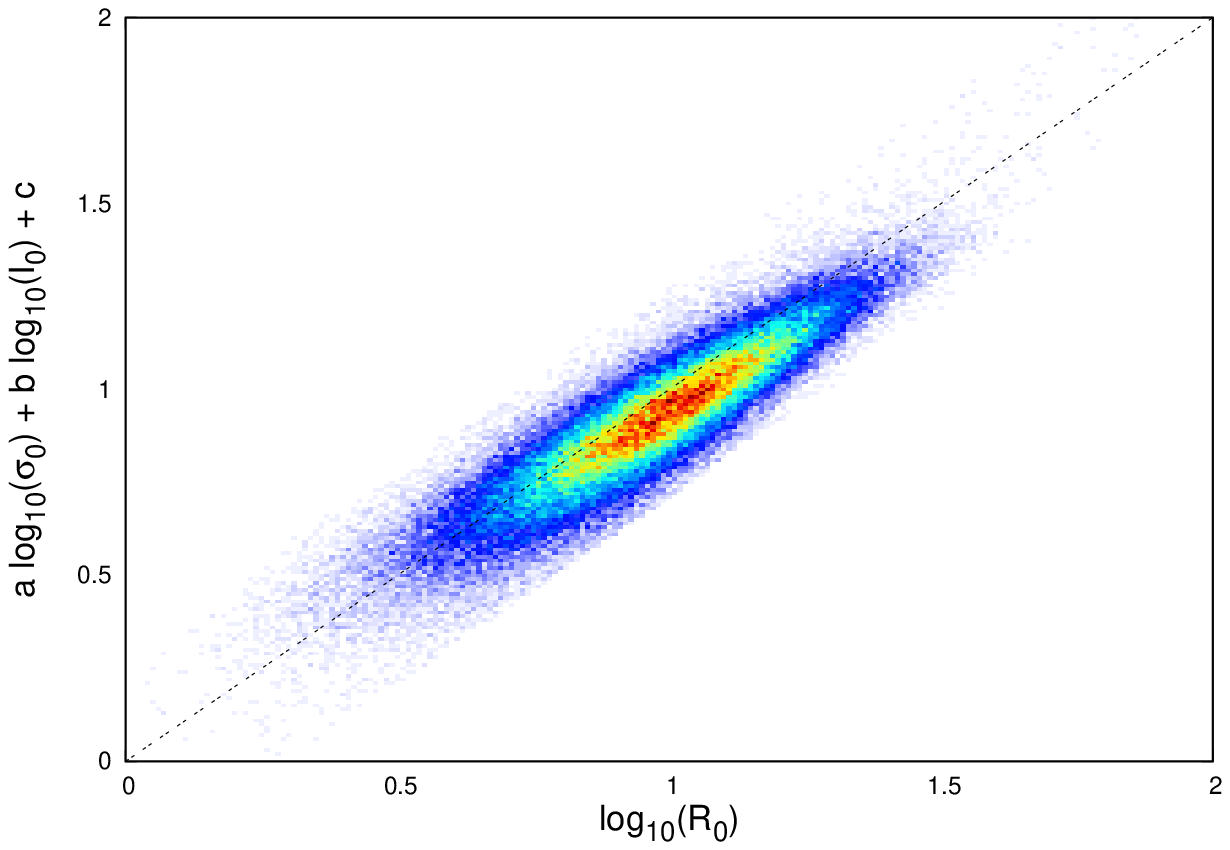}\\ 
\caption{Projection of the fundamental plane for the g band of the p model.}
\label{p_fp_g_poster}
\end{center}
\end{figure}
\begin{figure}[H]
\begin{center}
\includegraphics[width=0.45\textwidth]{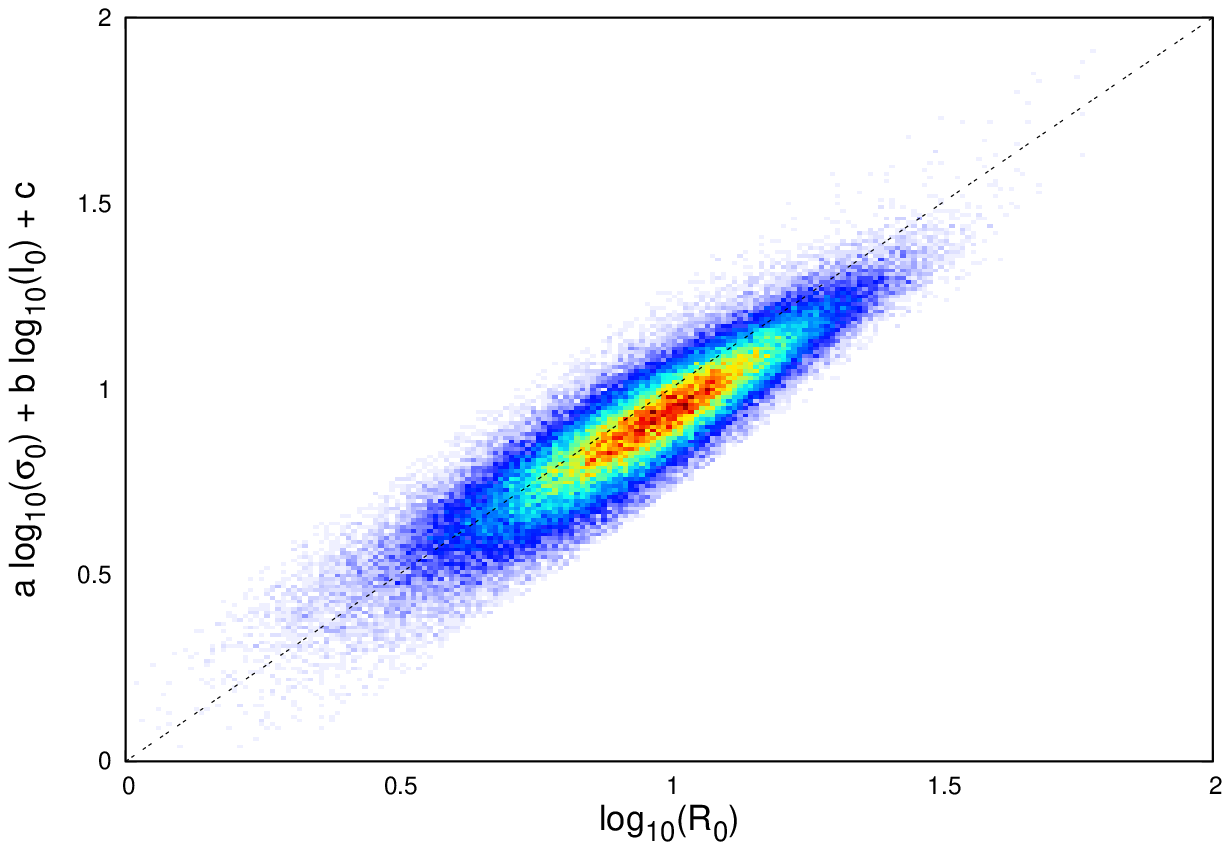}\\ 
\caption{Projection of the fundamental plane for the r band of the p model.}
\label{p_fp_r_poster}
\end{center}
\end{figure}
\begin{figure}[H]
\begin{center}
\includegraphics[width=0.45\textwidth]{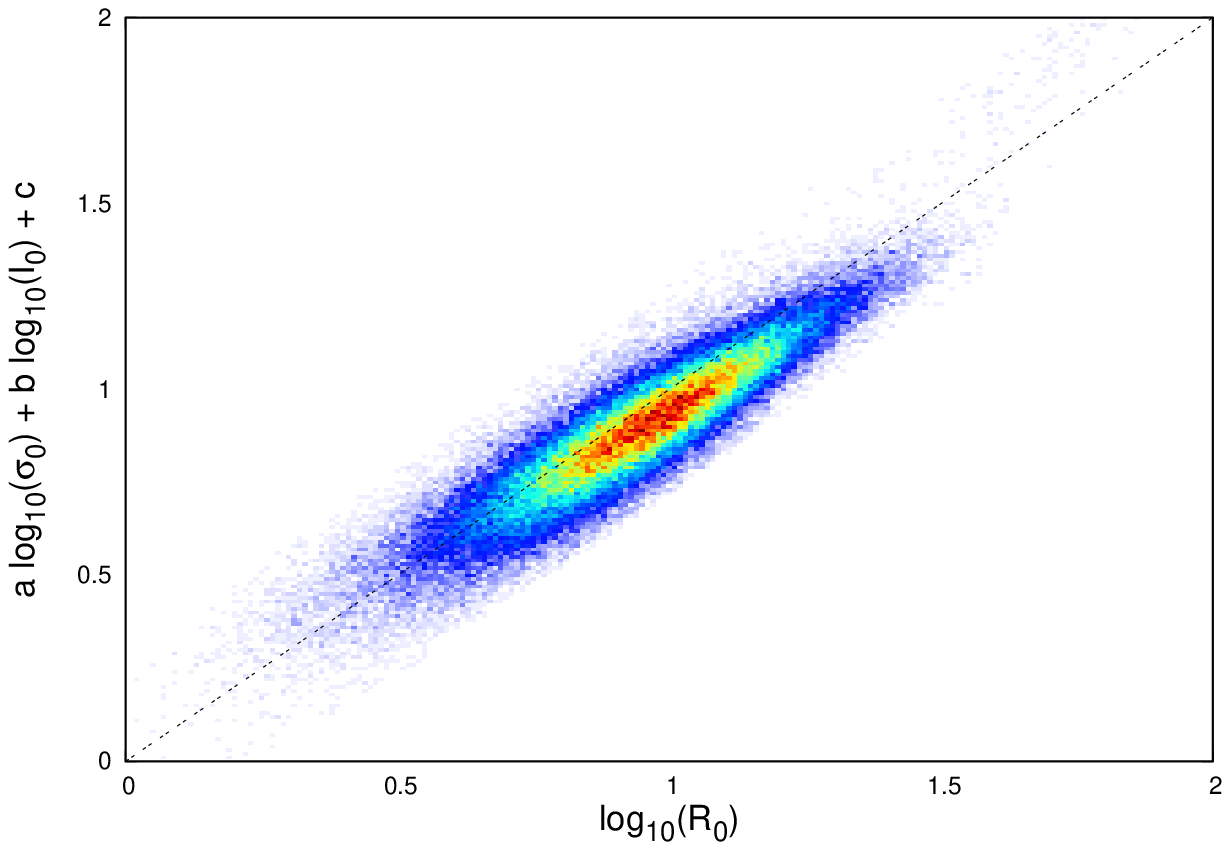}\\ 
\caption{Projection of the fundamental plane for the i band of the p model.}
\label{p_fp_i_poster}
\end{center}
\end{figure}
\begin{figure}[H]
\begin{center}
\includegraphics[width=0.45\textwidth]{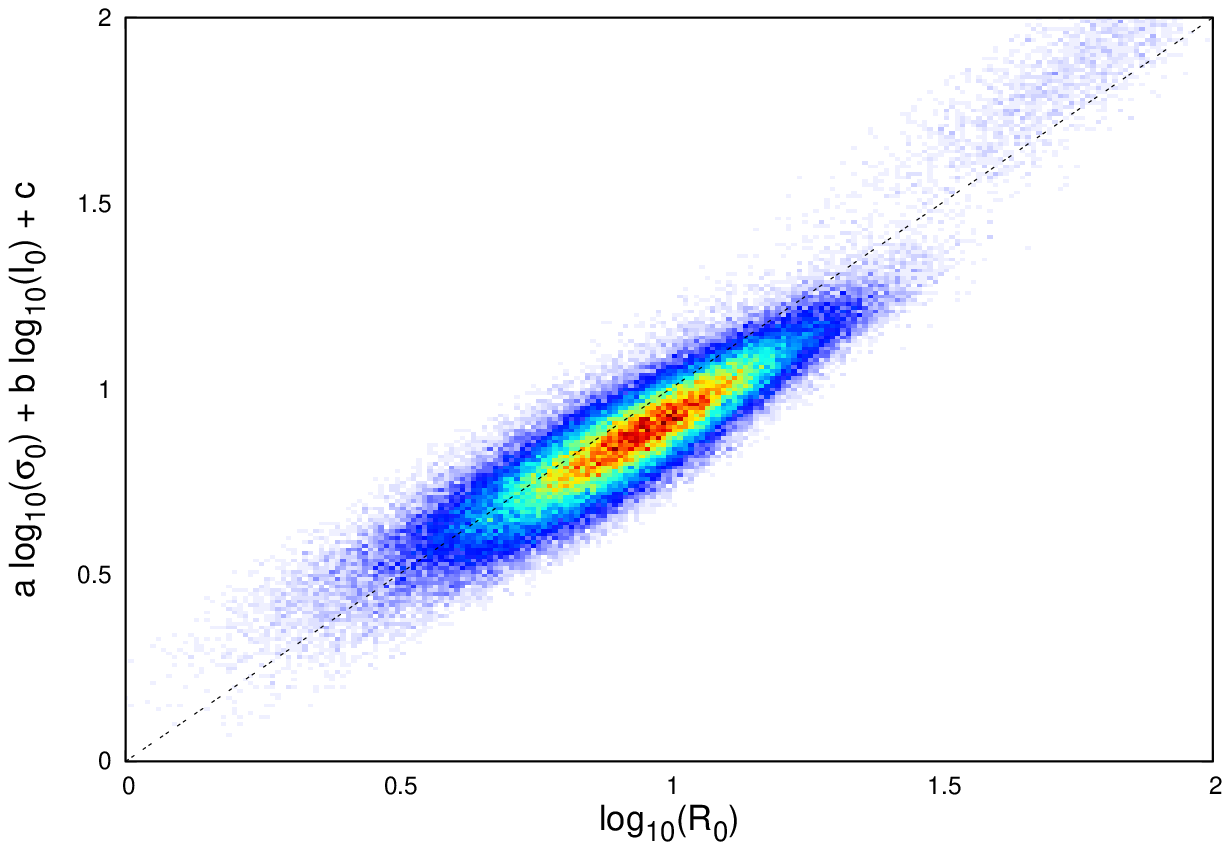}\\ 
\caption{Projection of the fundamental plane for the z band of the p model.}
\label{p_fp_z_poster}
\end{center}
\end{figure}

\begin{figure}[H]
\begin{center}
\includegraphics[width=0.45\textwidth]{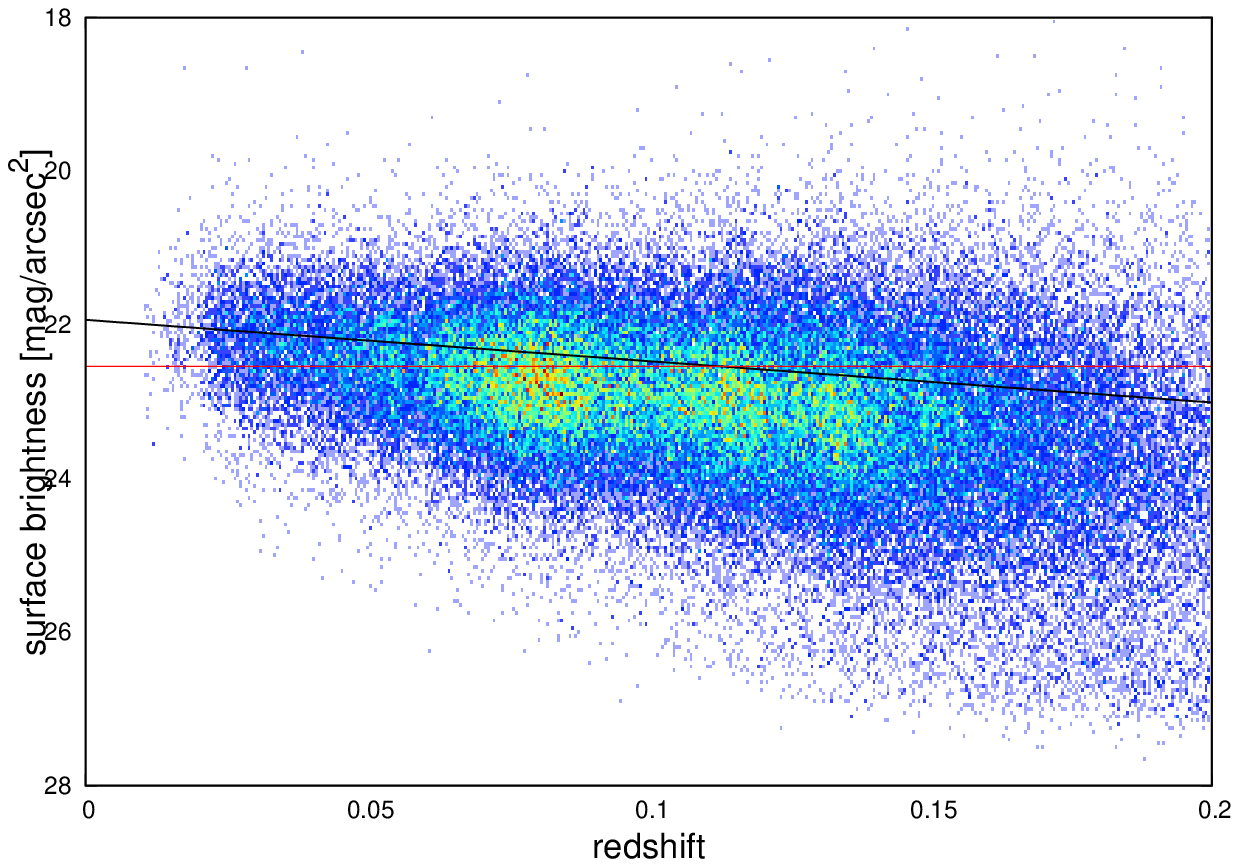}\\ 
\caption{Redshift evolution of the surface brightness in the u band of dV model indicated by the solid black line. The solid red line shows the Malmquist-bias-corrected average value of the surface brightness for this particular filter and model.}
\label{dV_z_dependence_sb_u}
\end{center}
\end{figure}
\begin{figure}[H]
\begin{center}
\includegraphics[width=0.45\textwidth]{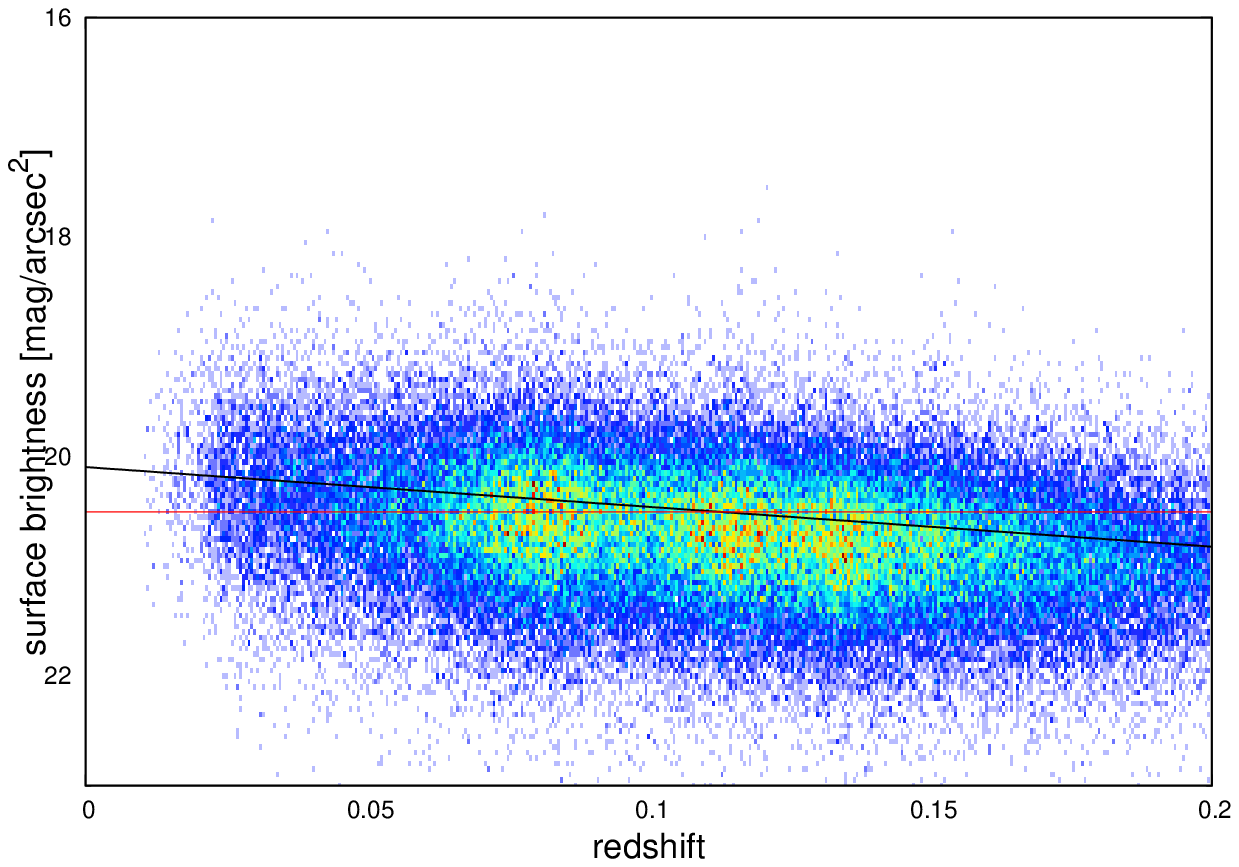}\\ 
\caption{Redshift evolution of the surface brightness in the g band of dV model indicated by the solid black line. The solid red line shows the Malmquist-bias-corrected average value of the surface brightness for this particular filter and model.}
\label{dV_z_dependence_sb_g}
\end{center}
\end{figure}
\begin{figure}[H]
\begin{center}
\includegraphics[width=0.45\textwidth]{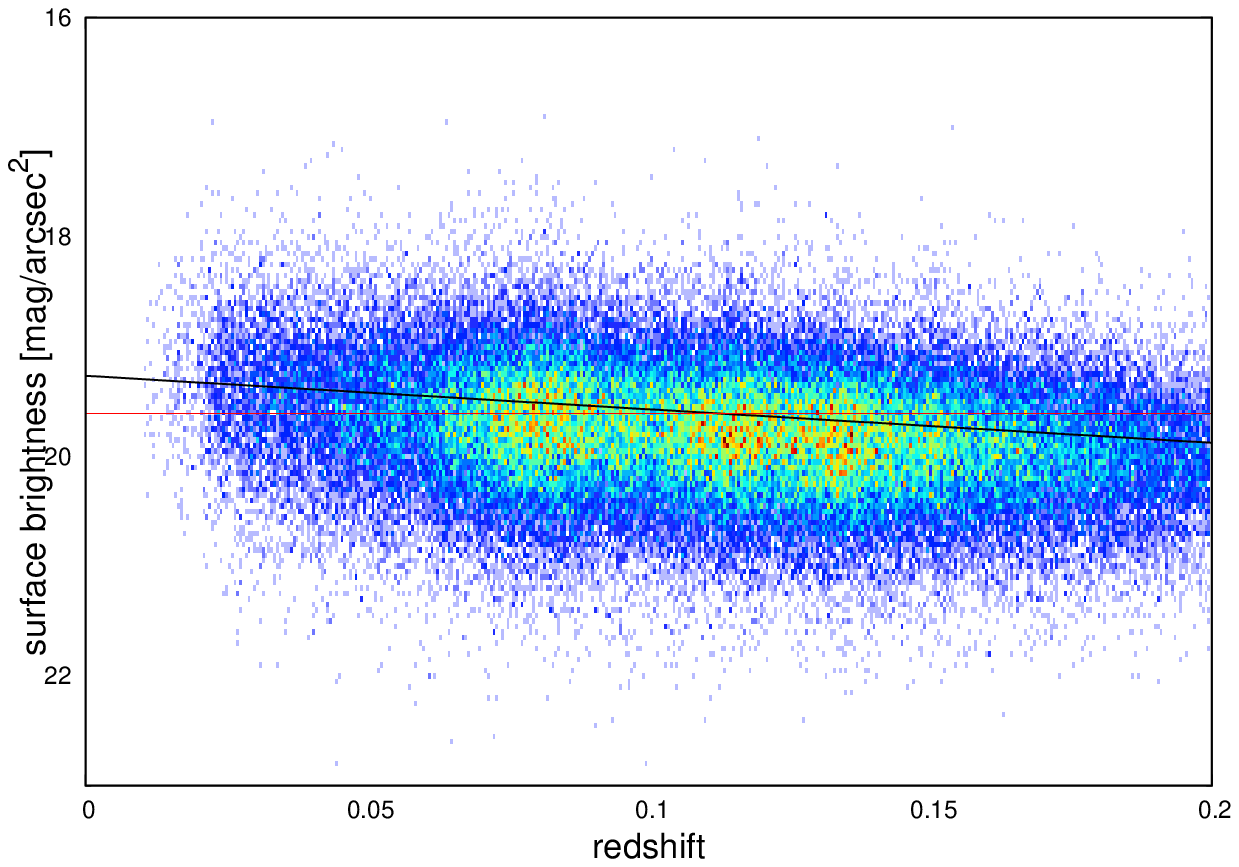}\\ 
\caption{Redshift evolution of the surface brightness in the r band of dV model indicated by the solid black line. The solid red line shows the Malmquist-bias-corrected average value of the surface brightness for this particular filter and model.}
\label{dV_z_dependence_sb_r}
\end{center}
\end{figure}
\begin{figure}[H]
\begin{center}
\includegraphics[width=0.45\textwidth]{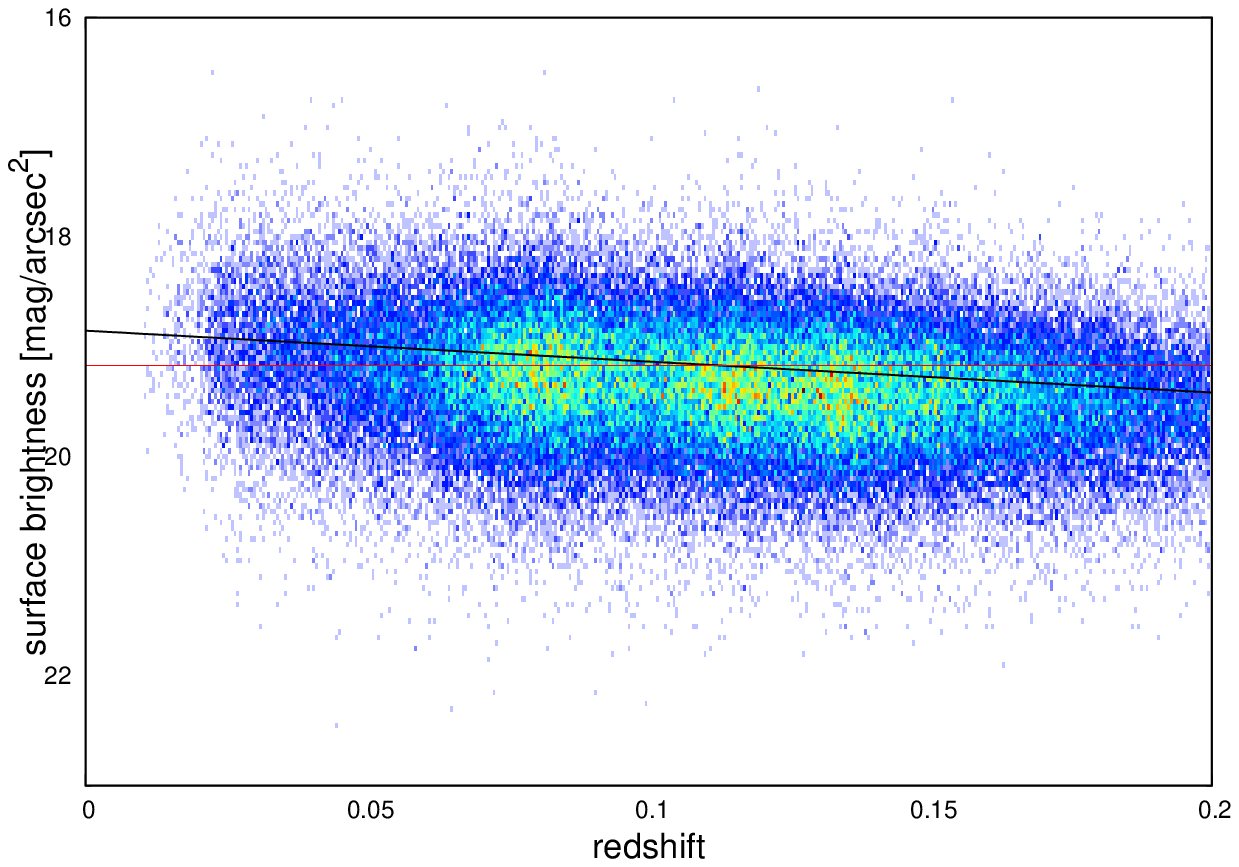}\\ 
\caption{Redshift evolution of the surface brightness in the i band of dV model indicated by the solid black line. The solid red line shows the Malmquist-bias-corrected average value of the surface brightness for this particular filter and model.}
\label{dV_z_dependence_sb_i}
\end{center}
\end{figure}
\begin{figure}[H]
\begin{center}
\includegraphics[width=0.45\textwidth]{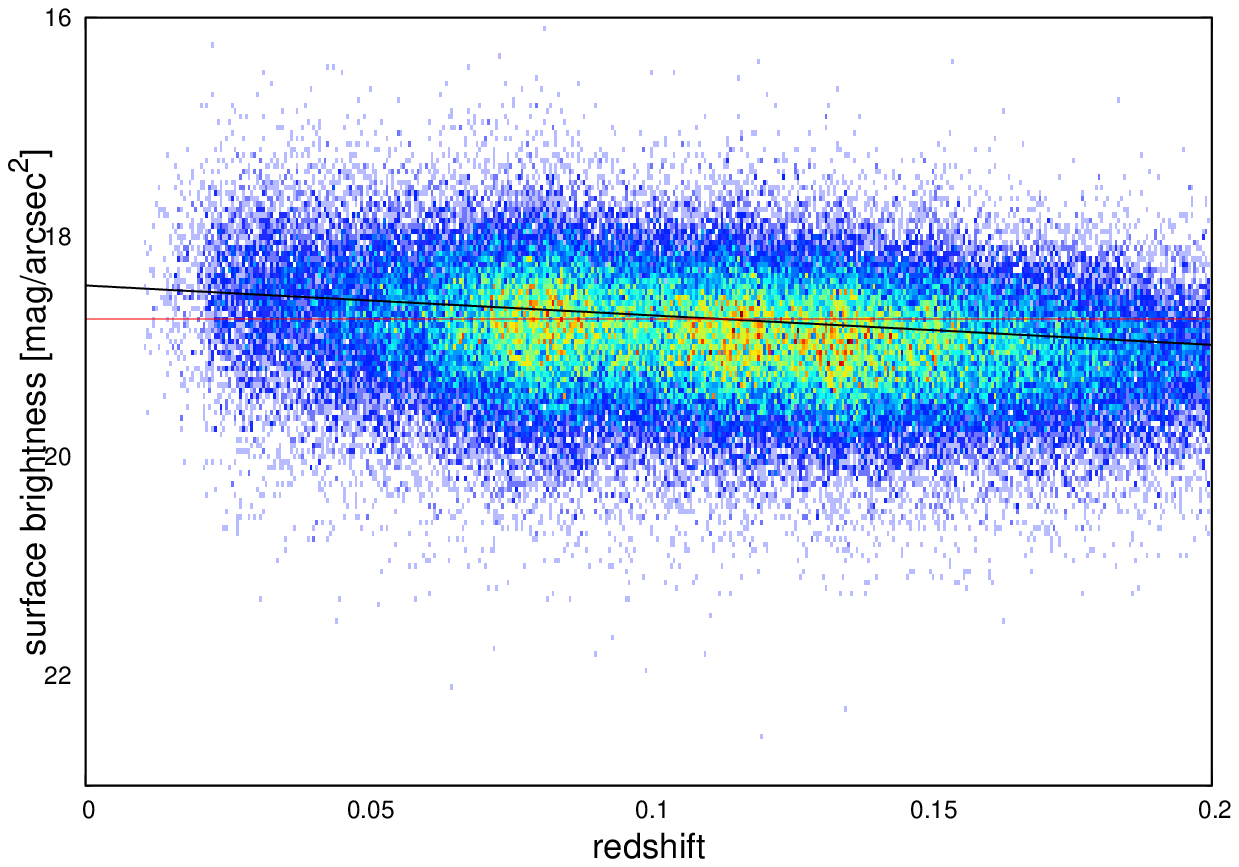}\\ 
\caption{Redshift evolution of the surface brightness in the z band of dV model indicated by the solid black line. The solid red line shows the Malmquist-bias-corrected average value of the surface brightness for this particular filter and model.}
\label{dV_z_dependence_sb_z}
\end{center}
\end{figure}

\begin{figure}[H]
\begin{center}
\includegraphics[width=0.45\textwidth]{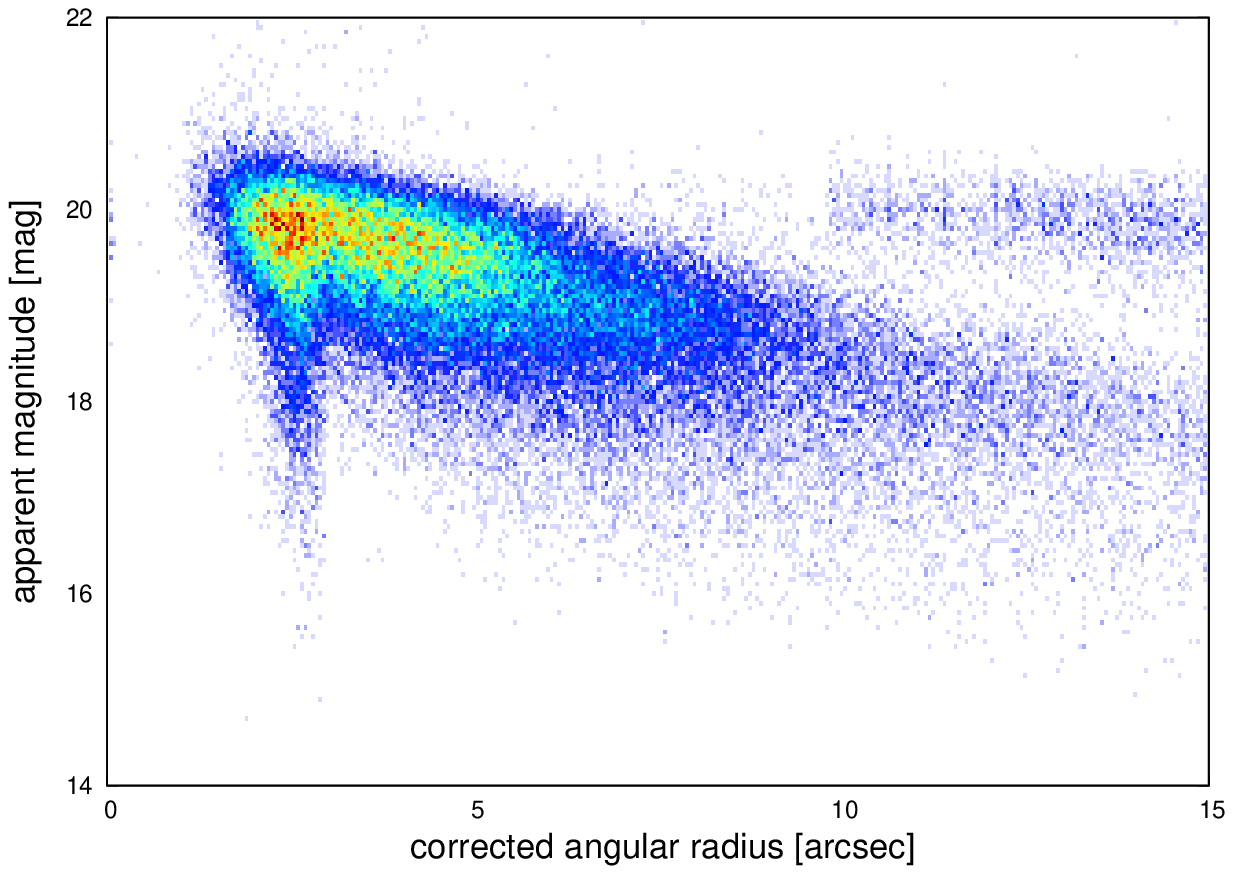}\\ 
\caption{Corrected angular radii plotted against the apparent magnitudes, showing some grouping in the u band for the p model.}
\label{p_r_cor_vs_m_app_u}
\end{center}
\end{figure}
\begin{figure}[H]
\begin{center}
\includegraphics[width=0.45\textwidth]{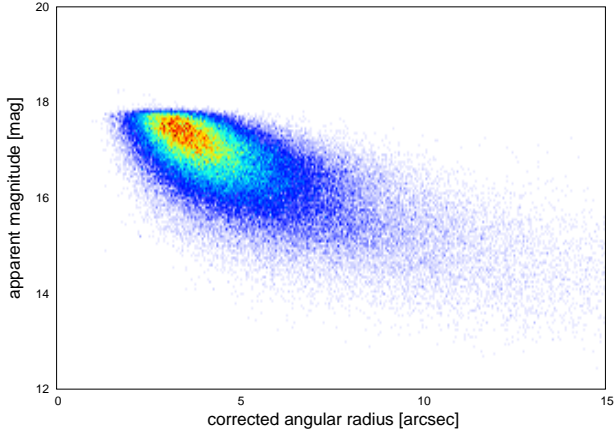}\\ 
\caption{Corrected angular radii plotted against the apparent magnitudes, not showing any peculiar features in the r band for the p model.}
\label{p_r_cor_vs_m_app_r}
\end{center}
\end{figure}
\begin{figure}[H]
\begin{center}
\includegraphics[width=0.45\textwidth]{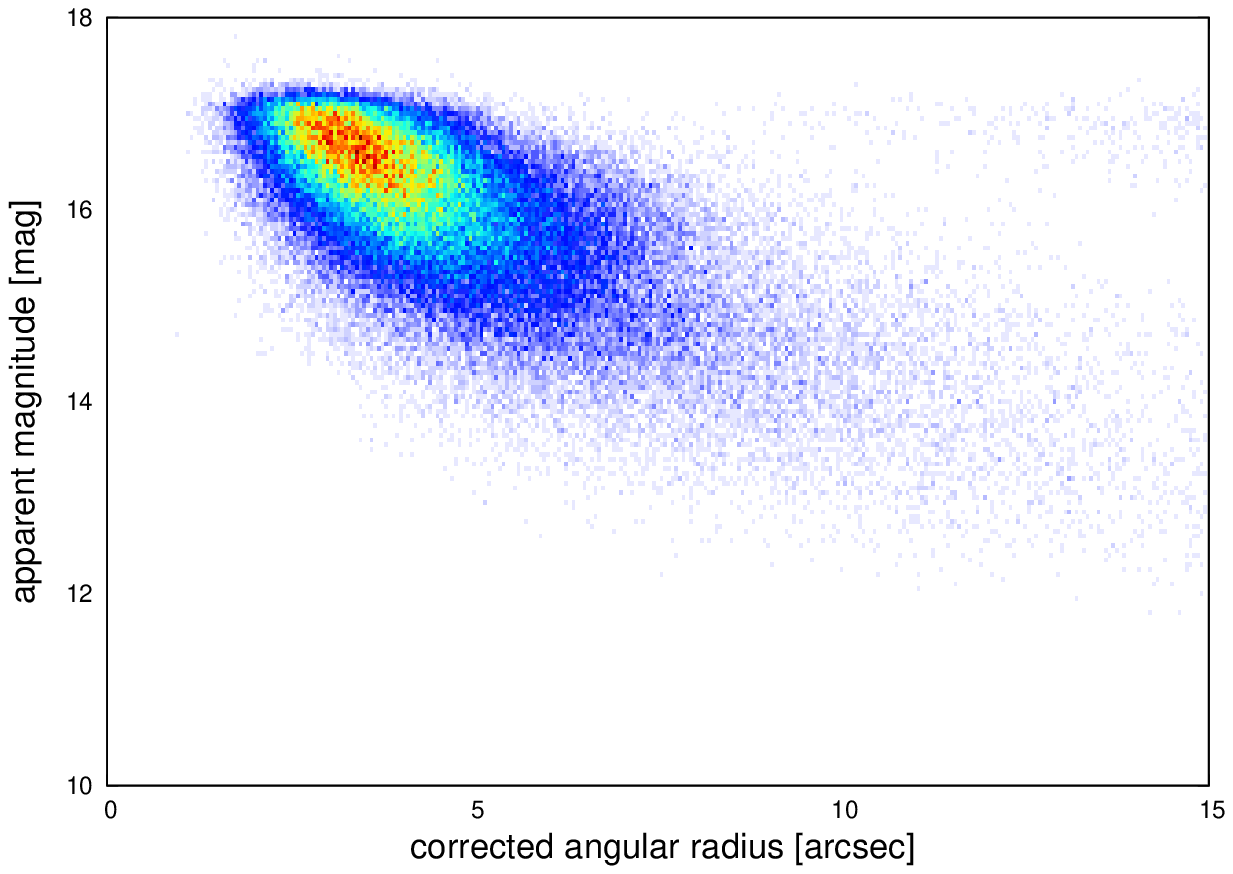}\\ 
\caption{Corrected angular radii plotted against the apparent magnitudes, showing some small grouping in the z band for the p model.}
\label{p_r_cor_vs_m_app_z}
\end{center}
\end{figure}

\begin{figure}[H]
\begin{center}
\includegraphics[width=0.45\textwidth]{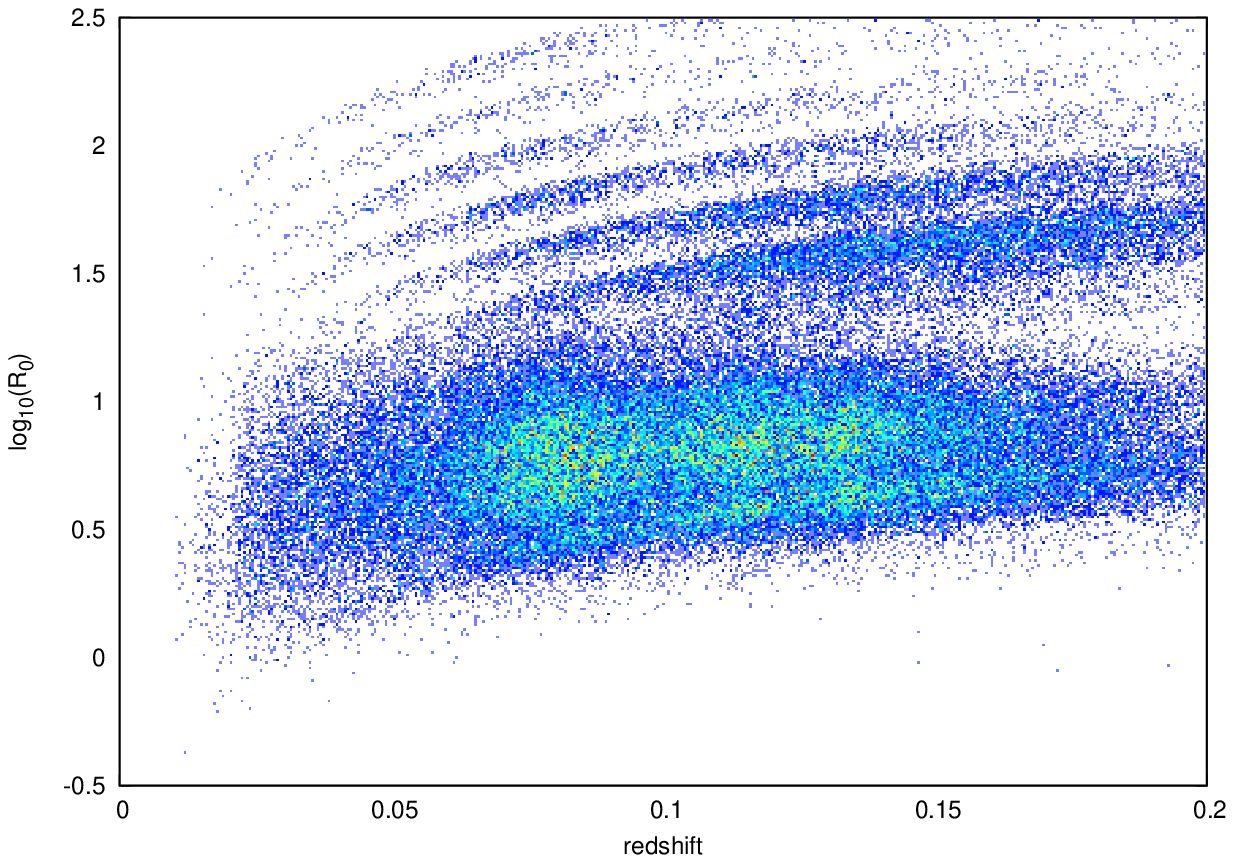}\\ 
\caption{Logarithm of the physical radii against the redshift, clearly showing band-like structures in the u band for the p model.}
\label{p_z_dependence_logR0_u}
\end{center}
\end{figure}
\begin{figure}[H]
\begin{center}
\includegraphics[width=0.45\textwidth]{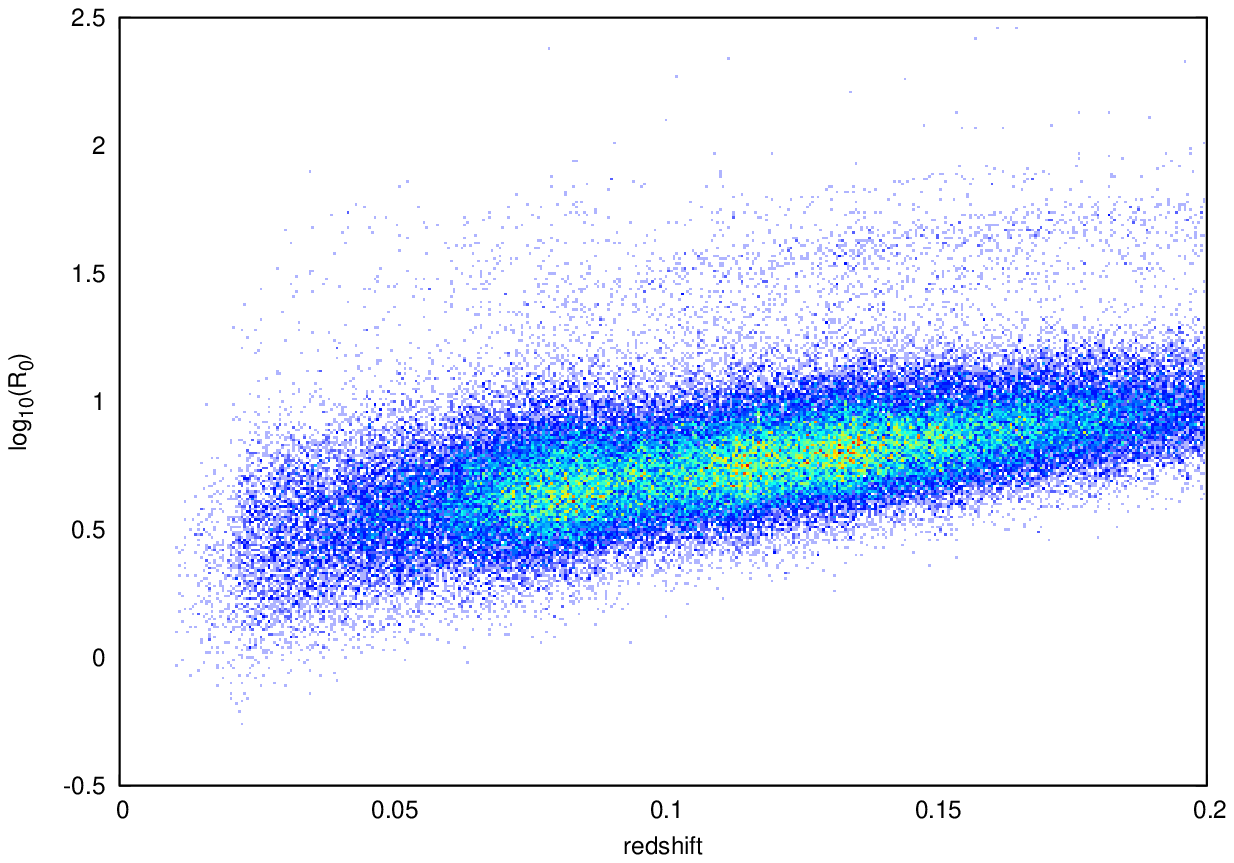}\\ 
\caption{Logarithm of the physical radii against the redshift, showing band-like structures in the z band for the p model.}
\label{p_z_dependence_logR0_z}
\end{center}
\end{figure}

\begin{figure}[H]
\begin{center}
\includegraphics[width=0.45\textwidth]{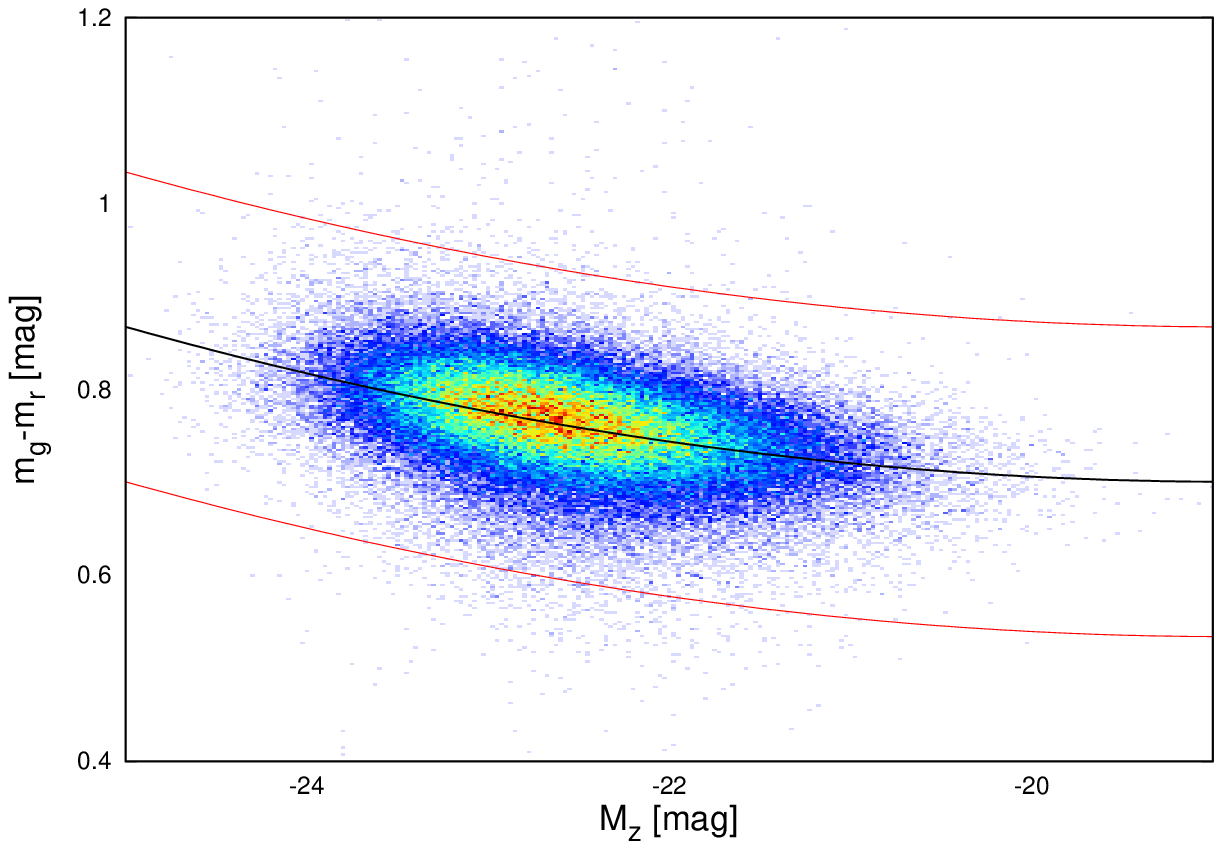}\\ 
\caption{Colour-magnitude diagram of the red sequence for the c model. The solid black line represents our best fit and the two solid red lines indicate the 3-$\sigma$ confidence limits beyond which we clipped the sample.}
\label{c_mz_vs_g-r_map}
\end{center}
\end{figure}
\begin{figure}[H]
\begin{center}
\includegraphics[width=0.45\textwidth]{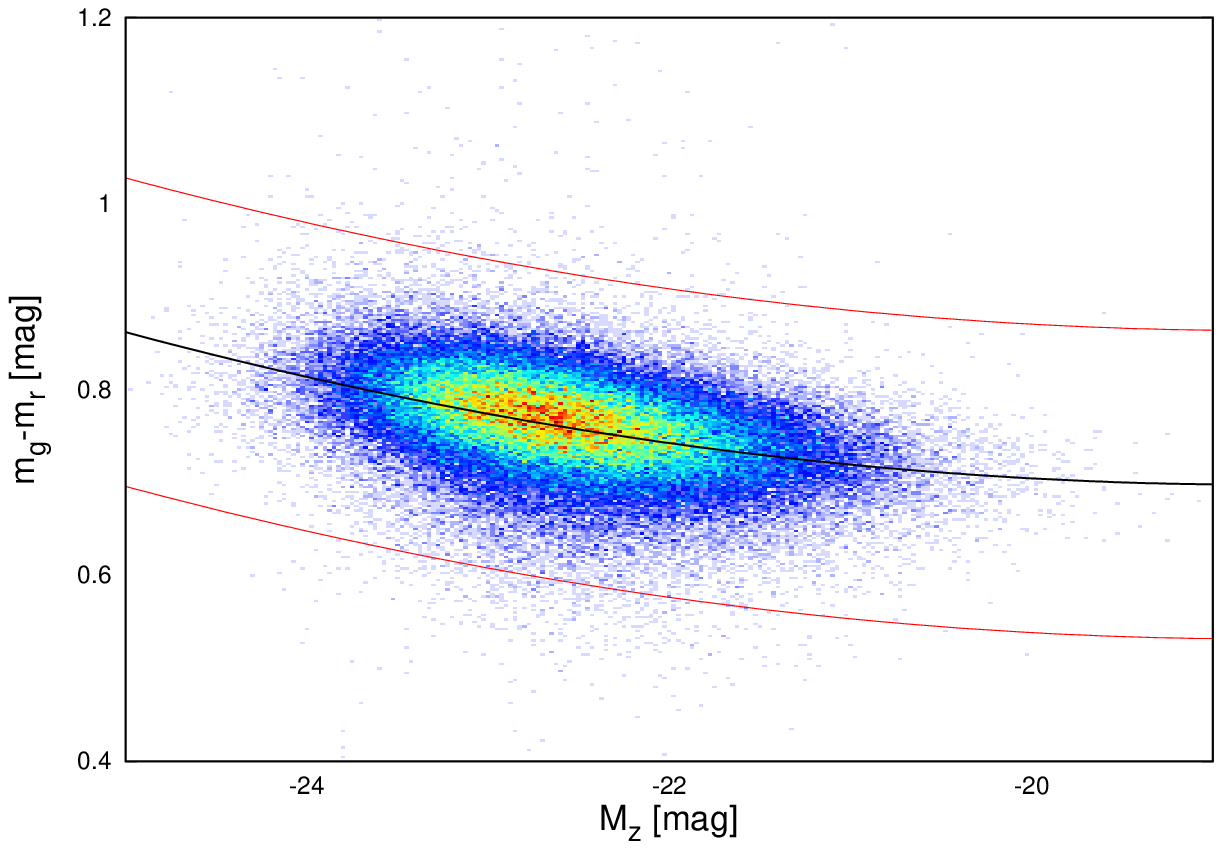}\\ 
\caption{Colour-magnitude diagram of the red sequence for the dV model. The solid black line represents our best fit and the two solid red lines indicate the 3-$\sigma$ confidence limits beyond which we clipped the sample.}
\label{dV_mz_vs_g-r_map}
\end{center}
\end{figure}
\begin{figure}[H]
\begin{center}
\includegraphics[width=0.45\textwidth]{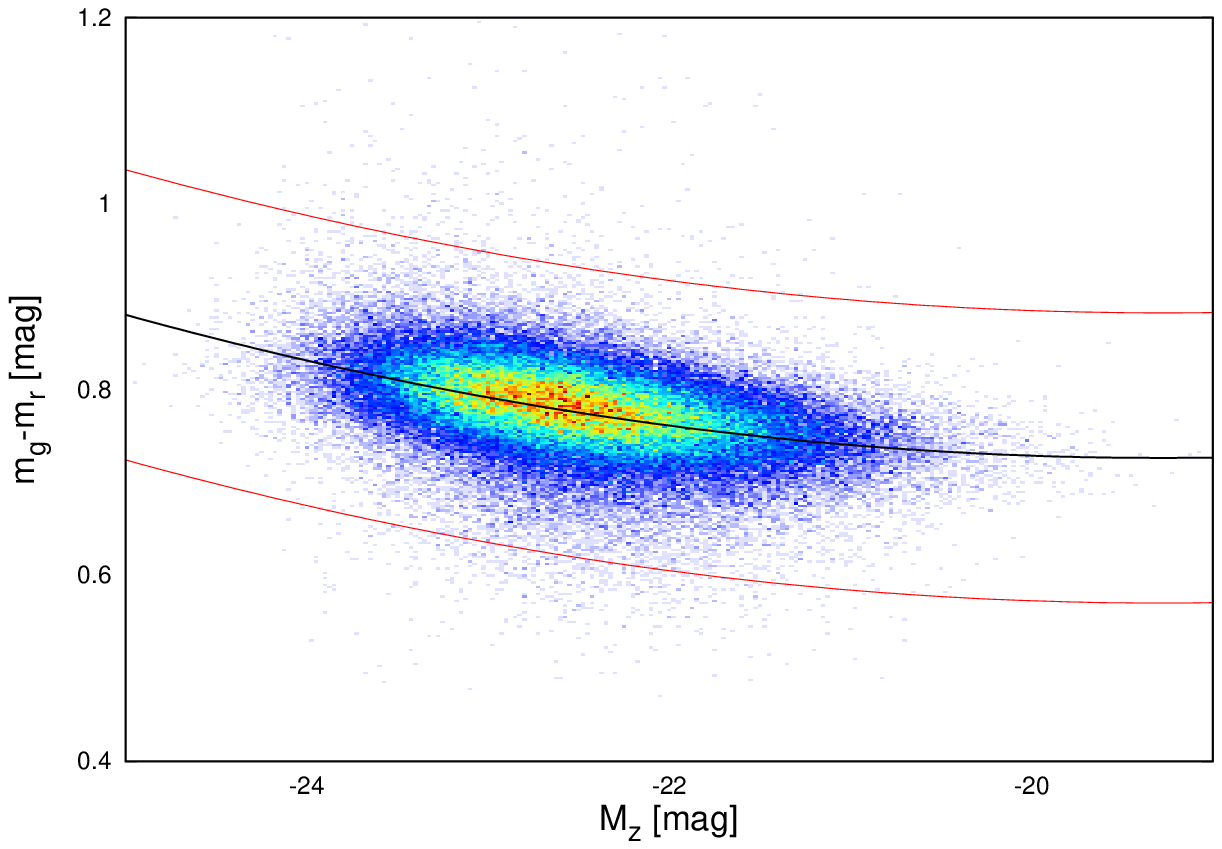}\\ 
\caption{Colour-magnitude diagram of the red sequence for the p model. The solid black line represents our best fit and the two solid red lines indicate the 3-$\sigma$ confidence limits beyond which we clipped the sample.}
\label{p_mz_vs_g-r_map}
\end{center}
\end{figure}

\begin{figure}[H]
\begin{center}
\includegraphics[width=0.45\textwidth]{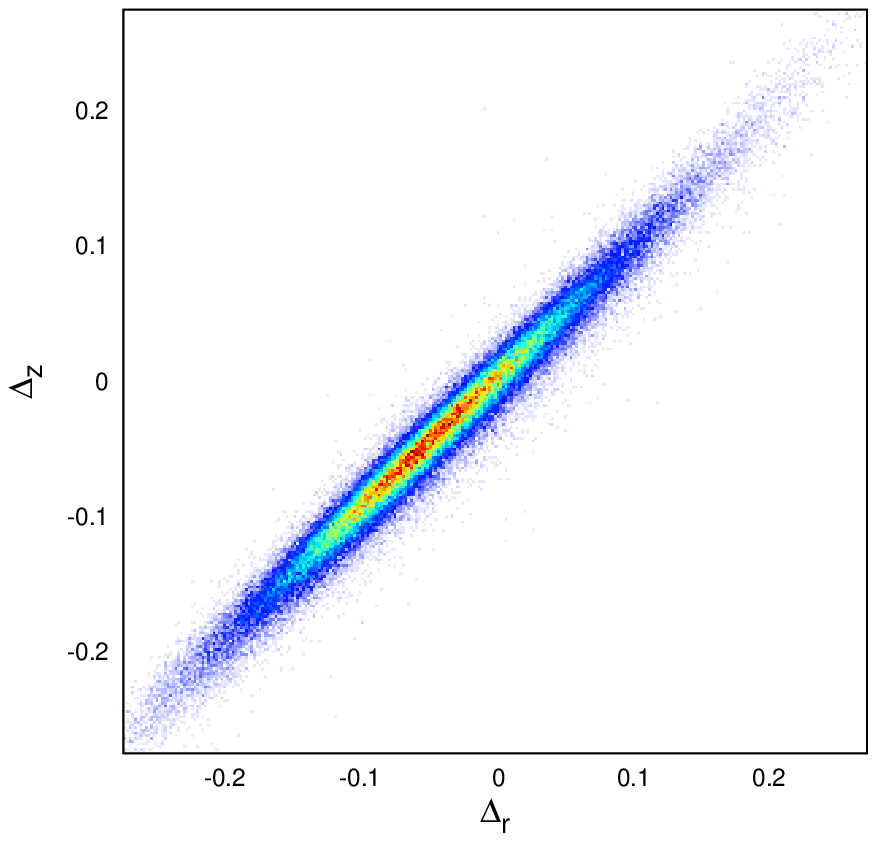}\\ 
\caption{Tight correlation between the residuals of the fundamental plane in the r band $\Delta_{r}$ and of those in the z band $\Delta_{z}$. This plot uses the fundamental-plane fit for the dV model.}
\label{dV_residuals_map_rz}
\end{center}
\end{figure}

\begin{figure}[H]
\begin{center}
\includegraphics[width=0.45\textwidth]{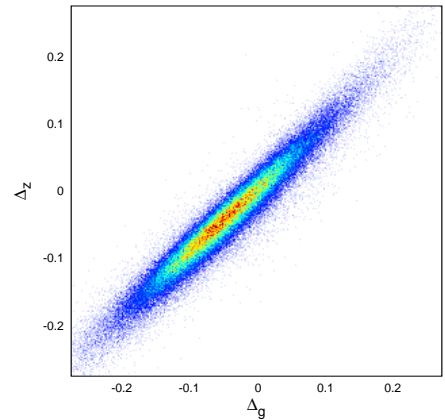}\\ 
\caption{Strong correlation between the residuals of the fundamental plane in the g band $\Delta_{g}$ and of those in the z band $\Delta_{z}$, however the correlation is visible weaker than for previous plots, due to the larger difference in the wavelength between the two filters. This plot uses the fundamental-plane fit for the dV model.}
\label{dV_residuals_map_gz}
\end{center}
\end{figure}

\begin{figure}[H]
\begin{center}
\includegraphics[width=0.45\textwidth]{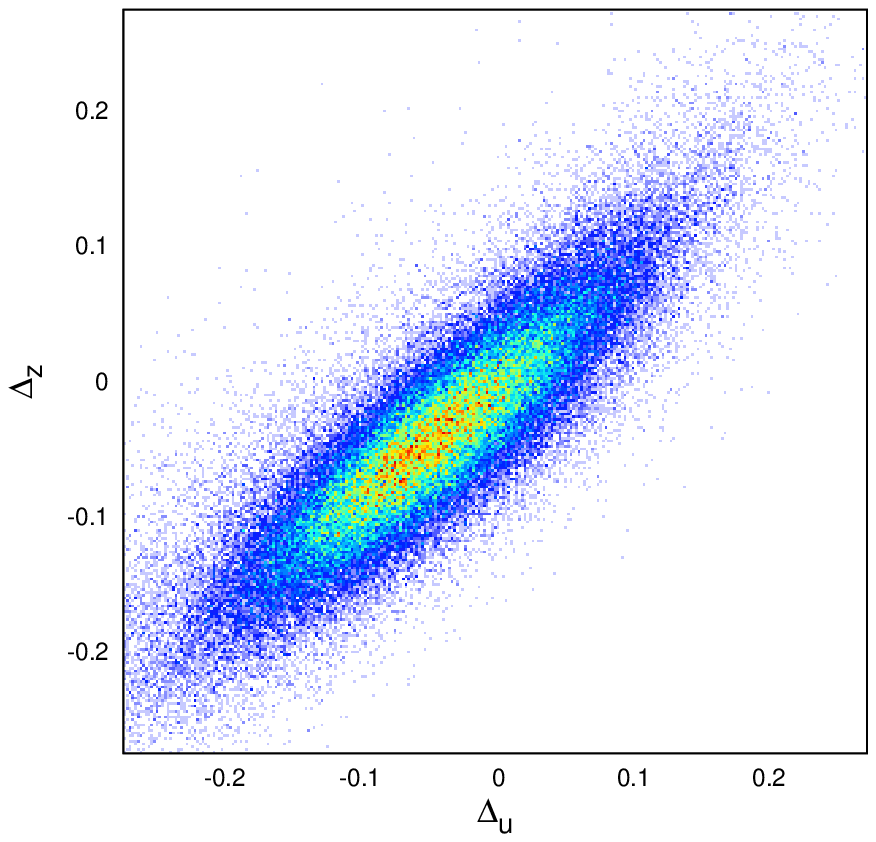}\\ 
\caption{Correlation between the residuals of the fundamental plane in the u band $\Delta_{u}$ and of those in the z band $\Delta_{z}$. Due to the larger scatter in the u band, the correlation is significantly weaker than for all other filters. This plot uses the fundamental-plane fit for the dV model.}
\label{dV_residuals_map_uz}
\end{center}
\end{figure}

\begin{figure}[H]
\begin{center}
\includegraphics[width=0.45\textwidth]{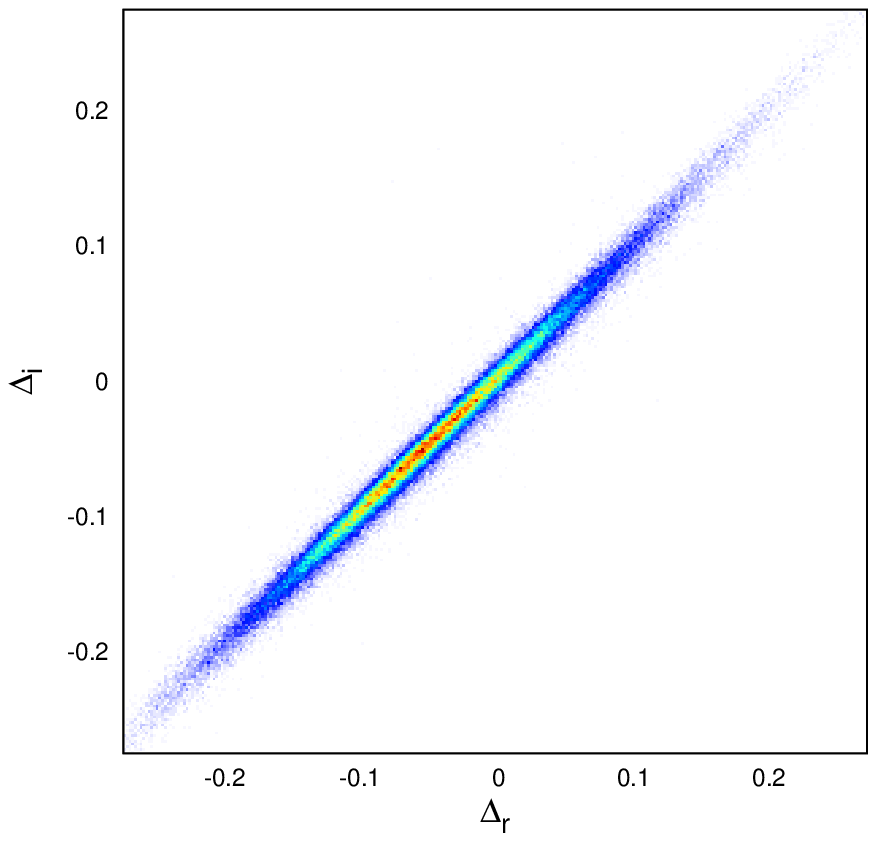}\\ 
\caption{Tight correlation between the residuals of the fundamental plane in the r band $\Delta_{r}$ and of those in the z band $\Delta_{i}$. This plot uses the fundamental-plane fit for the dV model.}
\label{dV_residuals_map_ri}
\end{center}
\end{figure}

\begin{figure}[H]
\begin{center}
\includegraphics[width=0.45\textwidth]{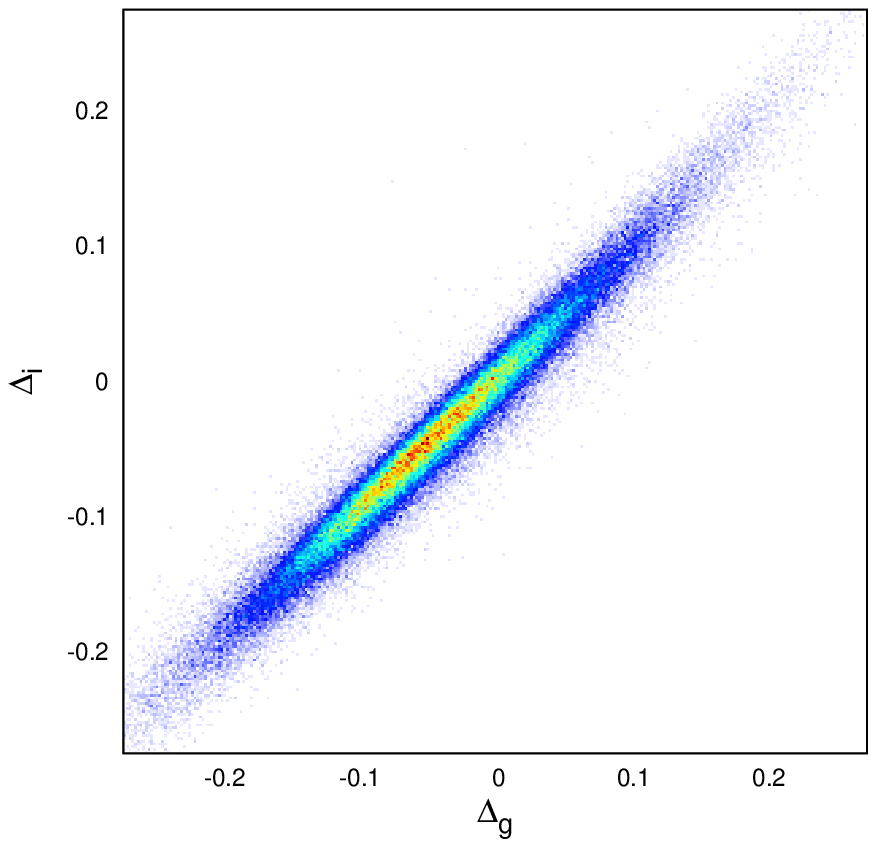}\\ 
\caption{Strong correlation between the residuals of the fundamental plane in the g band $\Delta_{g}$ and of those in the i band $\Delta_{i}$. This plot uses the fundamental-plane fit for the dV model.}
\label{dV_residuals_map_gi}
\end{center}
\end{figure}

\begin{figure}[H]
\begin{center}
\includegraphics[width=0.45\textwidth]{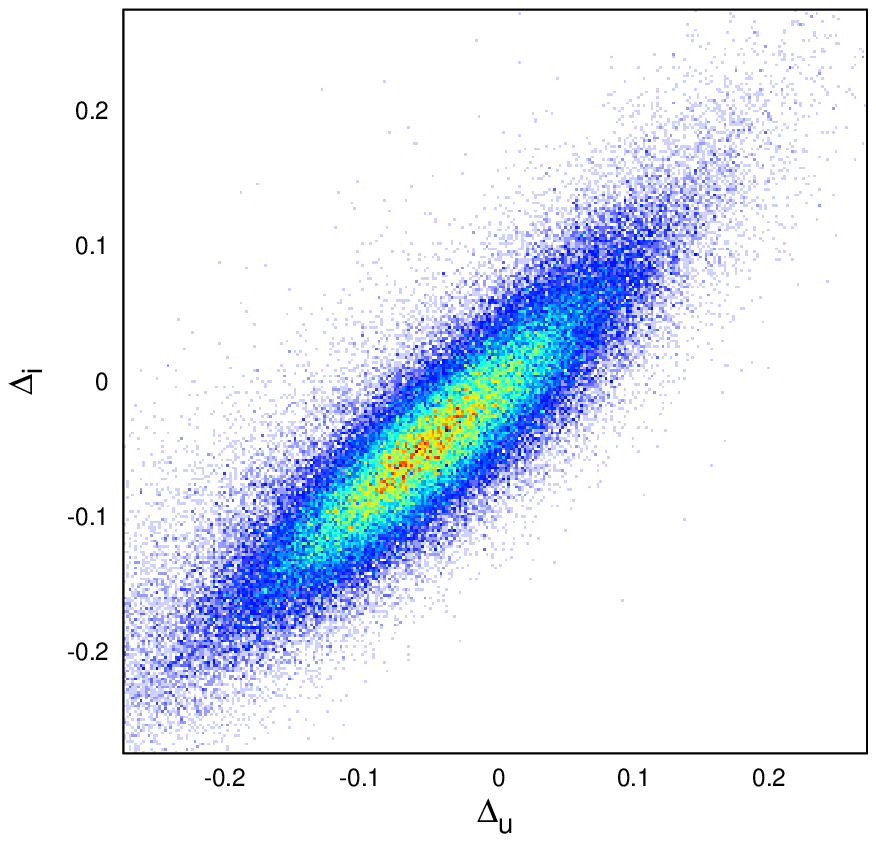}\\ 
\caption{Correlation between the residuals of the fundamental plane in the u band $\Delta_{u}$ and of those in the i band $\Delta_{i}$. Due to the larger scatter in the u band, the correlation is significantly weaker than for all other filters. This plot uses the fundamental-plane fit for the dV model.}
\label{dV_residuals_map_ui}
\end{center}
\end{figure}

\begin{figure}[H]
\begin{center}
\includegraphics[width=0.45\textwidth]{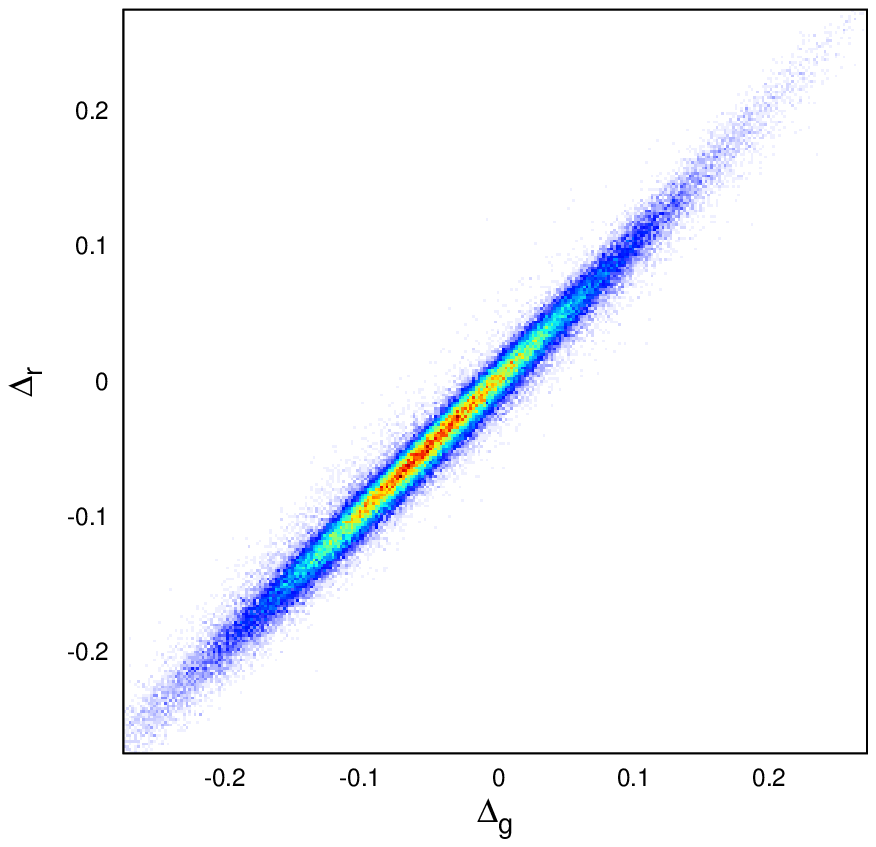}\\ 
\caption{Strong correlation between the residuals of the fundamental plane in the g band $\Delta_{g}$ and of those in the r band $\Delta_{r}$. This plot uses the fundamental-plane fit for the dV model.}
\label{dV_residuals_map_gr}
\end{center}
\end{figure}

\begin{figure}[H]
\begin{center}
\includegraphics[width=0.45\textwidth]{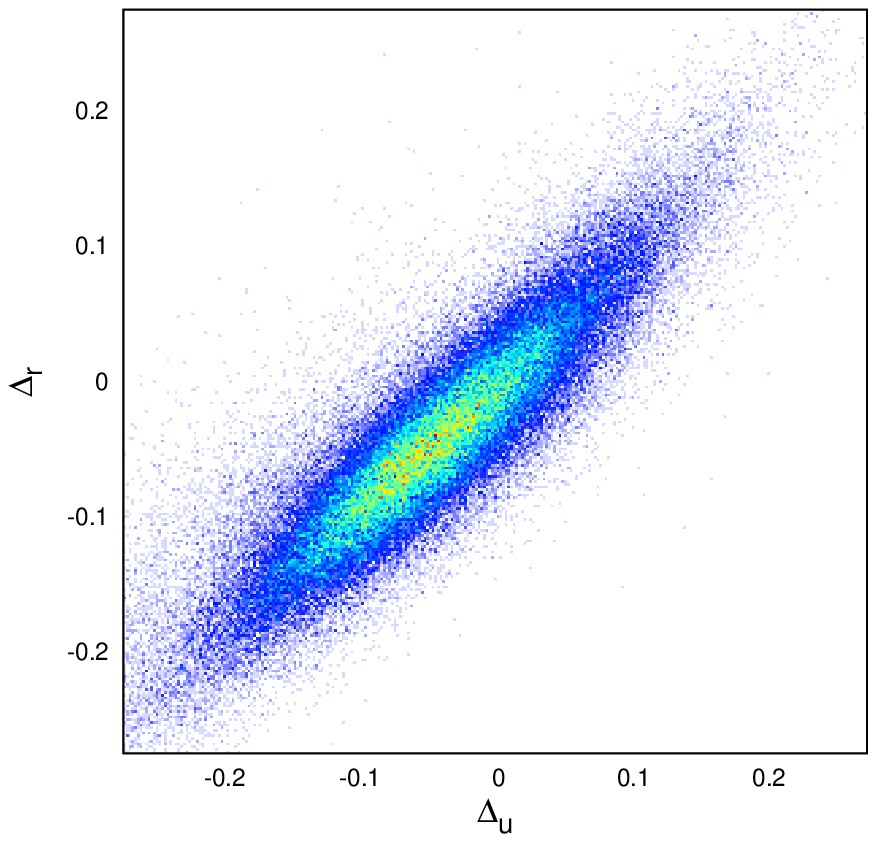}\\ 
\caption{Correlation between the residuals of the fundamental plane in the u band $\Delta_{u}$ and of those in the r band $\Delta_{r}$. Due to the larger scatter in the u band, the correlation is significantly weaker than for all other filters. This plot uses the fundamental-plane fit for the dV model.}
\label{dV_residuals_map_ur}
\end{center}
\end{figure}

\begin{figure}[H]
\begin{center}
\includegraphics[width=0.45\textwidth]{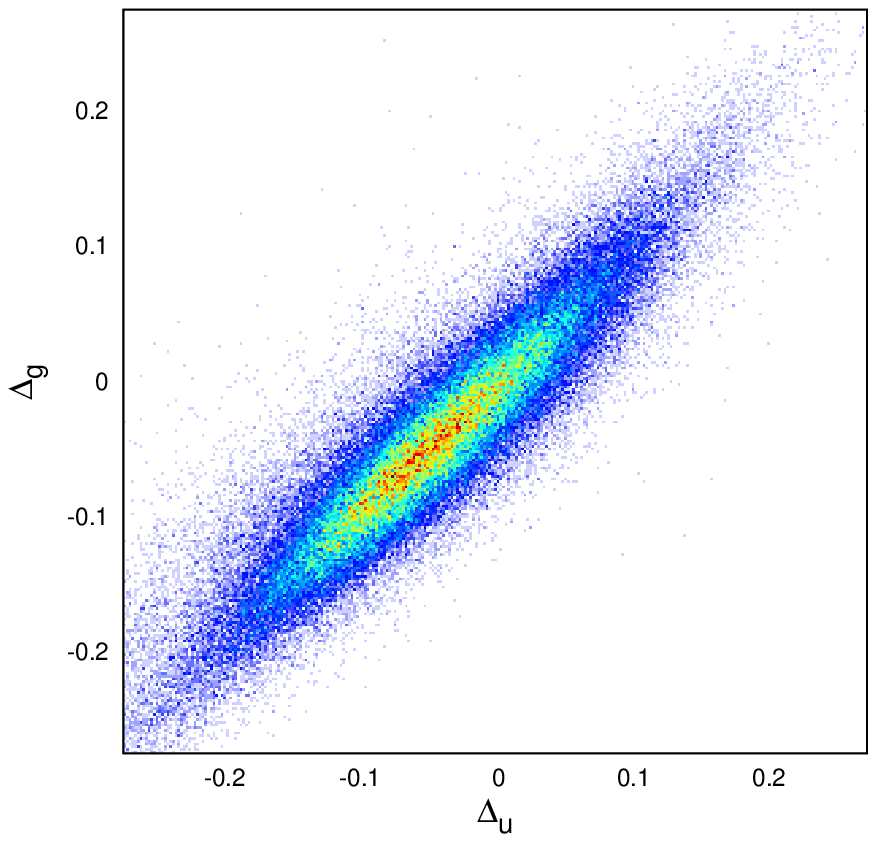}\\ 
\caption{Correlation between the residuals of the fundamental plane in the u band $\Delta_{u}$ and of those in the g band $\Delta_{g}$. Due to the larger scatter in the u band, the correlation is significantly weaker than for all other filters. This plot uses the fundamental-plane fit for the dV model.}
\label{dV_residuals_map_ug}
\end{center}
\end{figure}

\FloatBarrier
\section{Additional Tables}

\begin{table}[H]
\begin{center}
\begin{tabular}{c|cccc}
& (u-r)$^0$ & (u-r)$^1$ & (u-r)$^2$ & (u-r)$^3$ \\ \hline
$z^0$ &0 &0 &0 &0\\
$z^1$ &10.3686 &-6.12658 &2.58748 &-0.299322 \\
$z^2$ &-138.069& 45.0511& -10.8074 &0.95854 \\
$z^3$ &540.494&  -43.7644 &3.84259& 0 \\
$z^4$ &-1005.28& 10.9763 &0& 0\\
$z^5$ &710.482& 0 &0& 0 
\end{tabular}
\end{center}
\caption{Coefficients for the K-correction in the u band using u-r colours.}
\label{Kcor_u}
\end{table}

\begin{table}[H]
\begin{center}
\begin{tabular}{c|cccc}
& (g-r)$^0$ & (g-r)$^1$ & (g-r)$^2$ & (g-r)$^3$ \\ \hline
$z^0$ &0 &0 &0 &0\\
$z^1$ &-2.45204 & 4.10188 & 10.5258 & -13.5889 \\
$z^2$ &56.7969 & -140.913 &144.572 & 57.2155 \\
$z^3$ &-466.949 & 222.789 & -917.46& -78.0591\\
$z^4$ &2906.77& 1500.8 & 1689.97 &30.889\\
$z^5$ &-10453.7 &-4419.56  &-1011.01 &0\\
$z^6$ &17568 &3236.68 &0 &0\\
$z^7$ &-10820.7& 0& 0 &0
\end{tabular}
\end{center}
\caption{Coefficients for the K-correction in the g band using g-r colours.}
\label{Kcor_g}
\end{table}

\begin{table}[H]
\begin{center}
\begin{tabular}{c|cccc}
& (g-r)$^0$ & (g-r)$^1$ & (g-r)$^2$ & (g-r)$^3$ \\ \hline
$z^0$ &0& 0& 0& 0\\
$z^1$ &1.83285 &-2.71446 &4.97336 &-3.66864  \\
$z^2$ &-19.7595& 10.5033 &18.8196 & 6.07785 \\
$z^3$ &33.6059 &-120.713 &-49.299 &0\\
$z^4$ &144.371 &216.453 &0 &0 \\
$z^5$ &-295.39& 0 &0& 0 
\end{tabular}
\end{center}
\caption{Coefficients for the K-correction in the r band using g-r colours.}
\label{Kcor_r}
\end{table}

\begin{table}[H]
\begin{center}
\begin{tabular}{c|cccc}
& (g-i)$^0$ & (g-i)$^1$ & (g-i)$^2$ & (g-i)$^3$ \\ \hline
$z^0$ &0& 0& 0 &0 \\
$z^1$ &-2.21853 & 3.94007 & 0.678402 &-1.24751 \\
$z^2$ &-15.7929 &-19.3587& 15.0137 &2.27779 \\
$z^3$ &118.791 &-40.0709 &-30.6727& 0 \\
$z^4$ &-134.571 &125.799 &0& 0 \\
$z^5$ &-55.4483& 0& 0& 0
\end{tabular}
\end{center}
\caption{Coefficients for the K-correction in the i band using g-i colours.}
\label{Kcor_i}
\end{table}

\begin{table}[H]
\begin{center}
\begin{tabular}{c|cccc}
& (g-z)$^0$ & (g-z)$^1$ & (g-z)$^2$ & (g-z)$^3$ \\ \hline
$z^0$ &  0& 0 &0& 0 \\
$z^1$ &0.30146 &-0.623614 &1.40008 &-0.534053\\
$z^2$ &-10.9584& -4.515& 2.17456&  0.913877\\
$z^3$ &66.0541& 4.18323& -8.42098& 0\\
$z^4$ &-169.494 &14.5628 &0& 0\\
$z^5$ & 144.021& 0& 0& 0
\end{tabular}
\end{center}
\caption{Coefficients for the K-correction in the z band using g-z colours.}
\label{Kcor_z}
\end{table}

\begin{table}
\begin{center}
\begin{tabular}{c|cccc}
 models & $c_{2}$ & $c_{1}$ & $c_{0}$ & rms \\ \hline 
 c Model & 0.0045 & 0.1681 & 2.2878 & 0.0555\\
 dV Model & 0.0042 & 0.1555 & 2.1523 & 0.0553\\ 
 p Model & 0.0047 & 0.1816 & 2.4768 & 0.0520
\end{tabular}
\end{center}
\caption{Coefficients and the root mean square of the best fit for the red sequence using our sample. The polynomial for these coefficients is of the shape $(m_{g}-m_{r}) = c_{2} \cdot M_{z}^{2} + c_{1} \cdot M_{z} + c_{0} $ .}
\label{redsequence_fit}
\end{table}

\begin{table}[H]
\begin{center}
\begin{tabular}{c|cc}
 filters & $s_{\varepsilon}$ & $\bar{\sigma}_{\textrm{dist}}$ [\%] \\ \hline \hline
   \citet{Bernardi:2003c} &  & \\ \hline
 g & $0.1056$ & $16.8$\\
 r & $0.1054$ & $17.9$\\
 i & $0.1028$ & $16.8$\\
 z & $0.1134$ & $16.0$\\ \hline \hline
   \citet{Hyde:2009} &  & \\ \hline
 g & $0.1063$ & $18.8$ \\
 r & $0.1042$ & $17.5$ \\
 i & $0.1031$ & $17.4$ \\
 z & $0.1087$ & $18.6$ 
\end{tabular}
\end{center}
\caption{Quality of the fundamental plane as a distance indicator using our selected sample of 94922 elliptical galaxies, but with the direct-fit coefficients of \citet{Bernardi:2003c} or \citet{Hyde:2009}, respectively (the coefficients are listed in Table \ref{fp_coefficients_by_others} of this paper). We found that our best-fit coefficients (see Table \ref{fitparameters}) are better by a few  percent than those of our esteemed colleagues.}
\label{accuracy_Bernardi}
\end{table}

\begin{table*}
\begin{center}
\begin{tabular}{c|ccccc}
SDSS-filter & u & g & r & i & z \\ \hline \hline

 $\bar{r}_{\textrm{cor}}$ (c model) [arcsec] &    2.95       &    2.29       &    2.13       &    2.04       &    1.89      \\
  $\bar{r}_{\textrm{cor}}$ (dV model) [arcsec] &    2.93     &    2.28      &    2.13      &    2.03    &    1.88    \\
   $\bar{r}_{\textrm{cor}}$ (p model) [arcsec] &    9.35    &    4.67    &    4.44    &    4.37     &    4.54    \\
  
 $\sigma_{r_{\textrm{cor}}}$ (c model) [arcsec] &    2.13     &    1.26     &    1.16   &    1.12   &    1.02   \\
  $\sigma_{r_{\textrm{cor}}}$ (dV model) [arcsec] &    2.12  &    1.25    &    1.16    &    1.12     &    1.01    \\
   $\sigma_{r_{\textrm{cor}}}$ (p model) [arcsec] &    15.26    &    2.33    &    2.11    &    2.25     &    3.69    \\ \hline
   
 $\bar{m}_{\textrm{app}}$ (c model) [mag] &    19.02   &    17.33  &    16.58   &    16.24   &    15.98    \\
  $\bar{m}_{\textrm{app}}$ (dV model) [mag] &    19.01   &    17.32     &    16.58   &    16.24   &    15.97   \\
   $\bar{m}_{\textrm{app}}$ (p model) [mag] &    19.14    &    17.40    &    16.64     &    16.28   &    16.01  \\
 
 $\sigma_{m_{\textrm{app}}}$ (c model) [mag] &   0.92    &   0.87  &   0.87     &   0.87     &   0.87   \\
  $\sigma_{m_{\textrm{app}}}$ (dV model) [mag] &   0.92    &   0.88    &   0.87    &   0.87  &   0.87   \\
   $\sigma_{m_{\textrm{app}}}$ (p model) [mag] &   0.90     &   0.87    &   0.86   &   0.87    &   0.86  \\ \hline
 
 $\bar{\sigma}_{0}$ (c model) [km/s] &    180.1     &    177.7     &    176.8   &    176.8    &    177.0  \\
  $\bar{\sigma}_{0}$ (dV model) [km/s] &    180.4    &    177.9    &    176.9  &    176.9     &    177.1     \\
   $\bar{\sigma}_{0}$ (p model) [km/s] &    173.5     &    171.9     &    170.9     &    170.6     &    170.1  \\
 
 $\sigma_{\sigma_{0}}$ (c model) [km/s] &   46.0      &   45.3     &   45.1     &   45.1      &  45.4     \\
  $\sigma_{\sigma_{0}}$ (dV model) [km/s] &   46.0       &   45.4       &   45.1       &   45.1     &   45.4     \\
   $\sigma_{\sigma_{0}}$ (p model) [km/s] &  44.4       &   43.9      &   43.7       &   43.6      &   43.7     \\ \hline
   
 $\overline{\textrm{log}_{10}(R_{0})}$ (c model) [$\textrm{log}_{10}(\textrm{kpc})$] &   0.564  &   0.483  &   0.449  &  0.431 &  0.394      \\
  $\overline{\textrm{log}_{10}(R_{0})}$ (dV model) [$\textrm{log}_{10}(\textrm{kpc})$] &   0.562 &   0.482 &   0.448 &   0.430 &   0.394 \\
   $\overline{\textrm{log}_{10}(R_{0})}$ (p model) [$\textrm{log}_{10}(\textrm{kpc})$] &   0.950 &   0.801 &   0.777 &   0.770 &   0.765 \\
 
 $\sigma_{\textrm{log}_{10}(R_{0})}$ (c model) [$\textrm{log}_{10}(\textrm{kpc})$] &   0.305 &  0.235&   0.231 &   0.228 &   0.228 \\
  $\sigma_{\textrm{log}_{10}(R_{0})}$ (dV model) [$\textrm{log}_{10}(\textrm{kpc})$] &   0.304  &  0.235 &   0.231 &   0.228 &   0.228 \\
   $\sigma_{\textrm{log}_{10}(R_{0})}$ (p model) [$\textrm{log}_{10}(\textrm{kpc})$] &   0.399 &  0.229 &   0.225 &   0.226 &   0.250 \\ \hline
 
 $\bar{\mu}_{0}$ (c model) [mag/arcsec$^2$] &    22.57 &    20.52 &    19.62  &    19.18  &    18.75  \\
  $\bar{\mu}_{0}$ (dV model) [mag/arcsec$^2$] &    22.54 &    20.50  &    19.61  &    19.17   &    18.75  \\
     $\bar{\mu}_{0}$ (p model) [mag/arcsec$^2$] &    24.60  &    22.19  &    21.32 &    20.93  &    20.65 \\ 
 
 $\sigma_{\mu_{0}}$ (c model) [mag/arcsec$^2$] &   0.99   &   0.65    &   0.62   &   0.61    &   0.62     \\
  $\sigma_{\mu_{0}}$ (dV model) [mag/arcsec$^2$] &   0.98   &   0.64  &   0.62   &   0.61  &   0.61  \\
    $\sigma_{\mu_{0}}$ (p model) [mag/arcsec$^2$] &    1.72  &   0.64  &   0.62      &   0.64   &   0.83  \\ \hline
 
 $\overline{\textrm{log}_{10}(\sigma_{0})}$ (c model) [$\textrm{log}_{10}(\textrm{km/s})$] &  2.24  &    2.24 &    2.23 &    2.23   &    2.23    \\
 $\overline{\textrm{log}_{10}(\sigma_{0})}$ (dV model) [$\textrm{log}_{10}(\textrm{km/s})$] & 2.24 &    2.24 &    2.23 &    2.23 &   2.23\\
   $\overline{\textrm{log}_{10}(\sigma_{0})}$ (p model) [$\textrm{log}_{10}(\textrm{km/s})$] &    2.23&    2.22&    2.22 &    2.22 &    2.22\\ 
 
 $\sigma_{\textrm{log}_{10}(\sigma_{0})}$ (c model) [$\textrm{log}_{10}(\textrm{km/s})$] &   0.11 &   0.11 &   0.11 &   0.11 &   0.11 \\
 $\sigma_{\textrm{log}_{10}(\sigma_{0})}$ (dV model) [$\textrm{log}_{10}(\textrm{km/s})$] &   0.11 &   0.11 &   0.11 &   0.11 &   0.11 \\
 $\sigma_{\textrm{log}_{10}(\sigma_{0})}$ (p model) [$\textrm{log}_{10}(\textrm{km/s})$] &   0.11 &   0.11 &   0.11  &   0.11 &   0.11
\end{tabular}
\end{center}
\caption{Mean values and standard deviations of several different parameters that have to be calculated or measured for the calibration of the fundamental plane. These values are given for all models and all filters. $\bar{r}_{\textrm{cor}}$ stands for the mean value of the apparent corrected radius $r_{\textrm{cor}}$, and $\sigma_{r_{\textrm{cor}}}$ is the corresponding standard deviation. $\bar{m}_{\textrm{app}}$ denotes the mean value of the apparent magnitude $m_{\textrm{app}}$, and $\sigma_{m_{\textrm{app}}}$ its standard deviation. The mean value of the central velocity dispersion $\sigma_{0}$ is given by $\bar{\sigma}_{0}$ and the corresponding standard deviation by $\sigma_{\sigma_{0}}$. $\overline{\textrm{log}_{10}(R_{0})}$ denotes the mean value of logarithm of the physical radius $R_{0}$, and $\sigma_{\textrm{log}_{10}(R_{0})}$ the corresponding standard deviation. $\bar{\mu}_{0}$ is the mean value of the mean surface brightness $\mu_{0}$, and $\sigma_{\mu_{0}}$ is its standard deviation. The mean value of the logarithm of the central velocity dispersion $\sigma_{0}$ is given by $\overline{\textrm{log}_{10}(\sigma_{0})}$ and the corresponding standard deviation by $\sigma_{\textrm{log}_{10}(\sigma_{0})}$.}
\label{additional_parameters}
\end{table*}

\begin{table*}
\begin{center}
\begin{tabular}{c|ccccc}
 models and filters & $a$ & $b$ & $c$ & $s_{\varepsilon}$ & $\bar{\sigma}_{\textrm{dist}}$ [\%] \\ \hline \hline
 c model &  &  &  & \\ \hline
 u & $  0.806   \pm   0.029 $  & $ -0.683    \pm   0.008 $  & $  -7.41  \pm   0.10 $ & $  0.1026   $ & $   17.2    $\\
 g & $  0.972     \pm   0.030 $  & $ -0.722     \pm   0.012 $  & $  -7.62     \pm   0.12   $ & $  0.1011   $ & $   16.2     $\\
 r & $   1.038   \pm   0.030 $  & $ -0.738   \pm   0.013 $  & $  -7.66   \pm   0.12    $ & $  0.1010   $ & $   15.8    $\\
 i & $   1.065     \pm   0.030 $  & $ -0.744    \pm   0.013 $  & $  -7.66   \pm   0.13   $ & $  0.9979$ & $   15.5     $\\
 z & $   1.113   \pm  0.030$  & $ -0.753     \pm   0.013 $  & $  -7.74     \pm   0.13  $ & $  0.1026   $ & $   15.3  $\\
 \hline \hline
 dV model &  &  &  & \\ \hline
 u & $  0.823  \pm   0.029 $  & $ -0.669  \pm  0.008 $  & $  -7.31 \pm   0.10 $ & $  0.1079  $ & $   17.3   $\\
 g & $  0.964    \pm   0.030 $  & $ -0.727 \pm   0.012 $  & $  -7.63    \pm   0.12   $ & $  0.1012   $ & $   16.2     $\\
 r & $   1.031  \pm   0.030 $  & $ -0.742    \pm   0.013 $  & $  -7.67    \pm   0.13    $ & $  0.1011    $ & $   15.8  $\\
 i & $   1.059   \pm   0.030 $  & $ -0.747    \pm   0.013 $  & $  -7.66   \pm   0.13    $ & $  0.0998 $ & $   15.5    $\\
 z & $   1.107  \pm  0.030 $  & $ -0.757  \pm  0.013 $  & $  -7.76     \pm   0.13   $ & $  0.1079  $ & $   15.3   $\\
 \hline \hline
  p model &  &  &  & \\ \hline
 u & $  0.623     \pm   0.029 $  & $ -0.494 \pm  0.005 $  & $  -5.32    \pm   0.08 $ & $  0.1729     $ & $   23.1    $\\
 g & $  0.992    \pm   0.030 $  & $ -0.676     \pm   0.012$  & $  -7.41  \pm   0.13   $ & $  0.1054  $ & $   17.0      $\\
 r & $   1.058 \pm   0.030 $  & $ -0.713 \pm   0.013 $  & $  -7.66   \pm   0.13   $ & $  0.1029  $ & $   16.5    $\\
 i & $   1.081    \pm   0.030 $  & $ -0.670  \pm   0.012 $  & $  -7.24  \pm   0.13  $ & $  0.1048 $ & $   16.5   $\\
 z & $   1.106  \pm   0.030 $  & $ -0.621   \pm   0.009 $  & $  -6.817 \pm   0.110     $ & $  0.1729  $ & $   17.1$

\end{tabular}
\end{center}
\caption{Best fits for the fundamental-plane coefficients in all filters and for all models with the 3-$\sigma$ clipping disabled.}
\label{fitparameters_noclipping}
\end{table*}

\begin{table*}
\begin{center}
\begin{tabular}{c|ccccc}
 models and filters & $a$ & $b$ & $c$ & $s_{\varepsilon}$ & $\bar{\sigma}_{\textrm{dist}}$ [\%] \\ \hline \hline
 c model &  &  &  & \\ \hline
 u & $  0.720  \pm   0.031 $  & $ -0.681    \pm  0.007 $  & $  -7.14 \pm   0.10 $ & $  0.0859 $ & $   19.7  $\\
 g & $  0.890 \pm   0.032 $  & $ -0.701 \pm   0.013 $  & $  -7.20  \pm   0.13    $ & $  0.0800$ & $   18.4  $\\
 r & $  0.955  \pm   0.032$  & $ -0.724  \pm   0.014 $  & $  -7.31 \pm   0.13  $ & $  0.0787 $ & $   18.0  $\\
 i & $  0.977   \pm   0.032 $  & $ -0.735  \pm   0.014 $  & $  -7.34   \pm   0.13  $ & $  0.0768 $ & $   17.6   $\\
 z & $   1.003   \pm   0.032 $  & $ -0.738    \pm   0.014 $  & $  -7.33 \pm   0.13 $ & $  0.0859 $ & $   17.4  $\\
\hline \hline
 dV model &  &  &  & \\ \hline
 u & $  0.708  \pm   0.031 $  & $ -0.684  \pm   0.007 $  & $  -7.13 \pm   0.10 $ & $  0.0860 $ & $   19.8    $\\
 g & $  0.882  \pm   0.032 $  & $ -0.707   \pm   0.013 $  & $  -7.23  \pm   0.13    $ & $  0.0802 $ & $   18.4   $\\
 r & $  0.948 \pm   0.032 $  & $ -0.729 \pm   0.014 $  & $  -7.33  \pm   0.13  $ & $  0.0787 $ & $   18.0  $\\
 i & $  0.973  \pm   0.032 $  & $ -0.739 \pm   0.014 $  & $  -7.36   \pm   0.13 $ & $  0.0768 $ & $   17.6  $\\
 z & $  0.999 \pm   0.032 $  & $ -0.743 \pm  0.014$  & $  -7.36   \pm   0.14    $ & $  0.0860 $ & $   17.4 $\\
\hline \hline
  p model &  &  &  & \\ \hline
 u & $  0.796\pm   0.031 $  & $ -0.538\pm   0.004 $  & $  -6.04 \pm   0.09 $ & $  0.0996 $ & $   22.4   $\\
 g & $  0.903 \pm   0.031 $  & $ -0.659 \pm  0.013 $  & $  -6.99 \pm   0.14   $ & $  0.0842 $ & $   19.2  $\\
 r & $  0.973 \pm   0.031 $  & $ -0.690  \pm   0.014 $  & $  -7.21\pm   0.15 $ & $  0.0819 $ & $   18.7   $\\
 i & $  0.997  \pm   0.032 $  & $ -0.688 \pm   0.014 $  & $  -7.14 \pm   0.14     $ & $  0.0807 $ & $   18.5      $\\
 z & $  1.018 \pm   0.032 $  & $ -0.611 \pm   0.010 $  & $  -6.48   \pm   0.12   $ & $  0.0996 $ & $   18.9  $
\end{tabular}
\end{center}
\caption{Best fits for the fundamental-plane coefficients in all filters and for all models with the volume weights disabled. Consequently, these results suffer from a Malmquist bias.}
\label{fitparameters_novolumeweight}
\end{table*}

\begin{table*}
\begin{center}
\begin{tabular}{c|ccccc}
 models and filters & $a$ & $b$ & $c$ & $s_{\varepsilon}$ & $\bar{\sigma}_{\textrm{dist}}$ [\%] \\ \hline \hline
 c model &  &  &  & \\ \hline
 u & $  0.865 \pm   0.030 $  & $ -0.699 \pm   0.008 $  & $  -7.66 \pm   0.10$ & $  0.0998 $ & $   17.0     $\\
 g & $   1.036\pm   0.030 $  & $ -0.738 \pm   0.013 $  & $  -7.87\pm   0.13  $ & $  0.0995 $ & $   16.2 $\\
 r & $   1.103 \pm  0.030 $  & $ -0.751 \pm   0.013 $  & $  -7.89 \pm   0.13 $ & $  0.0997 $ & $   16.1  $\\
 i & $   1.131 \pm   0.030 $  & $ -0.757 \pm   0.014$  & $  -7.87 \pm   0.13 $ & $  0.0982 $ & $   15.7    $\\
 z & $   1.176  \pm  0.030 $  & $ -0.763 \pm   0.014$  & $  -7.93 \pm   0.13 $ & $  0.0998 $ & $   15.6$\\
 \hline \hline
 dV model &  &  &  & \\ \hline
 u & $  0.852\pm   0.030 $  & $ -0.704\pm   0.009 $  & $  -7.68 \pm   0.10 $ & $  0.0990 $ & $   16.9  $\\
 g & $   1.028 \pm   0.030 $  & $ -0.743\pm   0.013 $  & $  -7.89 \pm   0.13 $ & $  0.0995 $ & $   16.2$\\
 r & $   1.096 \pm   0.030 $  & $ -0.755\pm   0.013 $  & $  -7.90 \pm   0.13$ & $  0.0997 $ & $   16.0 $\\
 i & $   1.125\pm   0.030 $  & $ -0.759 \pm   0.014 $  & $  -7.88\pm   0.13$ & $  0.0981 $ & $   15.7 $\\
 z & $   1.171  \pm   0.030 $  & $ -0.766 \pm   0.014 $  & $  -7.94 \pm   0.13$ & $  0.0990 $ & $   15.5 $\\
 \hline \hline
  p model &  &  &  & \\ \hline
 u & $  0.893 \pm   0.030 $  & $ -0.552 \pm   0.005 $  & $  -6.45 \pm   0.08 $ & $  0.1149 $ & $   20.1 $\\
 g & $   1.042  \pm   0.030 $  & $ -0.705 \pm   0.013 $  & $  -7.74 \pm   0.14  $ & $  0.1028 $ & $   17.2$\\
 r & $   1.112 \pm   0.030 $  & $ -0.725 \pm   0.014 $  & $  -7.85 \pm   0.14    $ & $  0.1015$ & $   16.8   $\\
 i & $   1.137\pm   0.030 $  & $ -0.718 \pm   0.013 $  & $  -7.74   \pm   0.14 $ & $  0.1004 $ & $   16.6 $\\
 z & $   1.157   \pm   0.030 $  & $ -0.642 \pm   0.010 $  & $  -7.09 \pm   0.12$ & $  0.1149$ & $   17.5 $

\end{tabular}
\end{center}
\caption{Best fits for the fundamental-plane coefficients in all filters and for all models with the correction for redshift evolution completely disabled.}
\label{fitparameters_noevolution}
\end{table*}

\begin{table*}
\begin{center}
\begin{tabular}{c|ccccc}
 models and filters & $a$ & $b$ & $c$ & $s_{\varepsilon}$ & $\bar{\sigma}_{\textrm{dist}}$ [\%] \\ \hline \hline
 c model &  &  &  & \\ \hline
 u & $  0.643\pm   0.030 $  & $ -0.671\pm   0.008 $  & $  -7.01    \pm  0.09 $ & $  0.0850 $ & $   16.4 $\\
 g & $  0.880  \pm   0.030 $  & $ -0.720\pm   0.012 $  & $  -7.43\pm   0.12$ & $  0.08592 $ & $   15.2 $\\
 r & $  0.974\pm   0.030 $  & $ -0.742 \pm   0.013 $  & $  -7.58 \pm   0.12  $ & $  0.0875 $ & $   14.8$\\
 i & $   1.012 \pm   0.030 $  & $ -0.749\pm   0.013 $  & $  -7.60  \pm   0.13 $ & $  0.0872 $ & $   14.6 $\\
 z & $   1.061 \pm   0.030 $  & $ -0.755 \pm   0.013 $  & $  -7.66 \pm   0.13 $ & $  0.0850 $ & $   14.5 $\\
 \hline \hline
 dV model &  &  &  & \\ \hline
 u & $  0.626\pm   0.030 $  & $ -0.672\pm   0.008 $  & $  -6.98 \pm   0.09 $ & $  0.0840 $ & $   16.4$\\
 g & $  0.869\pm   0.030 $  & $ -0.725 \pm   0.012 $  & $  -7.45 \pm   0.12 $ & $  0.0853 $ & $   15.1 $\\
 r & $  0.965\pm   0.030 $  & $ -0.747 \pm   0.013 $  & $  -7.59 \pm   0.12 $ & $  0.0868 $ & $   14.8$\\
 i & $   1.005\pm  0.030 $  & $ -0.752\pm   0.013 $  & $  -7.61 \pm   0.13 $ & $  0.0865 $ & $   14.5$\\
 z & $   1.056 \pm   0.030 $  & $ -0.759\pm   0.013 $  & $  -7.68\pm   0.13$ & $  0.0840 $ & $   14.4 $\\
 \hline \hline
  p model &  &  &  & \\ \hline
 u & $  0.641\pm   0.030 $  & $ -0.536\ pm   0.004 $  & $  -5.84\pm   0.08 $ & $  0.0938 $ & $   17.7$\\
 g & $  0.840 \pm  0.030 $  & $ -0.661\pm   0.012 $  & $  -6.99\pm   0.12$ & $  0.0867 $ & $   15.9$\\
 r & $  0.928\pm   0.030 $  & $ -0.689\pm   0.013 $  & $  -7.21\pm   0.13$ & $  0.0867 $ & $   15.4$\\
 i & $  0.955 \pm   0.030 $  & $ -0.683\pm   0.012 $  & $  -7.12\pm   0.13$ & $  0.0861$ & $   15.4 $\\
 z & $  0.966 \pm   0.030 $  & $ -0.615 \pm   0.009 $  & $  -6.507\pm   0.11$ & $  0.0938 $ & $   15.7$

\end{tabular}
\end{center}
\caption{Best fits for the fundamental-plane coefficients in all filters and for all models with volume weights and 3-$\sigma$ clipping and a filter-dependend redshift evolution derived from the redshift evolution of the surface brightness (see Table \ref{z_evolution}).}
\label{fitparameters_filterevolution}
\end{table*}

\begin{table*}
\begin{center}
\begin{tabular}{c|ccccc}
models and filters  & $Q_{\textrm{u}} $[mag/arcsec$^2$] & $Q_{\textrm{g}}$ [mag/arcsec$^2$]& $Q_{\textrm{r}}$ [mag/arcsec$^2$]& $Q_{\textrm{i}}$ [mag/arcsec$^2$]& $Q_{\textrm{z}}$ [mag/arcsec$^2$]\\ \hline 
 c model & 4.26 & 2.69 & 2.18 & 1.96 & 1.87\\
 dV model & 4.40 & 2.72 & 2.20 & 1.97 & 1.88\\ 
  p model & 6.03 & 3.78 & 3.29 & 3.24 & 3.73
\end{tabular}
\end{center}
\caption{Redshift evolution derived from changes in the surface brightness using non-evolution-corrected magnitudes.}
\label{z_evolution}
\end{table*}

\begin{table*}
\begin{center}
\begin{tabular}{c|ccccc}
models and filters  & $a$ & $b$ & $c$ & $s_{\varepsilon}$ & $\bar{\sigma}_{\textrm{dist}}$ [\%] \\ \hline \hline
 c model &  &  &  & \\ \hline
 u & $  0.847 \pm   0.106 $  & $ -0.688 \pm   0.040 $  & $  -7.53 \pm   0.45$ & $  0.0989 $ & $   19.9 $\\
 g & $  0.994 \pm   0.107$  & $ -0.723\pm   0.046 $  & $  -7.67 \pm   0.46 $ & $  0.0942 $ & $   19.0$\\
 r & $   1.058 \pm   0.107$  & $ -0.742\pm   0.047 $  & $  -7.74 \pm   0.47 $ & $  0.0914 $ & $   18.4     $\\
 i & $   1.102\pm   0.107$  & $ -0.758 \pm   0.048 $  & $  -7.85\pm   0.47 $ & $  0.0895$ & $   18.0$\\
 z & $   1.126\pm   0.108$  & $ -0.762 \pm   0.048 $  & $  -7.84 \pm   0.47 $ & $  0.0989 $ & $   17.5  $\\
 \hline \hline
 dV model &  &  &  & \\ \hline
 u & $  0.832 \pm   0.106 $  & $ -0.705   \pm   0.041 $  & $  -7.64 \pm   0.46 $ & $  0.0988 $ & $   19.9   $\\
 g & $  0.987 \pm   0.107  $  & $ -0.731  \pm   0.047 $  & $  -7.72  \pm   0.47  $ & $  0.0939 $ & $   18.9  $\\
 r & $   1.052 \pm   0.107 $  & $ -0.747  \pm   0.048 $  & $  -7.76   \pm   0.47   $ & $  0.0911 $ & $   18.3  $\\
 i & $   1.097 \pm   0.107   $  & $ -0.763  \pm   0.048 $  & $  -7.87 \pm   0.47  $ & $  0.0892 $ & $   17.9     $\\
 z & $   1.122 \pm   0.108  $  & $ -0.768  \pm   0.049 $  & $  -7.87  \pm   0.47  $ & $  0.0988 $ & $   17.4   $\\
 \hline \hline
  p model &  &  &  & \\ \hline
 u & $  0.833      \pm   0.106      $  & $ -0.553     \pm   0.020 $  & $  -6.34  \pm   0.32     $ & $  0.1081 $ & $   22.0      $\\
 g & $  0.988   \pm   0.106    $  & $ -0.656   \pm   0.043 $  & $  -7.21 \pm   0.47   $ & $  0.0989 $ & $   20.1    $\\
 r & $   1.059  \pm   0.107  $  & $ -0.681 \pm   0.045 $  & $  -7.37  \pm   0.49  $ & $  0.0956 $ & $   19.4    $\\
 i & $   1.093   \pm   0.107$  & $ -0.688  \pm  0.044 $  & $  -7.40 \pm   0.47 $ & $  0.0941 $ & $   19.1     $\\
 z & $   1.111      \pm   0.109 $  & $ -0.640 \pm   0.040 $  & $  -6.99 \pm   0.45 $ & $  0.1081  $ & $   19.3      $

\end{tabular}
\end{center}
\caption{Fundamental-plane coefficients in all filters and for all models derived from the volume-limited subsample, which is to 95,45\% (2-$\sigma$) completed. This condition limits the sample to a redshift of 0.0513. Owing to its completeness, the Malmquist-bias correction was disabled.}
\label{fitparameters_volumelimited}
\end{table*}

\begin{table*}
\begin{center}
\begin{tabular}{c|ccccc}
models and filters  & $a$ & $b$ & $c$ & $s_{\varepsilon}$ & $\bar{\sigma}_{\textrm{dist}}$ [\%] \\ \hline \hline
 c model &  &  &  & \\ \hline
 u & $  0.820 \pm   0.029 $  & $ -0.697  \pm   0.008 $  & $  -7.56 \pm   0.10 $ & $  0.0952 $ & $   16.6$\\
 g & $  0.987  \pm   0.029 $  & $ -0.738     \pm   0.013 $  & $  -7.78      \pm   0.12   $ & $  0.0937 $ & $   15.6   $\\
 r & $   1.054       \pm   0.029 $  & $ -0.752  \pm  0.013 $  & $  -7.81  \pm   0.12      $ & $  0.0936 $ & $   15.3 $\\
 i & $   1.080    \pm   0.029$  & $ -0.757      \pm   0.013 $  & $  -7.79       \pm   0.12    $ & $  0.0922 $ & $   15.0 $\\
 z & $   1.124      \pm   0.029 $  & $ -0.762      \pm   0.013$  & $  -7.84  \pm   0.13 $ & $  0.0952 $ & $   14.8     $\\
 \hline \hline
 dV model &  &  &  & \\ \hline
 u & $  0.809  \pm   0.029 $  & $ -0.701 \pm   0.008 $  & $  -7.57 \pm   0.10 $ & $  0.0946 $ & $   16.6$\\
 g & $  0.979 \pm   0.029 $  & $ -0.742 \pm   0.012 $  & $  -7.79\pm   0.12    $ & $  0.0937$ & $   15.6   $\\
 r & $   1.047  \pm   0.029 $  & $ -0.755  \pm   0.013 $  & $  -7.81  \pm   0.12   $ & $  0.0936$ & $   15.3 $\\
 i & $   1.075  \pm   0.029 $  & $ -0.759 \pm   0.013 $  & $  -7.79 \pm   0.12   $ & $  0.0922 $ & $   15.0$\\
 z & $   1.120   \pm   0.029 $  & $ -0.766      \pm   0.013 $  & $  -7.85 \pm   0.13    $ & $  0.0946 $ & $   14.8      $\\
 \hline \hline
  p model &  &  &  & \\ \hline
 u & $  0.871 \pm  0.029 $  & $ -0.551 \pm   0.004 $  & $  -6.41 \pm   0.08 $ & $  0.11099  $ & $   19.7 $\\
 g & $   1.001 \pm  0.029 $  & $ -0.700  \pm   0.012 $  & $  -7.63  \pm   0.13  $ & $  0.0977 $ & $   16.6    $\\
 r & $   1.070 \pm   0.029 $  & $ -0.720 \pm   0.013 $  & $  -7.74 \pm   0.13  $ & $  0.0962$ & $   16.1   $\\
 i & $   1.095   \pm   0.029 $  & $ -0.713 \pm   0.013 $  & $  -7.63  \pm   0.13  $ & $  0.0952 $ & $   16.0  $\\
 z & $   1.121  \pm   0.029 $  & $ -0.639 \pm   0.010 $  & $  -6.99 \pm   0.11 $ & $  0.1110 $ & $   16.7 $
\end{tabular}
\end{center}
\caption{Fundamental-plane coefficients in all filters and for all models derived from an extended sample up to a redshift of 0.3. However, it already suffers from an additional bias beyond the Malmquist bias at these distances.}
\label{fitparameters_z03}
\end{table*}

\FloatBarrier
\addcontentsline{toc}{section}{References}
\bibliography{refer}\label{bib}

\end{document}